\documentclass[11pt]{article}
\pdfoutput=1
\usepackage{outlines}
\usepackage{jheppub}
\usepackage[utf8]{inputenc}
\usepackage{booktabs}
\usepackage{graphicx}
\usepackage{bbold}
\usepackage{subcaption}
\usepackage{slashed}
\usepackage{mathtools}
\usepackage{easybmat}
\usepackage{lmodern}
\usepackage{comment}

\input{widebar}
\usepackage{xkeyval}

\usepackage{tikz,fp}
\usetikzlibrary{external,fixedpointarithmetic,decorations.pathmorphing,positioning,calc,fadings,decorations.markings,decorations.pathreplacing,patterns,arrows,arrows.meta}

\tikzset{
  label distance=-3pt,
  inner sep=1.5pt,
  >=stealth,
  boson/.style={
    decoration={snake, segment length=2mm, amplitude=0.5mm},
    decorate,
  },
  graviton/.style={
  	line width=0.4pt,
    decoration={snake, segment length=2mm, amplitude=0.5mm},
    decorate,
  },
  quark/.style={
    postaction={decorate},
    decoration={markings,mark=at position .5 with {\draw[->] (-0.01,0) -- (0.07,0);}}
  },
  aquark/.style={
    postaction={decorate},
    decoration={markings,mark=at position .5 with {\draw[->] (0.01,0) -- (-0.07,0);}}
  },
  gluonOld/.style={
    line width=0.7pt,
    decorate,
    decoration={coil,aspect=-0.5,amplitude=-2pt,segment length=2.2pt}
  },
  gluon/.style={
    line width=0.5pt,
    decorate,
    decoration={coil,aspect=-0.75,amplitude=-1.5pt,segment length=2pt}
  },
  agluon/.style={
    line width=0.5pt,
    decorate,
    decoration={coil,aspect=0.75,amplitude=1.5pt,segment length=2pt}
  },
  scalar/.style={
    dashed
  },
  dscalar/.style={
    dashed,
    postaction={decorate},
    decoration={markings,mark=at position 0.65 with {\draw[->] (-0.01,0) -- (0.07,0);}}
  },
  adscalar/.style={
    dashed,
    postaction={decorate},
    decoration={markings,mark=at position 0.65 with {\draw[->] (-0.01,0) -- (0.07,0);}}
  },
  line/.style={
  },
  doubleline/.style={
    double
  },
  middleArrow/.style={
    draw=white,
    decoration={markings,mark=at position 0.60 with{\arrow[scale=0.16,line width=8pt,black]{Straight Barb}}},
    decorate
  },
  dot/.style={
  	postaction={decorate},
    decoration={markings,mark=between positions 0.5 and 0.5 step 1 with {\draw[fill] (0,0) circle [radius=1.5pt];}}  
  },
  dots/.style={
  	postaction={decorate},
    decoration={markings,mark=between positions 0.33 and 0.66 step 0.33 with {\draw[fill] (0,0) circle [radius=1.5pt];}}  
  },
  dotss/.style={
  	postaction={decorate},
    decoration={markings,mark=between positions 0.25 and 0.75 step 0.25 with {\draw[fill] (0,0) circle [radius=1.5pt];}}  
  },
  blob/.style={
  	pattern=north west lines,
    draw=black
  },
  blobBlack/.style={
    draw=black,
    fill=black
  },
  blobWhite/.style={
    draw=black,
    fill=white
  },
  xSource/.style={
    circle, minimum width=6pt, draw, inner sep=0pt, path picture={\draw (path picture bounding box.south east) -- (path picture bounding box.north west) (path picture bounding box.south west) -- (path picture bounding box.north east);}
  },
  source/.style={
    circle,fill,inner sep=1pt
  },
  mid/.style={
    circle,fill,inner sep=0.5pt
  }
}

\makeatletter
\def\gSetStdExtTypes#1{\presetkeys{gTypes}{eA=#1,eB=#1,eC=#1,eD=#1,eE=#1}{}}
\def\gSetStdIntTypes#1{\presetkeys{gTypes}{iA=#1,iB=#1,iC=#1,iD=#1,iE=#1,iF=#1,iG=#1,iH=#1,iI=#1,iJ=#1}{}}
\def\gSetStdTypes#1{\gSetStdExtTypes{#1}\gSetStdIntTypes{#1}}

\define@cmdkeys{general}{scale}
\define@cmdkeys{general}{rotate}
\define@cmdkeys{general}{font}
\define@cmdkeys{general}{yshift}
\presetkeys{general}{scale=1, rotate=0, font=\scriptsize, yshift=0pt}{}

\define@cmdkeys{gTypes}{eA,eB,eC,eD,eE,iA,iB,iC,iD,iE,iF,iG,iH,iI,iJ}
\gSetStdTypes{line}
\define@key{pre}{all}{\gSetStdTypes{#1}}
\define@key{pre}{allE}{\gSetStdExtTypes{#1}}
\define@key{pre}{allI}{\gSetStdIntTypes{#1}}

\define@cmdkeys{gLabels}{eLA,eLB,eLC,eLD,eLE,iLA,iLB,iLC,iLD,iLE,iLF,iLG,iLH,iLI,iLJ}
\presetkeys{gLabels}{eLA={},eLB={},eLC={},eLD={},eLE={},iLA={},iLB={},iLC={},iLD={},iLE={},iLF={},iLG={},iLH={},iLI={},iLJ={}}{}

\define@boolkeys{dots}{lDots,rDots}
\presetkeys{dots}{lDots=false, rDots=false}{}

\define@cmdkeys{blobs}{blobA,blobB,blobC}
\presetkeys{blobs}{blobA=blob,blobB=blob,blobC=blob}{}

\def\gTreeTri{\@ifnextchar[\@gTreeTri{\@gTreeTri[]}}
\def\@gTreeTri[#1]#2{%
  \setkeys*{pre}{#1}%
  \setrmkeys{general,gTypes,gLabels}%
  \begin{tikzpicture}
    [line width=1pt,
    baseline={([yshift=-0.5ex+\cmdKV@general@yshift]current bounding box.center)},
    scale=\cmdKV@general@scale,
    rotate=\cmdKV@general@rotate,
    font=\cmdKV@general@font]
    \draw[\cmdKV@gTypes@eA] (0.7,0.3) -- (0.4,0.3) node[label=(-180+\cmdKV@general@rotate):\cmdKV@gLabels@eLA] {};
    \draw[\cmdKV@gTypes@eB] (0.7,0.3) -- (0.9,0.6) node[label=(\cmdKV@general@rotate):\cmdKV@gLabels@eLB] {};
    \draw[\cmdKV@gTypes@eC] (0.7,0.3) -- (0.9,0) node[label=(\cmdKV@general@rotate):\cmdKV@gLabels@eLC] {};
    #2
  \end{tikzpicture}
  \gSetStdTypes{line}
}

\def\gTreeS{\@ifnextchar[\@gTreeS{\@gTreeS[]}}
\def\@gTreeS[#1]#2{%
  \setkeys*{pre}{#1}%
  \setrmkeys{general,gTypes,gLabels}%
  \begin{tikzpicture}
    [line width=1pt,
    baseline={([yshift=-0.5ex+\cmdKV@general@yshift]current bounding box.center)},
    scale=\cmdKV@general@scale,
    rotate=\cmdKV@general@rotate,
    font=\cmdKV@general@font]
    \draw[\cmdKV@gTypes@eA] (0.2,0.3) -- (0,0) node[label=(-180+\cmdKV@general@rotate):\cmdKV@gLabels@eLA] {};
    \draw[\cmdKV@gTypes@eB] (0.2,0.3) -- (0,0.6) node[label=(180+\cmdKV@general@rotate):\cmdKV@gLabels@eLB] {};
    \draw[\cmdKV@gTypes@eC] (0.7,0.3) -- (0.9,0.6) node[label=(\cmdKV@general@rotate):\cmdKV@gLabels@eLC] {};
    \draw[\cmdKV@gTypes@eD] (0.7,0.3) -- (0.9,0) node[label=(\cmdKV@general@rotate):\cmdKV@gLabels@eLD] {};    
    \draw[label distance=0,\cmdKV@gTypes@iA] (0.2,0.3) -- node[label=(90+\cmdKV@general@rotate):\cmdKV@gLabels@iLA] {} (0.7,0.3);
    #2
  \end{tikzpicture}
  \gSetStdTypes{line}
}

\def\gTreeT{\@ifnextchar[\@gTreeT{\@gTreeT[]}}
\def\@gTreeT[#1]#2{%
  \setkeys*{pre}{#1}%
  \setrmkeys{general,gTypes,gLabels}%
  \begin{tikzpicture}
    [line width=1pt,
    baseline={([yshift=-0.5ex+\cmdKV@general@yshift]current bounding box.center)},
    scale=\cmdKV@general@scale,
    rotate=\cmdKV@general@rotate,
    font=\cmdKV@general@font]
    \draw[\cmdKV@gTypes@eA] (0.45,0.1) -- (0,0) node[label=(-180+\cmdKV@general@rotate):\cmdKV@gLabels@eLA] {};
    \draw[\cmdKV@gTypes@eB] (0.45,0.5) -- (0,0.6) node[label=(180+\cmdKV@general@rotate):\cmdKV@gLabels@eLB] {};
    \draw[\cmdKV@gTypes@eC] (0.45,0.5) -- (0.9,0.6) node[label=(\cmdKV@general@rotate):\cmdKV@gLabels@eLC] {};
    \draw[\cmdKV@gTypes@eD] (0.45,0.1) -- (0.9,0) node[label=(\cmdKV@general@rotate):\cmdKV@gLabels@eLD] {};
    \draw[label distance=0,\cmdKV@gTypes@iA] (0.45,0.5) -- node[label=(\cmdKV@general@rotate):\cmdKV@gLabels@iLA] {} (0.45,0.1);
    #2
  \end{tikzpicture}
  \gSetStdTypes{line}
}

\def\gTreeU{\@ifnextchar[\@gTreeU{\@gTreeU[]}}
\def\@gTreeU[#1]#2{%
  \setkeys*{pre}{#1}%
  \setrmkeys{general,gTypes,gLabels}%
  \begin{tikzpicture}
    [line width=1pt,
    baseline={([yshift=-0.5ex+\cmdKV@general@yshift]current bounding box.center)},
    scale=\cmdKV@general@scale,
    rotate=\cmdKV@general@rotate,
    font=\cmdKV@general@font]
    \draw[\cmdKV@gTypes@eA] (0.45,0.5) -- (0,0) node[label=(-180+\cmdKV@general@rotate):\cmdKV@gLabels@eLA] {};
    \foreach \x in {5,...,14} \fill[opacity=0.12,white] (0.27,0.3) circle (\x/100);
    \draw[\cmdKV@gTypes@eB] (0.45,0.1) -- (0,0.6) node[label=(180+\cmdKV@general@rotate):\cmdKV@gLabels@eLB] {};
    \draw[\cmdKV@gTypes@eC] (0.45,0.5) -- (0.9,0.6) node[label=(\cmdKV@general@rotate):\cmdKV@gLabels@eLC] {};
    \draw[\cmdKV@gTypes@eD] (0.45,0.1) -- (0.9,0) node[label=(\cmdKV@general@rotate):\cmdKV@gLabels@eLD] {};
    \draw[label distance=0,\cmdKV@gTypes@iA] (0.45,0.5) -- node[label=(\cmdKV@general@rotate):\cmdKV@gLabels@iLA] {} (0.45,0.1);
    #2
  \end{tikzpicture}
  \gSetStdTypes{line}
}

\def\gTreeFourPt{\@ifnextchar[\@gTreeFourPt{\@gTreeFourPt[]}}
\def\@gTreeFourPt[#1]#2{%
  \setkeys*{pre}{#1}%
  \setrmkeys{general,gTypes,gLabels}%
  \begin{tikzpicture}
    [line width=1pt,
    baseline={([yshift=-0.5ex+\cmdKV@general@yshift]current bounding box.center)},
    scale=\cmdKV@general@scale,
    rotate=\cmdKV@general@rotate,
    font=\cmdKV@general@font]
    \draw[\cmdKV@gTypes@eA] (0,0) -- (-0.4,-0.4) node[label=(-180+\cmdKV@general@rotate):\cmdKV@gLabels@eLA] {};
    \draw[\cmdKV@gTypes@eB] (0,0) -- (-0.4,0.4) node[label=(180+\cmdKV@general@rotate):\cmdKV@gLabels@eLB] {};
    \draw[\cmdKV@gTypes@eC] (0,0) -- (0.4,0.4) node[label=(\cmdKV@general@rotate):\cmdKV@gLabels@eLC] {};
    \draw[\cmdKV@gTypes@eD] (0,0) -- (0.4,-0.4) node[label=(\cmdKV@general@rotate):\cmdKV@gLabels@eLD] {};
    #2
  \end{tikzpicture}
  \gSetStdTypes{line}
}

\def\gTadA{\@ifnextchar[\@gTadA{\@gTadA[]}}
\def\@gTadA[#1]#2{%
  \setkeys*{pre}{#1}%
  \setrmkeys{general,gTypes,gLabels}%
  \begin{tikzpicture}
    [line width=1pt,
    baseline={([yshift=-0.5ex+\cmdKV@general@yshift]current bounding box.center)},
    scale=\cmdKV@general@scale,
    rotate=\cmdKV@general@rotate,
    font=\cmdKV@general@font]
    \draw[\cmdKV@gTypes@eA] (-0.7,0.3) -- (-0.9,0) node[label=(180+\cmdKV@general@rotate):\cmdKV@gLabels@eLA] {};
    \draw[\cmdKV@gTypes@eB] (-0.7,0.3) -- (-0.9,0.6) node[label=(180+\cmdKV@general@rotate):\cmdKV@gLabels@eLB] {};
    \draw[\cmdKV@gTypes@eC] (0.15,0.45) -- (0.5,0.6) node[label=(\cmdKV@general@rotate):\cmdKV@gLabels@eLC] {};
    \draw[\cmdKV@gTypes@eD] (-0.3,0.3) -- (-0.2,0) node[label=(\cmdKV@general@rotate):\cmdKV@gLabels@eLD] {};
    \draw[label distance=0,\cmdKV@gTypes@iA] (-0.7,0.3) -- node[label=(90+\cmdKV@general@rotate):\cmdKV@gLabels@iLA] {} (-0.3,0.3);
    \draw[label distance=0,\cmdKV@gTypes@iB] (-0.3,0.3) -- node[label=(-80+\cmdKV@general@rotate):\cmdKV@gLabels@iLB] {} (0.15,0.45);
    \draw[label distance=0,\cmdKV@gTypes@iC] (0.15,0.45) -- node[label=(45+\cmdKV@general@rotate):\cmdKV@gLabels@iLC] {} (-0.18,0.69);
    \draw[label distance=-0.5,\cmdKV@gTypes@iD] (-0.18,0.69) .. controls ($(-0.18,0.69) + (-0.12,-0.12)$) and ($(-0.35,0.87) + (-0.12,-0.12)$) .. (-0.35,0.87) node[label=(180+\cmdKV@general@rotate):\cmdKV@gLabels@iLD] {} .. controls ($(-0.35,0.87) + (0.12,0.12)$) and ($(-0.18,0.69) + (0.12,0.12)$) .. (-0.18,0.69);
    #2
  \end{tikzpicture}
  \gSetStdTypes{line}
}

\def\gTadB{\@ifnextchar[\@gTadB{\@gTadB[]}}
\def\@gTadB[#1]#2{%
  \setkeys*{pre}{#1}%
  \setrmkeys{general,gTypes,gLabels}%
  \begin{tikzpicture}
    [line width=1pt,
    baseline={([yshift=-0.5ex+\cmdKV@general@yshift]current bounding box.center)},
    scale=\cmdKV@general@scale,
    rotate=\cmdKV@general@rotate,
    font=\cmdKV@general@font]
    \draw[\cmdKV@gTypes@eA] (0,0.3) -- (-0.2,0) node[label=(-180+\cmdKV@general@rotate):\cmdKV@gLabels@eLA] {};
    \draw[\cmdKV@gTypes@eB] (0,0.3) -- (-0.2,0.6) node[label=(180+\cmdKV@general@rotate):\cmdKV@gLabels@eLB] {};
    \draw[\cmdKV@gTypes@eC] (0.7,0.3) -- (0.9,0.6) node[label=(\cmdKV@general@rotate):\cmdKV@gLabels@eLC] {};
    \draw[\cmdKV@gTypes@eD] (0.7,0.3) -- (0.9,0) node[label=(\cmdKV@general@rotate):\cmdKV@gLabels@eLD] {};
    \draw[label distance=0,\cmdKV@gTypes@iA] (0,0.3) -- node[label=(-90+\cmdKV@general@rotate):\cmdKV@gLabels@iLA] {} (0.35,0.3);
    \draw[label distance=0,\cmdKV@gTypes@iB] (0.35,0.3) -- node[label=(\cmdKV@general@rotate):\cmdKV@gLabels@iLB] {} (0.35,0.6);
    \draw[label distance=-0.5,\cmdKV@gTypes@iC] (0.35,0.6) .. controls (0.18,0.6) and (0.18,0.85) .. (0.35,0.85) .. controls (0.52,0.85) and (0.52,0.6) .. node[label=(\cmdKV@general@rotate):\cmdKV@gLabels@iLC] {} (0.35,0.6);
    \draw[label distance=0,\cmdKV@gTypes@iD] (0.35,0.3) -- node[label=(-90+\cmdKV@general@rotate):\cmdKV@gLabels@iLD] {} (0.7,0.3);
    #2
  \end{tikzpicture}
  \gSetStdTypes{line}
}
\def\gBubA{\@ifnextchar[\@gBubA{\@gBubA[]}}
\def\@gBubA[#1]#2{%
  \setkeys*{pre}{#1}%
  \setrmkeys{general,gTypes,gLabels}%
  \begin{tikzpicture}
    [line width=1pt,
    baseline={([yshift=-0.5ex+\cmdKV@general@yshift]current bounding box.center)},
    scale=\cmdKV@general@scale,
    rotate=\cmdKV@general@rotate,
    font=\cmdKV@general@font]
    \draw[\cmdKV@gTypes@eA] (0.5,0.7) -- (0.2,0.7) node[label=(180+\cmdKV@general@rotate):\cmdKV@gLabels@eLA] {};
    \draw[\cmdKV@gTypes@eB] (0.9,0.7) -- (1.2,0.7) node[label=(\cmdKV@general@rotate):\cmdKV@gLabels@eLB] {};
    \draw[label distance=0,\cmdKV@gTypes@iA] (0.5,0.7) .. controls (0.5,0.95) and (0.9,0.95) .. node[label=(90+\cmdKV@general@rotate):\cmdKV@gLabels@iLA] {} (0.9,0.7);
    \draw[label distance=0,\cmdKV@gTypes@iB] (0.9,0.7) .. controls (0.9,0.45) and (0.5,0.45) .. node[label=(-90+\cmdKV@general@rotate):\cmdKV@gLabels@iLB] {} (0.5,0.7);
    #2
  \end{tikzpicture}
  \gSetStdTypes{line}
}
\def\gBubB{\@ifnextchar[\@gBubB{\@gBubB[]}}
\def\@gBubB[#1]#2{%
  \setkeys*{pre}{#1}%
  \setrmkeys{general,gTypes,gLabels}%
  \begin{tikzpicture}
    [line width=1pt,
    baseline={([yshift=-0.5ex+\cmdKV@general@yshift]current bounding box.center)},
    scale=\cmdKV@general@scale,
    rotate=\cmdKV@general@rotate,
    font=\cmdKV@general@font]
    \draw[\cmdKV@gTypes@eA] (0.2,0.7) -- (0,0.4) node[label=(-180+\cmdKV@general@rotate):\cmdKV@gLabels@eLA] {};
    \draw[\cmdKV@gTypes@eB] (0.2,0.7) -- (0,1) node[label=(180+\cmdKV@general@rotate):\cmdKV@gLabels@eLB] {};
    \draw[\cmdKV@gTypes@eC] (1.2,0.7) -- (1.4,1) node[label=(\cmdKV@general@rotate):\cmdKV@gLabels@eLC] {};
    \draw[\cmdKV@gTypes@eD] (1.2,0.7) -- (1.4,0.4) node[label=(\cmdKV@general@rotate):\cmdKV@gLabels@eLD] {};
    \draw[label distance=0,\cmdKV@gTypes@iA] (0.2,0.7) -- node[label=(90+\cmdKV@general@rotate):\cmdKV@gLabels@iLA] {} (0.5,0.7);
    \draw[label distance=0,\cmdKV@gTypes@iB] (0.5,0.7) .. controls (0.5,0.95) and (0.9,0.95) .. node[label=(90+\cmdKV@general@rotate):\cmdKV@gLabels@iLB] {} (0.9,0.7);
    \draw[label distance=0,\cmdKV@gTypes@iC] (0.9,0.7) .. controls (0.9,0.45) and (0.5,0.45) .. node[label=(-90+\cmdKV@general@rotate):\cmdKV@gLabels@iLC] {} (0.5,0.7);
    \draw[label distance=0,\cmdKV@gTypes@iD] (0.9,0.7) -- node[label=(90+\cmdKV@general@rotate):\cmdKV@gLabels@iLD] {} (1.2,0.7);
    #2
  \end{tikzpicture}
  \gSetStdTypes{line}
}

\def\gBubC{\@ifnextchar[\@gBubC{\@gBubC[]}}
\def\@gBubC[#1]#2{%
  \setkeys*{pre}{#1}%
  \setrmkeys{general,gTypes,gLabels}%
  \begin{tikzpicture}
    [line width=1pt,
    baseline={([yshift=-0.5ex+\cmdKV@general@yshift]current bounding box.center)},
    scale=\cmdKV@general@scale,
    rotate=\cmdKV@general@rotate,
    font=\cmdKV@general@font]
    \draw[\cmdKV@gTypes@eA] (-0.7,0.3) -- (-0.9,0) node[label=(180+\cmdKV@general@rotate):\cmdKV@gLabels@eLA] {};
    \draw[\cmdKV@gTypes@eB] (-0.7,0.3) -- (-0.9,0.6) node[label=(180+\cmdKV@general@rotate):\cmdKV@gLabels@eLB] {};
    \draw[\cmdKV@gTypes@eC] (0.3,0.5) -- (0.5,0.6) node[label=(\cmdKV@general@rotate):\cmdKV@gLabels@eLC] {};
    \draw[\cmdKV@gTypes@eD] (-0.3,0.3) -- (-0.2,0) node[label=(\cmdKV@general@rotate):\cmdKV@gLabels@eLD] {};
    \draw[label distance=0,\cmdKV@gTypes@iA] (-0.7,0.3) -- node[label=(90+\cmdKV@general@rotate):\cmdKV@gLabels@iLA] {} (-0.3,0.3);
    \draw[label distance=0,\cmdKV@gTypes@iB] (-0.3,0.3) -- node[label=(90+\cmdKV@general@rotate):\cmdKV@gLabels@iLB] {} (-0,0.4);
    \draw[label distance=0,\cmdKV@gTypes@iC] (0,0.4) .. controls (-0.03,0.6) and (0.22,0.65) .. node[label=(90+\cmdKV@general@rotate):\cmdKV@gLabels@iLC] {} (0.3,0.5);
    \draw[label distance=0,\cmdKV@gTypes@iD] (0.3,0.5) .. controls (0.38,0.3) and (0.08,0.21) .. node[label=(-90+\cmdKV@general@rotate):\cmdKV@gLabels@iLD] {} (0,0.4);
    #2
  \end{tikzpicture}
  \gSetStdTypes{line}
}
\def\gTriA{\@ifnextchar[\@gTriA{\@gTriA[]}}
\def\@gTriA[#1]#2{%
  \setkeys*{pre}{#1}%
  \setrmkeys{general,gTypes,gLabels}%
  \begin{tikzpicture}
    [line width=1pt,
    baseline={([yshift=-0.5ex+\cmdKV@general@yshift]current bounding box.center)},
    scale=\cmdKV@general@scale,
    rotate=\cmdKV@general@rotate,
    font=\cmdKV@general@font]
    \draw[\cmdKV@gTypes@eC] (0:0.3) -- (0:0.55) node[label=(\cmdKV@general@rotate):\cmdKV@gLabels@eLC] {};
    \draw[\cmdKV@gTypes@eA] (-120:0.3) -- (-120:0.55) node[label=(-180+\cmdKV@general@rotate):\cmdKV@gLabels@eLA] {};
    \draw[\cmdKV@gTypes@eB] (120:0.3) -- (120:0.55) node[label=(180+\cmdKV@general@rotate):\cmdKV@gLabels@eLB] {};    
    \draw[label distance=0,\cmdKV@gTypes@iA] (-120:0.3) -- node[label=(180+\cmdKV@general@rotate):\cmdKV@gLabels@iLA] {} (120:0.3);
    \draw[label distance=0,\cmdKV@gTypes@iB] (120:0.3) -- node[label=(60+\cmdKV@general@rotate):\cmdKV@gLabels@iLB] {} (0:0.3);
    \draw[label distance=0,\cmdKV@gTypes@iC] (0:0.3) -- node[label=(-60+\cmdKV@general@rotate):\cmdKV@gLabels@iLC] {} (-120:0.3);
    #2
  \end{tikzpicture}
  \gSetStdTypes{line}
}
\def\gTriB{\@ifnextchar[\@gTriB{\@gTriB[]}}
\def\@gTriB[#1]#2{%
  \setkeys*{pre}{#1}%
  \setrmkeys{general,gTypes,gLabels}%
  \begin{tikzpicture}
    [line width=1pt,
    baseline={([yshift=-0.5ex+\cmdKV@general@yshift]current bounding box.center)},
    scale=\cmdKV@general@scale,
    rotate=\cmdKV@general@rotate,
    font=\cmdKV@general@font]
    \draw[\cmdKV@gTypes@eA] (0.3,0.45) -- (0,0.3) node[label=(-180+\cmdKV@general@rotate):\cmdKV@gLabels@eLA] {};
    \draw[\cmdKV@gTypes@eB] (0.3,0.95) -- (0,1.1) node[label=(180+\cmdKV@general@rotate):\cmdKV@gLabels@eLB] {};
    \draw[\cmdKV@gTypes@eC] (1,0.7) -- (1.3,1.1) node[label=(\cmdKV@general@rotate):\cmdKV@gLabels@eLC] {};
    \draw[\cmdKV@gTypes@eD] (1,0.7) -- (1.3,0.3) node[label=(\cmdKV@general@rotate):\cmdKV@gLabels@eLD] {};
    \draw[label distance=0,\cmdKV@gTypes@iA] (0.3,0.45) -- node[label=(180+\cmdKV@general@rotate):\cmdKV@gLabels@iLA] {} (0.3,0.95);
    \draw[label distance=0,\cmdKV@gTypes@iB] (0.3,0.95) -- node[label=(70+\cmdKV@general@rotate):\cmdKV@gLabels@iLB] {} (0.7,0.7);
    \draw[label distance=0,\cmdKV@gTypes@iC] (0.7,0.7) -- node[label=(-70+\cmdKV@general@rotate):\cmdKV@gLabels@iLC] {} (0.3,0.45);
    \draw[label distance=0,\cmdKV@gTypes@iD] (1,0.7) -- node[label=(90+\cmdKV@general@rotate):\cmdKV@gLabels@iLD] {} (0.7,0.7);
    #2
  \end{tikzpicture}
  \gSetStdTypes{line}
}
\def\gBox{\@ifnextchar[\@gBox{\@gBox[]}}
\def\@gBox[#1]#2{%
  \setkeys*{pre}{#1}%
  \setrmkeys{general,gTypes,gLabels}%
  \begin{tikzpicture}
    [line width=1pt,
    baseline={([yshift=-0.5ex+\cmdKV@general@yshift]current bounding box.center)},
    scale=\cmdKV@general@scale,
    rotate=\cmdKV@general@rotate,
    font=\cmdKV@general@font]
    \draw[\cmdKV@gTypes@eA] (0.25,0.25) -- (0,0) node[label=(-180+\cmdKV@general@rotate):\cmdKV@gLabels@eLA] {};
    \draw[\cmdKV@gTypes@eB] (0.25,0.75) -- (0,1) node[label=(180+\cmdKV@general@rotate):\cmdKV@gLabels@eLB] {};
    \draw[\cmdKV@gTypes@eC] (0.75,0.75) -- (1,1) node[label=(\cmdKV@general@rotate):\cmdKV@gLabels@eLC] {};
    \draw[\cmdKV@gTypes@eD] (0.75,0.25) -- (1,0) node[label=(\cmdKV@general@rotate):\cmdKV@gLabels@eLD] {};
    \draw[label distance=0,\cmdKV@gTypes@iA] (0.25,0.25) -- node[label=(180+\cmdKV@general@rotate):\cmdKV@gLabels@iLA] {} (0.25,0.75);
    \draw[label distance=0,\cmdKV@gTypes@iB] (0.25,0.75) -- node[label=(90+\cmdKV@general@rotate):\cmdKV@gLabels@iLB] {} (0.75,0.75);
    \draw[label distance=0,\cmdKV@gTypes@iC] (0.75,0.75) -- node[label=(\cmdKV@general@rotate):\cmdKV@gLabels@iLC] {} (0.75,0.25);
    \draw[label distance=0,\cmdKV@gTypes@iD] (0.75,0.25) -- node[label=(-90+\cmdKV@general@rotate):\cmdKV@gLabels@iLD] {} (0.25,0.25);
    #2
  \end{tikzpicture}
  \gSetStdTypes{line}
}
\def\gPenta{\@ifnextchar[\@gPenta{\@gPenta[]}}
\def\@gPenta[#1]#2{%
  \setkeys*{pre}{#1}%
  \setrmkeys{general,gTypes,gLabels}%
  \begin{tikzpicture}
    [line width=1pt,
    baseline={([yshift=-0.5ex+\cmdKV@general@yshift]current bounding box.center)},
    scale=\cmdKV@general@scale,
    rotate=\cmdKV@general@rotate,
    font=\cmdKV@general@font]
    \draw[\cmdKV@gTypes@eA] (-144:0.35) -- (-144:0.75) node[label=(-144+\cmdKV@general@rotate):\cmdKV@gLabels@eLA] {};
    \draw[\cmdKV@gTypes@eB] (144:0.35) -- (144:0.75) node[label=(144+\cmdKV@general@rotate):\cmdKV@gLabels@eLB] {};
    \draw[\cmdKV@gTypes@eC] (72:0.35) -- (72:0.75) node[label=(72+\cmdKV@general@rotate):\cmdKV@gLabels@eLC] {};
    \draw[\cmdKV@gTypes@eD] (0:0.35) -- (0:0.75) node[label=(\cmdKV@general@rotate):\cmdKV@gLabels@eLD] {};
    \draw[\cmdKV@gTypes@eE] (-72:0.35) -- (-72:0.75) node[label=(-72+\cmdKV@general@rotate):\cmdKV@gLabels@eLE] {};
    \draw[label distance=0,\cmdKV@gTypes@iA] (-144:0.35) -- node[label=(180+\cmdKV@general@rotate):\cmdKV@gLabels@iLA] {} (144:0.35);
    \draw[label distance=0,\cmdKV@gTypes@iB] (144:0.35) -- node[label=(108+\cmdKV@general@rotate):\cmdKV@gLabels@iLB] {} (72:0.35);
    \draw[label distance=0,\cmdKV@gTypes@iC] (72:0.35) -- node[label=(36+\cmdKV@general@rotate):\cmdKV@gLabels@iLC] {} (0:0.35);
    \draw[label distance=0,\cmdKV@gTypes@iD] (0:0.35) -- node[label=(-36+\cmdKV@general@rotate):\cmdKV@gLabels@iLD] {} (-72:0.35);
    \draw[label distance=0,\cmdKV@gTypes@iE] (-72:0.35) -- node[label=(-108+\cmdKV@general@rotate):\cmdKV@gLabels@iLE] {} (-144:0.35);
    #2
  \end{tikzpicture}
  \gSetStdTypes{line}
}

\def\gBoxBox{\@ifnextchar[\@gBoxBox{\@gBoxBox[]}}
\def\@gBoxBox[#1]#2{%
  \setkeys*{pre}{#1}%
  \setrmkeys{general,gTypes,gLabels}%
  \begin{tikzpicture}
    [line width=1pt,
    baseline={([yshift=-0.5ex+\cmdKV@general@yshift]current bounding box.center)},
    scale=\cmdKV@general@scale,
    rotate=\cmdKV@general@rotate,
    font=\cmdKV@general@font]
    \draw[\cmdKV@gTypes@eA] (0.3,0.3) -- (0,0) node[label=(-180+\cmdKV@general@rotate):\cmdKV@gLabels@eLA] {};
    \draw[\cmdKV@gTypes@eB] (0.3,0.9) -- (0,1.2) node[label=(180+\cmdKV@general@rotate):\cmdKV@gLabels@eLB] {};
    \draw[\cmdKV@gTypes@eC] (1.5,0.9) -- (1.8,1.2) node[label=(\cmdKV@general@rotate):\cmdKV@gLabels@eLC] {};
    \draw[\cmdKV@gTypes@eD] (1.5,0.3) -- (1.8,0) node[label=(\cmdKV@general@rotate):\cmdKV@gLabels@eLD] {};
    \draw[label distance=0,\cmdKV@gTypes@iA] (0.3,0.3) -- node[label=(180+\cmdKV@general@rotate):\cmdKV@gLabels@iLA] {} (0.3,0.9);
    \draw[label distance=0,\cmdKV@gTypes@iB] (0.3,0.9) -- node[label=(90+\cmdKV@general@rotate):\cmdKV@gLabels@iLB] {} (0.9,0.9);
    \draw[label distance=0,\cmdKV@gTypes@iC] (0.9,0.9) -- node[label=(90+\cmdKV@general@rotate):\cmdKV@gLabels@iLC] {} (1.5,0.9);
    \draw[label distance=0,\cmdKV@gTypes@iD] (1.5,0.9) -- node[label=(\cmdKV@general@rotate):\cmdKV@gLabels@iLD] {} (1.5,0.3);
    \draw[label distance=0,\cmdKV@gTypes@iE] (1.5,0.3) -- node[label=(-90+\cmdKV@general@rotate):\cmdKV@gLabels@iLE] {} (0.9,0.3);
    \draw[label distance=0,\cmdKV@gTypes@iF] (0.9,0.3) -- node[label=(-90+\cmdKV@general@rotate):\cmdKV@gLabels@iLF] {} (0.3,0.3);
    \draw[label distance=0,\cmdKV@gTypes@iG] (0.9,0.9) -- node[label=(\cmdKV@general@rotate):\cmdKV@gLabels@iLG] {} (0.9,0.3);
    #2
  \end{tikzpicture}
  \gSetStdTypes{line}
}
\def\gBoxBoxNP{\@ifnextchar[\@gBoxBoxNP{\@gBoxBoxNP[]}}
\def\@gBoxBoxNP[#1]#2{%
  \setkeys*{pre}{#1}%
  \setrmkeys{general,gTypes,gLabels}%
  \begin{tikzpicture}
    [line width=1pt,
    baseline={([yshift=-0.5ex+\cmdKV@general@yshift]current bounding box.center)},
    scale=\cmdKV@general@scale,
    rotate=\cmdKV@general@rotate,
    font=\cmdKV@general@font]
    \draw[\cmdKV@gTypes@eA] (-1.9,0.3) -- (-2.2,0) node[label=(180+\cmdKV@general@rotate):\cmdKV@gLabels@eLA] {};
    \draw[\cmdKV@gTypes@eB] (-1.9,1.1) -- (-2.2,1.4) node[label=(180+\cmdKV@general@rotate):\cmdKV@gLabels@eLB] {};
    \draw[\cmdKV@gTypes@eC] (-0.3,0.7) -- (0,0.7) node[label=(\cmdKV@general@rotate):\cmdKV@gLabels@eLC] {};
    \draw[\cmdKV@gTypes@eD] (-1.1,0.7) -- (-1.4,0.7) node[label=(180+\cmdKV@general@rotate):\cmdKV@gLabels@eLD] {};
    \draw[label distance=0,\cmdKV@gTypes@iA] (-1.9,0.3) -- node[label=(180+\cmdKV@general@rotate):\cmdKV@gLabels@iLA] {} (-1.9,1.1);
    \draw[label distance=0,\cmdKV@gTypes@iB] (-1.9,1.1) -- node[label=(90+\cmdKV@general@rotate):\cmdKV@gLabels@iLB] {} (-0.7,1.1);
    \draw[label distance=0,\cmdKV@gTypes@iC] (-0.7,1.1) -- node[label=(45+\cmdKV@general@rotate):\cmdKV@gLabels@iLC] {} (-0.3,0.7);
    \draw[label distance=0,\cmdKV@gTypes@iD] (-0.3,0.7) -- node[label=(-45+\cmdKV@general@rotate):\cmdKV@gLabels@iLD] {} (-0.7,0.3);
    \draw[label distance=0,\cmdKV@gTypes@iE] (-0.7,0.3) -- node[label=(-90+\cmdKV@general@rotate):\cmdKV@gLabels@iLE] {} (-1.9,0.3);
    \draw[label distance=0,\cmdKV@gTypes@iF] (-0.7,1.1) -- node[label=(180+\cmdKV@general@rotate):\cmdKV@gLabels@iLF] {} (-1.1,0.7);
    \draw[label distance=0,\cmdKV@gTypes@iG] (-1.1,0.7) -- node[label=(-180+\cmdKV@general@rotate):\cmdKV@gLabels@iLG] {} (-0.7,0.3);
    #2
  \end{tikzpicture}
  \gSetStdTypes{line}
}

\def\gBoxBoxNPB{\@ifnextchar[\@gBoxBoxNPB{\@gBoxBoxNPB[]}}
\def\@gBoxBoxNPB[#1]#2{%
  \setkeys*{pre}{#1}%
  \setrmkeys{general,gTypes,gLabels}%
  \begin{tikzpicture}
    [line width=1pt,
    baseline={([yshift=-0.5ex+\cmdKV@general@yshift]current bounding box.center)},
    scale=\cmdKV@general@scale,
    rotate=\cmdKV@general@rotate,
    font=\cmdKV@general@font]
    \draw[\cmdKV@gTypes@eA] (0.3,0.3) -- (0,0) node[label=(-180+\cmdKV@general@rotate):\cmdKV@gLabels@eLA] {};
    \draw[\cmdKV@gTypes@eB] (0.3,0.9) -- (0,1.2) node[label=(180+\cmdKV@general@rotate):\cmdKV@gLabels@eLB] {};
    \draw[\cmdKV@gTypes@eC] (1.5,0.9) -- (1.8,1.2) node[label=(\cmdKV@general@rotate):\cmdKV@gLabels@eLC] {};
    \draw[\cmdKV@gTypes@eD] (1.5,0.3) -- (1.8,0) node[label=(\cmdKV@general@rotate):\cmdKV@gLabels@eLD] {};
    \draw[label distance=0,\cmdKV@gTypes@iA] (0.3,0.3) -- node[label=(180+\cmdKV@general@rotate):\cmdKV@gLabels@iLA] {} (0.3,0.9);
    \draw[label distance=0,\cmdKV@gTypes@iB] (0.3,0.9) -- node[label=(90+\cmdKV@general@rotate):\cmdKV@gLabels@iLB] {} (0.9,0.9);
    \draw[label distance=0,\cmdKV@gTypes@iC] (0.9,0.9) -- node[label=(90+\cmdKV@general@rotate):\cmdKV@gLabels@iLC] {} (1.5,0.3);
    \foreach \x in {5,...,14} \fill[opacity=0.12,white] (1.2,0.6) circle (\x/100);
    \draw[label distance=0,\cmdKV@gTypes@iD] (1.5,0.3) -- node[label=(\cmdKV@general@rotate):\cmdKV@gLabels@iLD] {} (1.5,0.9);
    \draw[label distance=0,\cmdKV@gTypes@iE] (1.5,0.9) -- node[label=(-90+\cmdKV@general@rotate):\cmdKV@gLabels@iLE] {} (0.9,0.3);
    \draw[label distance=0,\cmdKV@gTypes@iF] (0.9,0.3) -- node[label=(-90+\cmdKV@general@rotate):\cmdKV@gLabels@iLF] {} (0.3,0.3);
    \draw[label distance=-1,\cmdKV@gTypes@iG] (0.9,0.9) -- node[label=(180+\cmdKV@general@rotate):\cmdKV@gLabels@iLG] {} (0.9,0.3);
    #2
  \end{tikzpicture}
  \gSetStdTypes{line}
}

\def\gBoxBoxNPC{\@ifnextchar[\@gBoxBoxNPC{\@gBoxBoxNPC[]}}
\def\@gBoxBoxNPC[#1]#2{%
  \setkeys*{pre}{#1}%
  \setrmkeys{general,gTypes,gLabels}%
  \begin{tikzpicture}
    [line width=1pt,
    baseline={([yshift=-0.5ex+\cmdKV@general@yshift]current bounding box.center)},
    scale=\cmdKV@general@scale,
    rotate=\cmdKV@general@rotate,
    font=\cmdKV@general@font]
    \draw[\cmdKV@gTypes@eA] (0.3,0.3) -- (0,0) node[label=(-180+\cmdKV@general@rotate):\cmdKV@gLabels@eLA] {};
    \draw[\cmdKV@gTypes@eB] (0.3,0.9) -- (0,1.2) node[label=(180+\cmdKV@general@rotate):\cmdKV@gLabels@eLB] {};
    \draw[\cmdKV@gTypes@eC] (1.5,0.9) -- (1.8,1.2) node[label=(\cmdKV@general@rotate):\cmdKV@gLabels@eLC] {};
    \draw[\cmdKV@gTypes@eD] (1.5,0.3) -- (1.8,0) node[label=(\cmdKV@general@rotate):\cmdKV@gLabels@eLD] {};
    \draw[label distance=0,\cmdKV@gTypes@iA] (0.3,0.3) -- node[label=(180+\cmdKV@general@rotate):\cmdKV@gLabels@iLA] {} (0.3,0.9);
    \draw[label distance=0,\cmdKV@gTypes@iB] (0.3,0.9) -- node[label=(90+\cmdKV@general@rotate):\cmdKV@gLabels@iLB] {} (0.9,0.9);
    \draw[label distance=0,\cmdKV@gTypes@iC] (0.9,0.9) -- node[label=(90+\cmdKV@general@rotate):\cmdKV@gLabels@iLC] {} (1.5,0.9);
    \draw[label distance=0,\cmdKV@gTypes@iD] (1.5,0.9) -- node[label=(\cmdKV@general@rotate):\cmdKV@gLabels@iLD] {} (0.9,0.3);
    \foreach \x in {5,...,14} \fill[opacity=0.12,white] (1.2,0.6) circle (\x/100);
    \draw[label distance=0,\cmdKV@gTypes@iE] (1.5,0.3) -- node[label=(-90+\cmdKV@general@rotate):\cmdKV@gLabels@iLE] {} (0.9,0.3);
    \draw[label distance=0,\cmdKV@gTypes@iF] (0.9,0.3) -- node[label=(-90+\cmdKV@general@rotate):\cmdKV@gLabels@iLF] {} (0.3,0.3);
    \draw[label distance=-1,\cmdKV@gTypes@iG] (0.9,0.9) -- node[label=(180+\cmdKV@general@rotate):\cmdKV@gLabels@iLG] {} (1.5,0.3);
    #2
  \end{tikzpicture}
  \gSetStdTypes{line}
}

\def\gBoxTriNPA{\@ifnextchar[\@gBoxTriNPA{\@gBoxTriNPA[]}}
\def\@gBoxTriNPA[#1]#2{%
  \setkeys*{pre}{#1}%
  \setrmkeys{general,gTypes,gLabels}%
  \begin{tikzpicture}
    [line width=1pt,
    baseline={([yshift=-0.5ex+\cmdKV@general@yshift]current bounding box.center)},
    scale=\cmdKV@general@scale,
    rotate=\cmdKV@general@rotate,
    font=\cmdKV@general@font]
    \draw[\cmdKV@gTypes@eA] (0.3,0.6) -- (0,0) node[label=(-180+\cmdKV@general@rotate):\cmdKV@gLabels@eLA] {};
    \draw[\cmdKV@gTypes@eB] (0.3,0.6) -- (0,1.2) node[label=(180+\cmdKV@general@rotate):\cmdKV@gLabels@eLB] {};
    \draw[\cmdKV@gTypes@eC] (1.5,0.9) -- (1.8,1.2) node[label=(\cmdKV@general@rotate):\cmdKV@gLabels@eLC] {};
    \draw[\cmdKV@gTypes@eD] (1.5,0.3) -- (1.8,0) node[label=(\cmdKV@general@rotate):\cmdKV@gLabels@eLD] {};
    \draw[label distance=0,\cmdKV@gTypes@iA] (0.3,0.6) -- node[label=(90+\cmdKV@general@rotate):\cmdKV@gLabels@iLA] {} (0.9,0.9);
    \draw[label distance=0,\cmdKV@gTypes@iB] (0.9,0.9) -- node[label=(90+\cmdKV@general@rotate):\cmdKV@gLabels@iLB] {} (1.5,0.9);
    \draw[label distance=-1,\cmdKV@gTypes@iC] (0.9,0.9) -- node[label=(180+\cmdKV@general@rotate):\cmdKV@gLabels@iLC] {} (1.5,0.3);
    \foreach \x in {5,...,14} \fill[opacity=0.12,white] (1.2,0.6) circle (\x/100);
    \draw[label distance=0,\cmdKV@gTypes@iD] (1.5,0.3) -- node[label=(-90+\cmdKV@general@rotate):\cmdKV@gLabels@iLD] {} (0.9,0.3);
    \draw[label distance=0,\cmdKV@gTypes@iE] (0.9,0.3) -- node[label=(-90+\cmdKV@general@rotate):\cmdKV@gLabels@iLE] {} (0.3,0.6);
    \draw[label distance=-1,\cmdKV@gTypes@iF] (1.5,0.9) -- node[label=(180+\cmdKV@general@rotate):\cmdKV@gLabels@iLF] {} (0.9,0.3);
    #2
  \end{tikzpicture}
  \gSetStdTypes{line}
}

\def\gTriPenta{\@ifnextchar[\@gTriPenta{\@gTriPenta[]}}
\def\@gTriPenta[#1]#2{%
  \setkeys*{pre}{#1}%
  \setrmkeys{general,gTypes,gLabels}%
  \begin{tikzpicture}
    [line width=1pt,
    baseline={([yshift=-0.5ex+\cmdKV@general@yshift]current bounding box.center)},
    scale=\cmdKV@general@scale,
    rotate=\cmdKV@general@rotate,
    font=\cmdKV@general@font]
    \draw[\cmdKV@gTypes@eA] (-108:0.5) -- (-135:1) node[label=(145+\cmdKV@general@rotate):\cmdKV@gLabels@eLA] {};
    \draw[\cmdKV@gTypes@eB] (180:0.5) -- (180:1) node[label=(180+\cmdKV@general@rotate):\cmdKV@gLabels@eLB] {};
    \draw[\cmdKV@gTypes@eC] (108:0.5) -- (135:1) node[label=(-145+\cmdKV@general@rotate):\cmdKV@gLabels@eLC] {};
    \draw[\cmdKV@gTypes@eD] (0:1.2) -- (0:1.5) node[label=(\cmdKV@general@rotate):\cmdKV@gLabels@eLD] {};
    \draw[label distance=0,\cmdKV@gTypes@iA] (-108:0.5) -- node[label=(-135+\cmdKV@general@rotate):\cmdKV@gLabels@iLA] {} (-180:0.5);
    \draw[label distance=0,\cmdKV@gTypes@iB] (180:0.5) -- node[label=(135+\cmdKV@general@rotate):\cmdKV@gLabels@iLB] {} (108:0.5);
    \draw[label distance=0,\cmdKV@gTypes@iC] (108:0.5) -- node[label=(72+\cmdKV@general@rotate):\cmdKV@gLabels@iLC] {} (36:0.5);
    \draw[label distance=0,\cmdKV@gTypes@iD] (36:0.5) -- node[label=(72+\cmdKV@general@rotate):\cmdKV@gLabels@iLD] {} (0:1.2);
    \draw[label distance=2,\cmdKV@gTypes@iE] (0:1.2) -- node[label=(-90+\cmdKV@general@rotate):\cmdKV@gLabels@iLE] {} (-36:0.5);
    \draw[label distance=2,\cmdKV@gTypes@iF] (-36:0.5) -- node[label=(-90+\cmdKV@general@rotate):\cmdKV@gLabels@iLF] {} (-108:0.5);
    \draw[label distance=0,\cmdKV@gTypes@iG] (36:0.5) -- node[label=(180+\cmdKV@general@rotate):\cmdKV@gLabels@iLG] {} (-36:0.5);
    #2
  \end{tikzpicture}
  \gSetStdTypes{line}
}
\def\gTriTri{\@ifnextchar[\@gTriTri{\@gTriTri[]}}
\def\@gTriTri[#1]#2{%
  \setkeys*{pre}{#1}%
  \setrmkeys{general,gTypes,gLabels}%
  \begin{tikzpicture}
    [line width=1pt,
    baseline={([yshift=-0.5ex+\cmdKV@general@yshift]current bounding box.center)},
    scale=\cmdKV@general@scale,
    rotate=\cmdKV@general@rotate,
    font=\cmdKV@general@font]
    \draw[\cmdKV@gTypes@eA] (0.3,0.3) -- (0,0) node[label=(-180+\cmdKV@general@rotate):\cmdKV@gLabels@eLA] {};
    \draw[\cmdKV@gTypes@eB] (0.3,1.1) -- (0,1.4) node[label=(180+\cmdKV@general@rotate):\cmdKV@gLabels@eLB] {};
    \draw[\cmdKV@gTypes@eC] (1.9,1.1) -- (2.2,1.4) node[label=(\cmdKV@general@rotate):\cmdKV@gLabels@eLC] {};
    \draw[\cmdKV@gTypes@eD] (1.9,0.3) -- (2.2,0) node[label=(\cmdKV@general@rotate):\cmdKV@gLabels@eLD] {};
    \draw[label distance=0,\cmdKV@gTypes@iA] (0.3,0.3) -- node[label=(180+\cmdKV@general@rotate):\cmdKV@gLabels@iLA] {} (0.3,1.1);
    \draw[label distance=0,\cmdKV@gTypes@iB] (0.3,1.1) -- node[label=(45+\cmdKV@general@rotate):\cmdKV@gLabels@iLB] {} (0.9,0.7);
    \draw[label distance=0,\cmdKV@gTypes@iC] (0.9,0.7) -- node[label=(-45+\cmdKV@general@rotate):\cmdKV@gLabels@iLC] {} (0.3,0.3);
    \draw[label distance=0,\cmdKV@gTypes@iD] (0.9,0.7) -- node[label=(90+\cmdKV@general@rotate):\cmdKV@gLabels@iLD] {} (1.3,0.7);
    \draw[label distance=0,\cmdKV@gTypes@iE] (1.3,0.7) -- node[label=(135+\cmdKV@general@rotate):\cmdKV@gLabels@iLE] {} (1.9,1.1);
    \draw[label distance=0,\cmdKV@gTypes@iF] (1.9,1.1) -- node[label=(\cmdKV@general@rotate):\cmdKV@gLabels@iLF] {} (1.9,0.3);
    \draw[label distance=0,\cmdKV@gTypes@iG] (1.9,0.3) -- node[label=(-135+\cmdKV@general@rotate):\cmdKV@gLabels@iLG] {} (1.3,0.7);
    #2
  \end{tikzpicture}
  \gSetStdTypes{line}
}

\def\gBoxBubA{\@ifnextchar[\@gBoxBubA{\@gBoxBubA[]}}
\def\@gBoxBubA[#1]#2{%
  \setkeys*{pre}{#1}%
  \setrmkeys{general,gTypes,gLabels}%
  \begin{tikzpicture}
    [line width=1pt,
    baseline={([yshift=-0.5ex+\cmdKV@general@yshift]current bounding box.center)},
    scale=\cmdKV@general@scale,
    rotate=\cmdKV@general@rotate,
    font=\cmdKV@general@font]
    \draw[\cmdKV@gTypes@eA] (0.1,0.1) -- (-0.1,-0.1) node[label=(-180+\cmdKV@general@rotate):\cmdKV@gLabels@eLA] {};
    \draw[\cmdKV@gTypes@eB] (0.1,0.9) -- (-0.1,1.1) node[label=(180+\cmdKV@general@rotate):\cmdKV@gLabels@eLB] {};
    \draw[\cmdKV@gTypes@eC] (0.9,0.9) -- (1.1,1.1) node[label=(\cmdKV@general@rotate):\cmdKV@gLabels@eLC] {};
    \draw[\cmdKV@gTypes@eD] (0.9,0.1) -- (1.1,-0.1) node[label=(\cmdKV@general@rotate):\cmdKV@gLabels@eLD] {};
    \draw[label distance=0,\cmdKV@gTypes@iA] (0.1,0.1) -- node[label=(180+\cmdKV@general@rotate):\cmdKV@gLabels@iLA] {} (0.1,0.9);
    \draw[label distance=0,\cmdKV@gTypes@iB] (0.1,0.9) -- node[label=(90+\cmdKV@general@rotate):\cmdKV@gLabels@iLB] {} (0.32,0.9);
    \draw[label distance=0,\cmdKV@gTypes@iC] (0.32,0.9) .. controls (0.32,1.15) and (0.68,1.15) .. node[label=(90+\cmdKV@general@rotate):\cmdKV@gLabels@iLC] {} (0.68,0.9);
    \draw[label distance=0,\cmdKV@gTypes@iD] (0.68,0.9) .. controls (0.68,0.65) and (0.32,0.65) .. node[label=(-90+\cmdKV@general@rotate):\cmdKV@gLabels@iLD] {} (0.32,0.9);
    \draw[label distance=0,\cmdKV@gTypes@iE] (0.68,0.9) -- node[label=(90+\cmdKV@general@rotate):\cmdKV@gLabels@iLE] {} (0.9,0.9);
    \draw[label distance=0,\cmdKV@gTypes@iF] (0.9,0.9) -- node[label=(\cmdKV@general@rotate):\cmdKV@gLabels@iLF] {} (0.9,0.1);
    \draw[label distance=0,\cmdKV@gTypes@iG] (0.9,0.1) -- node[label=(-90+\cmdKV@general@rotate):\cmdKV@gLabels@iLG] {} (0.1,0.1);
    #2
  \end{tikzpicture}
  \gSetStdTypes{line}
}

\def\gBoxBubB{\@ifnextchar[\@gBoxBubB{\@gBoxBubB[]}}
\def\@gBoxBubB[#1]#2{%
  \setkeys*{pre}{#1}%
  \setrmkeys{general,gTypes,gLabels}%
  \begin{tikzpicture}
    [line width=1pt,
    baseline={([yshift=-0.5ex+\cmdKV@general@yshift]current bounding box.center)},
    scale=\cmdKV@general@scale,
    rotate=\cmdKV@general@rotate,
    font=\cmdKV@general@font]
    \draw[label distance=-1,\cmdKV@gTypes@eA] (0,-0.4) -- (0,-0.7) node[label=(180+\cmdKV@general@rotate):\cmdKV@gLabels@eLA] {};
    \draw[\cmdKV@gTypes@eB] (-0.4,0) -- (-0.7,0) node[label=(180+\cmdKV@general@rotate):\cmdKV@gLabels@eLB] {};
    \draw[label distance=-1,\cmdKV@gTypes@eC] (0,0.4) -- (0,0.7) node[label=(180+\cmdKV@general@rotate):\cmdKV@gLabels@eLC] {};
    \draw[\cmdKV@gTypes@eD] (1.1,0) -- (1.4,0) node[label=(\cmdKV@general@rotate):\cmdKV@gLabels@eLD] {};
    \draw[label distance=0,\cmdKV@gTypes@iA] (0,-0.4) -- node[label=(-135+\cmdKV@general@rotate):\cmdKV@gLabels@iLA] {} (-0.4,0);
    \draw[label distance=0,\cmdKV@gTypes@iB] (-0.4,0) -- node[label=(135+\cmdKV@general@rotate):\cmdKV@gLabels@iLB] {} (0,0.4);
    \draw[label distance=0,\cmdKV@gTypes@iC] (0,0.4) -- node[label=(45+\cmdKV@general@rotate):\cmdKV@gLabels@iLC] {} (0.4,0);
    \draw[label distance=0,\cmdKV@gTypes@iD] (0.4,0) -- node[label=(-45+\cmdKV@general@rotate):\cmdKV@gLabels@iLD] {} (0,-0.4);
    \draw[label distance=0,\cmdKV@gTypes@iE] (0.4,0) --  node[label=(90+\cmdKV@general@rotate):\cmdKV@gLabels@iLE] {} (0.7,0);
    \draw[label distance=0,\cmdKV@gTypes@iF] (0.7,0) .. controls (0.7,0.26) and (1.1,0.26) .. node[label=(90+\cmdKV@general@rotate):\cmdKV@gLabels@iLF] {} (1.1,0);
    \draw[label distance=0,\cmdKV@gTypes@iG] (1.1,0) .. controls (1.1,-0.26) and (0.7,-0.26) ..  node[label=(-90+\cmdKV@general@rotate):\cmdKV@gLabels@iLG] {} (0.7,0);
    #2
  \end{tikzpicture}
  \gSetStdTypes{line}
}

\def\gPentaTad{\@ifnextchar[\@gPentaTad{\@gPentaTad[]}}
\def\@gPentaTad[#1]#2{%
  \setkeys*{pre}{#1}%
  \setrmkeys{general,gTypes,gLabels}%
  \begin{tikzpicture}
    [line width=1pt,
    baseline={([yshift=-0.5ex+\cmdKV@general@yshift]current bounding box.center)},
    scale=\cmdKV@general@scale,
    rotate=\cmdKV@general@rotate,
    font=\cmdKV@general@font]
    \draw[label distance=-0.5,\cmdKV@gTypes@eA] (108:0.35) -- (108:0.75) node[label=(180+\cmdKV@general@rotate):\cmdKV@gLabels@eLA] {};
    \draw[\cmdKV@gTypes@eB] (36:0.35) -- (36:0.75) node[label=(36+\cmdKV@general@rotate):\cmdKV@gLabels@eLB] {};
    \draw[\cmdKV@gTypes@eC] (-36:0.35) -- (-36:0.75) node[label=(-36+\cmdKV@general@rotate):\cmdKV@gLabels@eLC] {};
    \draw[label distance=-0.5,\cmdKV@gTypes@eD] (-108:0.35) -- (-108:0.75) node[label=(-180+\cmdKV@general@rotate):\cmdKV@gLabels@eLD] {};
    \draw[label distance=0,\cmdKV@gTypes@iA] (108:0.35) -- node[label=(72+\cmdKV@general@rotate):\cmdKV@gLabels@iLA] {} (36:0.35);
    \draw[label distance=0,\cmdKV@gTypes@iB] (36:0.35) -- node[label=(\cmdKV@general@rotate):\cmdKV@gLabels@iLB] {} (-36:0.35);
    \draw[label distance=0,\cmdKV@gTypes@iC] (-36:0.35) -- node[label=(-72+\cmdKV@general@rotate):\cmdKV@gLabels@iLC] {} (-108:0.35);
    \draw[label distance=0,\cmdKV@gTypes@iD] (-108:0.35) -- node[label=(-144+\cmdKV@general@rotate):\cmdKV@gLabels@iLD] {} (-180:0.35);
    \draw[label distance=0,\cmdKV@gTypes@iE] (180:0.35) -- node[label=(144+\cmdKV@general@rotate):\cmdKV@gLabels@iLE] {} (108:0.35);
    \draw[label distance=0,\cmdKV@gTypes@iF] (180:0.35) -- node[label=(90+\cmdKV@general@rotate):\cmdKV@gLabels@iLF] {} (180:0.75);
    \draw[label distance=-0.5,\cmdKV@gTypes@iG] (180:0.75) .. controls ($(180:0.75)+(0,-0.18)$) and ($(180:1)+(0,-0.18)$) .. (180:1) node[label=(180+\cmdKV@general@rotate):\cmdKV@gLabels@iLG] {} .. controls ($(180:1)+(0,0.18)$) and ($(180:0.75)+(0,0.18)$) .. (180:0.75);
    #2
  \end{tikzpicture}
  \gSetStdTypes{line}
}

\def\gBoxTadA{\@ifnextchar[\@gBoxTadA{\@gBoxTadA[]}}
\def\@gBoxTadA[#1]#2{%
  \setkeys*{pre}{#1}%
  \setrmkeys{general,gTypes,gLabels}%
  \begin{tikzpicture}
    [line width=1pt,
    baseline={([yshift=-0.5ex+\cmdKV@general@yshift]current bounding box.center)},
    scale=\cmdKV@general@scale,
    rotate=\cmdKV@general@rotate,
    font=\cmdKV@general@font]
    \draw[label distance=-0.5,\cmdKV@gTypes@eA] (0,0.4) -- (0,0.7) node[label=(\cmdKV@general@rotate):\cmdKV@gLabels@eLA] {};
    \draw[\cmdKV@gTypes@eB] (0.4,0) -- (0.7,0) node[label=(\cmdKV@general@rotate):\cmdKV@gLabels@eLB] {};
    \draw[label distance=-0.5,\cmdKV@gTypes@eC] (0,-0.4) -- (0,-0.7) node[label=(\cmdKV@general@rotate):\cmdKV@gLabels@eLC] {};
    \draw[\cmdKV@gTypes@eD] (-0.7,0) -- (-0.7,-0.3) node[label=(-90+\cmdKV@general@rotate):\cmdKV@gLabels@eLD] {};
    \draw[label distance=0,\cmdKV@gTypes@iA] (0,0.4) -- node[label=(45+\cmdKV@general@rotate):\cmdKV@gLabels@iLA] {} (0.4,0);
    \draw[label distance=0,\cmdKV@gTypes@iB] (0.4,0) -- node[label=(-45+\cmdKV@general@rotate):\cmdKV@gLabels@iLB] {} (0,-0.4);
    \draw[label distance=0,\cmdKV@gTypes@iC] (0,-0.4) -- node[label=(-135+\cmdKV@general@rotate):\cmdKV@gLabels@iLC] {} (-0.4,0);
    \draw[label distance=0,\cmdKV@gTypes@iD] (-0.4,0) -- node[label=(135+\cmdKV@general@rotate):\cmdKV@gLabels@iLD] {} (0,0.4);
    \draw[label distance=0,\cmdKV@gTypes@iE] (-0.4,0) --  node[label=(90+\cmdKV@general@rotate):\cmdKV@gLabels@iLE] {} (-0.7,0);
    \draw[label distance=0,\cmdKV@gTypes@iF] (-0.7,0) -- node[label=(90+\cmdKV@general@rotate):\cmdKV@gLabels@iLF] {} (-1.1,0);
    \draw[label distance=-0.5,\cmdKV@gTypes@iG] (-1.1,0) .. controls (-1.1,-0.18) and (-1.35,-0.18) .. (-1.35,0) node[label=(180+\cmdKV@general@rotate):\cmdKV@gLabels@iLG] {} .. controls (-1.35,0.18) and (-1.1,0.18) .. (-1.1,0);
    #2
  \end{tikzpicture}
  \gSetStdTypes{line}
}

\def\gSunsetA{\@ifnextchar[\@gSunsetA{\@gSunsetA[]}}
\def\@gSunsetA[#1]#2{%
  \setkeys*{pre}{#1}%
  \setrmkeys{general,gTypes,gLabels}%
  \begin{tikzpicture}
    [line width=1pt,
    baseline={([yshift=-0.5ex+\cmdKV@general@yshift]current bounding box.center)},
    scale=\cmdKV@general@scale,
    rotate=\cmdKV@general@rotate,
    font=\cmdKV@general@font]
    \draw[\cmdKV@gTypes@eA] (0.3,0.6) -- (0,0) node[label=(-180+\cmdKV@general@rotate):\cmdKV@gLabels@eLA] {};
    \draw[\cmdKV@gTypes@eB] (0.3,0.6) -- (0,1.2) node[label=(180+\cmdKV@general@rotate):\cmdKV@gLabels@eLB] {};
    \draw[\cmdKV@gTypes@eC] (1.5,0.6) -- (1.8,1.2) node[label=(\cmdKV@general@rotate):\cmdKV@gLabels@eLC] {};
    \draw[\cmdKV@gTypes@eD] (1.5,0.6) -- (1.8,0) node[label=(\cmdKV@general@rotate):\cmdKV@gLabels@eLD] {};
    \draw[label distance=0,\cmdKV@gTypes@iA] (0.3,0.6) to[bend left=90] node[label=(90+\cmdKV@general@rotate):\cmdKV@gLabels@iLA] {} (1.5,0.6);
    \draw[label distance=0,\cmdKV@gTypes@iB] (1.5,0.6) to[bend left=90] node[label=(-90+\cmdKV@general@rotate):\cmdKV@gLabels@iLB] {} (0.3,0.6);
    \draw[label distance=-2,\cmdKV@gTypes@iC] (0.3,0.6) -- node[label=(90+\cmdKV@general@rotate):\cmdKV@gLabels@iLC] {} (1.5,0.6);
    #2
  \end{tikzpicture}
  \gSetStdTypes{line}
}

\def\gTriBubA{\@ifnextchar[\@gTriBubA{\@gTriBubA[]}}
\def\@gTriBubA[#1]#2{%
  \setkeys*{pre}{#1}%
  \setrmkeys{general,gTypes,gLabels}%
  \begin{tikzpicture}
    [line width=1pt,
    baseline={([yshift=-0.5ex+\cmdKV@general@yshift]current bounding box.center)},
    scale=\cmdKV@general@scale,
    rotate=\cmdKV@general@rotate,
    font=\cmdKV@general@font]
    \draw[\cmdKV@gTypes@eA] (0.4,0.3) -- (0,0) node[label=(-180+\cmdKV@general@rotate):\cmdKV@gLabels@eLA] {};
    \draw[\cmdKV@gTypes@eB] (0.4,0.9) -- (0,1.2) node[label=(180+\cmdKV@general@rotate):\cmdKV@gLabels@eLB] {};
    \draw[\cmdKV@gTypes@eC] (1.4,0.6) -- (1.8,1.2) node[label=(\cmdKV@general@rotate):\cmdKV@gLabels@eLC] {};
    \draw[\cmdKV@gTypes@eD] (1.4,0.6) -- (1.8,0) node[label=(\cmdKV@general@rotate):\cmdKV@gLabels@eLD] {};
    \draw[label distance=0,\cmdKV@gTypes@iA] (0.4,0.3) to[bend left=40] node[label=(180+\cmdKV@general@rotate):\cmdKV@gLabels@iLA] {} (0.4,0.9);
    \draw[label distance=0,\cmdKV@gTypes@iB] (0.4,0.9) -- node[label=(90+\cmdKV@general@rotate):\cmdKV@gLabels@iLB] {} (1.4,0.6);
    \draw[label distance=0,\cmdKV@gTypes@iC] (1.4,0.6) -- node[label=(-90+\cmdKV@general@rotate):\cmdKV@gLabels@iLC] {} (0.4,0.3);
    \draw[label distance=-1,\cmdKV@gTypes@iD] (0.4,0.9) to[bend left=40] node[label=(\cmdKV@general@rotate):\cmdKV@gLabels@iLD] {} (0.4,0.3);
    #2
  \end{tikzpicture}
  \gSetStdTypes{line}
}

\def\gBubBubA{\@ifnextchar[\@gBubBubA{\@gBubBubA[]}}
\def\@gBubBubA[#1]#2{%
  \setkeys*{pre}{#1}%
  \setrmkeys{general,gTypes,gLabels}%
  \begin{tikzpicture}
    [line width=1pt,
    baseline={([yshift=-0.5ex+\cmdKV@general@yshift]current bounding box.center)},
    scale=\cmdKV@general@scale,
    rotate=\cmdKV@general@rotate,
    font=\cmdKV@general@font]
    \draw[\cmdKV@gTypes@eA] (0.3,0.6) -- (0,0) node[label=(-180+\cmdKV@general@rotate):\cmdKV@gLabels@eLA] {};
    \draw[\cmdKV@gTypes@eB] (0.3,0.6) -- (0,1.2) node[label=(180+\cmdKV@general@rotate):\cmdKV@gLabels@eLB] {};
    \draw[\cmdKV@gTypes@eC] (1.5,0.6) -- (1.8,1.2) node[label=(\cmdKV@general@rotate):\cmdKV@gLabels@eLC] {};
    \draw[\cmdKV@gTypes@eD] (1.5,0.6) -- (1.8,0) node[label=(\cmdKV@general@rotate):\cmdKV@gLabels@eLD] {};
    \draw[label distance=0,\cmdKV@gTypes@iA] (0.3,0.6) to[bend left=90] node[label=(90+\cmdKV@general@rotate):\cmdKV@gLabels@iLA] {} (0.9,0.6);
    \draw[label distance=0,\cmdKV@gTypes@iB] (0.9,0.6) to[bend left=90] node[label=(90+\cmdKV@general@rotate):\cmdKV@gLabels@iLB] {} (1.5,0.6);
    \draw[label distance=0,\cmdKV@gTypes@iC] (1.5,0.6) to[bend left=90] node[label=(-90+\cmdKV@general@rotate):\cmdKV@gLabels@iLC] {} (0.9,0.6);
    \draw[label distance=0,\cmdKV@gTypes@iD] (0.9,0.6) to[bend left=90] node[label=(-90+\cmdKV@general@rotate):\cmdKV@gLabels@iLD] {} (0.3,0.6);
    #2
  \end{tikzpicture}
  \gSetStdTypes{line}
}

\def\gBoxBubC{\@ifnextchar[\@gBoxBubC{\@gBoxBubC[]}}
\def\@gBoxBubC[#1]#2{%
  \setkeys*{pre}{#1}%
  \setrmkeys{general,gTypes,gLabels}%
  \begin{tikzpicture}
    [line width=1pt,
    baseline={([yshift=-0.5ex+\cmdKV@general@yshift]current bounding box.center)},
    scale=\cmdKV@general@scale,
    rotate=\cmdKV@general@rotate,
    font=\cmdKV@general@font]
    \draw[\cmdKV@gTypes@eA] (0.1,0.1) -- (-0.1,-0.1) node[label=(-180+\cmdKV@general@rotate):\cmdKV@gLabels@eLA] {};
    \draw[\cmdKV@gTypes@eB] (0.1,0.9) -- (-0.1,1.1) node[label=(180+\cmdKV@general@rotate):\cmdKV@gLabels@eLB] {};
    \draw[\cmdKV@gTypes@eC] (0.9,0.9) -- (1.1,1.1) node[label=(\cmdKV@general@rotate):\cmdKV@gLabels@eLC] {};
    \draw[\cmdKV@gTypes@eD] (0.9,0.1) -- (1.1,-0.1) node[label=(\cmdKV@general@rotate):\cmdKV@gLabels@eLD] {};
    \draw[label distance=0,\cmdKV@gTypes@iA] (0.1,0.1) -- node[label=(180+\cmdKV@general@rotate):\cmdKV@gLabels@iLA] {} (0.1,0.9);
    \draw[label distance=0,\cmdKV@gTypes@iB] (0.1,0.9) to[bend left] node[label=(90+\cmdKV@general@rotate):\cmdKV@gLabels@iLB] {} (0.9,0.9);
    \draw[label distance=0,\cmdKV@gTypes@iC] (0.9,0.9) to[bend left] node[label=(-90+\cmdKV@general@rotate):\cmdKV@gLabels@iLC] {} (0.1,0.9);
    \draw[label distance=0,\cmdKV@gTypes@iD] (0.9,0.9) -- node[label=(\cmdKV@general@rotate):\cmdKV@gLabels@iLD] {} (0.9,0.1);
    \draw[label distance=0,\cmdKV@gTypes@iE] (0.9,0.1) -- node[label=(-90+\cmdKV@general@rotate):\cmdKV@gLabels@iLE] {} (0.1,0.1);
    #2
  \end{tikzpicture}
  \gSetStdTypes{line}
}

\def\gBoxZ{\@ifnextchar[\@gBoxZ{\@gBoxZ[]}}
\def\@gBoxZ[#1]#2{%
  \setkeys*{pre}{#1}%
  \setrmkeys{general,gTypes,gLabels}%
  \begin{tikzpicture}
    [line width=1pt,
    baseline={([yshift=-0.5ex+\cmdKV@general@yshift]current bounding box.center)},
    scale=\cmdKV@general@scale,
    rotate=\cmdKV@general@rotate,
    font=\cmdKV@general@font]
    \draw[\cmdKV@gTypes@eA] (0.15,0.15) -- (-0.1,-0.1) node[label=(-180+\cmdKV@general@rotate):\cmdKV@gLabels@eLA] {};
    \draw[\cmdKV@gTypes@eB] (0.15,0.85) -- (-0.1,1.1) node[label=(180+\cmdKV@general@rotate):\cmdKV@gLabels@eLB] {};
    \draw[\cmdKV@gTypes@eC] (0.85,0.85) -- (1.1,1.1) node[label=(\cmdKV@general@rotate):\cmdKV@gLabels@eLC] {};
    \draw[\cmdKV@gTypes@eD] (0.85,0.15) -- (1.1,-0.1) node[label=(\cmdKV@general@rotate):\cmdKV@gLabels@eLD] {};
    \draw[label distance=0,\cmdKV@gTypes@iA] (0.15,0.15) -- node[label=(180+\cmdKV@general@rotate):\cmdKV@gLabels@iLA] {} (0.15,0.85);
    \draw[label distance=0,\cmdKV@gTypes@iB] (0.15,0.85) -- node[label=(90+\cmdKV@general@rotate):\cmdKV@gLabels@iLB] {} (0.85,0.85);
    \draw[label distance=0,\cmdKV@gTypes@iC] (0.85,0.85) -- node[label=(\cmdKV@general@rotate):\cmdKV@gLabels@iLC] {} (0.85,0.15);
    \draw[label distance=0,\cmdKV@gTypes@iD] (0.85,0.15) -- node[label=(-90+\cmdKV@general@rotate):\cmdKV@gLabels@iLD] {} (0.15,0.15);
    \draw[label distance=-4,\cmdKV@gTypes@iE] (0.15,0.15) -- node[label=(135+\cmdKV@general@rotate):\cmdKV@gLabels@iLE] {} (0.85,0.85);
    #2
  \end{tikzpicture}
  \gSetStdTypes{line}
}

\def\gBoxBoxBoxA{\@ifnextchar[\@gBoxBoxBoxA{\@gBoxBoxBoxA[]}}
\def\@gBoxBoxBoxA[#1]#2{%
  \setkeys*{pre}{#1}%
  \setrmkeys{general,gTypes,gLabels}%
  \begin{tikzpicture}
    [line width=1pt,
    baseline={([yshift=-0.5ex+\cmdKV@general@yshift]current bounding box.center)},
    scale=\cmdKV@general@scale,
    rotate=\cmdKV@general@rotate,
    font=\cmdKV@general@font]
    \draw[\cmdKV@gTypes@eA] (-0.3,0.3) -- (-0.6,0) node[label=(-180+\cmdKV@general@rotate):\cmdKV@gLabels@eLA] {};
    \draw[\cmdKV@gTypes@eB] (-0.3,0.9) -- (-0.6,1.2) node[label=(180+\cmdKV@general@rotate):\cmdKV@gLabels@eLB] {};
    \draw[\cmdKV@gTypes@eC] (1.5,0.9) -- (1.8,1.2) node[label=(\cmdKV@general@rotate):\cmdKV@gLabels@eLC] {};
    \draw[\cmdKV@gTypes@eD] (1.5,0.3) -- (1.8,0) node[label=(\cmdKV@general@rotate):\cmdKV@gLabels@eLD] {};
    \draw[label distance=0,\cmdKV@gTypes@iA] (-0.3,0.3) -- node[label=(180+\cmdKV@general@rotate):\cmdKV@gLabels@iLA] {} (-0.3,0.9);
    \draw[label distance=0,\cmdKV@gTypes@iB] (-0.3,0.9) -- node[label=(90+\cmdKV@general@rotate):\cmdKV@gLabels@iLB] {} (0.3,0.9);
    \draw[label distance=0,\cmdKV@gTypes@iC] (0.3,0.9) -- node[label=(90+\cmdKV@general@rotate):\cmdKV@gLabels@iLC] {} (0.9,0.9);
    \draw[label distance=0,\cmdKV@gTypes@iD] (0.9,0.9) -- node[label=(90+\cmdKV@general@rotate):\cmdKV@gLabels@iLD] {} (1.5,0.9);
    \draw[label distance=0,\cmdKV@gTypes@iE] (1.5,0.9) -- node[label=(\cmdKV@general@rotate):\cmdKV@gLabels@iLE] {} (1.5,0.3);
    \draw[label distance=0,\cmdKV@gTypes@iF] (1.5,0.3) -- node[label=(-90+\cmdKV@general@rotate):\cmdKV@gLabels@iLF] {} (0.9,0.3);
    \draw[label distance=0,\cmdKV@gTypes@iG] (0.9,0.3) -- node[label=(-90+\cmdKV@general@rotate):\cmdKV@gLabels@iLG] {} (0.3,0.3);
    \draw[label distance=0,\cmdKV@gTypes@iH] (0.3,0.3) -- node[label=(-90+\cmdKV@general@rotate):\cmdKV@gLabels@iLH] {} (-0.3,0.3);
    \draw[label distance=0,\cmdKV@gTypes@iI] (0.3,0.9) -- node[label=(\cmdKV@general@rotate):\cmdKV@gLabels@iLI] {} (0.3,0.3);
    \draw[label distance=0,\cmdKV@gTypes@iJ] (0.9,0.9) -- node[label=(\cmdKV@general@rotate):\cmdKV@gLabels@iLJ] {} (0.9,0.3);
    #2
  \end{tikzpicture}
  \gSetStdTypes{line}
}

\def\gBoxBoxBoxB{\@ifnextchar[\@gBoxBoxBoxB{\@gBoxBoxBoxB[]}}
\def\@gBoxBoxBoxB[#1]#2{%
  \setkeys*{pre}{#1}%
  \setrmkeys{general,gTypes,gLabels}%
  \begin{tikzpicture}
    [line width=1pt,
    baseline={([yshift=-0.5ex+\cmdKV@general@yshift]current bounding box.center)},
    scale=\cmdKV@general@scale,
    rotate=\cmdKV@general@rotate,
    font=\cmdKV@general@font]
    \draw[\cmdKV@gTypes@eA] (0.3,0.3) -- (0,0) node[label=(-180+\cmdKV@general@rotate):\cmdKV@gLabels@eLA] {};
    \draw[\cmdKV@gTypes@eB] (0.3,1.5) -- (0,1.8) node[label=(180+\cmdKV@general@rotate):\cmdKV@gLabels@eLB] {};
    \draw[\cmdKV@gTypes@eC] (1.5,1.5) -- (1.8,1.8) node[label=(\cmdKV@general@rotate):\cmdKV@gLabels@eLC] {};
    \draw[\cmdKV@gTypes@eD] (1.5,0.3) -- (1.8,0) node[label=(\cmdKV@general@rotate):\cmdKV@gLabels@eLD] {};
    \draw[label distance=0,\cmdKV@gTypes@iA] (0.3,0.3) -- node[label=(180+\cmdKV@general@rotate):\cmdKV@gLabels@iLA] {} (0.3,0.9);
    \draw[label distance=0,\cmdKV@gTypes@iB] (0.3,0.9) -- node[label=(180+\cmdKV@general@rotate):\cmdKV@gLabels@iLB] {} (0.3,1.5);
    \draw[label distance=0,\cmdKV@gTypes@iC] (0.3,1.5) -- node[label=(90+\cmdKV@general@rotate):\cmdKV@gLabels@iLC] {} (0.9,1.5);
    \draw[label distance=0,\cmdKV@gTypes@iD] (0.9,1.5) -- node[label=(90+\cmdKV@general@rotate):\cmdKV@gLabels@iLD] {} (1.5,1.5);
    \draw[label distance=0,\cmdKV@gTypes@iE] (1.5,1.5) -- node[label=(\cmdKV@general@rotate):\cmdKV@gLabels@iLE] {} (1.5,0.3);
    \draw[label distance=0,\cmdKV@gTypes@iF] (1.5,0.3) -- node[label=(-90+\cmdKV@general@rotate):\cmdKV@gLabels@iLF] {} (0.9,0.3);
    \draw[label distance=0,\cmdKV@gTypes@iG] (0.9,0.3) -- node[label=(-90+\cmdKV@general@rotate):\cmdKV@gLabels@iLG] {} (0.3,0.3);
    \draw[label distance=0,\cmdKV@gTypes@iH] (0.9,1.5) -- node[label=(\cmdKV@general@rotate):\cmdKV@gLabels@iLH] {} (0.9,0.9);
    \draw[label distance=0,\cmdKV@gTypes@iI] (0.9,0.9) -- node[label=(\cmdKV@general@rotate):\cmdKV@gLabels@iLI] {} (0.9,0.3);
    \draw[label distance=0,\cmdKV@gTypes@iJ] (0.9,0.9) -- node[label=(90+\cmdKV@general@rotate):\cmdKV@gLabels@iLJ] {} (0.3,0.9);
    #2
  \end{tikzpicture}
  \gSetStdTypes{line}
}

\def\gEFTTree{\@ifnextchar[\@gEFTTree{\@gEFTTree[]}}
\def\@gEFTTree[#1]#2{%
  \setkeys*{pre}{#1}%
  \setrmkeys{general,gTypes,gLabels}%
  \begin{tikzpicture}
    [line width=1pt,
    baseline={([yshift=-0.5ex+\cmdKV@general@yshift]current bounding box.center)},
    scale=\cmdKV@general@scale,
    rotate=\cmdKV@general@rotate,
    font=\cmdKV@general@font]
    \draw[\cmdKV@gTypes@iA] (0,0.5) node[label=(90+\cmdKV@general@rotate):\cmdKV@gLabels@eLA] {} -- node[label=(\cmdKV@general@rotate):\cmdKV@gLabels@iLA] {} (0,-0.5) node[label=(-90+\cmdKV@general@rotate):\cmdKV@gLabels@eLB] {};
    \draw[fill] (0,0.5) circle (0.02);
    \draw[fill] (0,-0.5) circle (0.02);
    #2
  \end{tikzpicture}
  \gSetStdTypes{line}
}

\def\gEFTV{\@ifnextchar[\@gEFTV{\@gEFTV[]}}
\def\@gEFTV[#1]#2{%
  \setkeys*{pre}{#1}%
  \setrmkeys{general,gTypes,gLabels}%
  \begin{tikzpicture}
    [line width=1pt,
    baseline={([yshift=-0.5ex+\cmdKV@general@yshift]current bounding box.center)},
    scale=\cmdKV@general@scale,
    rotate=\cmdKV@general@rotate,
    font=\cmdKV@general@font]
    \draw[\cmdKV@gTypes@iA] (0,0.5) node[label=(90+\cmdKV@general@rotate):\cmdKV@gLabels@eLA] {} -- node[label=(200+\cmdKV@general@rotate):\cmdKV@gLabels@iLA] {} (0.3,-0.5) node[label=(-90+\cmdKV@general@rotate):\cmdKV@gLabels@eLC] {};
    \draw[\cmdKV@gTypes@iB] (0.6,0.5) node[label=(90+\cmdKV@general@rotate):\cmdKV@gLabels@eLB] {} -- node[label=(-20+\cmdKV@general@rotate):\cmdKV@gLabels@iLB] {} (0.3,-0.5);
    \draw[fill] (0,0.5) circle (0.02);
    \draw[fill] (0.3,-0.5) circle (0.02);
    \draw[fill] (0.6,0.5) circle (0.02);
    #2
  \end{tikzpicture}
  \gSetStdTypes{line}
}

\def\gEFTY{\@ifnextchar[\@gEFTY{\@gEFTY[]}}
\def\@gEFTY[#1]#2{%
  \setkeys*{pre}{#1}%
  \setrmkeys{general,gTypes,gLabels}%
  \begin{tikzpicture}
    [line width=1pt,
    baseline={([yshift=-0.5ex+\cmdKV@general@yshift]current bounding box.center)},
    scale=\cmdKV@general@scale,
    rotate=\cmdKV@general@rotate,
    font=\cmdKV@general@font]
    \draw[\cmdKV@gTypes@iA] (0,0.5) node[label=(90+\cmdKV@general@rotate):\cmdKV@gLabels@eLA] {} -- node[label=(225+\cmdKV@general@rotate):\cmdKV@gLabels@iLA] {} (0.3,0);
    \draw[\cmdKV@gTypes@iB] (0.6,0.5) node[label=(90+\cmdKV@general@rotate):\cmdKV@gLabels@eLB] {} -- node[label=(-45+\cmdKV@general@rotate):\cmdKV@gLabels@iLB] {} (0.3,0);
    \draw[\cmdKV@gTypes@iC] (0.3,0) -- node[label=(\cmdKV@general@rotate):\cmdKV@gLabels@iLC] {} (0.3,-0.5) node[label=(-90+\cmdKV@general@rotate):\cmdKV@gLabels@eLC] {};
    \draw[fill] (0,0.5) circle (0.02);
    \draw[fill] (0.6,0.5) circle (0.02);
    \draw[fill] (0.3,-0.5) circle (0.02);
    #2
  \end{tikzpicture}
  \gSetStdTypes{line}
}

\def\gEFTPsi{\@ifnextchar[\@gEFTPsi{\@gEFTPsi[]}}
\def\@gEFTPsi[#1]#2{%
  \setkeys*{pre}{#1}%
  \setrmkeys{general,gTypes,gLabels}%
  \begin{tikzpicture}
    [line width=1pt,
    baseline={([yshift=-0.5ex+\cmdKV@general@yshift]current bounding box.center)},
    scale=\cmdKV@general@scale,
    rotate=\cmdKV@general@rotate,
    font=\cmdKV@general@font]
    \draw[\cmdKV@gTypes@iA] (0,0.5) node[label=(90+\cmdKV@general@rotate):\cmdKV@gLabels@eLA] {} -- node[label=(225+\cmdKV@general@rotate):\cmdKV@gLabels@iLA] {} (0.5,0);
    \draw[\cmdKV@gTypes@iB] (0.5,0.5) node[label=(90+\cmdKV@general@rotate):\cmdKV@gLabels@eLB] {} -- node[label=(\cmdKV@general@rotate):\cmdKV@gLabels@iLB] {} (0.5,0);
    \draw[\cmdKV@gTypes@iC] (1,0.5) node[label=(90+\cmdKV@general@rotate):\cmdKV@gLabels@eLC] {} -- node[label=(-45+\cmdKV@general@rotate):\cmdKV@gLabels@iLC] {} (0.5,0);
    \draw[\cmdKV@gTypes@iD] (0.5,0) -- node[label=(\cmdKV@general@rotate):\cmdKV@gLabels@iLD] {} (0.5,-0.5) node[label=(-90+\cmdKV@general@rotate):\cmdKV@gLabels@eLD] {};
    \draw[fill] (0,0.5) circle (0.02);
    \draw[fill] (0.5,0.5) circle (0.02);
    \draw[fill] (1,0.5) circle (0.02);
    \draw[fill] (0.5,-0.5) circle (0.02);
    #2
  \end{tikzpicture}
  \gSetStdTypes{line}
}

\def\gEFTVinY{\@ifnextchar[\@gEFTVinY{\@gEFTVinY[]}}
\def\@gEFTVinY[#1]#2{%
  \setkeys*{pre}{#1}%
  \setrmkeys{general,gTypes,gLabels}%
  \begin{tikzpicture}
    [line width=1pt,
    baseline={([yshift=-0.5ex+\cmdKV@general@yshift]current bounding box.center)},
    scale=\cmdKV@general@scale,
    rotate=\cmdKV@general@rotate,
    font=\cmdKV@general@font]
    \draw[\cmdKV@gTypes@iA] (0,0.5) node[label=(90+\cmdKV@general@rotate):\cmdKV@gLabels@eLA] {} -- node[label=(225+\cmdKV@general@rotate):\cmdKV@gLabels@iLA] {} (0.25,0.25);
    \draw[\cmdKV@gTypes@iE] (0.25,0.25) -- node[label=(225+\cmdKV@general@rotate):\cmdKV@gLabels@iLE] {} (0.5,0);
    \draw[\cmdKV@gTypes@iB] (0.5,0.5) node[label=(90+\cmdKV@general@rotate):\cmdKV@gLabels@eLB] {} -- node[label=(\cmdKV@general@rotate):\cmdKV@gLabels@iLB] {} (0.25,0.25);
    \draw[\cmdKV@gTypes@iC] (1,0.5) node[label=(90+\cmdKV@general@rotate):\cmdKV@gLabels@eLC] {} -- node[label=(-45+\cmdKV@general@rotate):\cmdKV@gLabels@iLC] {} (0.5,0);
    \draw[\cmdKV@gTypes@iD] (0.5,0) -- node[label=(\cmdKV@general@rotate):\cmdKV@gLabels@iLD] {} (0.5,-0.5) node[label=(-90+\cmdKV@general@rotate):\cmdKV@gLabels@eLD] {};
    \draw[fill] (0,0.5) circle (0.02);
    \draw[fill] (0.5,0.5) circle (0.02);
    \draw[fill] (1,0.5) circle (0.02);
    \draw[fill] (0.5,-0.5) circle (0.02);
    #2
  \end{tikzpicture}
  \gSetStdTypes{line}
}

\def\gEFTYinvY{\@ifnextchar[\@gEFTYinvY{\@gEFTYinvY[]}}
\def\@gEFTYinvY[#1]#2{%
  \setkeys*{pre}{#1}%
  \setrmkeys{general,gTypes,gLabels}%
  \begin{tikzpicture}
    [line width=1pt,
    baseline={([yshift=-0.5ex+\cmdKV@general@yshift]current bounding box.center)},
    scale=\cmdKV@general@scale,
    rotate=\cmdKV@general@rotate,
    font=\cmdKV@general@font]
    \draw[\cmdKV@gTypes@iA] (0,0.5) node[label=(90+\cmdKV@general@rotate):\cmdKV@gLabels@eLA] {} -- node[label=(225+\cmdKV@general@rotate):\cmdKV@gLabels@iLA] {} (0.3,0.166);
    \draw[\cmdKV@gTypes@iB] (0.6,0.5) node[label=(90+\cmdKV@general@rotate):\cmdKV@gLabels@eLB] {} -- node[label=(-45+\cmdKV@general@rotate):\cmdKV@gLabels@iLB] {} (0.3,0.166);
    \draw[\cmdKV@gTypes@iC] (0.3,-0.166) -- node[label=(135+\cmdKV@general@rotate):\cmdKV@gLabels@iLC] {} (0,-0.5) node[label=(-90+\cmdKV@general@rotate):\cmdKV@gLabels@eLC] {}; 
    \draw[\cmdKV@gTypes@iD] (0.3,-0.166) -- node[label=(45+\cmdKV@general@rotate):\cmdKV@gLabels@iLD] {} (0.6,-0.5) node[label=(-90+\cmdKV@general@rotate):\cmdKV@gLabels@eLD] {};
    \draw[\cmdKV@gTypes@iE] (0.3,0.166) -- node[label=(\cmdKV@general@rotate):\cmdKV@gLabels@iLE] {} (0.3,-0.166);
    \draw[fill] (0,0.5) circle (0.02);
    \draw[fill] (0.6,0.5) circle (0.02);
    \draw[fill] (0,-0.5) circle (0.02);
    \draw[fill] (0.6,-0.5) circle (0.02);
    #2
  \end{tikzpicture}
  \gSetStdTypes{line}
}

\def\gEFTX{\@ifnextchar[\@gEFTX{\@gEFTX[]}}
\def\@gEFTX[#1]#2{%
  \setkeys*{pre}{#1}%
  \setrmkeys{general,gTypes,gLabels}%
  \begin{tikzpicture}
    [line width=1pt,
    baseline={([yshift=-0.5ex+\cmdKV@general@yshift]current bounding box.center)},
    scale=\cmdKV@general@scale,
    rotate=\cmdKV@general@rotate,
    font=\cmdKV@general@font]
    \draw[\cmdKV@gTypes@iA] (0,0.5) node[label=(90+\cmdKV@general@rotate):\cmdKV@gLabels@eLA] {} -- node[label=(225+\cmdKV@general@rotate):\cmdKV@gLabels@iLA] {} (0.3,0);
    \draw[\cmdKV@gTypes@iB] (0.6,0.5) node[label=(90+\cmdKV@general@rotate):\cmdKV@gLabels@eLB] {} -- node[label=(-45+\cmdKV@general@rotate):\cmdKV@gLabels@iLB] {} (0.3,0);
    \draw[\cmdKV@gTypes@iC] (0.3,0) -- node[label=(135+\cmdKV@general@rotate):\cmdKV@gLabels@iLC] {} (0,-0.5) node[label=(-90+\cmdKV@general@rotate):\cmdKV@gLabels@eLC] {}; 
    \draw[\cmdKV@gTypes@iD] (0.3,0) -- node[label=(45+\cmdKV@general@rotate):\cmdKV@gLabels@iLD] {} (0.6,-0.5) node[label=(-90+\cmdKV@general@rotate):\cmdKV@gLabels@eLD] {};
    \draw[fill] (0,0.5) circle (0.02);
    \draw[fill] (0.6,0.5) circle (0.02);
    \draw[fill] (0,-0.5) circle (0.02);
    \draw[fill] (0.6,-0.5) circle (0.02);
    #2
  \end{tikzpicture}
  \gSetStdTypes{line}
}

\def\gEFTH{\@ifnextchar[\@gEFTH{\@gEFTH[]}}
\def\@gEFTH[#1]#2{%
  \setkeys*{pre}{#1}%
  \setrmkeys{general,gTypes,gLabels}%
  \begin{tikzpicture}
    [line width=1pt,
    baseline={([yshift=-0.5ex+\cmdKV@general@yshift]current bounding box.center)},
    scale=\cmdKV@general@scale,
    rotate=\cmdKV@general@rotate,
    font=\cmdKV@general@font]
    \draw[\cmdKV@gTypes@iA] (0,0.5) node[label=(90+\cmdKV@general@rotate):\cmdKV@gLabels@eLA] {} -- node[label=(180+\cmdKV@general@rotate):\cmdKV@gLabels@iLA] {} (0,0);
    \draw[\cmdKV@gTypes@iB] (0.6,0.5) node[label=(90+\cmdKV@general@rotate):\cmdKV@gLabels@eLB] {} -- node[label=(\cmdKV@general@rotate):\cmdKV@gLabels@iLB] {} (0.6,0);
    \draw[\cmdKV@gTypes@iC] (0.0,0) -- node[label=(180+\cmdKV@general@rotate):\cmdKV@gLabels@iLC] {} (0,-0.5) node[label=(-90+\cmdKV@general@rotate):\cmdKV@gLabels@eLC] {}; 
    \draw[\cmdKV@gTypes@iD] (0.6,0) -- node[label=(\cmdKV@general@rotate):\cmdKV@gLabels@iLD] {} (0.6,-0.5) node[label=(-90+\cmdKV@general@rotate):\cmdKV@gLabels@eLD] {};
    \draw[\cmdKV@gTypes@iE] (0,0) -- node[label=(90+\cmdKV@general@rotate):\cmdKV@gLabels@iLE] {} (0.6,0);
    \draw[fill] (0,0.5) circle (0.02);
    \draw[fill] (0.6,0.5) circle (0.02);
    \draw[fill] (0,-0.5) circle (0.02);
    \draw[fill] (0.6,-0.5) circle (0.02);
    #2
  \end{tikzpicture}
  \gSetStdTypes{line}
}

\def\cTree{\@ifnextchar[\@cTree{\@cTree[]}}
\def\@cTree[#1]#2{%
  \setkeys*{pre}{#1}%
  \setrmkeys{general,gTypes,gLabels,blobs}%
  \begin{tikzpicture}
    [line width=1pt,
    baseline={([yshift=-0.5ex+\cmdKV@general@yshift]current bounding box.center)},
    scale=\cmdKV@general@scale,
    rotate=\cmdKV@general@rotate,
    font=\cmdKV@general@font]
    \fill[\cmdKV@blobs@blobA] (0,0) ellipse (0.25 and 0.25);
    \draw[\cmdKV@gTypes@eA] ($ (0,0) + (-135:0.25) $) -- ($ (0,0) + (-135:0.6) $) node[label=(-180+\cmdKV@general@rotate):\cmdKV@gLabels@eLA] {};
    \draw[\cmdKV@gTypes@eB] ($ (0,0) + (135:0.25) $) -- ($ (0,0) + (135:0.6) $) node[label=(180+\cmdKV@general@rotate):\cmdKV@gLabels@eLB] {};
    \draw[\cmdKV@gTypes@eC] ($ (0,0) + (45:0.25) $) -- ($ (0,0) + (45:0.6) $) node[label=(\cmdKV@general@rotate):\cmdKV@gLabels@eLC] {};
    \draw[\cmdKV@gTypes@eD] ($ (0,0) + (-45:0.25) $) -- ($ (0,0) + (-45:0.6) $) node[label=(\cmdKV@general@rotate):\cmdKV@gLabels@eLD] {};
    \node[label=(90+\cmdKV@general@rotate):\cmdKV@gLabels@iLA] at (0,0.3) {};
    \node[label=(\cmdKV@general@rotate):\cmdKV@gLabels@iLB] at (0.3,0) {};
    #2
  \end{tikzpicture}
  \gSetStdTypes{line}
}

\def\cBoxA{\@ifnextchar[\@cBoxA{\@cBoxA[]}}
\def\@cBoxA[#1]#2{%
  \setkeys*{pre}{#1}%
  \setrmkeys{general,gTypes,gLabels,dots,blobs}%
  \begin{tikzpicture}
    [line width=1pt,
    baseline={([yshift=-0.5ex+\cmdKV@general@yshift]current bounding box.center)},
    scale=\cmdKV@general@scale,
    rotate=\cmdKV@general@rotate,
    font=\cmdKV@general@font]
    \fill[\cmdKV@blobs@blobA] (-0.4,0) ellipse (0.2 and 0.4);
    \fill[\cmdKV@blobs@blobB] (0.4,0) ellipse (0.2 and 0.4);
    \draw[\cmdKV@gTypes@eA] ($ (-0.4,0) + (-135:0.26) $) -- ($ (-0.4,0) + (-135:0.7) $) node[label=(-180+\cmdKV@general@rotate):\cmdKV@gLabels@eLA] {};
    \draw[\cmdKV@gTypes@eB] ($ (-0.4,0) + (135:0.26) $) -- ($ (-0.4,0) + (135:0.7) $) node[label=(180+\cmdKV@general@rotate):\cmdKV@gLabels@eLB] {};
    \draw[\cmdKV@gTypes@eC] ($ (0.4,0) + (45:0.26) $) -- ($ (0.4,0) + (45:0.7) $) node[label=(\cmdKV@general@rotate):\cmdKV@gLabels@eLC] {};
    \draw[\cmdKV@gTypes@eD] ($ (0.4,0) + (-45:0.26) $) -- ($ (0.4,0) + (-45:0.7) $) node[label=(\cmdKV@general@rotate):\cmdKV@gLabels@eLD] {};
    \draw[label distance=1,\cmdKV@gTypes@iA] ($ (-0.4,0) + (45:0.26) $) -- node[label=(90+\cmdKV@general@rotate):\cmdKV@gLabels@iLA] {}($ (0.4,0) + (135:0.26) $);
    \draw[label distance=1,\cmdKV@gTypes@iB] ($ (0.4,0) + (-135:0.26) $) -- node[label=(-90+\cmdKV@general@rotate):\cmdKV@gLabels@iLB] {}($ (-0.4,0) + (-45:0.26) $);
    \node[label=(180+\cmdKV@general@rotate):\cmdKV@gLabels@iLC] at (-0.6,0) {};
    \node[label=(\cmdKV@general@rotate):\cmdKV@gLabels@iLD] at (0.6,0) {};
    \ifKV@dots@lDots
      \draw[dotted] ($(-0.4,0) + (-150:0.7)$) to[bend left] ($(-0.4,0) + (150:0.7)$);%
    \fi
    \ifKV@dots@rDots
      \draw[dotted] ($(0.4,0) + (30:0.7)$) to[bend left] ($(0.4,0) + (-30:0.7)$);%
    \fi
    #2
  \end{tikzpicture}
  \gSetStdTypes{line}
}

\def\cBoxAA{\@ifnextchar[\@cBoxAA{\@cBoxAA[]}}
\def\@cBoxAA[#1]#2{%
  \setkeys*{pre}{#1}%
  \setrmkeys{general,gTypes,gLabels,dots,blobs}%
  \begin{tikzpicture}
    [line width=1pt,
    baseline={([yshift=-0.5ex+\cmdKV@general@yshift]current bounding box.center)},
    scale=\cmdKV@general@scale,
    rotate=\cmdKV@general@rotate,
    font=\cmdKV@general@font]
    \fill[\cmdKV@blobs@blobA] (-0.4,0) ellipse (0.2 and 0.4);
    \fill[\cmdKV@blobs@blobB] (0.4,0) ellipse (0.2 and 0.4);
    \draw[\cmdKV@gTypes@eA] ($ (-0.4,0) + (-135:0.26) $) -- ($ (-0.4,0) + (-135:0.7) $) node[label=(-180+\cmdKV@general@rotate):\cmdKV@gLabels@eLA] {};
    \draw[\cmdKV@gTypes@eB] ($ (-0.4,0) + (135:0.26) $) -- ($ (-0.4,0) + (135:0.7) $) node[label=(180+\cmdKV@general@rotate):\cmdKV@gLabels@eLB] {};
    \draw[\cmdKV@gTypes@eC] ($ (0.4,0) + (45:0.26) $) -- ($ (0.4,0) + (45:0.7) $) node[label=(\cmdKV@general@rotate):\cmdKV@gLabels@eLC] {};
    \draw[\cmdKV@gTypes@eD] ($ (0.4,0) + (-45:0.26) $) -- ($ (0.4,0) + (-45:0.7) $) node[label=(\cmdKV@general@rotate):\cmdKV@gLabels@eLD] {};
    \draw[label distance=1,\cmdKV@gTypes@iA] ($ (-0.4,0) + (60:0.32) $) to[bend left] node[label=(90+\cmdKV@general@rotate):\cmdKV@gLabels@iLA] {}($ (0.4,0) + (120:0.32) $);
    \draw[label distance=1,\cmdKV@gTypes@iB] ($ (0.4,0) + (-120:0.32) $) to[bend left]  node[label=(-90+\cmdKV@general@rotate):\cmdKV@gLabels@iLB] {}($ (-0.4,0) + (-60:0.32) $);
    \draw[dotted] (0,0.25) -- (0,-0.25);
    \node[label=(180+\cmdKV@general@rotate):\cmdKV@gLabels@iLC] at (-0.6,0) {};
    \node[label=(\cmdKV@general@rotate):\cmdKV@gLabels@iLD] at (0.6,0) {};
    \ifKV@dots@lDots
      \draw[dotted] ($(-0.4,0) + (-150:0.7)$) to[bend left] ($(-0.4,0) + (150:0.7)$);%
    \fi
    \ifKV@dots@rDots
      \draw[dotted] ($(0.4,0) + (30:0.7)$) to[bend left] ($(0.4,0) + (-30:0.7)$);%
    \fi
    #2
  \end{tikzpicture}
  \gSetStdTypes{line}
}

\def\cBoxB{\@ifnextchar[\@cBoxB{\@cBoxB[]}}
\def\@cBoxB[#1]#2{%
  \setkeys*{pre}{#1}%
  \setrmkeys{general,gTypes,gLabels,blobs}%
  \begin{tikzpicture}
    [line width=1pt,
    baseline={([yshift=-0.5ex+\cmdKV@general@yshift]current bounding box.center)},
    scale=\cmdKV@general@scale,
    rotate=\cmdKV@general@rotate,
    font=\cmdKV@general@font]
    \fill[\cmdKV@blobs@blobA] (0,-0.3) ellipse (0.4 and 0.2);
    \fill[\cmdKV@blobs@blobB] (0,0.3) ellipse (0.4 and 0.2);
    \draw[\cmdKV@gTypes@eA] ($ (0,-0.3) + (180:0.4) $) -- ($ (0,-0.3) + (-170:0.7) $) node[label=(180+\cmdKV@general@rotate):\cmdKV@gLabels@eLA] {};
    \draw[\cmdKV@gTypes@eB] ($ (0,0.3) + (180:0.4) $) -- ($ (0,0.3) + (170:0.7) $) node[label=(180+\cmdKV@general@rotate):\cmdKV@gLabels@eLB] {};
    \draw[\cmdKV@gTypes@eC] ($ (0,0.3) + (0:0.4) $) -- ($ (0,0.3) + (10:0.7) $) node[label=(\cmdKV@general@rotate):\cmdKV@gLabels@eLC] {};
    \draw[\cmdKV@gTypes@eD] ($ (0,-0.3) + (0:0.4) $) -- ($ (0,-0.3) + (-10:0.7) $) node[label=(\cmdKV@general@rotate):\cmdKV@gLabels@eLD] {};
    \draw[label distance=1,\cmdKV@gTypes@iA] ($ (0,-0.3) + (135:0.26) $) -- node[label=(-180+\cmdKV@general@rotate):\cmdKV@gLabels@iLA] {}($ (0,0.3) + (-135:0.26) $);
    \draw[label distance=1,\cmdKV@gTypes@iB] ($ (0,0.3) + (-45:0.26) $) -- node[label=(\cmdKV@general@rotate):\cmdKV@gLabels@iLB] {}($ (0,-0.3) + (45:0.26) $);
    \node[label=(90+\cmdKV@general@rotate):\cmdKV@gLabels@iLC] at (0,0.5) {};
    \node[label=(-90+\cmdKV@general@rotate):\cmdKV@gLabels@iLD] at (0,-0.5) {};
    #2
  \end{tikzpicture}
  \gSetStdTypes{line}
}

\def\cBoxBoxA{\@ifnextchar[\@cBoxBoxA{\@cBoxBoxA[]}}
\def\@cBoxBoxA[#1]#2{%
  \setkeys*{pre}{#1}%
  \setrmkeys{general,gTypes,gLabels,blobs}%
  \begin{tikzpicture}
    [line width=1pt,
    baseline={([yshift=-0.5ex+\cmdKV@general@yshift]current bounding box.center)},
    scale=\cmdKV@general@scale,
    rotate=\cmdKV@general@rotate,
    font=\cmdKV@general@font]
    \fill[\cmdKV@blobs@blobA] (-0.4,0) ellipse (0.2 and 0.4);
    \fill[\cmdKV@blobs@blobB] (0.4,0) ellipse (0.2 and 0.4);
    \fill[\cmdKV@blobs@blobC] (1.2,0) ellipse (0.2 and 0.4);
    \draw[\cmdKV@gTypes@eA] ($ (-0.4,0) + (-135:0.26) $) -- ($ (-0.4,0) + (-135:0.7) $) node[label=(-180+\cmdKV@general@rotate):\cmdKV@gLabels@eLA] {};
    \draw[\cmdKV@gTypes@eB] ($ (-0.4,0) + (135:0.26) $) -- ($ (-0.4,0) + (135:0.7) $) node[label=(180+\cmdKV@general@rotate):\cmdKV@gLabels@eLB] {};
    \draw[\cmdKV@gTypes@eC] ($ (1.2,0) + (45:0.26) $) -- ($ (1.2,0) + (45:0.7) $) node[label=(\cmdKV@general@rotate):\cmdKV@gLabels@eLC] {};
    \draw[\cmdKV@gTypes@eD] ($ (1.2,0) + (-45:0.26) $) -- ($ (1.2,0) + (-45:0.7) $) node[label=(\cmdKV@general@rotate):\cmdKV@gLabels@eLD] {};
    \draw[label distance=1,\cmdKV@gTypes@iA] ($ (-0.4,0) + (45:0.26) $) -- node[label=(90+\cmdKV@general@rotate):\cmdKV@gLabels@iLA] {}($ (0.4,0) + (135:0.26) $);
    \draw[label distance=1,\cmdKV@gTypes@iB] ($ (0.4,0) + (45:0.26) $) -- node[label=(90+\cmdKV@general@rotate):\cmdKV@gLabels@iLB] {}($ (1.2,0) + (135:0.26) $);
    \draw[label distance=1,\cmdKV@gTypes@iC] ($ (1.2,0) + (-135:0.26) $) -- node[label=(-90+\cmdKV@general@rotate):\cmdKV@gLabels@iLC] {}($ (0.4,0) + (-45:0.26) $);
    \draw[label distance=1,\cmdKV@gTypes@iD] ($ (0.4,0) + (-135:0.26) $) -- node[label=(-90+\cmdKV@general@rotate):\cmdKV@gLabels@iLD] {}($ (-0.4,0) + (-45:0.26) $);
    \node[label=(180+\cmdKV@general@rotate):\cmdKV@gLabels@iLE] at (-0.6,0) {};
    \node[label=(\cmdKV@general@rotate):\cmdKV@gLabels@iLF] at (0.6,0) {};
    \node[label=(\cmdKV@general@rotate):\cmdKV@gLabels@iLG] at (1.4,0) {};
    #2
  \end{tikzpicture}
  \gSetStdTypes{line}
}

\def\cBoxBoxB{\@ifnextchar[\@cBoxBoxB{\@cBoxBoxB[]}}
\def\@cBoxBoxB[#1]#2{%
  \setkeys*{pre}{#1}%
  \setrmkeys{general,gTypes,gLabels}%
  \begin{tikzpicture}
    [line width=1pt,
    baseline={([yshift=-0.5ex+\cmdKV@general@yshift]current bounding box.center)},
    scale=\cmdKV@general@scale,
    rotate=\cmdKV@general@rotate,
    font=\cmdKV@general@font]
    \fill[\cmdKV@blobs@blobA] (0,0.3) ellipse (0.4 and 0.2);
    \fill[\cmdKV@blobs@blobB] (0,-0.3) ellipse (0.4 and 0.2);
    \fill[\cmdKV@blobs@blobC] (1,0) ellipse (0.2 and 0.4);
    \draw[\cmdKV@gTypes@eA] ($ (0,-0.3) + (180:0.4) $) -- ($ (0,-0.3) + (-170:0.7) $) node[label=(180+\cmdKV@general@rotate):\cmdKV@gLabels@eLA] {};
    \draw[\cmdKV@gTypes@eB] ($ (0,0.3) + (180:0.4) $) -- ($ (0,0.3) + (170:0.7) $) node[label=(180+\cmdKV@general@rotate):\cmdKV@gLabels@eLB] {};
    \draw[\cmdKV@gTypes@eC] ($ (1,0) + (45:0.26) $) -- ($ (1,0) + (45:0.7) $) node[label=(\cmdKV@general@rotate):\cmdKV@gLabels@eLC] {};
    \draw[\cmdKV@gTypes@eD] ($ (1,0) + (-45:0.26) $) -- ($ (1,0) + (-45:0.7) $) node[label=(\cmdKV@general@rotate):\cmdKV@gLabels@eLD] {};
    \draw[label distance=1,\cmdKV@gTypes@iA] ($ (0,-0.3) + (135:0.26) $) -- node[label=(-180+\cmdKV@general@rotate):\cmdKV@gLabels@iLA] {}($ (0,0.3) + (-135:0.26) $);
    \draw[label distance=1,\cmdKV@gTypes@iB] ($ (0,0.3) + (0:0.4) $) -- node[label=(90+\cmdKV@general@rotate):\cmdKV@gLabels@iLB] {} ($ (1,0) + (135:0.26) $);
    \draw[label distance=1,\cmdKV@gTypes@iC] ($ (1,0) + (-135:0.26) $) -- node[label=(-90+\cmdKV@general@rotate):\cmdKV@gLabels@iLC] {} ($ (0,-0.3) + (0:0.4) $);
    \draw[label distance=1,\cmdKV@gTypes@iD] ($ (0,0.3) + (-45:0.26) $) -- node[label=(\cmdKV@general@rotate):\cmdKV@gLabels@iLD] {}($ (0,-0.3) + (45:0.26) $);
    \node[label=(90+\cmdKV@general@rotate):\cmdKV@gLabels@iLE] at (0,0.5) {};
    \node[label=(\cmdKV@general@rotate):\cmdKV@gLabels@iLF] at (1.2,0) {};
    \node[label=(-90+\cmdKV@general@rotate):\cmdKV@gLabels@iLG] at (0,-0.5) {};
    #2
  \end{tikzpicture}
  \gSetStdTypes{line}
}

\def\cNPBox{\@ifnextchar[\@cNPBox{\@cNPBox[]}}
\def\@cNPBox[#1]#2{%
  \setkeys*{pre}{#1}%
  \setrmkeys{general,gTypes,gLabels,blob}%
  \begin{tikzpicture}
    [line width=1pt,
    baseline={([yshift=-0.5ex+\cmdKV@general@yshift]current bounding box.center)},
    scale=\cmdKV@general@scale,
    rotate=\cmdKV@general@rotate,
    font=\cmdKV@general@font]
    \fill[\cmdKV@blobs@blobA] (-0.6,0) ellipse (0.2 and 0.4);
    \fill[\cmdKV@blobs@blobB] (0.4,0) ellipse (0.2 and 0.4);
    \draw[\cmdKV@gTypes@eA] ($ (-0.6,0) + (-180:0.2) $) -- ($ (-0.6,0) + (-180:0.6) $) node[label=(-180+\cmdKV@general@rotate):\cmdKV@gLabels@eLA] {};
    \draw[\cmdKV@gTypes@eB] ($ (-0.6,0) + (0:0.2) $) -- (-0.18,0) node[label=(\cmdKV@general@rotate):\cmdKV@gLabels@eLB] {};
    \draw[\cmdKV@gTypes@eC] ($ (0.4,0) + (45:0.26) $) -- ($ (0.4,0) + (45:0.7) $) node[label=(\cmdKV@general@rotate):\cmdKV@gLabels@eLC] {};
    \draw[\cmdKV@gTypes@eD] ($ (0.4,0) + (-45:0.26) $) -- ($ (0.4,0) + (-45:0.7) $) node[label=(\cmdKV@general@rotate):\cmdKV@gLabels@eLD] {};
    \draw[label distance=1,\cmdKV@gTypes@iA] ($ (-0.6,0) + (60:0.3) $) -- node[label=(90+\cmdKV@general@rotate):\cmdKV@gLabels@iLA] {}($ (0.4,0) + (135:0.26) $);
    \draw[label distance=1,\cmdKV@gTypes@iB] ($ (0.4,0) + (-135:0.26) $) -- node[label=(-90+\cmdKV@general@rotate):\cmdKV@gLabels@iLB] {}($ (-0.6,0) + (-60:0.3) $);
    \node[label=(\cmdKV@general@rotate):\cmdKV@gLabels@iLD] at (0.6,0) {};
    #2
  \end{tikzpicture}
  \gSetStdTypes{line}
}

\def\cFourPtRec{\@ifnextchar[\@cFourPtRec{\@cFourPtRec[]}}
\def\@cFourPtRec[#1]#2{%
  \setkeys*{pre}{#1}%
  \setrmkeys{general,gTypes,gLabels,blobs}%
  \begin{tikzpicture}
    [line width=1pt,
    baseline={([yshift=-0.5ex+\cmdKV@general@yshift]current bounding box.center)},
    scale=\cmdKV@general@scale,
    rotate=\cmdKV@general@rotate,
    font=\cmdKV@general@font]
    \fill[\cmdKV@blobs@blobA] (-0.5,0) ellipse (0.2 and 0.3);
    \fill[\cmdKV@blobs@blobB] (0.5,0) ellipse (0.3 and 0.4);
    \filldraw[fill=white,draw=black] (0.55,0.15) circle (0.1);
    \filldraw[fill=white,draw=black] (0.45,-0.1) circle (0.1);
    \draw[\cmdKV@gTypes@eA] ($(-0.5,0) + (135:0.23)$) -- ($(-0.5,0) + (135:0.5)$) node[label=(180+\cmdKV@general@rotate):\cmdKV@gLabels@eLA] {};
    \draw[\cmdKV@gTypes@eB] ($(-0.5,0) + (-135:0.23)$) -- ($(-0.5,0) + (-135:0.5)$) node[label=(180+\cmdKV@general@rotate):\cmdKV@gLabels@eLB] {};
    \draw[\cmdKV@gTypes@eC] ($(0.5,0) + (45:0.35)$) -- ($(0.5,0) + (45:0.7)$) node[label=(\cmdKV@general@rotate):\cmdKV@gLabels@eLC] {};
    \draw[\cmdKV@gTypes@eD] ($(0.5,0) + (-45:0.35)$) -- ($(0.5,0) + (-45:0.7)$) node[label=(\cmdKV@general@rotate):\cmdKV@gLabels@eLD] {};
    \draw[label distance=1,\cmdKV@gTypes@iA] ($(-0.5,0) + (30:0.23)$) -- node[label=(90+\cmdKV@general@rotate):\cmdKV@gLabels@iLA] {} ($(0.5,0) + (157:0.3)$);
    \draw[label distance=1,\cmdKV@gTypes@iB] ($(0.5,0) + (-157:0.3)$) -- node[label=(-90+\cmdKV@general@rotate):\cmdKV@gLabels@iLB] {} ($(-0.5,0) + (-30:0.23)$);
    #2
 \end{tikzpicture}
  \gSetStdTypes{line}
}

\makeatother

\numberwithin{equation}{section}

\makeatletter
\g@addto@macro\bfseries{\boldmath}
\makeatother

\DeclareMathOperator{\arcsinh}{arcsinh}

\DeclareMathOperator{\Arctan}{arctan}
\DeclareMathOperator{\arccosh}{arccosh}

\DeclareMathOperator{\adj}{adj}

\DeclareMathSymbol{\shortminusN}{\mathbin}{AMSa}{"39}
\DeclareMathOperator{\shortminusS}{\shortminusN\hspace{-2pt}}
\DeclareMathOperator{\shortminus}{\hspace{-1pt}\shortminusS}

\newcommand{\nn}{\nonumber}

\newcommand\beq{\begin{equation}}
\newcommand\eeq{\end{equation}}
\newcommand{\cD}{\mathcal{D}}

\newcommand\bea{\begin{eqnarray}}
\newcommand\eea{\end{eqnarray}}

\newcommand\dd{{\mathrm d}}

\newcommand\Mp{M_{\rm Pl}}

\newcommand{\bell}{{\boldsymbol \ell}}

\newcommand{\bk}{{\boldsymbol k}}

\newcommand{\bn}{{\boldsymbol n}}
\newcommand{\bp}{{\boldsymbol p}}
\newcommand{\bP}{{\boldsymbol P}}

\newcommand{\bq}{{\boldsymbol q}}
\newcommand{\br}{{\boldsymbol r}}

\newcommand{\bx}{{\boldsymbol x}}

\newcommand{\bD}{{\boldsymbol D}}

\newcommand\cO{\mathcal{O}}
\newcommand\cM{\mathcal{M}}
\newcommand\cS{\mathcal{S}}
\newcommand\cP{\mathcal{P}}
\newcommand\cF{\mathcal{F}}
\newcommand\cU{\mathcal{U}}
\newcommand\cI{\mathcal{I}}

\newcommand\cL{\mathcal{L}}

\makeatletter
\newcommand{\Biggg}{\bBigg@{3.5}}
\makeatother

\definecolor{desycyan}{rgb}{0.00,0.68,0.93}
\definecolor{desyorange}{rgb}{0.93,0.58,0.16}
\definecolor{internationalorange}{rgb}{1.0, 0.31, 0.0}

\newcommand\cdotnew{\!\cdot\!}

\newcommand\vecbf[1]{{\boldsymbol #1}}

\tikzset{
  graviton/.style={
  	line width=0.1pt,
    color=desycyan,
    decoration={snake, segment length=0.7mm, amplitude=0.2mm},
    decorate,
  }
}
\usetikzlibrary{matrix}

\begin{document}
\preprint{\texttt{DESY\,23-041}}

\title{ Bootstrapping the relativistic two-body problem} 
\author[a]{\large Christoph Dlapa,}
\author[a]{\large Gregor K\"alin,}
\author[b]{\large Zhengwen Liu,}
\author[a,c,d]{\large and Rafael A. Porto}

\affiliation[a]{Deutsches Elektronen-Synchrotron DESY\\  Notkestr.\,85, 22607 Hamburg,\,Germany}
\affiliation[b]{Niels Bohr International Academy, Niels Bohr Institute, University of Copenhagen\\
Blegdamsvej 17, 2100 Copenhagen \O{}, Denmark}
\affiliation[c]{Instituto de F\'isica, Facultad de Ingenier\'ia, Universidad de la Rep\'ublica\\ J.H.y Reissig 565, 11000 Montevideo, Uruguay}
\affiliation[d]{Instituto de F\'isica, Facultad de Ciencias, Universidad de la Rep\'ublica\\ Igua 4225, 11400 Montevideo, Uruguay}
 
\emailAdd{christoph.dlapa@desy.de}
\emailAdd{gregor.kaelin@desy.de}
\emailAdd{zhengwen.liu@nbi.ku.dk}
\emailAdd{rafael.porto@desy.de}
\abstract{We describe the formalism to compute gravitational-wave observables for compact binaries via the effective field theory framework in combination with modern tools from collider physics. We put particular emphasis on solving the `multi-loop' integration problem via the methodology of differential equations and expansion by regions. This allows us to {\it bootstrap} the two-body relativistic dynamics in the Post-Minkowskian (PM) expansion from boundary data evaluated in the near-static ({\it soft}) limit. We illustrate the procedure with the derivation of the total spacetime impulse in the scattering of non-spinning bodies to 4PM (three-loop) order, i.e.~${\cal O}(G^4)$, including conservative and dissipative effects.}
\maketitle
\newpage

\section{Introduction} \label{sec:introduction}

The direct observation of gravitational waves (GWs) by the Ligo-Virgo-Kagra collaboration~\cite{LIGOScientific:2021djp} opens a window onto the universe that can shed light on long-standing issues in astro, particle and gravitational physics \cite{buosathya,tune,music,Maggiore:2019uih,Barausse:2020rsu,Bernitt:2022aoa}. Yet, searching for the minute imprints of the traveling ripples of spacetime on earth- and space-based detectors, and at the same time be able to unravel the nature of the sources, requires accurate waveform models. Moreover, next-generation GW observatories will be sensitive to even earlier phases of the two-body dynamics of compact objects, covering many more cycles in the detectors' band. The expected empirical reach is thus inaccessible to current numerical simulations, often limited to the near-merger phase of comparable masses and slowly spinning bodies~\cite{Ajith:2012az,Szilagyi:2015rwa,Dietrich:2018phi}. Therefore, analytic techniques~\cite{Damour:2008yg,blanchet,Schafer:2018kuf,Barack:2018yvs,walterLH,iragrg,foffa,review,Goldberger:2022ebt,Buonanno:2022pgc} remain an essential tool toward the construction of high-precision template banks for GW searches and parameter estimation with present \cite{LIGOScientific:2021djp} and---more critically---future networks of GW interferometers such as the Laser interferometer Space Antenna (LISA) \cite{LISA}, the Cosmic Explorer \cite{CE} and the Einstein Telescope (ET) \cite{ET}.\vskip 4pt For many years the weapon of choice to study the binary problem has been the Post-Newtonian (PN) expansion in small velocities and weak fields, following either traditional techniques in general relativity \cite{Damour:2008yg,blanchet,Schafer:2018kuf}, modern effective field theory (EFT) methods \cite{walterLH,iragrg,foffa,review,Goldberger:2022ebt}, or combinations thereof  (see e.g.\,\cite{4pndjs,4pnbla,4pnbla2,Marchand:2017pir,damour3n,binidam1,binidam2,binit,Bini:2021gat,Bini:2021qvf,Bini:2022yrk,Bini:2022xpp,Bini:2022enm,Damour:2020tta,Marchand:2020fpt,Larrouturou:2021dma,Larrouturou:2021gqo,nrgr,nrgrs,dis1,dis2,prl,nrgrss,nrgrs2,nrgrso,andirad,andirad2,amps,srad,chadRR,chadbr2,tail,natalia1,natalia2,dis3,withchad,apparent,nrgr4pn1,nrgr4pn2,5pn1,5pn2,hered1,hered2,tail3,blum,blum2,Blumlein:2020pyo,Blumlein:2021txe,radnrgr,pardo,Almeida:2022jrv,Cho:2021mqw,Cho2022,Cho:1,Cho:2,Kim:2022bwv,Kim:2022pou,Mandal:2022nty,Mandal:2022ufb,Damour:2022ybd}). More recently, motivated by the effective-one-body (EOB) formalism \cite{damour1,damour2,Damour:2019lcq} and the boundary-to-bound (B2B) dictionary \cite{paper1,paper2,b2b3}, the study of scattering processes within the Post-Minkowskian (PM) approximation, featuring an expansion in Newton's constant but to all orders in the velocity, has experienced a {\it renaissance}, rapidly achieving new results using both EFT-based, e.g.~\cite{pmeft,3pmeft,tidaleft,pmefts,eftrad,4pmeft,4pmeft2,4pmeftot,janmogul,janmogul2,Jakobsen:2021zvh,Jakobsen:2022fcj,Mougiakakos:2021ckm,Riva:2021vnj,Mougiakakos:2022sic,Riva:2022fru,Jakobsen:2022psy,Jinno:2022sbr,Damgaard:2019lfh,Brandhuber:2023hhy}, as well as amplitude-based approaches, e.g.~\cite{ira1,Vaidya:2014kza,Walter,Goldberger:2016iau,cheung,bohr,Guevara:2018wpp,cristof1,donal,donalvines,zvi1,Haddad:2020que,Aoude:2022thd,Bjerrum-Bohr:2021din,andres2,4pmzvi,4pmzvi2,Gabriele,Gabriele2,Parra,Parra3,Brandhuber:2021eyq,Cristofoli:2021vyo,FebresCordero:2022jts,Manohar:2022dea,DiVecchia:2022nna}. Notably, using the EFT methodology introduced in \cite{pmeft,eftrad}, the total change of the relativistic momentum (a.k.a.~the impulse), total radiated energy and GW flux, have been obtained to the fourth PM (4PM) order for non-spinning bodies~\cite{4pmeftot}, with several aspects of the complete results independently confirmed by partial calculations in the PN \cite{Bini:2021gat,Bini:2022enm} and PM  \cite{4pmeft,4pmeft2,4pmzvi,4pmzvi2,Manohar:2022dea,DiVecchia:2022nna} literature, as well as numerical simulations~\cite{Damour:2022ybd}. The purpose of this paper is therefore  to elaborate on the various technical details behind the derivations in  \cite{4pmeft,4pmeft2,4pmeftot}, in particular regarding the integration problem.\vskip 4pt 

When dealing with dissipative systems, the field equations must be solved using causality-preserving (retarded) Green's functions. As emphasized in \cite{chadRR}, this is implemented through the  Schwinger-Keldysh (``in-in") formalism. The in-in framework also allows for a natural identification of conservative and dissipative effects \cite{chadprl,eftrad}. The former are obtained via the more standard ``in-out" approach, using Feynman propagators, and retaining the real part of the answer \cite{4pmeft,4pmeft2}; whereas the latter is derived either by subtracting the conservative part from the full in-in solution with retarded propagators~\cite{4pmeftot}, or by explicitly accounting for the mismatch between Green's functions~\cite{eftrad}. After the dust settles, the binary problem transforms into computing a series of `multi-loop'-type integrals---featuring Feynman and retarded propagators---similar to those in collider physics, even though in a purely classical~setting.\vskip 4pt 

As in other paradigmatic examples in particle physics \cite{iraTasi}, it is often useful to split the integration problem into {\it regions} \cite{beneke}. For the two-body problem we encounter potential (off-shell) and radiation (on-shell) modes \cite{nrgr}. The resulting {\it mode factorization} can be done either at the level of the integrals, by performing an asymptotic expansion in momentum space, or directly separating modes of the gravitational field~\cite{iraTasi}. We can then {\it integrate out} the (near-zone) potential degrees of freedom by {\it matching} to an effective (far-zone) theory featuring only radiation fields coupled to a (source) stress-energy tensor. The latter can be used, for instance, to directly compute the instantaneous contribution to the radiated (source) energy. See e.g.\,\cite{Mougiakakos:2021ckm,Riva:2021vnj,Mougiakakos:2022sic,Riva:2022fru} for some recent developments in this direction.
In the realm of the PN approximation, it is further possible to perform a multipole expansion, introducing a series of (time-dependent) mass- and current-type moments \cite{andirad,andirad2}. This allows to conveniently integrate out the radiation fields, systematically incorporating nonlinear corrections in the far zone, such as tail effects \cite{tail}. See e.g.\,\cite{apparent,nrgr4pn1,nrgr4pn2,5pn1,5pn2,hered1,hered2,tail3,blum,blum2,Blumlein:2020pyo,Blumlein:2021txe,radnrgr,pardo,Almeida:2022jrv,Cho:2021mqw,Cho2022,Cho:1,Cho:2,Kim:2022bwv,Kim:2022pou,Mandal:2022nty,Mandal:2022ufb} for some recent results. 

\vskip 4pt
This {\it one-at-a-time} integration technique is extremely useful to handle slowly-moving sources, yielding a natural separation between far and near zones and simplifying the PN integration problem by reducing it to the computation of three-dimensional (mass-independent and static) integrals, e.g.\,\cite{nrgrG5,radnrgr,Cho2022}. However, once we enter the PM regime, and moreover once nonlinear and hereditary radiation-reaction effects start to contribute, unless one is able to resum an infinite tower of velocity corrections, it seems hopeless to try to implement this PN-type strategy for the relativistic case. As we shall see in detail, however, the key new idea is to replace the resummation problem with solving a set of differential equations \cite{Parra,3pmeft}---somewhat resembling a {\it renormalization group} flow---in the (incoming) relative velocity.\vskip 4pt It is straightforward to show that, at a given PM order, there is a single relevant scale in the (classical) relativistic scattering problem. Therefore, up to overall factors dictated by dimensional analysis, the resulting (multi-loop) integrals can only depend on the scalar product of the initial velocities, often denoted as $\gamma$.  By using the methodology of differential equations, the fully relativistic dynamics is then reduced to a set of boundary constants, which can be conveniently evaluated in the $\gamma \to 1^+$ limit. This is often referred in the literature as the near-static or {\it soft} limit, which then becomes one of the main actors in the PM regime. Furthermore, modulo the implementation of `symmetry relations' connecting different (master) integrals,  these differential equations are insensitive to the choice of Green's functions, so that we simultaneously incorporate conservative and dissipative effects alike.~The distinction is simply translated to the boundary conditions \cite{eftrad}. This then promotes the method of regions to a novel role, namely, the computation of boundary integrals in the soft limit featuring retarded and Feynman propagators. Not only---by linking the number of radiation modes to linear, nonlinear and/or hereditary radiation-reaction effects---we isolate the various conservative and dissipative (radiative) contributions, this approach also allows us to make direct contact with the vast literature in the EFT for the PN regime \cite{Goldberger:2022ebt}. The complete solution to the relativistic two-body problem is thus {\it bootstrapped} to all orders in the velocity from, often recycled, PN-type integrals.\vskip 4pt Fortunately, the resulting (mass-independent) integrals involved in the (classical) scattering problem turn out to be simpler than those arising in full-fledged (quantum) scattering amplitudes, such that various other powerful techniques used in collider physics~\cite{Smirnov:2012gma,Weinzierl:2022eaz,Kotikov:1991pm,Remiddi:1997ny,Henn:2013pwa,Prausa:2017ltv,Lee:2020zfb,Lee:2014ioa,Adams:2018yfj,Chetyrkin:1981qh,Tkachov:1981wb,Smirnov:2019qkx,Smirnov:2020quc,Lee:2012cn,Lee:2013mka,Beneke:1997zp,Jantzen:2012mw,Smirnov:2015mct,Meyer:2016zeb,Meyer:2016slj,Broedel:2019kmn,Primo:2017ipr,Hidding:2020ytt,Goncharov:2001iea,Chen:1977oja,Duhr:2014woa,Duhr:2019tlz,Dlapa:2020cwj,Smirnov:2021rhf,Lee:2019zop,Blumlein:2021pgo} can have an even bigger impact in PM computations. In addition to the use of dimensional regularization (dim. reg.), in combination with advanced methods for handling the differential equations \cite{Henn:2013pwa,Prausa:2017ltv,Lee:2020zfb,Lee:2014ioa}, the reduction to master integrals using integration-by-parts (IBP) relations \cite{Chetyrkin:1981qh, Tkachov:1981wb,Smirnov:2019qkx,Smirnov:2020quc,Lee:2012cn,Lee:2013mka} has emerged as an indispensable tool to tackle the PM integration problem, allowing us to swiftly move forward in the perturbative expansion. Unsurprisingly, these techniques have featured prominently in the solution to the two-body problem to 4PM~\cite{4pmeft,4pmeft2,4pmeftot}, which is akin of a (single-scale) `three-loop' calculation in particle physics. \vskip 4pt In~the remaining of this paper we go over several details behind our derivations in \cite{4pmeft,4pmeft2,4pmeftot} of the state-of-the-art at three-loop order. We have organized it as follows: In \S\ref{sec:EFT} we briefly review the worldline EFT approach for PM dynamics \cite{pmeft}, and its extension to the in-in formalism needed to incorporate dissipative effects~\cite{eftrad}. (For simplicity we consider structureless non-spinning bodies, see \cite{natalia1,natalia2,pmefts,Riva:2022fru} and \cite{pmeft,Mougiakakos:2022sic} for the inclusion of rotational degrees of freedom and tidal deformations in the worldline PM theory, respectively.) We then illustrate how the standard method of regions, extensively used in PN calculations~\cite{walterLH,iragrg,foffa,review,Goldberger:2022ebt}, is promoted to a new role in a PM scheme: To tackle boundary (master) integrals in the soft limit. We~provide a few examples of the computation of  some paradigmatic integrals involving both Feynman and retarded propagators.\vskip 4pt In \S\ref{sec:integrand} we discuss the construction of the integrand for the total impulse. We~review the Feynman rules and simplifications thereof, as well as the general structure. The application of various integration tools is discussed in \S\ref{sec:integrals}, where we also go over the 3PM and 4PM cases in more detail. Section \S\ref{sec:de} is devoted to the topic of differential equations as the main methodology used to obtain the full velocity dependence of the master integrals. After introducing the general ideas we discuss the specific examples at two- and three-loop orders, emphasizing their {\it canonical} and {\it non-canonical} structures.\vskip 4pt In \S\ref{sec:bc} we discuss the computation of the boundary integrals needed for the total impulse to 4PM order. We discuss both the full in-in derivation involving retarded propagators, as well as the conservative part constructed via Feynman's~$i0$-prescription. We collect in \S\ref{sec:data} results for the conservative and dissipative impulses, total radiated spacetime momentum and GW energy flux. The latter can be readily used to compute observables for generic (un)bound orbits. We also provide results for the relative scattering angle and discuss Firsov's resummation. We conclude in \S\ref{sec:conc} with  overall remarks, challenges, and future directions. Additional details are relegated to various appendices. Explicit results are summarized in an ancillary file containing ready-to-use expressions. 
 \newpage
\paragraph{Notation and conventions.} We use the mostly minus signature: $\eta_{\mu\nu} = {\rm diag}(+,-,-,-)$. The Minkowski product between four-vectors is denoted as $k \cdot x = \eta_{\mu\nu} k^\mu x^\nu$, while we use $\bk \cdot \bx = \delta^{ij} \bk^i \bx^j$ for the Euclidean version, with bold letters representing ${\bf 3}$-vectors. We use $\bk_\perp$ for vectors in the plane perpendicular to the direction of the scattering particles.  We implement the shorthand notation  
\begin{equation}\label{int-mesaure-def-ep}
\begin{aligned}
\int_\ell &\equiv \int \frac{e^{\epsilon\gamma_E}\dd^{d+1}\ell}{\pi^{d/2}}\,,\quad
\int_\bell & \equiv \int\frac{e^{\epsilon\gamma_E}\dd^{d}\bell}{\pi^{d/2}}\,,
\end{aligned}
\end{equation}
for the integral measure, with $\gamma_E$ the Euler constant. We use dim.\ reg.\ in $d=3-2\epsilon$ dimensions. The Planck mass is given by $\Mp^{-1} \equiv \sqrt{32\pi G}$ in $\hbar=c=1$ units. 
We~denote  $M=m_1+m_2$ the total mass, $\mu = m_1m_2/M$ the reduced mass, and $\nu \equiv \mu/M$ the symmetric mass ratio. We also introduce the variables \beq \Delta_m \equiv \frac{m_1-m_2}{M}\,,\quad \Gamma\equiv \frac{E}{M}\,,\quad  \xi \equiv \frac{E_1E_2}{E^2}\,,\eeq with $E=E_1+E_2$ the total incoming energy.\vskip 4pt 

For the space spanned by the external vectors we use the impact parameter~$b \equiv b_1-b_2$ and the 4-velocities, $u_a$, of the incoming point-particles (with $a \in \{1,2\}$ the particle's index)
\begin{align}\label{eq:extVecs}
  \hat{b}^\mu &\equiv \frac{b^\mu}{\sqrt{-b^2}}\,, & \check{u}_1^\mu &\equiv \frac{\gamma u_2^\mu-u_1^\mu}{\gamma^2-1}\,, & \check{u}_2^\mu &\equiv \frac{\gamma u_1^\mu-u_2^\mu}{\gamma^2-1}\,,
\end{align}
where
\beq  u_a^2=1\,,\quad \gamma \equiv  u_1\cdot u_2  
\,,\eeq
such that $\check{u}_a\cdot \check{u}_b = \delta_{ab}$. We will also extensively use the quantities $v_\infty$ and $x$ defined through
\begin{align}
v_{\infty} &\equiv \sqrt{\gamma^2-1}\,,
\qquad \gamma = \frac{x^2-1}{2x}\,,
\end{align}
when performing near-static ($v_\infty^2 \ll 1$) expansions and solving differential equations.
We denote $J \equiv p_\infty b$ the total angular momentum (without spin), $p_\infty \equiv M \nu\, v_\infty/\Gamma$ the incoming momentum at infinity, and $j \equiv J/(GM^2\nu)$ the reduced angular momentum.

\newpage

\section{Effective field theory approach} \label{sec:EFT}

 
\vskip 4pt

We review here the main ideas behind the EFT formalism in a PM scheme \cite{pmeft,eftrad}, notably the somewhat new role of the method of regions \cite{nrgr}. In what follows we restrict ourselves to non-spinning bodies (see e.g.\,\cite{pmefts,natalia1,natalia2} for spin effects). 

\subsection{Schwinger-Keldysh}

The in-in effective action is obtained by performing a {\it closed-time-path} integral \cite{eftrad} 
\beq
{\rm exp}  \big(i\cS_{\rm eff}[x_{a(1)}, x_{a(2)}]\big) = \!\int\! \cD h_1 \cD h_2\, {\rm exp} \big( i S_{\rm EH}[h_1] - iS_{\rm EH}[h_2] + i S_{\rm pp}[h_1,x_{a,1}] - iS_{\rm pp}[h_2,x_{a ,2}]\big)\label{seff},
\eeq
where we integrate only over the metric degrees of freedom in a saddle-point approximation, 
\beq
S_{\rm EH} = - 2 \Mp \int \dd^4 x \sqrt{-g} \, R
\eeq
is the standard Einstein-Hilbert action, and the (point-like) particles are described by \cite{pmeft}
\beq
S_{\rm pp} = -\sum_a \frac{m_a}{2} \int \dd \tau_a \left( v_a^2 + \frac{ h_{\mu\nu}}{\Mp}\,v_a^\mu v_a^\nu\right)\,,
\eeq
with $h_{\mu\nu} \equiv g_{\mu\nu} - \eta_{\mu\nu}$ the metric perturbation. It is convenient to use the Keldysh basis,
\begin{align}
&
\begin{aligned}
    h^-_{\mu\nu} &= \frac{1}{2}(h_{1\mu\nu}  +h_{2\mu\nu} ) \\
    h^+_{\mu\nu} &= h_{1\mu\nu}  -h_{2\mu\nu} 
\end{aligned}
\quad\Longleftrightarrow\quad
\begin{aligned}
    h_{1\mu\nu} &= h^-_{\mu\nu}+\frac{1}{2} h^+_{\mu\nu}\\
     h_{2\mu\nu} &= h^-_{\mu\nu}-\frac{1}{2} h^+_{\mu\nu}\,,
\end{aligned}
\\[0.35 em]
&
\begin{aligned}
    x^\alpha_{a,+} &= \frac{1}{2}(x^\alpha_{a,1}+x^\alpha_{a,2}) \\
    x^\alpha_{a,-} &= x^\alpha_{a,1}-x^\alpha_{a,2}
\end{aligned}
\quad\Longleftrightarrow\quad
\begin{aligned}
    x^\alpha_{a,1} &= x^\alpha_{a,+}+\frac{1}{2}x^\alpha_{a,-} \\
    x^\alpha_{a,2} &= x^\alpha_{a,+}-\frac{1}{2}x^\alpha_{a,-}\,,
\end{aligned}
\end{align}
for which the matrix of (classical) propagators for the metric field becomes 
   \begin{equation}
  K^{AB}(x-y) = i \begin{pmatrix} 0 & -\Delta_{\textnormal{adv}}(x-y) \\ -\Delta_{\textnormal{ret}}(x-y) & 0 \end{pmatrix}\,,\label{kmatrix}
\end{equation}
with $A,B \in\{+,-\}$ and $\Delta_{\rm ret/adv}$ the standard retarded/advanced Green's functions. The~equations of motion for the classical worldline follow as usual,  
 \begin{equation}\label{eq:inineom}
  m_b \frac{d}{d\tau} {v}_{b}^\mu(\tau) = \left. -\eta^{\mu\nu}\frac{\delta\cS_\textrm{eff}[x_{a,+},x_{a,-}]}{\delta x_{b,-}^\nu(\tau)}\right|_{\rm PL}\,,
\end{equation}
where `PL' stands for the {\it Physical Limit}: $\{x_{a,-}\to~0,\, x_{a,+}\to x_a\}$ \cite{chadprl}. The total impulse is then given by \beq
  \begin{aligned}
  \label{impulse}
    \Delta p_a^\mu &= 
   \left.-\eta^{\mu\nu} \int_{-\infty}^\infty \dd\tau_1 \frac{\delta\cS_\textrm{eff}[x_{b,\pm}]}{\delta x_{a,-}^\nu(\tau_a)}\right|_{\rm PL}\,.
  \end{aligned}
\eeq 
\vskip 4pt The derivation of the impulse entails a series of Feynman diagrams constructed using the rules that follow from \eqref{seff} and the relationship in \eqref{impulse} \cite{eftrad}.\footnote{Having the equations that follow from~\eqref{eq:inineom} at hand, we can in principle compute other observables. For instance, it is straightforward to construct an integrand describing the total change of angular momentum.} The total impulse in \eqref{impulse} must be evaluated perturbatively on solutions to the equations of motion. This entails, for the $(n+1)$PM impulse, the addition of {\it iterations} involving the trajectories to $n$PM order, 
\beq
\begin{aligned}
x^\mu_{a} (\tau_a) &= b_{a}^\mu + u_{a}^\mu \tau_a + \sum_{k=1}^n \delta^{(k)} x^\mu_{a}(\tau_a)\,,\quad
v^\mu_{a} (\tau_a) =  u_{a}^\mu  + \sum_{k=1}^n \delta^{(k)} v^\mu_{a}(\tau_a)\,.\label{traj}
\end{aligned}
\eeq
We return to the construction of the PM integrand in more detail momentarily, see also \cite{eftrad}.
\vskip 4pt

\subsection{Method of regions}\label{sec:MoR}

We will discuss the implementation of the methodology of differential equations in gory detail in \S\ref{sec:de}. In order to illustrate the prominent role of the method of regions in our EFT framework in the PM scheme, we illustrate the procedure below with a few paradigmatic cases of the (boundary) integration problem. We will discuss this in a bit more detail again in \S\ref{sec:bc}.\vskip 4pt
Since the kinematical variables obey (with $q$ the transfer momentum) 
\beq\label{kinematics} q\cdot u_a=0,\, \quad u_a^2=1\,,\eeq
prior to the last Fourier transform into impact parameter space, the result for the integrals can only depend on $q^2$ and $u_1\cdot u_2 = 
\gamma$. Moreover, in the soft limit we can set $q =(0,\bq)$, and the overall $\bq^2$-dependence can be easily determined from dimensional analysis. For simplicity, we will set $\bq^2=1$ on some of the expressions below and reinstall overall factors at the end.

\subsubsection{3PM example}\label{3pmexample}
Let us consider the following integral which, as we shall see, appears in the derivation of the impulse at 3PM order \cite{3pmeft,eftrad},
\beq\label{region-example-1}
I_{1}=
\begin{tikzpicture}
    [
        baseline={([yshift=-0.5ex]current bounding box.center)}
    ]
    \draw[dashed] (-1,0) -- (1,0);
    \draw[boson] (-0.8,0) --node[left] {$\ell_1\!\uparrow$} (-0.8,-1.5);
    \draw[boson] (0.8,0) --node[right] {$\uparrow\!\ell_2$} (0.8,-1.5);
    \draw[boson] (-0.8,-1.5) -- (0.8,0);
    \draw[dashed] (-1,-1.5) -- (1,-1.5);
  \end{tikzpicture}=
\int_{\ell_1\ell_2} \, \frac{ \delta (\ell_1 \!\cdot\! u_1)\, \delta(\ell_2\!\cdot\! u_2)} {\ell_1^2\, \ell_2^2\, (\ell_1+\ell_2-q)^2}\,.
\eeq
The diagram is meant to illustrate the type of scalar integral involved, as well as the momentum routing.
The~Dirac-$\delta$ functions are described by dashed lines for each particle (on the top and bottom),\footnote{ Later on we will also introduce straight lines to represent linear propagators, i.e. $(\pm \ell\cdot u_a  +i0)^{-1}$,  which enter through iterations of solutions of lower order equations of motion \cite{pmeft,eftrad}.}  whereas the wavy line(s) represent the graviton propagator(s). We do not make any assumption about a particular $i0$-prescription, but we will emphasize at which point the result with Feynman and/or causal Green's function begins to differ.\vskip 4pt
\textbf{\textit{Parameter space.}}
Let us start with Feynman propagators, such that $\ell_a^2 \to (\ell_a^2+i0)$. This integral then contributes to the conservative sector. In order to systematically isolate the relevant regions of integration, it is convenient to use the following  parametrization 
\begin{equation}
 I_{1,\rm Fey} = - e^{2\epsilon \gamma_E}\Gamma(3-d) \Bigg(\prod_{i=1}^3\int_0^\infty \dd\alpha_i\Bigg) \,
 \delta\big(1-\alpha_{123}\big)\,
{\mathcal{U}^{(7-3d)/2}\, \mathcal{U}_\delta^{-1/2} \over \mathcal{F}^{3-d}}\,,
\end{equation}
with $\alpha_{123} = \alpha_1  + \alpha_2 + \alpha_3$. (See App.\,\ref{sec:param} for more generic cases.) The polynomials $\cU$, $\cF$, and $\cU_\delta$ are given by \begin{align}
  \cF =\alpha_1 \alpha_2 \alpha_3,
  \qquad
  \cU = \alpha_1 \alpha_2+\alpha_2 \alpha_3+\alpha_1 \alpha_3,
  \qquad
  \cU_\delta = \cU - (\gamma^2-1) \alpha_3^2.
\end{align}
While $\mathcal{U}$ and $\mathcal{F}$ are the standard first and second Symanzik polynomials, $\mathcal{U}_\delta$ encodes the information of the Dirac-$\delta$ constraints on the topology. Notice that $\mathcal{U}_\delta$ reduces to $\mathcal{U}$ when $v_\infty \to 0 $, such that the integral becomes of the standard parameterized form, with \,$\mathcal{U}^{3-3d/2}/\mathcal{F}^{3-d}$.
This is the case, for instance, when radiation is ignored.
\vskip 4pt
 
An integration region in the soft limit is represented by a vector,~$r_i$, which corresponds to a rescaling of the Feynman parameters, (recall $v_{\infty}^2 \equiv \gamma^2-1$) 
\beq \alpha_i \rightarrow v_\infty^{2r_i} \alpha_i\,,\eeq
 yielding a hierarchical structure for each given region. Notably, all except a finite number of rescalings turn into scaleless integrals which in dim. reg. are set to zero.
A~proof via a systematic procedure to identify non-trivial regions can be found in~\cite{Pak:2010pt,Jantzen:2012mw}.\vskip 4pt

It is straightforward to see that the above integral has only two non-vanishing regions, which we isolate by means of the \texttt{asy2.m} code included in the \texttt{FIESTA} package \cite{Jantzen:2012mw,Smirnov:2015mct,Smirnov:2021rhf},
\begin{equation}
  \begin{aligned}
    \br_\textrm{pot} &= (0,0,0) \,,\quad
    \br_\textrm{rad} &= (0,0,-1) \,.
  \end{aligned}
\end{equation}
Performing the  above rescaling we find the following expressions in the soft limit:\footnote{Needless to say the $i0$-prescription is inconsequential in the potential region.}
\begin{align}\label{eq:paramRegions}
 I_{1, \rm Fey}^\text{pot} &= - e^{2\epsilon\gamma_E}\Gamma(3 -d)\,  \Bigg(\prod_{i=1}^3\int_0^\infty \dd\alpha_i\Bigg) \,
 \delta\big(1-\alpha_{123}\big)\,
{\mathcal{U}^{3-3d/2} \over \mathcal{F}^{3-d}}
+ \mathcal{O}(v_\infty^2),
\\[0.35 em]
 I_{1,\rm Fey}^\text{rad} &= - e^{2\epsilon\gamma_E}\Gamma(3-d)\,v_\infty^{d-2}\,  \Bigg(\prod_{i=1}^3\int_0^\infty \dd\alpha_i\Bigg) \,
{\delta(1-\alpha_{123})\,   (\alpha_1\alpha_2)^{d-3} \alpha_3^{-d/2}   \over (\alpha_1 + \alpha_2)^{(3d-7)/2} \sqrt{\alpha_1 + \alpha_2 - \alpha_3}}
+ \mathcal{O}(v_\infty^d).
\nonumber
\end{align}
The reader will immediately notice the factor of $v_\infty^{-2\epsilon}$, which is a trademark for radiation modes \cite{4pmeft,4pmeft2,4pmeftot}. This feature plays a key role, both via consistency conditions through the differential equations as well as helping us identify the various different contributions to generic boundary conditions  \cite{4pmeft,4pmeft2,4pmeftot}.\vskip 4pt

\vskip 4pt
The above integrals are straightforward to compute. However, the parameterization in \eqref{eq:paramRegions} is only valid for the case of Feynman propagators. In other words, it can be used to compute conservative contributions, but not for the full case. The reason is the lack of a similar representation with retarded Green's functions, for which we must follow a different strategy. As we show next, this is achieved by performing an expansion in momentum space, which also provides us with a direct connection to  EFT derivations in the PN regime.\vskip 4pt
 
\textbf{\textit{Momentum space.}} It is easy to see that, due to the Dirac-$\delta$ functions, only the $(\ell_1 {+} \ell_2 {-} q)^2$ propagator can go on-shell. It is then useful to introduce new loop momenta,  $k\equiv\ell_1+\ell_2-q$ and $\ell\equiv \ell_2$, such that we can directly perform an expansion in $k$. Furthermore, it is also convenient to evaluate the integral in the rest frame, say of particle 1, where
\beq\label{eq:restFrame1}
u_1=(1,0,0,0)\,,\quad u_2=(\gamma, 0,0,v_\infty)\,,
\eeq
so that the product of Dirac-$\delta$ functions in \eqref{region-example-1} becomes
\begin{align}\label{region-exmple-1-cut}
\delta (\ell_1 \!\cdot\! u_1)\, \delta(\ell_2\!\cdot\! u_2)
\,\longrightarrow\, \delta (\ell^0 - k^0)\, \delta\big(\gamma \ell^0 - v_\infty \ell^z\big)\,.
\end{align}
We are now in position to follow the EFT methodology in the PN scheme \cite{nrgr}, by identifying the regions of integration in the soft limit according to the $v_\infty$-scaling of the components of the loop momenta. The number of regions remains the same, as expected, and we find\begin{equation}
  \begin{aligned}
    &\mathrm{potential:} & \ell&\sim (v_\infty,1)|{\bq}|\,, & k&\sim(v_\infty,1)|{\bq}|\,,\\
    &\mathrm{radiation:} & \ell&\sim (v_\infty,1)|{\bq}|\,, & k&\sim(v_\infty,v_\infty)|{\bq}|\,.
  \end{aligned}
\end{equation}
After using \eqref{region-exmple-1-cut} to resolve the temporal components, these two regions can be isolated via the following rescaling of the {\it spatial} momenta\footnote{Although technically speaking the temporal component must also be rescaled, the product of the time measure and Dirac-$\delta$ function remains invariant.} \begin{equation}
\begin{aligned}
&\mathrm{potential:} \quad (\bell\to \bell, \bk\to \bk)\,,\\ 
&\mathrm{radiation:}  \quad (\bell\to \bell, \bk\to v_\infty \tilde\bk)\,,\quad \tilde\bk\sim \bq\,,\label{res3pm}
\end{aligned}\end{equation} 
respectively. We then find
\begin{align}
\begin{aligned}
I_{1}^\mathrm{pot} &= -\int_{\bell\bk} \frac{1}{[(\bk-\bell+\bq)^2]\, [\bell^2]\, [\bk^2]} + \cO(v_\infty^2)\,,
\\[0.35 em]
I_{1}^\mathrm{rad} &= - \int_{\bell}  \frac{1}{  [(\bell - \vecbf{q})^2]\,[\bell^2] }\,
\int_{\tilde \bk} \frac{v_\infty^{d-2} }{  [\tilde\bk^2 - (\ell^z)^2] }
+\cO(v_\infty^{d})\,,
\end{aligned}
\end{align}
where all momenta scale homogeneously. Notice the choice of $i0$-prescription is translated to the $(\ell^z)^2$ term. Moreover, it is also  transparent how the factors of $v_\infty^{-2\epsilon}$ come about, with the measure contributing a $d$-dependent factor in the radiation region. For the sake of notation, from now on we will remove the tildes in the rescaled momenta.\vskip 4pt
So far we have been purposely agnostic about the choice of propagators. While the potential region is trivially independent, the radiation region does depend on the choice of Green's function. Let us reinsert now its explicit $i0$ dependence, yielding 
\begin{align}
\begin{aligned}\label{eq:paramBoundary3PM}
I_{1,\rm Fey}^\mathrm{rad} &= - \int_{\bell}  {1  \over  [(\bell - \vecbf{q})^2]\,[\bell^2] }\,
\int_{\bk} {v_\infty^{d-2} \over  [\bk^2 - (\ell^z)^2 - i0] }
+\cO(v_\infty^{d})\,,
\\[0.35 em]
I_{1,\rm ret}^\mathrm{rad} &= - \int_{\bell}  {1  \over  [(\bell - \vecbf{q})^2]\,[\bell^2] }\,
\int_{\bk} {v_\infty^{d-2} \over  [\bk^2 - (\ell^z + i0)^2] }
+\cO(v_\infty^{d})\,.
\end{aligned}
\end{align}
These are most conveniently evaluated as a nest, {\it loop by loop}, integral. The inner one, in $\bk$, is a simple tadpole which evaluates to \cite{tail}
\begin{align}\label{eq:tadpoles}
\begin{aligned}
\int_{\bk} \frac{1 }{  [\bk^2 - (\ell^z)^2 \pm i0]} 
&= \frac{e^{\epsilon\gamma_E}\Gamma\big(1 - \tfrac{d}{2}\big) }{ [-(\ell^z)^2 \pm i0]^{1- d/2}},
\\[0.35 em]
\int_{\bk}\frac{1 }{  [\bk^2 - (\ell^z \pm i0)^2]} 
&= \frac{e^{\epsilon\gamma_E}\Gamma\big(1 - \tfrac{d}{2}\big)}{ [ - (\ell^z \pm i0)^2]^{1 - d/2}}
= {e^{\mp i \pi (d/2 - 1)+\epsilon\gamma_E} \Gamma\big(1 - \tfrac{d}{2}\big) \over  (\ell^z \pm i0)^{2 - d}}.
\end{aligned}
\end{align}
For the leftover one-loop integral over $\bell$, the case of retarded propagators naturally falls into the category of a 2PM (potential) boundary integral. The reader will immediately notice that the term proportional to $\ell^z$ resembles the presence of a linear propagator (due to a 1PM iteration \cite{pmeft}) evaluated in the static limit; this can be easily seen upon rewriting $\ell^z = \bell\cdot \bn$, with $\bn \equiv (0,0,1)$.  There is, however, an important caveat, which is manifest in the $d$-dependent power in \eqref{eq:tadpoles}. Nevertheless, these type of one-loop integrals are known, which allows us to directly obtain the answer using standard tools.\vskip 4pt 

The Feynman version is a tad more involved. It can  be computed, in this particular case, by integration over a parametric representation. The result agrees with the direct computation in \eqref{eq:paramRegions}. The final answers read (up to an overall $(\bq^2)^{d-3}$) 
\beq
\begin{aligned}
I_{1}^\textrm{pot} &= -e^{2\epsilon\gamma_E}\,{\Gamma(3 - d)\, \Gamma^3\big(\frac{d}{2}-1\big)  \over  \Gamma\big(\frac{3 d}{2} - 3\big)}\,,
\\
I_{1,\rm Fey}^\textrm{rad} &= e^{i\pi d/2 + 2\epsilon\gamma_E} \frac{\Gamma(3-d)\, \Gamma\big(1-\frac{d}{2}\big)\, \Gamma(d-2)\,
\Gamma\big(\frac{d-1}{2}\big)}{\Gamma\big(d-\frac{3}{2}\big)}\, v_\infty^{d-2}
+\cO(v_\infty^{d})\,,
\\
I_{1,\rm ret}^\textrm{rad} &= 
e^{2\epsilon\gamma_E}\frac{8^{2-d}\,\sqrt{\pi}\, \Gamma^2\big(1-\frac{d}{2}\big)\, \Gamma (d-1)}{\Gamma\big(d - \frac{3}{2}\big)}\, v_{\infty }^{d-2}
+ \cO(v_\infty^{d-1})\,.
\end{aligned}
\eeq
Notice that, as expected, the Feynman and retarded answer differ due to the presence of an imaginary part (in $d=3$) for the former, while the latter is real.
Hence, the conservative sector does not include radiation modes at 3PM, and it is uniquely captured by the potential-only part at this order \cite{3pmeft}. As we shall see, this will not be the case at 4PM~\cite{4pmeft,4pmeft2,4pmeftot}.\vskip 4pt

It is instructive to compare the above manipulations with the EFT philosophy in the PN scheme. In principle, although the type of integrals are clearly linked, the above momentum expansion is done in {\it reversed} order. In the case of slowly moving sources, one starts by integrating out the potential modes first, which in the above example would correspond to performing the $\bell$-integration, followed by the integral over $\bk$, the radiation modes. Pictorially, this process would be represented as
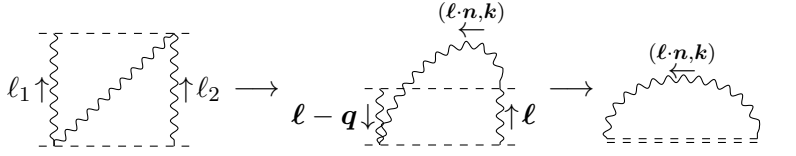
\begin{equation}
\nn
  \begin{tikzpicture}
    [
        baseline={([yshift=-0.5ex]current bounding box.center)}
    ]
    \draw[dashed] (-1,0) -- (1,0);
    \draw[boson] (-0.8,0) --node[left] {$\ell_1\!\uparrow$} (-0.8,-1.5);
    \draw[boson] (0.8,0) --node[right] {$\uparrow\!\ell_2$} (0.8,-1.5);
    \draw[boson] (-0.8,-1.5) -- (0.8,0);
    \draw[dashed] (-1,-1.5) -- (1,-1.5);
  \end{tikzpicture}
  \longrightarrow
  \begin{tikzpicture}
    [
        baseline={([yshift=-2.05ex]current bounding box.center)}
    ]
    \draw[dashed] (-1,0) -- (1,0);
    \draw[boson] (-0.8,0) --node[left] {$\bell-\bq\!\downarrow$} (-0.8,-0.75);
    \draw[boson] (0.8,0) --node[right] {$\uparrow\!\bell$} (0.8,-0.75);
    \draw[boson] (-0.8,-0.75) .. controls (-0.5,0.5) and (0.8,1) .. node[above=0.1,pos=0.65] {$\overset{(\bell\cdot\bn,\bk)}{\leftarrow}$} (0.8,0);
    \draw[dashed] (-1,-0.75) -- (1,-0.75);
  \end{tikzpicture}
  \longrightarrow
  \begin{tikzpicture}
    [
        baseline={([yshift=-0.5ex]current bounding box.center)}
    ]
    \draw[dashed] (0,0) -- (2,0);
    \draw[dashed] (0,-0.05) -- (2,-0.05);
    \draw[boson] (0,0) .. controls (0,1) and (2,1) .. node[above] {$\overset{(\bell\cdot\bn,\bk)}{\leftarrow}$} (2,0);
  \end{tikzpicture}\quad\,,
\end{equation}
where the last diagram corresponds to the (one-loop) radiative contribution after the potential modes have been matched into a series of (source) multipole moments, see e.g.\,\cite{tail}. In the PM scheme, on the other hand, we have found it more convenient to perform these steps in reversed order, first performing the (spatial) integral over the radiation region, and subsequently over the potential modes. At 3PM this distinction is perhaps only of academic interest. However, as we show next, we find the order does alter the complexity of the product at higher orders.
\vskip 4pt

\subsubsection{4PM example}

We move now into a three-loop example, which we hope will make the general pattern somewhat obvious. 
Using the same diagrammatic rules, we consider the following master integral
\begin{equation}
  I_{2}=\begin{tikzpicture}
    [
        baseline={([yshift=-0.5ex]current bounding box.center)}
    ]
    \draw[dashed] (-1.5,0) -- (1.5,0);
    \draw[boson] (-1.2,0) --node[left] {\small$\ell_1\!\uparrow$} (-1.2,-0.75);
    \draw[boson] (-1.2,-0.75) --node[left] {\small$\ell_3\!\uparrow$} (-1.2,-1.5);
    \draw[boson] (1.2,0) --node[right] {\small$\downarrow\!\ell_2{-}q$} (1.2,-0.75);
    \draw[boson] (1.2,-0.75) --node[right] {\small$\downarrow\!\ell_3{-}q$} (1.2,-1.5);
    \draw[boson] (0,-0.75) --node[below] {\small$\overset{\leftarrow}{\ell_2{-}\ell_3}$} (1.2,-0.75);
    \draw[boson] (-1.2,-0.75) --node[below] {\small$\overset{\rightarrow}{\ell_3{-}\ell_1}$} (0,-0.75);
    \draw[boson] (0,0) --node[right] {\small$\downarrow\!\ell_1{-}\ell_2$} (0,-0.75);
    \draw[dashed] (-1.5,-1.5) -- (1.5,-1.5);
  \end{tikzpicture}
 = \int_{\ell_1\ell_2\ell_3}  \frac{ \delta (\ell_1\!\cdot\! u_1)\, \delta (\ell_2\!\cdot\! u_1)\, \delta(\ell_3\!\cdot\! u_2)} {\ell_1^2\,\ell_3^2\,(\ell_2 {-}q)^2\, (\ell_3 {-} q)^2\, (\ell_1 {-} \ell_2)^2\, (\ell_2 {-} \ell_3)^2\, (\ell_3{-}\ell_1)^2}\,.
\end{equation}

\textbf{\textit{Parameter space.}} In order to isolate the (non-vanishing) regions we once again use the Feynman parameterization and the \texttt{asy2.m} package, which identifies the following four:
\begin{equation}\label{example-4pm-asy2}
  \begin{aligned}
    \br_\textrm{pot} &= (0,0,0,0,0,0,0) \,,\\
    \br_\textrm{1rad}^{(1)} &= (0,0,0,0,0,0,-1) \,,\\
    \br_\textrm{1rad}^{(2)} &= (0,0,0,0,0,-1,0) \,,\\
    \br_\textrm{2rad} &= (0,0,0,0,-1,-1,-1) \,,\\
  \end{aligned}
\end{equation}
Although, in principle, we need to rescale three different parameters, we refer to the last region as `2rad'.\footnote{In this particular example, the other possible 2rad-type regions $\br = (0,0,0,0,-1,-1)$, $\br = (0,0,0,-1,-1,0)$, and $\br = (0,0,0,-1,0,-1)$, are absent, although they feature in other integrals.}  This will be clear momentarily, from the scaling in powers of $v_\infty^2$, and more directly in momentum space (see below).
\vskip 4pt

The explicit expressions for the $\alpha_i$ integrals is not particularly illuminating. (Moreover, for simplicity we also omit the result for the combined rad1 regions, which only contributes an imaginary part with Feynman propagators.) Hence, using the methods described in App.~\ref{sec:app-bc}, we find (up to an overall $(\vecbf{q}^2)^{3d/2 - 6}$)
\beq\label{2fey}
\begin{aligned}
I_{2}^\textrm{pot} &= e^{3\epsilon\gamma_E}\frac{\pi ^{5/2} 2^{6-d} \Gamma \big(\frac{d}{2}-1\big)^2}{3 (d-4)^4 \Gamma \big(\frac{d-1}{2}\big)^2}\begin{multlined}[t]
\left[
3 \csc \big(\tfrac{3 \pi  d}{2}\big) \left(\frac{5\ 2^{10-3 d} \Gamma (d-2)}{\Gamma \big(d-\frac{7}{2}\big) \Gamma \big(\frac{3 d}{2}-5\big)}
+\frac{\pi  \csc \big(\frac{\pi  d}{2}\big)}{\Gamma \big(\frac{3}{2}-\frac{d}{2}\big) \Gamma (2 d-8)}\right)\right.\\
\left.+\frac{\sqrt{\pi } 2^{3-d} (d-3) \csc
\big(\frac{\pi  d}{2}\big) \csc (\pi  d)}{\Gamma \big(\frac{3 d}{2}-6\big)}
\right] + \cO(v_\infty^2)\,,
\end{multlined}\\
I_{2,\rm Fey}^\textrm{2rad} &= e^{3\epsilon\gamma_E-i\pi d}\frac{9\ 2^{7-2 d} (d-4) \cos \big(\frac{\pi  d}{2}\big)  \Gamma (3-d) \Gamma \big(1-\frac{d}{2}\big) \Gamma \big(d-\frac{5}{2}\big) \Gamma \big(\frac{d-1}{2}\big)}{\pi  \sin \big(\frac{3 \pi  d}{2}\big) \Gamma \big(\frac{3 d}{2}-\frac{7}{2}\big)}v_\infty^{2d-6}+\cO(v_\infty^{2d-4})\,.
\end{aligned}
\eeq

As discussed in \cite{4pmeft,4pmeft2}, the potential and (real part of the) 2rad pieces are building blocks for the conservative contributions.\vskip 4pt

\textbf{\textit{Momentum space.}} Let us translate now our knowledge from parameter to momentum space. From the scaling of $\alpha_i$-parameters one could have thought that all three propagators could be on-shell. However, it is straightforward to show that the first two Dirac-$\delta$ functions make it impossible for the third to last to have support on the on-shell condition. This is easily seen by again choosing the rest frame of particle 1, in which case the vector $\ell_1-\ell_2$ does not have a time component. Therefore, in this example, only two propagators can be on-shell, with the third turning into a `long-wavelength' potential mode, reminiscent of tail-type interactions~\cite{andirad}. We perform the following relabeling, $k_1=\ell_3-\ell_1$ and $k_2=\ell_2-\ell_3$, and upon rewriting $\ell=\ell_3$, the $I_2$ integral can be expressed as
\begin{equation}
I_{2} = \int_{\ell\, k_1 k_2} \frac{ \delta(k_1^0-\ell^0)\delta(k_2^0 + \ell^0)\delta(\gamma\ell^0 - v_\infty\ell^z)}
{(\ell-k_1)^2\,\ell^2\, (k_2+\ell-q)^2\, (\ell-q)^2\, (k_1+k_2)^2\, k_2^2\, k_1^2}\,.
\end{equation}
Hence, applying the scalings rules for potential and radiation modes we find 
the regions\begin{equation}
  \begin{aligned}
  &\mathrm{pot:} & k_1&\sim(v_\infty,1)|\bq|\,, & k_2&\sim(v_\infty,1)|\bq|\,, & \ell&\sim(v_\infty,1)|\bq|\,,\\
  &\mathrm{1rad}^{(1)}: & k_1&\sim(v_\infty,v_\infty)|\bq|\,, & k_2&\sim(v_\infty,1)|\bq|\,, & \ell&\sim(v_\infty,1)|\bq|\,,\\
  &\mathrm{1rad}^{(2)}: & k_1&\sim(v_\infty,1)|\bq|\,, & k_2&\sim(v_\infty,v_\infty)|\bq|\,, & \ell&\sim(v_\infty,1)|\bq|\,,\\
  &\mathrm{2rad:} & k_1&\sim(v_\infty,v_\infty)|\bq|\,, & k_2&\sim(v_\infty,v_\infty)|\bq|\,, & \ell&\sim(v_\infty,1)|\bq|\,,
  \end{aligned}
\end{equation}
which, performing similar rescaling as in \eqref{res3pm}, are
in one-to-one correspondence with those in \eqref{example-4pm-asy2}. Expanding around each region we have\beq
\begin{aligned}
\label{example-4pm}
I_{2}^\textrm{pot} &= \int_{\bell\,\bk_1\bk_2} {1 \over [(\bell-\bk_1)^2]\,[\bell^2]\, [(\bk_2+\bell-\bq)^2]\, [(\bell-\bq)^2]\, [(\bk_1+\bk_2)^2]\, [\bk_2^2]\, [\bk_1^2]} +\cO(v_\infty^2)\,,
\\[0.35 em]
I_{2}^\textrm{1rad,(1)} &= \int_{\bell\,\bk_2} {1 \over [\bell^2]\,[\bell^2]\, [(\bk_2+\bell-\bq)^2]\, [(\bell-\bq)^2]\, [\bk_2^2]^2}
\int_{\bk_1} {v_\infty^{d-2}  \over \bk_1^2 - (\ell^z)^2} +\cO(v_\infty^{d})\,,
\\[0.35 em]
I_{2}^\textrm{1rad,(2)} &= \int_{\bell\,\bk_1} {1 \over [(\bell-\bk_1)^2]\,[\bell^2]\, [(\bell-\bq)^2]\, [(\bell-\bq)^2]\, [\bk_1^2]^2}
\int_{\bk_2} {v_\infty^{d-2}  \over \bk_2^2 - (\ell^z)^2} +\cO(v_\infty^{d})\,,
\\[0.35 em]
I_{2}^\textrm{2rad} &=  \int_\bell {1  \over  [\bell^2]\,[\bell^2]\, [(\bell-\bq)^2]^2}
\int_{\bk_1\bk_2} {v_\infty^{2d-6} \over  [(\bk_1+\bk_2)^2]\, [\bk_2^2 - (\ell^z)^2]\, [\bk_1^2 - (\ell^z)^2]} +\cO(v_\infty^{2d-4})\,.
\end{aligned}
\eeq
The analytic result for the potential-only modes is straightforward using the nested one-loop integral, with the answer given in \eqref{2fey}. For the others, let us concentrate on the case of retarded propagators.  For the 1rad contribution(s) in \eqref{example-4pm}, the inner integral over $\bk_1$ is shown in \eqref{eq:tadpoles}, and the remaining two-loop integral becomes straightforward.  Furthermore, using IBP relations, we find that the two-fold inner integral in the 2rad contribution in \eqref{example-4pm} can be fully factorized, 
\begin{align}\label{eq:inner2rad}
\int_{\bk_1\bk_2} {1 \over  [(\bk_1+\bk_2)^2]\, [\bk_2^2 - (\ell^z)^2]\, [\bk_1^2 - (\ell^z)^2]}
\,=\, \frac{d-2}{2 (d-3)\, [(\ell^z)^2]}
\bigg[\int_{\bk} {1 \over \bk^2 - (\ell^z)^2}\bigg]^2,
\end{align}
such that it can be readily evaluated following our previous 3PM example. The explicit results are, up to an overall factor of $(\vecbf{q}^2)^{3d/2 - 6}$, 
\beq
\begin{aligned}
\label{example-4pmres}
I_{2,\rm ret}^\textrm{1rad} &= -e^{3\epsilon\gamma_E}\frac{3 \pi ^2 2^{6-d}  \cot \big(\frac{\pi  d}{2}\big) \csc \big(\frac{3 \pi  d}{2}\big) \Gamma (d-2)}{(d-6) (d-2) \Gamma \big(\frac{5 d}{2}-8\big)}v_\infty^{d-2} + \cO(v_\infty^d) \,,\\
I_{2,\rm ret}^\textrm{2rad} &= e^{3\epsilon\gamma_E}\frac{9 \sqrt{\pi } 32^{3-d} (d-2) \sin (\pi  d)  \Gamma (5-d) \Gamma \big(1-\frac{d}{2}\big)^2 \Gamma (2 d-6)}{\sin \big(\frac{3 \pi  d}{2}\big)\Gamma \big(\frac{3 d}{2}-\frac{7}{2}\big)}v_\infty^{2 d-6} + \cO(v_\infty^{2d-4})\,,
\end{aligned}
\eeq
where we combined the results for the two rad1-regions.\vskip 4pt

The alert reader will immediately notice that the inner integral for the $I_2^{\rm 2rad}$ contribution in \eqref{example-4pm}  has a celebrated counterpart in the PN regime. After identifying $\ell^z$ with $\omega$, the frequency, the integral turns out to be equivalent to the tail-type computation in the long-wavelength EFT approach \cite{tail}, diagrammatically represented as 
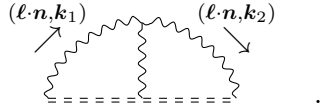
\begin{equation*}
\begin{aligned}
  \begin{tikzpicture}
    \draw[dashed] (0,0) -- (2.5,0);
    \draw[dashed] (0,-0.05) -- (2.5,-0.05);
    \draw[boson] (0,0) .. controls (0,0.5) and (0.75,1) .. node[above left=-0.2] {$\overset{(\bell\cdot\bn,\bk_1)}{\nearrow}$} (1.25,1);
    \draw[boson] (2.5,0) .. controls (2.5,0.5) and (1.75,1) .. node[above right=-0.2] {$\overset{(\bell\cdot\bn,\bk_2)}{\searrow}$} (1.25,1);
    \draw[boson] (1.25,0) -- (1.25,1);
  \end{tikzpicture}\quad\,.
\end{aligned}
\end{equation*}
Notice also that all of the momenta in the two-loop radiation region descend from modes rescaled by a factor of $v_{\infty}$, which are brought up to the numerator in \eqref{example-4pm}. However, due to the Dirac-$\delta$ functions, the one involving $k_1+k_2$, i.e. the middle line in the above diagram, has zero temporal component, thus corresponding to a {\it soft} potential mode.\vskip 4pt

 Let us conclude this section with a few remarks.
 The simplest way to identify regions in a given kinematic limit is in parametrized form, namely due to the existence of an algorithmic procedure, implemented through the \texttt{asy2.m} code \cite{Jantzen:2012mw,Smirnov:2015mct,Smirnov:2021rhf}.
 Unfortunately, retarded Green's functions do not enjoy the same type of parametrization as Feynman propagators, which is one of the reasons we performed the above manipulations in momentum space.
 Yet, for a given integral, using Feynman propagators allows us to find the relevant momentum-space rescalings, which can then be readily transported to the case of causal Green's functions.
 We will discuss more details of the two- and three-loop derivations in \S\ref{sec:bc}.

\section{Building the integrand}  \label{sec:integrand}

Our goal is to derive an integrand for the impulse  defined in \eqref{impulse}, say of particle~$1$, expanded to a given PM order,
\begin{equation}
  \Delta p_1^\mu = \sum_{n=1}^\infty G^n\, \Delta^{\!(n)}p_1^\mu\,.
\end{equation}
We discuss here the generic procedure, using the variation of the effective in-in action and Feynman rules described in \cite{eftrad}. Throughout this section we also elaborate on some technical details regarding the overall construction.
\subsection{Feynman rules}

We compute the (variation of the) effective action in \eqref{seff} by a saddle point approximation. Diagrammatically, this corresponds to a set of tree-level Feynman-like diagrams built out of graviton vertices and causally-directed edges (from $S_\textrm{EH}$ and $S_\textrm{GF}$) coupled to static worldline sources (from $S_\textrm{pp}$) in the form of \emph{source}- and \emph{sink}-type vertices, according to the causal flow~\cite{eftrad}. The computational complexity can be significantly decreased by optimizing the Feynman rules via a clever choice of the gauge fixing terms, and adding total time-derivatives to the action~\cite{pmeft}. In principle, we are allowed field redefinitions, however, for the case of non-spinning structureless bodies we focus on, this would lead to higher-point worldline (seagull-type) couplings in $S_\textrm{pp}$, hence to more diagrams. Therefore, we chose to keep the worldline coupling at linear order in $h_{\mu\nu}$. We fix the $\cO(h^2)$ Lagrangian to agree with the commonly used de Donder gauge
\begin{equation}
  \cL_{hh} = \frac{1}{2}\partial_\alpha h^{\mu\nu}\partial^\alpha h_{\mu\nu}-\frac{1}{4}\partial_\mu h\partial^\mu h\,,
\end{equation}
where $h\equiv {h_\alpha}^\alpha$. Following ideas described in~\cite{safi}, we optimize the  weak-field expansion of the Einstein-Hilbert action to have the fewer number of terms up to $\cO(h^5)$. This is achieved through \texttt{FORM}~\cite{Ruijl:2017dtg} and the Mathematica package \texttt{xAct}~\cite{xactpackage}. We find a version of the  Lagrangian that has following number of terms (per order in $h$):
\begin{align}
  \cL_{h^3}:\, 6\,, \qquad \cL_{h^4}:\, 18\,, \qquad \cL_{h^5}:\,36\,. \nn
\end{align}
Extracting Feynman rules for the graviton vertices and edges follows in the standard fashion using the in-in action in the Keldysh representation of \eqref{kmatrix}. Since we are ultimately interested in the impulse, there is only a single worldline vertex hit by the variation in \eqref{impulse} that must be distinguished from the others.
Such a \emph{sink} vertex has the rule
\begin{equation}\label{eq:WLOutV}
  \begin{tikzpicture}
    [baseline={([yshift=-0.5ex]current bounding box.center)}]
    \node[xSource] (A) at (0,1) {};
    \draw (0,0) -- node[left=0.1] {$\downarrow\! k$}  (A);
    \draw[middleArrow] (0,0) -- (A);
  \end{tikzpicture}
  \,
  = -\frac{i m}{2\Mp}  \,e^{i\,k\cdot x} \left[ i\, k^\alpha v^\mu v^\nu-i\,k\cdot v\, \eta^{\mu\alpha}v^\nu-\eta^{\mu\alpha}\dot{v}^\nu-i\,k\cdot v\, \eta^{\nu\alpha}v^\mu-\eta^{\nu\alpha}\dot{v}^\mu\right]\,.
\end{equation}
All other worldline vertices take the same form as in the in-out formalism, represented by
\begin{equation}
  \begin{tikzpicture}
    [baseline={([yshift=-0.5ex]current bounding box.center)}]
    \draw (0,1) -- node[left=0.1] {$\downarrow\! k$} (0,0);
    \draw[middleArrow] (0,1) -- (0,0);
    \draw[fill] (0,1) circle (0.05);
  \end{tikzpicture}
  \,
  =-\frac{i  m}{2\Mp}\int\dd\tau \, e^{i\, k\cdot x} v^\mu v^\nu\,.
\end{equation}
\vskip 4pt

From these rules we then compute the variation of the effective action. For instance, to 4PM order we have
\begingroup
\allowdisplaybreaks
\input{Seff_graphs.tex}
\endgroup
where, for simplicity, at three-loop order we have omitted all the \emph{self-energy} diagrams (``\emph{selfies}"), where the graviton lines attach to only one of the particles.    

\subsection{Integrand structure}\label{subsec:integrandStructure}
 
 At a fixed $n$PM order, we start from the variation of the effective action including terms up to $\cO{(
 G^n)}$, evaluate them on solutions of the equations of motion up to order ${\cO}(G^{n-1})$, see \eqref{traj}, and extract the ${\cO}(G^n)$ contribution, yielding
\begin{equation}
  \Delta^{\!(n)} p_1^\mu = -\eta^{\mu\nu}\int_{-\infty}^\infty \dd\tau_1\left(\left.\sum_{\ell=1}^{n}\frac{\delta \cS^{\ell\textrm{PM}}_\textrm{eff}}{\delta x_{1-}^\nu(\tau_1)}\right|_{\rm PL}\left[x_{a+}(\tau_a) \to b_a + u_a \tau_a + \sum_{k=1}^{n-1}\delta^{(k)}x_a(\tau_a)\right]\right)_{\cO(G^n)}\,.
\end{equation}
In addition to the propagators for the gravitational field,  the (recursive) solution for the trajectories introduces also a new type of Green's function, which can be associated with the (time) propagation of the worldline sources. Since these are given by factors of $(p\cdot u_a \pm i0)^{-1}$ in Fourier space,  with the $i0$-prescription determined by causality \cite{pmeft}, we will loosely refer to them as \emph{linear} propagators. Performing the integrals over the proper times yields a series of Dirac-$\delta$ functions which, as we discussed before, ultimately constrain the integration regions.\vskip 4pt 
After Lorentz contractions, and various algebraic manipulations, the impulse acquires the following simple tensorial structure 
\begin{equation}
  \Delta^{\!(n)} p^\mu_1 = \int_q \, e^{-i\, b\cdot q} \delta(q\cdot u_1)\delta(q\cdot u_2)
  \Bigg(\prod_{i=1}^{n-1}\int_{\ell_i}\Bigg)\Bigg[R_{q}(\ell_i,q,u_a) q^\mu + \sum_{j=1}^{n-1} R_{\ell_j}(\ell_i,q,u_a) \ell_j^\mu\Bigg]\,.
\end{equation}
The scalar $(R_q,R_{\ell_j})$ integrands are rational functions whose denominators are products of linear and quadratic propagators, and carry numerators involving polynomials in scalar products of loop, external momenta, as well as kinematical invariants. We will treat the transfer momentum~$q$ as an external variable, and only perform the Fourier transform at the end, after the $\ell_i$ integrals are obtained.\vskip 4pt Following the standard Passarino-Veltman reduction \cite{Passarino:1978jh}, we bring all tensor integrals proportional to $\ell_i^\mu$ to linear combinations of scalar integrals. This is due to the simple fact that the tensor structure in the final result must be expressible in terms of the external data, see e.g.~App.~B of \cite{pmefts}. This procedure then amounts to the following replacement rule at the level of the integrand (see~\eqref{eq:extVecs}),
\begin{equation}
  \ell_i^\mu \longrightarrow  \frac{\ell_i\cdot q}{q^2}\,q^\mu + \ell_i\cdot u_2\, \check{u}_2^\mu + \ell_i\cdot u_1\,\check{u}_1^\mu\,,
\end{equation}
(Notice that, due to the presence of Dirac-$\delta$ functions, only one of the two last terms can appear in a given decomposition at a time.) The integrand for the impulse then becomes \begin{align}\label{eq:intStructure}
  \Delta^{\!(n)} p_1^\mu
  &= \int_q \frac{\delta(q\cdotnew u_{1})\delta(q\cdotnew u_2)\,e^{-i\,q\cdot b}}{(-q^2)^{(d+(2-d)n)/2}}\left[-i q^\mu\,\cI_q(\gamma)+ \sqrt{-q^2}\left(\cI_{\check{u}_1}(\gamma) \check{u}_1^\mu+ \cI_{\check{u}_2}(\gamma) \check{u}_2^\mu\right)\right]\,,
 \end{align}
 where the coefficients in terms of the transfer momentum are simply determined by dimensional analysis. Using the formula, with $b_\perp \equiv b -(b\cdot u_1) \check{u}_1+(b\cdot u_2)\check{u}_2\,$,
\begin{equation}
   \int_q\, e^{-i\,b\cdot q}\delta(q\cdot u_1)\delta(q\cdot u_2)(-q^2)^\alpha
  = \frac{e^{\epsilon \gamma_E}}{\sqrt{\pi}}
  \frac{2^{d-1+2\alpha}}{\sqrt{\gamma^2-1}(-b_\perp^2)^{(d-1)/2+\alpha}}
    \frac{\Gamma\left(\frac{d-1}{2}+\alpha\right)}{\Gamma(-\alpha)}\,,
\end{equation}
the Fourier transform over $q$ is straightforward, leaving behind the $\cI_{q,\check{u}_a}$'s, containing linear combinations of scalar integrals which will be the subject of the next section.

\subsection{Constraints}\label{subsec:integrandConstraints} Let us conclude with a few comments on the various contributions to the impulse. In~general, we will parameterize it as follows,
 \beq
\label{impulsetot}
\Delta^{(n)} p_1^\mu =\frac{1}{|b|^n}\left(c^{(n)}_{1b}\, \hat b^\mu +\sum_{a} c^{(n)}_{1\check{u}_a}\, \check{u}_a^\mu \right)\,.
\eeq
Because of the conservation of the on-shell condition (for the case of non-absorptive dynamics), $(p_a+\Delta p_a)^2 = p_a^2$, or the conservation of energy-momentum for non-dissipative terms, $\sum_a \left(\Delta p_a\right)_{\rm cons}=0$, not all the coefficients turn out to be independent. To exploit this, it is convenient to make the mass dependence manifest, and write,
\beq
 c_{1b}^{(n)} = \sum_{i=1}^{n} m_1^{n-i+1} m_2^i\, c^{(n,i)}_{b}\,, \quad
 c^{(n)}_{1\check{u}_a}=  \sum_{i=1}^{n} m_1^{n-i+1} m_2^i \, c^{(n,i)}_{\check{u}_a} \,,
\eeq
where the $c^{(n,i)}_{1X}$'s, with $i=1, \cdots,n$,  then encapsulate all the $\gamma$-dependence of the impulse. It~is then straightforward to show that the  $c^{(n,i)}_{\check{u}_1}$ coefficients can be recursively determined by the on-shell condition through lower order contributions. We find, 
\begin{equation}
\sum_{i=1}^n m_1^{n-i+2}m_2^i \,c_{\check{u}_1}^{(n,i)} = |b|^n \sum_{k=1}^{n-1} \Delta^{(k)}p_1\cdot\Delta^{(n-k)}p_1\,.
\end{equation}
\vskip 4pt
For conservative contributions, on the other hand, the fact that $\Delta p_2$ can be obtained from $\Delta p_1$ by a simple relabelling $b\rightarrow -b$, $u_1\leftrightarrow u_2$, together with $m_1\leftrightarrow m_2$, we immediately have additional constraints
\begin{equation}\label{eq:momConservation}
  \begin{aligned}
    c_{1b,\textrm{cons}}^{(n,i)} &= c_{1b,\textrm{cons}}^{(n,n-i+1)}\,,\\
    c_{1\check{u}_1,\textrm{cons}}^{(n,i)} &= -c_{1\check{u}_2,\textrm{cons}}^{(n,n-i+1)}\,,\,\, {\rm or} \,\,
    c_{1\check{u}_2,\textrm{cons}}^{(n,i)} &= -c_{1\check{u}_1,\textrm{cons}}^{(n,n-i+1)}\,.
  \end{aligned}
\end{equation}
This implies that the $c_{1\check{u}_a,\textrm{cons}}^{(n,i)}$ coefficients are uniquely determined by lower order values. The above conditions allow us to conveniently write the conservative part of the impulse in the $b$-direction as,
\begin{equation}\label{eq:DpConsStructure}
   \frac{1}{M\nu} c_{1b}^{(n)\textrm{cons}} = M^n \sum_{i=1}^{\lceil n/2 \rceil} c^{(n,i)}_{1b,\textrm{cons}}\, \nu^{i-1} N^{(n,i)}(\nu)\,,
\end{equation}
with\footnote{The reader will notice that the $i$-integer also counts the order in a {\it self-force} expansion, with the $i=1$ case corresponding to the test-body (or Schwarzschild) limit.} 
\begin{equation}
  N^{(n,i)}(\nu) \equiv 2^{n-2i}\sum_{k=0}^{\lfloor\frac{n-2i+1}{2} \rfloor}\binom{n-2i+1}{2k}(1-4\nu)^k\,,
\end{equation}
so that the $c^{(n)}_{1b}$ coefficients have the expected symmetry between the $m_1^p m_2^q$ and $m_1^q m_2^p$ contributions \cite{paper1,Damour:2019lcq}. Although the above manipulations streamline the derivation of the impulse, particularly in the conservative sector, e.g.~\cite{3pmeft}, we will use them instead as nontrivial consistency checks of the complete results.

\section{Master integrals}\label{sec:integrals}

As in many standard computations in collider physics, our remaining task to obtain an observable quantity, to a given order in the coupling (Newton's) constant, consists on evaluating a number of loop-type integrals; this time featuring linear and quadratic propagators, with either Feynman or retarded $i0$-prescriptions, as well as Dirac-$\delta$ functions. We give here a few additional details regarding the separation into families and reduction to a small(er) set of {\it master} integrals. As key examples we illustrate the procedure for the two- and three-loop computations, at 3PM and 4PM order, respectively.

\subsection{Integral families}\label{sec:PMints}
After performing the manipulations described before, the computation of the impulse to $(n+1)$PM order can be reduced to the following family of ($n$-loop) integrals:
\begin{align}\label{PMint}
\mathcal{M}^{(a_1 \cdots a_n; \pm\cdots\pm)}_{\,\,\alpha_1 \cdots \alpha_n; \beta_1 \cdots \beta_m}(\gamma) \,=\, 
\Bigg(\prod_{i=1}^{n}\int_{\ell_i} \, \frac{ \delta(\ell_i \cdotnew u_{a_i})}{(\pm \ell_i\cdot u_{\slashed{a}_i} \! - i0)^{\alpha_i}} \Bigg) 
\frac{(-q^2)^{\sigma -nd/2}}{D_1^{\beta_1} D_2^{\beta_2} \cdots D_m^{\beta_m}}\,,
\end{align}
where $\{\alpha_i, \beta_r\} \in\mathbb{Z}$, $\sigma=(\alpha_1 + \cdots + \alpha_n)/2 + \beta_1 + \cdots + \beta_m$, and $a_i\in\{1,2\}$ are particle labels with $\slashed{1}=2$, $\slashed{2}=1$. At $n$-loop order we have  $m=n(n+3)/2$, which follows by counting independent scalar products involving the loop momenta. The factors of $D_i$ capture graviton propagators, as well as any other irreducible scalar product, that may appear in the numerator, i.e.
\begin{align}\label{}
D_i = -P_i^2
\quad\text{with}\quad
P_i = \lambda_{ij} \ell_j + \lambda_i q, \quad \lambda_{ij}, \lambda_i \in\{0,\pm 1\} \quad
\textrm{for } 1\leqslant i \leqslant m\,.
\end{align}
The associated Green's functions can be Feynman or retarded ones, namely
\begin{align}\label{}
P^2
~~\textrm{is either}~~
\begin{cases}
P^2 + i0 & \textrm{for Feynman, or} \\
(P^0 \pm i0)^2 - \vecbf{P}^2 & \textrm{for retarded/advanced}\,.
\end{cases}
\end{align}
The external kinematical variables satisfy \eqref{kinematics}. Because the overall scaling is dictated by dimensional analysis, we set $-q^2=1$ thus rendering the integrals in \eqref{PMint} dimensionless functions of a single kinematical invariant, $\gamma$, namely the product of the incoming velocities. 
\vskip 4pt

A notable feature is that each loop integral is constrained by a Dirac-$\delta$ functions, \,$\delta(\ell_i\cdotnew u_a)$, linear in the loop momentum and velocity. For the purpose of utilizing powerful collider-physics tools, it is convenient to resort to the so-called  `reverse unitarity' trick~\cite{Cutkosky:1960sp, Anastasiou:2002yz, Anastasiou:2003ds},
\begin{align}\label{reverse-unitarity}
  {\delta}(\ell_i\cdotnew u_a) \,=\, {1 \over 2\pi i} \left({1 \over \ell_i\cdotnew u_a - i0} +  {1 \over -\ell_i\cdotnew u_a - i0}\right)\,,
\end{align}
which can then be interpreted as `cut' linear propagators. This allows us to directly apply the strategy of IBP reductions (and ultimately differential equations) to our integrals.\footnote{Let us emphasize that higher powers of cut linear propagators will appear in (intermediate steps of) the IBP reduction, corresponding to derivatives of the Dirac-$\delta$ functions. On the other hand, negative powers in cut propagators automatically vanish, which leads to simplifications in the IBP relations.} The arguments of the cut linear propagators also provide a natural classification within the set in \eqref{PMint}.
We will denote as $(a_1\cdots a_n)$ a given configuration featuring the product $\delta(\ell_1 \cdot u_{a_1})\cdots\delta(\ell_n \cdot u_{a_n})$ in the numerator.\footnote{The reader will immediately recognize $(11\cdots1)$ and $(22\cdots2)$ as the test-body, or Schwarzschild configuration. In this case we can readily solve all the constraints by going to the rest frame of the associated particle, which resolves all of the $\ell^0_i$ integrals.}
\vskip 4pt Breaking it further down, a set of cut, linear, and square propagators defines an \emph{integral family}.
Each family can be treated independently within the IBP reduction and differential equations method.
For a given family, we organize integrals into \emph{sectors}~$S_A$, which are the collections of integrals that share the same set of propagators with $\alpha_i > 0$ and $\beta_r > 0$.
Similarly, a sector $S_B$ is said to be a sub-sector of $S_A$ if the set of positive $\{\alpha_i,\beta_r\}_B$ of sector $S_B$ is a subset of positive $\{\alpha_i,\beta_r\}_A$ of sector $S_A$.

\subsection{IBP reduction}
\label{sec:IBP-red}

At the heart of the IBP reduction lies the fact that, in principle, there are many identities among the integrals in \eqref{PMint} for different indices $\alpha_i$ and $\beta_r$. Schematically, IBP identities can be obtained from the condition \cite{Tkachov:1981wb,Chetyrkin:1981qh, Anastasiou:2004vj}
\begin{align}\label{IBP}
0 \,=\, \int_{\{\ell_i\}}
{\partial \over \partial \ell_j^\mu}
\bigg({v^\mu  \over (\ell_1 \cdotnew u_{a})^{\alpha_1} \cdots D_1^{\beta_1} D_{2}^{\beta_2} \cdots}\bigg),
\end{align}
where  the $\ell_j$'s are any loop momenta, and $v^\mu$ is a linear combination of loop and external data. In general, the IBP relations generate, within a given family, a large system of homogeneous (linear) equations. There are, however, various nontrivial {\it symmetry relations}, often found by performing linear shifts of the loop momenta (with unit determinant), which can simplify the system. As we shall see, the $i0$-prescription of squared propagators, namely whether we use Feynman or retarded Green's functions, will play a key role in the validity of these relations. (That is the case because Feynman propagators are invariant under a flip of the momentum, $p \to -p$, whereas causal propagators are not.)\vskip 4pt
After using all of these relations, we may then express any member of the integral family as a linear combination of a finite number of basis integrals, referred to as {\it master} integrals. Due to the conspicuous relevance of Feynman integrals in particle physics, a large number of methods have been developed, following Laporta's algorithm \cite{Laporta:2000dsw}, utilizing either algebraic properties of the IBP system \cite{Smirnov:2006wh, Smirnov:2006tz, Lee:2008tj}, finite field techniques \cite{vonManteuffel:2014ixa,Peraro:2016wsq, Klappert:2019emp,Klappert:2020aqs,Peraro:2019svx}, 
module intersection methods \cite{Boehm:2018fpv}, and techniques for optimizations of the master integral basis \cite{Boehm:2020ijp,Usovitsch:2020jrk,Smirnov:2020quc,Bendle:2019csk}. On the practical side, these state-of-the-art tools have been implemented in many publicly available computer programs, such as \texttt{LiteRed} \cite{Lee:2012cn,Lee:2013mka}, \texttt{FIRE6} \cite{Smirnov:2019qkx,Smirnov:2020quc} and \texttt{Kira2} \cite{Maierhofer:2017gsa,Klappert:2020nbg}. We have used a combination of \texttt{LiteRed} and \texttt{FIRE6} in all our computations.
 
\subsection{Two loops}

\begin{figure}[h]
\centering
\vspace{0.2cm}
\begin{tikzpicture}[scale=1]
  \draw[boson] (0,0) -- (0,1);
  \draw[boson] (1,0) -- (1,1);
  \draw[boson] (2,0) -- (2,1);
  \draw (0,0) -- (2,0);
  \draw[dashed] (0,1) -- (2,1);
\end{tikzpicture}
\hspace{0.3cm}
\begin{tikzpicture}[scale=1]
  \draw[boson] (0,0) -- (0,1);
  \draw[boson] (1,0) -- (1,1);
  \draw[boson] (2,0) -- (2,1);
  \draw[dashed] (0,0) -- (1,0);
  \draw[dashed] (1,1) -- (2,1);
  \draw (1,0) -- (2,0);
  \draw (0,1) -- (1,1);
\end{tikzpicture}
\hspace{0.3cm}
\begin{tikzpicture}[scale=1]
  \draw[boson] (0,0) -- (0,1);
  \draw[boson] (0.66,1) .. controls (0.66,0.5) and (1.33,0.5) .. (1.33,1);
  \draw[boson] (2,0) -- (2,1);
  \draw[dashed] (0.66,1) -- (1.33,1);
  \draw (0,1) -- (0.66,1);
  \draw (1.33,1) -- (2,1);
  \draw[dashed] (0,0) -- (2,0);
\end{tikzpicture}
\hspace{0.3cm}
\begin{tikzpicture}[scale=1]
  \draw[boson] (0.8,0) -- (0.8,0.33);
  \draw[boson] (0.8,0.33) -- (1.6,1);
  \draw[boson] (0.8,0.33) -- (0.4,0.66);
  \draw[boson] (0.4,0.66) -- (0,1);
  \draw[boson] (0.4,0.66) -- (0.8,1);
  \draw[dashed] (0,1) -- (1.6,1);
\end{tikzpicture}
\hspace{0.3cm}
\begin{tikzpicture}[scale=1]
  \draw[boson] (0,0) -- (0,1);
  \draw[boson] (1.3,0) -- (1.3,0.5);
  \draw[boson] (1.3,0.5) -- (0.8,1);
  \draw[boson] (1.3,0.5) -- (1.8,1);
  \draw[dashed] (0,0) -- (1.3,0);
  \draw (0,1) -- (0.8,1);
  \draw[dashed] (0.8,1) -- (1.8,1);
\end{tikzpicture}
\hspace{0.3cm}
\begin{tikzpicture}[scale=1]
  \draw[boson] (0,0) -- (0,0.5);
  \draw[boson] (1,0) -- (1,0.5);
  \draw[boson] (0,0.5) -- (1,0.5);
  \draw[boson] (0,0.5) -- (0,1);
  \draw[boson] (1,.5) -- (1,1);
  \draw[dashed] (0,0) -- (1,0);
  \draw[dashed] (0,1) -- (1,1);
\end{tikzpicture}\\
\caption{Sample integral topologies at 3PM. (See \S\ref{3pmexample} for the diagrammatic rules.)}\label{feynGraphs-3pm}
\end{figure}
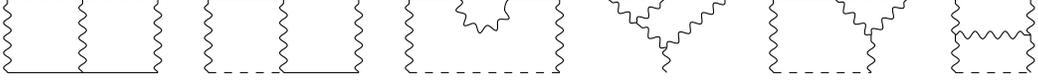

For the integral topologies at two-loop order, see Fig.\,\ref{feynGraphs-3pm}, there are at most five propagators with $\alpha_i>0$ and $\beta_r>0$. For instance, the ``H-diagram'' has five quadratic and no linear propagator(s), while others may feature some linear propagators but fewer quadratic ones. At this order it is straightforward to show that the full 3PM impulse (including also spin and finite-size effects) can be immersed into the following set of square propagators \cite{3pmeft,tidaleft}
\begin{align}\label{eq:intFam3PM}
\{D_i\}&= \{-\ell_1^2, -\ell_2^2, -(\ell_1 + \ell_2 - q)^2, -(\ell_1 - q)^2, -(\ell_2 - q)^2\}\,.
\end{align}
Due to the invariance under $u_1\leftrightarrow u_2$ only eight families of integrals, for this single set of square propagators, need to be considered:
\begin{align}
 (11;--)\,, & & (11;+-)\,, & & (11;-+)\,, & & (11;++)\,, & & (12;--)\,, & & (12;+-)\,, & & (12,-+)\,, & & (12;++)\,,
\end{align}
where we denoted a family by $(a_1a_2;\pm\pm)$, corresponding to the superscript in \eqref{PMint}.\vskip 4pt

For the case of Feynman-only propagators, we can use e.g.~\texttt{LiteRed} to detect symmetry relations. After performing an IBP reduction, using for instance \texttt{FIRE6}, we find a total of 21 master integrals appearing in the (spin-independent) impulse, 
\begin{equation}
\begin{aligned}
&\cM^{(11;--)}_{11;00111} & & \cM^{(11;+-)}_{01;00111} & & 
\cM^{(11;+-)}_{11;00111} & & \cM^{(11;++)}_{00;00111} & & 
\cM^{(11;++)}_{01;00111} & & \cM^{(11;++)}_{11;00111} & &
\cM^{(12;--)}_{11;00111} \\
&\cM^{(12;+-)}_{01;00111} & &
\cM^{(12;+-)}_{11;00111} & & \cM^{(12;++)}_{00;00111} & & 
\cM^{(12;++)}_{00;00112} & & \cM^{(12;++)}_{00;00211} & &
\cM^{(12;++)}_{00;01101} & & \cM^{(12;++)}_{00;11011} \\
&\cM^{(12;++)}_{00;11111} & & \cM^{(12;++)}_{00;11211} &&
\cM^{(12;++)}_{01;00111} & & \cM^{(12;++)}_{11;00111} & &
\cM^{(12;++)}_{\shortminusS10;00111} & & \cM^{(12;++)}_{\shortminusS 10;01101} & & 
\cM^{(12;++)}_{\shortminusS10;11111}\,.
\end{aligned}
\end{equation}
For the case of causal propagators, on the other hand, the lack of symmetry relations yields a larger number. In particular, a reduction with \texttt{FIRE6} returns 74 masters for the spin-independent case.\vskip 4pt
At this point, this result applies to unspecified signs of the $i0$'s in the causal propagators. Since we have not used any symmetry relations, which could exchange the order of propagators, these signs will trivially go along for the ride through the IBP machinery. This means, however, that the 74 master integrals need to be {\it dressed}, in principle, by $2^5$ combinations of different $i0$-prescriptions for the five square propagators.
Fortunately, the number is drastically reduced by realizing that the method of regions implies that only the $i0$-prescription for the $D_3$ propagator ever matters at 3PM.

\subsection{Three loops}

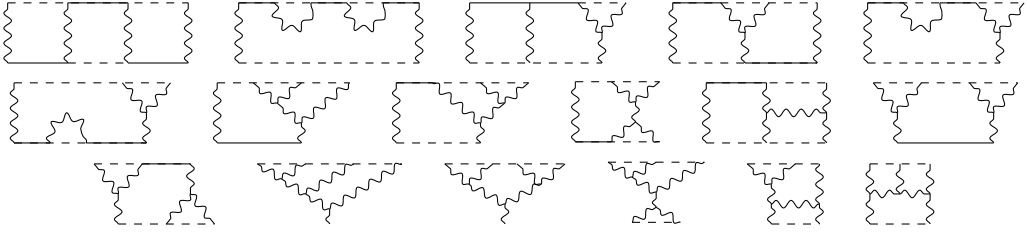
\begin{figure}[h]
\centering
\begin{tikzpicture}[scale=0.8]
  \draw[boson] (0,0) -- (0,1);
  \draw[boson] (1,0) -- (1,1);
  \draw[boson] (2,0) -- (2,1);
  \draw[boson] (3,0) -- (3,1);
  \draw (0,0) -- (1,0);
  \draw[dashed] (1,0) -- (2,0);
  \draw (2,0) -- (3,0);
  \draw[dashed] (0,1) -- (3,1);
  \draw (1,1) -- (2,1);
\end{tikzpicture}
\hspace{0.3cm}
\begin{tikzpicture}[scale=0.8]
  \draw[boson] (0,0) -- (0,1);
  \draw[boson] (0.6,1) .. controls (0.6,0.5) and (1.2,0.5) .. (1.2,1);
  \draw[boson] (1.8,1) .. controls (1.8,0.5) and (2.4,0.5) .. (2.4,1);
  \draw[boson] (3,0) -- (3,1);
  \draw[dashed] (0,1) -- (3,1);
  \draw (0,1) -- (0.6,1);
  \draw (1.2,1) -- (1.8,1);
  \draw (2.4,1) -- (3,1);
  \draw[dashed] (0,0) -- (3,0);
\end{tikzpicture}
\hspace{0.3cm}
\begin{tikzpicture}[scale=0.8]
  \draw[boson] (0,0) -- (0,1);
  \draw[boson] (1,0) -- (1,1);
  \draw[boson] (2.2,0) -- (2.2,0.5);
  \draw[boson] (2.2,0.5) -- (1.8,1);
  \draw[boson] (2.2,0.5) -- (2.6,1);
  \draw[dashed] (0,0) -- (2.2,0);
  \draw (0,1) -- (1.8,1);
  \draw[dashed] (1.8,1) -- (2.6,1);
\end{tikzpicture}
\hspace{0.3cm}
\begin{tikzpicture}[scale=0.8]
  \draw[boson] (0,0) -- (0,1);  
  \draw[boson] (1.2,0) -- (1.2,0.5);
  \draw[boson] (1.2,0.5) -- (0.8,1);
  \draw[boson] (1.2,0.5) -- (1.6,1);
  \draw[boson] (2.4,0) -- (2.4,1);
  \draw[dashed] (0,0) -- (2.4,0);
  \draw (1.2,0) -- (2.4,0);
  \draw[dashed] (0,1) -- (2.4,1);
  \draw (0,1) -- (0.8,1);
\end{tikzpicture}
\hspace{0.3cm}
\begin{tikzpicture}[scale=0.8]
  \draw[boson] (0,0) -- (0,1);  
  \draw[boson] (2.2,0) -- (2.2,0.5);
  \draw[boson] (2.2,0.5) -- (1.8,1);
  \draw[boson] (2.2,0.5) -- (2.6,1);
  \draw[boson] (0.6,1) .. controls (0.6,0.5) and (1.2,0.5) .. (1.2,1);
  \draw[dashed] (0,0) -- (2.2,0);
  \draw[dashed] (0,1) -- (2.6,1);
  \draw (0,1) -- (0.6,1);
  \draw (1.2,1) -- (1.8,1);
\end{tikzpicture}\\
\vspace{0.2cm}
\begin{tikzpicture}[scale=0.8]
  \draw[boson] (0,0) -- (0,1);  
  \draw[boson] (2.2,0) -- (2.2,0.5);
  \draw[boson] (2.2,0.5) -- (1.8,1);
  \draw[boson] (2.2,0.5) -- (2.6,1);
  \draw[boson] (0.6,0) .. controls (0.6,0.5) and (1.2,0.5) .. (1.2,0);
  \draw[dashed] (0,0) -- (2.2,0);
  \draw (0,0) -- (0.6,0);
  \draw (1.2,0) -- (2.2,0);
  \draw[dashed] (0,1) -- (2.6,1);
\end{tikzpicture}
\hspace{0.3cm}
\begin{tikzpicture}[scale=0.8]
  \draw[boson] (0,0) -- (0,1);
  \draw[boson] (1.4,0) -- (1.4,0.33);
  \draw[boson] (1.4,0.33) -- (2.2,1);
  \draw[boson] (1.4,0.33) -- (1,0.66);
  \draw[boson] (1,0.66) -- (0.6,1);
  \draw[boson] (1,0.66) -- (1.4,1);
  \draw (0,0) -- (1.4,0);
  \draw[dashed] (0,1) -- (2.2,1);
\end{tikzpicture}
\hspace{0.3cm}
\begin{tikzpicture}[scale=0.8]
  \draw[boson] (0,0) -- (0,1);
  \draw[boson] (1.4,0) -- (1.4,0.33);
  \draw[boson] (1.4,0.33) -- (0.6,1);
  \draw[boson] (1.4,0.33) -- (1.8,0.66);
  \draw[boson] (1.8,0.66) -- (2.2,1);
  \draw[boson] (1.8,0.66) -- (1.4,1);
  \draw[dashed] (0,0) -- (1.4,0);
  \draw[dashed] (0,1) -- (2.2,1);
  \draw (0,1) -- (0.6,1);
\end{tikzpicture}
\hspace{0.3cm}
\begin{tikzpicture}[scale=0.8]
  \draw[boson] (0,0) -- (0,1);
  \draw[boson] (0.6,0) -- (1,0.33);
  \draw[boson] (1.4,0) -- (1,0.33);
  \draw[boson] (1,0.33) -- (1,0.66);
  \draw[boson] (1,0.66) -- (0.6,1);
  \draw[boson] (1,0.66) -- (1.4,1);
  \draw[dashed] (0,0) -- (1.4,0);
  \draw (0,0) -- (0.6,0);
  \draw[dashed] (0,1) -- (1.4,1);
\end{tikzpicture}
\hspace{0.3cm}
\begin{tikzpicture}[scale=0.8]
  \draw[boson] (0,0) -- (0,1);
  \draw[boson] (1,0) -- (1,0.5);
  \draw[boson] (2,0) -- (2,0.5);
  \draw[boson] (1,0.5) -- (2,0.5);
  \draw[boson] (1,0.5) -- (1,1);
  \draw[boson] (2,0.5) -- (2,1);
  \draw[dashed] (0,0) -- (2,0);
  \draw[dashed] (0,1) -- (2,1);
  \draw (0,1) -- (1,1);
\end{tikzpicture}
\hspace{0.3cm}
\begin{tikzpicture}[scale=0.8]
  \draw[boson] (0,0) -- (0,0.5);
  \draw[boson] (0,0.5) -- (-0.4,1);
  \draw[boson] (0,0.5) -- (0.4,1);
  \draw[boson] (1.6,0) -- (1.6,0.5);
  \draw[boson] (1.6,0.5) -- (1.2,1);
  \draw[boson] (1.6,0.5) -- (2,1);
  \draw (0,0) -- (1.6,0);
  \draw[dashed] (-0.4,1) -- (2,1);
\end{tikzpicture}\\
\vspace{0.2cm}
\begin{tikzpicture}[scale=0.8]
  \draw[boson] (0,0) -- (0,0.5);
  \draw[boson] (0,0.5) -- (-0.4,1);
  \draw[boson] (0,0.5) -- (0.4,1);
  \draw[boson] (1.2,1) -- (1.2,0.5);
  \draw[boson] (1.2,0.5) -- (0.8,0);
  \draw[boson] (1.2,0.5) -- (1.6,0);
  \draw[dashed] (0,0) -- (1.6,0);
  \draw[dashed] (-0.4,1) -- (1.2,1);
  \draw (0.4,1) -- (1.2,1);
\end{tikzpicture}
\hspace{0.3cm}
\begin{tikzpicture}[scale=0.8]
  \draw[boson] (0,0) -- (0,0.25);
  \draw[boson] (0,0.25) -- (-0.4,0.5);
  \draw[boson] (-0.4,0.5) -- (-0.8,0.75);
  \draw[boson] (-0.8,0.75) -- (-1.2,1);
  \draw[boson] (-0.8,0.75) -- (-0.4,1);
  \draw[boson] (-0.4,0.5) -- (0.4,1);
  \draw[boson] (0,0.25) -- (1.2,1);
  \draw[dashed] (-1.2,1) -- (1.2,1);
\end{tikzpicture}
\hspace{0.3cm}
\begin{tikzpicture}[scale=0.8]
  \draw[boson] (0,0) -- (0,0.33);
  \draw[boson] (0,0.33) -- (-0.6,0.66);
  \draw[boson] (-0.6,0.66) -- (-1,1);
  \draw[boson] (-0.6,0.66) -- (-0.2,1);
  \draw[boson] (0,0.33) -- (0.6,0.66);
  \draw[boson] (0.6,0.66) -- (0.2,1);
  \draw[boson] (0.6,0.66) -- (1,1);
  \draw[dashed] (-1,1) -- (1,1);
\end{tikzpicture}
\hspace{0.3cm}
\begin{tikzpicture}[scale=0.8]
  \draw[boson] (-0.4,0) -- (0,0.25);
  \draw[boson] (0.4,0) -- (0,0.25);
  \draw[boson] (0,0.25) -- (0,0.5);
  \draw[boson] (0,0.5) -- (-0.4,0.75);
  \draw[boson] (-0.4,0.75) -- (-0.8,1);
  \draw[boson] (-0.4,0.75) -- (0,1);
  \draw[boson] (0,0.5) -- (0.8,1);
  \draw[dashed] (-0.4,0) -- (0.4,0);
  \draw[dashed] (-0.8,1) -- (0.8,1);
\end{tikzpicture}
\hspace{0.3cm}
\begin{tikzpicture}[scale=0.8]
  \draw[boson] (-0.4,0) -- (-0.4,0.33);
  \draw[boson] (0.4,0) -- (0.4,0.33);
  \draw[boson] (-0.4,0.33) -- (0.4,0.33);
  \draw[boson] (-0.4,0.33) -- (-0.4,0.66);
  \draw[boson] (-0.4,0.66) -- (-0.8,1);
  \draw[boson] (-0.4,0.66) -- (0,1);
  \draw[boson] (0.4,0.33) -- (0.4,1);
  \draw[dashed] (-0.4,0) -- (0.4,0);
  \draw[dashed] (-0.8,1) -- (0.4,1);
\end{tikzpicture}
\hspace{0.3cm}
\begin{tikzpicture}[scale=0.8]
  \draw[boson] (-0.5,0) -- (-0.5,0.5);
  \draw[boson] (0.5,0) -- (0.5,0.5);
  \draw[boson] (-0.5,0.5) -- (0,0.5);
  \draw[boson] (0,0.5) -- (0.5,0.5);
  \draw[boson] (-0.5,0.5) -- (-0.5,1);
  \draw[boson] (0.5,0.5) -- (0.5,1);
  \draw[boson] (0,0.5) -- (0,1);
  \draw[dashed] (-0.5,0) -- (0.5,0);
  \draw[dashed] (-0.5,1) -- (0.5,1);
\end{tikzpicture}
\caption{Sample scalar integral topologies at 4PM. (See \S\ref{3pmexample} for the diagrammatic rules.)}\label{feynGraphs-4pm}
\end{figure}

At 4PM order, there are at most seven propagators, see Fig.~\ref{feynGraphs-4pm}.
There are also 15 independent scalar products in total, involving the three loop momenta, the incoming velocity or the transfer momentum.
We use the freedom in the choice of scalar products to complete the $\{D_i\}$ set, such that our integrals can be collected into two sets of square propagators.
We~found that all scalar integrals needed for computing the spin-independent 4PM impulse can be immersed into the following two sets of square propagators \cite{4pmeft,4pmeft2,4pmeftot}
\begin{equation}\label{eq:props4PM}
\begin{aligned}
\{D_i^\textrm{I}\} &= \big\{
-\ell_1^2,
-\ell_2^2,
-\ell_3^2,
-(\ell_1 {-} q)^2,
-(\ell_2 {-} q)^2,
-(\ell_3 {-} q)^2
-(\ell_1 {-} \ell_2)^2,
-(\ell_2 {-} \ell_3)^2,
-(\ell_1 {-} \ell_3)^2
\big\},
\\[0.35 em]
\{D_i^\textrm{II}\} &= \big\{
-\ell_1^2,
-\ell_2^2,
-\ell_3^2,
-(\ell_1 {-} q)^2,
-(\ell_2 {-} q)^2,
-(\ell_3 {-} q)^2,
-(\ell_{12} {-} q)^2,
-(\ell_{23} {-} q)^2,
-(\ell_{123} {-} q)^2
\big\},
\end{aligned}
\end{equation}
with $\ell_{i\ldots j} \equiv \ell_i {+} \cdots {+} \ell_j$.\vskip 4pt There is clearly a large overlap of integrals between these two sets. In general, we give priority to the first set due to a larger number of symmetries. Using the invariance under $u_1\leftrightarrow u_2$, we have for each of these sets 16 families in which all integrals can be embedded. After an IBP reduction, allowing for symmetry relations of Feynman-type propagators, we find a total of 576 master integrals,\footnote{Let us point out that, by allowing various identities between families, it is in principle possible to reduce the total number by around a factor of three. However, such identities, which we only worked out at the level of master integrals, make the process of solving the differential equations (we discuss next) somewhat harder. We have therefore decided not to use such relations, thus keeping the full set of 576 master integrals.}   of which 266 are needed for the determination of the  $c_{1b}^{(4),\mathrm{cons}}$ coefficient, yielding the full 4PM conservative sector, see \S\ref{subsec:integrandStructure}. As a consistency check, we have also computed the remaining 310 master integrals, along the incoming velocities, which served as a powerful cross-check.\vskip 4pt On the other hand, for causal propagators, not allowing for symmetry relations, \texttt{FIRE6} returns a total of 1094 master integrals. Once again, all of these master integrals have to be dressed with various $i0$-prescriptions for the given square propagators. Luckily, by the method of regions, only the $i0$'s of the last three propagators in \eqref{eq:props4PM}, and moreover not concurrently, contribute at 4PM order.

\section{Differential equations}\label{sec:de}

 The method of ordinary differential equations is a powerful way to determine the dependency of loop integrals on kinematical variables. In the previous sections, we have shown that in the worldline EFT formalism the key to solving the relativistic two-body problem boils down to the evaluation of the $\gamma$-dependence of a number of master integrals. Therefore, this method provides us with a novel way to {\it bootstrap} relativistic information using boundary data computed in the near-static limit.\vskip 4pt

For a set of master integrals in a family, collected in a vector~$\vec{f}$, we take the derivative of the basis integrals with respect to the kinematic variable and use IBP identities to write the result again in terms of master integrals.
This yields a system of first-order differential equations of the form
\begin{align}
\label{eq:des-non-can}
 \frac{\partial}{\partial x}\vec{f}(x,\epsilon)=A(x,\epsilon)\,\vec{f}(x,\epsilon)\,,
\end{align}
where the coefficient matrix $A(x,\epsilon)$ is rational in $x$ and $\epsilon$, and the variable $x$ is introduced in order to rationalize the square root $\sqrt{\gamma^2 {-} 1}$ \cite{Parra}:
\begin{align}\label{}
\sqrt{\gamma^2 - 1} = {1 - x^2 \over 2x},
~~\Longrightarrow~~
d\gamma =  {x^2 - 1 \over 2x^2}\,dx,
\quad 1 \le \gamma,\quad 0< x \le 1.
\end{align}
 
\subsection{General method}

We start by discussing the general structure of the differential equations before we show how to solve them through a particular choice of basis. We also cover the case of integrals involving elliptic functions, appearing for the first time at 4PM order.

\subsubsection{Block-triangular form}
\label{sec:block-triangular}
For a given family, we order its masters integrals in the vector $\vec{f}$ as follows: 
\begin{equation}
 \vec{f} = \begin{pmatrix}
          \vec{f}_{S_1} \\
          \vdots \\
          \vec{f}_{S_s}
         \end{pmatrix},
\end{equation}
where $\vec{f}_{S_k}$ denotes the master integrals in sector $S_k$ and we ordered the sectors such that the number of denominators (i.e.~the number of positive indices $\alpha_i$ and $\beta_r$) of $S_j$ is bigger or equal to that of $S_k$ when $j>k$. As a result, the differential equations will take the following lower-block-triangular form:
\begin{equation}
\label{eq:block}
\frac{\partial}{\partial x}\begin{pmatrix}
          \vec{f}_{S_1} \\
          \vec{f}_{S_2} \\
          \vdots \\
          \vec{f}_{S_s}
         \end{pmatrix}=\begin{pmatrix}
                D_{1} & 0 & \ldots & 0 \\
                C_{2,1} & D_{2} & \ldots & 0 \\
                \vdots & \vdots & \ddots & \vdots \\
                C_{s,1} & C_{s,2} & \ldots & D_{s}
               \end{pmatrix}\begin{pmatrix}
          \vec{f}_{S_1} \\
          \vec{f}_{S_2} \\
          \vdots \\
          \vec{f}_{S_s}
         \end{pmatrix}\,
\end{equation}
The diagonal blocks, $D_i$, and off-diagonal blocks, $C_{i,j}$, are matrices of size $n_i\times n_i$ and $n_i\times n_j$, respectively, with $n_i$ the number of master integrals in $S_i$.
In practice, because not all integral sectors $S_j$ with $j<i$ are truly sub-sectors of $S_i$, the matrix in \eqref{eq:block} is usually very sparse, in the sense that many of the off-diagonal blocks vanish. Note that, in the context of differential equations, it is also natural to refer to the diagonal blocks themselves as sectors. 

\subsubsection{The \texorpdfstring{$\epsilon$}{eps}-form}
\label{sec:eps-form}

A convenient way to solve the differential equations in \eqref{eq:des-non-can} is by finding a transformation, $T$, and basis,
\begin{align}
 \vec{g}(x,\epsilon) =T(x,\epsilon)\, \vec{f}(x,\epsilon),
\end{align}
that bring the system into $\epsilon$-form \cite{Henn:2013pwa}:
\begin{align}
\label{eq:des-can}
 \frac{\partial}{\partial x}\vec{g}(x,\epsilon)=\epsilon \tilde{A}(x)\,\vec{g}(x,\epsilon)\,,
\end{align}
where the coefficient matrix now
depends on $\epsilon$ only through the overall factor, and moreover it has only logarithmic singularities. The latter criterion means that the differential equations have only regular and no essential singularities. The form in \eqref{eq:des-can} is useful since it is then straightforward to write down the solution for the integrals $\vec{g}$ in an expansion in $\epsilon$,
concretely, with
\begin{align}
\label{eq:gcan-exp}
 \vec{g}(x,\epsilon)=\sum_{k}\epsilon^k\,\vec{g}^{(k)}(x)\,,
\end{align}
one finds from \eqref{eq:des-can} that
\begin{align}
\label{eq:iterated-result}
 \vec{g}^{(k)}(x)=\int_{x_0}^x\tilde{A}(x^\prime)\,\vec{g}^{(k-1)}(x^\prime)\mathrm{d}x^\prime+\vec{g}_0^{(k)}\,,
\end{align}
where we expand also the boundary constants $\vec{g}_0(\epsilon)=\sum_{k} \epsilon^k\vec{g}_0^{(k)}$.\vskip 4pt

The difficulty of computing master integrals (in an $\epsilon$-expanded form) is hence reduced to finding the transformation $T$. The latter task can be further simplified by utilizing the lower-block-triangular form described in \S\ref{sec:block-triangular}, see also \cite{Lee:2014ioa}.\vskip 4pt As an example, let us assume that there are only two sectors $S_1$ and $S_2$. Then, one first uses transformations of the form
\begin{equation}
\label{eq:diag-transf}
 \begin{pmatrix}
  \vec{g}_{S_1}^\text{\,diag}\\
  \vec{g}_{S_2}^\text{\,diag}
 \end{pmatrix}=
\begin{pmatrix}
  T_1 & 0 \\
  0 & T_2
 \end{pmatrix}
 \begin{pmatrix}
  \vec{f}_{S_1} \\
  \vec{f}_{S_2}
 \end{pmatrix},
\end{equation}
to bring the $D_i$ blocks into $\epsilon$-form
\begin{equation}
 \frac{\partial}{\partial x}\begin{pmatrix}
  \vec{g}_{S_1}^\text{\,diag}\\
  \vec{g}_{S_2}^\text{\,diag}
 \end{pmatrix}=
\begin{pmatrix}
  \epsilon\tilde{D}_1(x) & 0 \\
  \hat{C}_{2,1}(\epsilon,x) & \epsilon\tilde{D}_2(x)
 \end{pmatrix}
 \begin{pmatrix}
  \vec{g}_{S_1}^\text{\,diag}\\
  \vec{g}_{S_2}^\text{\,diag}
 \end{pmatrix}.
\end{equation}
After doing this for all diagonal blocks, one can then use transformations
\begin{equation}
\label{eq:od-transf}
 \begin{pmatrix}
  \vec{g}_{S_1}\\
  \vec{g}_{S_2}
 \end{pmatrix}=
\begin{pmatrix}
  \mathbb{1} & 0 \\
  T_{2,1} & \mathbb{1}
 \end{pmatrix}
 \begin{pmatrix}
  \vec{g}_{S_1}^\text{\,diag}\\
  \vec{g}_{S_2}^\text{\,diag}
 \end{pmatrix},
\end{equation}
to do the same for the off-diagonal blocks, such that\
\begin{equation}
 \frac{\partial}{\partial x}\begin{pmatrix}
  \vec{g}_{S_1}\\
  \vec{g}_{S_2}
 \end{pmatrix}=
\begin{pmatrix}
  \epsilon\tilde{D}_1(x) & 0 \\
  \epsilon\tilde{C}_{2,1}(x) & \epsilon\tilde{D}_2(x)
 \end{pmatrix}
 \begin{pmatrix}
  \vec{g}_{S_1}\\
  \vec{g}_{S_2}
 \end{pmatrix}.
\end{equation}
Note that \eqref{eq:diag-transf} amounts to finding a transformation on a homogeneous system, whereas \eqref{eq:od-transf} only concerns transformations on the inhomogeneous part. For this reason, finding the off-diagonal transformations $T_{i,j}$ is usually much easier than finding the $T_i$'s.

\subsubsection{The canonical form and multiple polylogarithms}

In the majority of cases, the matrices $\tilde{D}_i(x)$ and $\tilde{C}_{i,j}(x)$ can be written in the form
\begin{align}
\label{eq:dlog-form}
 \sum_l M_l\, \partial_x\log \alpha_l(x)\,,
\end{align}
where $M_l$ are constant matrices. If this is the case, the differential equation or the block is said to be in \emph{canonical} form \cite{Henn:2013pwa}. The algebraic functions $\alpha_l(x)$ are then called the \emph{letters} and the set of all independent letters is called the \emph{alphabet}.

\vskip 4pt
As described in \cite{Lee:2014ioa}, finding a canonical form can be done in three steps: First, one removes essential singularities, after which the differential equations matrix is in Fuchsian form, $\sum_l \hat{M}_l(\epsilon)\partial_x \log\alpha_l(x)$. Then, one normalizes the eigenvalues of the coefficient-matrices $\hat{M}_l(\epsilon)$ such that\ they are proportional to $\epsilon$. Lastly, one finds an $x$-independent transformation that simultaneously factorizes the $\epsilon$-dependence from all coefficient matrices.
We note that this is, in general, the simplest of all steps. Furthermore, for off-diagonal blocks, the normalization of eigenvalues is not necessary and the {\it Fuchsification} is usually much easier.
\vskip 4pt
There exist multiple implementations of this algorithm, of which we use the publicly available \texttt{Mathematica} package \texttt{Libra} \cite{Lee:2020zfb} and the $\texttt{C}{+}{+}$ code \texttt{epsilon} \cite{Prausa:2017ltv}.
We note, however, that the algorithm of \cite{Lee:2014ioa} can only find transformations that are rational in $x$ and $\epsilon$.
Whenever that is the case, the result in \eqref{eq:iterated-result} can be expressed in terms of \emph{multiple polylogarithms} (MPLs) \cite{Chen:1977oja,Goncharov:2001iea}:
\begin{align}\label{MPL-def}
G(a_1,\ldots, a_n; z) = \int_0^z {dt \over t - a_1}\, G(a_2,\ldots, a_{n}; t), \quad
G(~;z)=1,
\end{align}
where $a_i,z\in\mathbb{C}$. For the special case where all $a_i=0$, one defines
\begin{align}\label{MPL-def-log}
G(\vec{0}_n; z) = {1 \over n!} \log^n (z),
\end{align}
with $\vec{a}_n = (\underbrace{a,\ldots,a}_{n})$.
MPLs contain the ordinary logarithm,
\begin{equation}
 G(\vec{a}_n;z)=\frac{1}{n!}\log^n\left(1-\frac{z}{a}\right),\qquad a\neq 0
\end{equation}
and the well-known classical polylogarithm
\begin{align}\label{MPL-Li}
\text{Li}_{n}(z) \equiv - G(\vec{0}_{n-1},1; z).
\end{align}
Depending on the boundary point, we will also use the following relation recursively to convert MPLs with argument $1-x$ to those with argument $x$:\footnote{For $a_1=1$, this formula is to be understood in a regularized way similar to \eqref{MPL-def-log}.}
\begin{equation}
\label{eq:MPL-arg-shift}
 G(a_1,a_2,\ldots,a_n;1-x)=G(a_1,a_2,\ldots,a_n;1)+\int_0^x\frac{\mathrm{d}t}{t-(1-a_1)}G(a_2,\ldots,a_n;1-t)\,.
\end{equation}
For more details on MPLs, see e.g.\,\cite{Duhr:2014woa, Duhr:2019tlz}.

\subsubsection{Elliptic integrals}
\label{sec:elliptic-integrals}

As stated above, the algorithm in \cite{Lee:2014ioa} can find a canonical form whenever the transformation involves only rational functions. At three-loop order the algorithm fails due to the presence of elliptic integrals.
In the following we describe a systematic procedure to bring sectors $S_\textrm{ell}$ containing elliptic integrals into $\epsilon$-form. We concentrate here on the diagonal blocks.
The procedure for the off-diagonal blocks is conceptually easier, and is discussed in App.~\ref{sec:app-ell-off-diag}.\vskip 4pt

Let us assume that there is a diagonal block, $D_{\text{ell}}$, where the above algorithm fails. Specifically, it fails to normalize the eigenvalues through rational transformations. In order to find an alternative path forward, we multiply the master integrals in $\vec{f}_{S_{\textrm{ell}}}$ by different $\epsilon$-dependent factors, such that the diagonal block starts at $\mathcal{O}(\epsilon^0)$:
\begin{align}
 D_{\text{ell}}(x,\epsilon)=\sum_{k=0}^\infty \epsilon^k D_{\text{ell}}^{(k)}(x)\,.
\end{align}
The main idea now is to remove the constant piece $D_{\text{ell}}^{(0)}(x)$ through the transformation $\vec{f}_{S_{\text{ell}}}^\prime=T_{\text{ell}}^{(0)}\vec{f}_{S_{\text{ell}}}$, defined by
\begin{align}
\label{eq:T0}
 \frac{\partial}{\partial x}[T_{\text{ell}}^{(0)}(x)]^{-1}=D_{\text{ell}}^{(0)}[T_{\text{ell}}^{(0)}(x)]^{-1}\,.
\end{align}
This system of differential equations is independent of $\epsilon$, and therefore it can often be solved directly in \texttt{Mathematica} by converting it to a higher-order differential (Picard-Fuchs) equation.
After implementing the transformation $T_{\text{ell}}^{(0)}$, the diagonal block becomes
\begin{equation}
 \hat{D}_{\text{ell}}(x,\epsilon)=\sum_{k=1}^\infty \epsilon^k \hat{D}_{\text{ell}}^{(k)}(x)\,.
\end{equation}
Although $T_{\text{ell}}^{(0)}$ does not quite transform the diagonal block into $\epsilon$-form, it is safe to assume that the functions entering in $T_{\text{ell}}^{(0)}$ are nonetheless involved in the transformation that  produces an $\epsilon$-form. That is because the step of removing $D_{\text{ell}}^{(0)}$ is very similar to the normalization of eigenvalues mentioned above, with the difference that we could still encounter essential poles. In most cases, one finds that only rational functions, logarithms and possibly square roots appear.
For square roots, we can use rationalization through a change of variables, as we did with the introduction of the $x$ variable.
However, it may happen that more complicated functions appear, especially for second and third-order differential equations. An important class of such functions is given by the complete elliptic integrals of the first and second kind, i.e.\
\begin{align}
 \mathrm{K}(z) &= \int_0^1\frac{\mathrm{d}t}{\sqrt{(1-t^2)(1-zt^2)}}\,,\quad {\rm and} \quad
\mathrm{E}(z) = \int_0^1\mathrm{d}t\frac{\sqrt{1-zt^2}}{\sqrt{1-t^2}}\,,
\end{align}
respectively.
Even though, in principle, we can identify the class of functions needed to achieve an $\epsilon$-form, in practice, we cannot simply go back and modify the above-mentioned algorithms to incorporate such cases. (At the time of this writing we are not aware of an existent algorithm in the literature.)  Instead, one can
try to make an ansatz for the $\epsilon$-form, and subsequently fix the coefficients through suitable constraints, see e.g.~\cite{Meyer:2017joq,Adams:2018yfj} or \cite{Dlapa:2020cwj,Dlapa:2022wdu}. For our derivations, we have resorted to the \texttt{INITIAL} algorithm introduced in \cite{Dlapa:2020cwj,Dlapa:2022wdu}, see also \S\ref{sec:DE-examples-4PM} for a few more details.

\subsection{Two loops}
\label{sec:3PM-de-example}

Let us start with the integrals in \eqref{PMint} at 3PM order, and consider master integrals of the sector $\mathcal{M}^{(12;++)}_{00;\beta_1\dots\beta_5}$, with $\beta_1,\dots,\beta_5>0$, and its subsectors. For brevity, we omit in what follows some of the indices, and write\, $\mathcal{M}_{\beta_1\dots\beta_5}\equiv\mathcal{M}^{(12;++)}_{00;\beta_1\dots\beta_5}$. There are seven master integrals, which we choose as
\begin{equation}
 \vec{f}=
 \big(
          \mathcal{M}_{01101},
          \mathcal{M}_{00111},
          \mathcal{M}_{00112},
          \mathcal{M}_{00211},
          \mathcal{M}_{11011},
          \mathcal{M}_{11111},
          \mathcal{M}_{11211}
\big)^{\mathrm T}\,.
\end{equation}
As advertised, the differential equations have lower-block-triangular structure:
\begin{equation}\label{eq:3PM-de-form}
 \partial_x\vec{f}=
\begin{tikzpicture}[baseline]
\matrix (m)[
    matrix of math nodes,
    left delimiter=(,right delimiter=),
  ] {
    \star & 0 & 0 & 0 & 0 & 0 & 0 \\
    0 & \star & \star & \star & 0 & 0 & 0 \\
    0 & \star & \star & \star & 0 & 0 & 0 \\
    0 & \star & \star & \star & 0 & 0 & 0 \\
    0 & 0 & 0 & 0 & 0 & 0 & 0 \\
    \star & \star & \star & \star & \star & \star & \star \\
    \star & \star & \star & \star & \star & \star & \star \\
  } ;
  \draw ($1*(m-1-1.south west)$) rectangle ($1*(m-1-1.north east)$);
  \draw ($1*(m-4-2.south west)+(0.015,0)$) rectangle ($1*(m-2-4.north east)+(0,0.055)$);
  \draw ($1*(m-5-5.south west)+(0.014,0)$) rectangle ($1*(m-5-5.north east)+(0,-0.014)$);
  \draw ($1*(m-7-6.south west)+(0.014,0)$) rectangle ($1*(m-6-7.north east)+(0,-0.013)$);
\end{tikzpicture}
\vec{f}\,,
\end{equation}
where the $\star$ indicates a non-zero element, and we have marked the diagonal blocks (sectors). By analyzing the elements of \eqref{eq:3PM-de-form}, we find that the differential equations in $x$ have the singular points $0,1,-1,\infty$, as well as $i$ and $-i$. However, the latter two points are spurious, because the eigenvalues of their coefficient matrices do not depend on $\epsilon$, and therefore will be removed during the second step of the canonicalization procedure. Both, \texttt{Libra} and \texttt{epsilon} readily find a canonical basis for the diagonal and off-diagonal blocks. The resulting differential equations become
\begin{equation}
 \partial_x\vec{g}=\epsilon\left(\frac{\tilde{M}_0}{x}+\frac{\tilde{M}_1}{x-1}+\frac{\tilde{M}_{-1}}{x+1}\right)\vec{g},
\end{equation}
with
\begin{equation}
 \tilde{M}_0=\begin{pmatrix}
      2 & 0 & 0 & 0 & 0 & 0 & 0  \\
      0 & 2 & 2 & 0 & 0 & 0 & 0  \\
      0 & -2 & 0 & -3 & 0 & 0 & 0  \\
      0 & 0 & 4 & -6 & 0 & 0 & 0  \\
      0 & 0 & 0 & 0 & 0 & 0 & 0  \\
      0 & 0 & 0 & 0 & 0 & 2 & -2  \\
      0 & 0 & 0 & 0 & 0 & 2 & -2  
     \end{pmatrix},\qquad
 \tilde{M}_{\pm 1}=\begin{pmatrix}
      -2 & 0 & 0 & 0 & 0 & 0 & 0  \\
      0 & -2 & 0 & 0 & 0 & 0 & 0  \\
      0 & 0 & 0 & 0 & 0 & 0 & 0  \\
      0 & 0 & 0 & 6 & 0 & 0 & 0  \\
      0 & 0 & 0 & 0 & 0 & 0 & 0  \\
      -1 & 1 & -1 & 1 & 0 & -2 & 0  \\
      2 & 0 & 0 & -1 & \pm 1 & 0 & 2  
     \end{pmatrix}.
\end{equation}
Following \eqref{eq:iterated-result}, we can write down the solution as an expansion in $\epsilon$ to any desired order. As an example, the result for the fourth integral to first order in $\epsilon$ is given by 
\begin{equation}
 g_4(x,\epsilon)=g_{0,4}^{(0)}+\epsilon\left[g_{0,4}^{(1)}+4g_{0,3}^{(0)}G(0;x)+6g_{0,4}^{(0)}(G(-1;x)-G(0;x)+G(1;x)-\log 2)\right]+\mathcal{O}(\epsilon^2)\,,
\end{equation}
where $g_{0,j}^{(k)}$ are the boundary constants in \eqref{eq:iterated-result} at ${\cal O}(\epsilon^k)$ and we take the integrals' normalization such that their $\epsilon$-expansion starts at $\mathcal{O}(\epsilon^0)$. We return in \S\ref{sec:bc} to the computation of the boundary constants.

\subsection{Three loops}
\label{sec:DE-examples-4PM}

Blocks that do not contain elliptic integrals can be brought into canonical form using either \texttt{epsilon} or \texttt{Libra}. However, we encounter a sector where both programs fail to normalize all the eigenvalues. A possible choice for the three master integrals in this sector is given by \begin{equation}
 \vec{f}(x,\epsilon)=\begin{pmatrix}
          \mathcal{M}^{(112;+++)}_{000;010101110} \\
          \mathcal{M}^{(112;+++)}_{000;010101120} \\
          \mathcal{M}^{(112;+++)}_{000;010102110}
         \end{pmatrix}\, .
\end{equation}
 Multiplying the second and third integral with one power of $\epsilon^{-1}$ makes the diagonal block of the differential equation matrix finite as $\epsilon\to0$: 
\begin{equation}
 A(x,\epsilon)=\begin{pmatrix}
       \frac{1-x^2}{2x(1+x^2)} & \frac{1+x^2}{4x(1-x^2)} & \frac{3x}{(1-x^2)(1+x^2)} \\
       -\frac{1-x^2}{x(1+x^2)} & \frac{3(1+x^2)}{2x(1-x^2)} & -\frac{6x}{(1-x^2)(1+x^2)} \\
       \frac{1-x^2}{x(1+x^2)} & -\frac{1+x^2}{2x(1-x^2)} & -\frac{1-4x^2+x^4}{x(1-x^2)(1+x^2)}
      \end{pmatrix}+\mathcal{O}(\epsilon)\, .
\end{equation}
Truncating at $\cO(\epsilon)$, the system of first-order differential equations is equivalent to a single third-order differential equation (for the first integral $f_1(x,\epsilon)$):
\begin{equation}
\label{eq:PF-explicit}
 \left[\partial_x^3-\frac{6x}{1-x^2}\partial_x^2-\frac{1-4x^2+7x^4}{x^2(1-x^2)^2}\partial_x-\frac{1+x^2}{x^3(1-x^2)}+\mathcal{O}(\epsilon)\right]f_1(x,\epsilon)=0\,.
\end{equation}
As it turns out, the solutions can be found using \texttt{Mathematica}, yielding
\begin{align}
\label{eq:PF-solutions}
 &x\mathrm{K}^2(1-x^2)\,,& &x\mathrm{K}(1-x^2)\mathrm{K}(x^2)\,,& &x\mathrm{K}^2(x^2)\,.
\end{align}
Alternatively, we can solve \eqref{eq:PF-explicit} by noticing that it is the symmetric square of a second-order differential equation, see e.g.\ \cite{Primo:2017ipr}.\vskip 4pt

The solutions in \eqref{eq:PF-solutions} confirm that the we must extend the class of functions used in the transformation to $\epsilon$-form. As described at the end of \S\ref{sec:elliptic-integrals}, we can find the transformation, e.g.\ through the \texttt{INITIAL} algorithm \cite{Dlapa:2020cwj,Dlapa:2022wdu}, after making an ansatz for the $\epsilon$-form. Following the analysis in \cite{Dlapa:2022wdu}, as well as other known examples, see e.g.\ \cite{Adams:2018yfj}, we take
\begin{equation}
 \tilde{D}_{\text{ell}}=\frac{\pi^2}{x(1-x^2)\mathrm{K}^2(1-x^2)}\sum_{k,l\geq0}\hat{M}_{k,l}\,x^k\left(\frac{\mathrm{K}^{2}(1-x^2)}{\pi^{2}}\right)^l,
\end{equation}
subject to the condition that essential singularities are not present. In order to build constraints that determine the unknown constant matrices $\hat{M}_{k,l}$, the \texttt{INITIAL} algorithm requires as a starting point an integral that is not changed by the transformation. We find that
${\pi^2f_1(x,\epsilon)/(x\mathrm{K}^2(1-x^2))}$ is sufficient for this purpose.
The \texttt{INITIAL} algorithm is then able to find the $\epsilon$-form of the diagonal block,
\begin{equation}\label{eq:ellEformDiag}
  \frac{\partial}{\partial x} \vec{g}_{\textrm{ell}}(x,\epsilon) = \epsilon \tilde{D}_{\textrm{ell}}(x) \vec{g}_{\textrm{ell}}(x,\epsilon) + \ldots,
\end{equation}
where 
\begin{equation}
 \tilde{D}_{\text{ell}}=\begin{pmatrix}
  -\frac{4(1+x^2)}{3x(1-x^2)} & \frac{\pi^2}{x(1-x^2)\mathrm{K}^2(1-x^2)} & 0 \\
  \frac{2(1+110x^2+x^4)\mathrm{K}^2(1-x^2)}{3\pi^2x(1-x^2)} & -\frac{4(1+x^2)}{3x(1-x^2)} & \frac{\pi^2}{x(1-x^2)\mathrm{K}^2(1-x^2)} \\
  \frac{16(1+x^2)(1-18x+x^2)(1+18x+x^2)\mathrm{K}^4(1-x^2)}{27\pi^2x(1-x^2)} & \frac{2(1+110x^2+x^4)\mathrm{K}^2(1-x^2)}{3\pi^2x(1-x^2)} & -\frac{4(1+x^2)}{3x(1-x^2)}
 \end{pmatrix}\,,
\end{equation}
and the ellipsis indicate off-diagonal contributions. See App.~\ref{sec:app-ell-off-diag} and \ref{sec:app-4PM-final-DE} for the treatment of the elliptic off-diagonal blocks as well as the final differential equations in $\epsilon$-form, respectively.

\section{{\it Soft} boundary conditions} \label{sec:bc}

After having found a general solution to the differential equations, in order to fully solve the problem we still need the boundary values around some (often singular) point, which we take as the near-static limit, $x \to 1^-$,\footnote{Since this is a singular point of the differential equations, we need to regularize \eqref{eq:iterated-result}, see \cite{Lee:2019zop}.} corresponding to a small-velocity expansion in \beq v_\infty\simeq(1-x)+\mathcal{O}((1-x)^2) \ll 1\,.\eeq For this purpose, and to reduce the number of independent constants, it is useful to compare the general solution to the specific values of the associated master integrals computed in this particular (soft) limit. A systematic procedure for doing so can be found in \cite{Lee:2019zop}, and is implemented in \texttt{Libra}.
However, since the latter package can only handle rational differential equations, we implement a slightly modified version of the procedure, which we briefly outline momentarily. We also provide here a more in-depth discussion on how to systematically compute the small-velocity expansion of the master integrals in the soft limit using the method of regions discussed in \S\ref{sec:MoR}. See App.~\ref{sec:app-bc} for more details.

\subsection{Boundary relations}\label{sec:bc-exp}

Using Wasow's method, see e.g.~\cite{wasow1965asymptotic,Bruser:2018jnc}, the master integrals can be written as
\begin{align}
\label{eq:Wasow}
 \vec{f}(v_\infty,\epsilon)\simeq T^{-1}P(v_\infty,\epsilon)v_\infty^{\epsilon\tilde{M}_{1}}\vec{g}_0(\epsilon)\,,
\end{align}
where $\tilde{A}(v_\infty)=\tilde{M}_{1}/v_\infty+\mathcal{O}(v_\infty^0)$, and $T$ is the transformation to the basis $\vec{g}$ which brings the differential equations to $\epsilon$-form.
The matrix $P(v_\infty,\epsilon)=\mathbb{1}+\sum_{i=1}^\infty v_\infty^iP^{(i)}(\epsilon)$ can be computed recursively by plugging \eqref{eq:Wasow} back into the differential equations. After evaluating the matrix exponential in \eqref{eq:Wasow}, we have
\begin{equation}
\label{eq:DE-bcs}
 \begin{aligned}
  \vec{f}(v_\infty,\epsilon)&\simeq \sum_{n_1,n_2,k}v_\infty^{n_1+n_2\epsilon}\log^kv_\infty\,H_{n_1,n_2,k}(\epsilon)\vec{g}_0(\epsilon)\,,
 \end{aligned}
\end{equation}
which becomes the solution of the master integrals near $v_\infty \simeq 0$, obtained through the differential equations, where $H_{n_1,n_2,k}(\epsilon)$ is a matrix of coefficients with $n_1,n_2$ and $k$ all integers.\footnote{If the canonical form involves square-roots, $n_1$ and $n_2$ can also be half-integers. Note that the possible values for $n_2$ can be inferred from the eigenvalues of $\tilde{M}_{1}$.}\vskip 4pt
For the specific evaluation of the boundary integrals, on the other hand, we can resort to the method of regions (see~\S\ref{sec:MoR}), yielding an expansion of the sort\footnote{We find that all of the (explicit) $\log^k v_\infty$ in \eqref{eq:asy-bcs} with $k>0$ do not contribute to 4PM. Notice that this does not happen if one chooses $x_0=0$ as a boundary point, in which case one would encounter ill-defined boundary integrals coming from collinear regions, see e.g.~\cite{Becher:2014oda}.}
\begin{equation}
\label{eq:asy-bcs}
 \begin{aligned}
  \vec{f}(v_\infty,\epsilon)&\simeq \sum_{n_1,n_2,k}v_\infty^{n_1+n_2\epsilon}\log^k v_\infty\,\vec{h}_{n_1,n_2,k}(\epsilon)\,.
 \end{aligned}
\end{equation}
Comparing \eqref{eq:DE-bcs} and \eqref{eq:asy-bcs} we have,
\begin{align}
\label{eq:bc-DE-asy}
 \vec{h}_{n_1,n_2,k}(\epsilon)=H_{n_1,n_2,k}(\epsilon)\vec{g}_0(\epsilon),\qquad\text{for all }\, n_1,n_2,k\,.
\end{align}
It is then clear that not all boundary integral coefficients are linearly independent. In fact, the expression in~\eqref{eq:bc-DE-asy} allows us to find relations between them, such that there are exactly as many independent coefficients as there are boundary constants in $\vec{g}_0$, which, in turn, is equal to the number of master integrals.\footnote{The number of independent boundary integrals can be reduced even further by performing an additional IBP reduction in the soft limit.}\vskip 4pt
Let us assume that we have identified a set of linearly independent coefficients through the linearly independent rows of the collection of all of the $H_{n_1,n_2,k}(\epsilon)$'s. Denoting as $H_{\text{indep}}(\epsilon)$ the square-matrix built from the corresponding linearly independent rows, and $\vec{h}_{\rm indep}(\epsilon)$ the set of independent coefficients in \eqref{eq:asy-bcs}, the vector of boundary constants, $\vec{g}_0(\epsilon)$, is simply 
\begin{equation}
 \label{eq:indep-bound-comp}
 \vec{g}_0(\epsilon)=H_{\text{indep}}(\epsilon)^{-1}\vec{h}_{\text{indep}}(\epsilon)\,.\
\end{equation}
An explicit example leading to  \eqref{eq:indep-bound-comp} at two-loop order is discussed in App.~\ref{sec:bcrels-3PM}.\vskip 4pt

The expansion of every master integral around each region leads to a ($\gamma$-independent) boundary integral which contributes to $\vec{h}_\textrm{indep}(\epsilon)$ according to its scaling behavior in $v_\infty$.
In~what follows we discuss various relevant regions to 4PM order in more detail.

\subsection{Potential modes}\label{sec:potReg}

In momentum space the potential region corresponds to the following scaling of the loop momenta $\ell_i\sim(v_\infty,1)|\bq|$. By going to the rest frame of particle~1, see \eqref{eq:restFrame1},  we can find simple rules for the expansion of a generic integral of the type in~\eqref{PMint}. The combination of measure, Dirac-$\delta$ function, and linear propagator behaves  as, where $\bn\equiv(0,0 ,1)$,
\begin{equation}
\begin{aligned}
  \dd^{d+1}\ell_i\frac{\delta(\ell_i\cdot u_1)}{(\pm \ell_i\cdot u_2)^{\alpha_i}} &= \dd^{d+1}\ell_i \, v_\infty^{-\alpha_i} \frac{\delta(\ell_i^0)}{(\mp\bell_i\cdot\bn)^{\alpha_i}}\,,\\
  \dd^{d+1}\ell_j \frac{\delta(\ell_j\cdot u_2)}{(\pm\ell_j\cdot u_1)^{\alpha_j}} &= \dd^{d+1}\ell_j\,v_\infty\frac{\delta(\gamma v_\infty \ell_j^0 - \ell_j^z v_\infty)}{(\pm\gamma v_\infty\ell_j^0)^{\alpha_j}} = \dd^{d+1}\ell_j\,v_\infty^{-\alpha_j} \frac{\delta(\ell_j^0-\bell_j\cdot\bn)}{(\mp\bell_j\cdot\bn)^{\alpha_j}} + \cO(v_\infty^{-\alpha_j+1})\,.
\end{aligned}
\end{equation}
Resolving these Dirac-$\delta$ conditions, together with the rescaling of the energy component, reduces all square propagators to their Euclidean $d$-dimensional version:
\begin{equation}
  D_i=-(\lambda_{ij}\ell_j+\lambda_i q)^2 \rightarrow (\lambda_{ij}\bell_j+\lambda_i \bq)^2 + \cO(v_\infty) \equiv \bD_i +\cO(v_\infty)\,.
\end{equation}
Hence, a generic~$\cM$ in \eqref{PMint} can be simply expanded in the potential region as
\begin{equation}\label{eq:potBCgeneric}
  \left.\mathcal{M}^{(a_1 \cdots a_n; \pm\cdots\pm)}_{\,\alpha_1 \cdots \alpha_n; \beta_1 \cdots \beta_m}\right|_{v_\infty\rightarrow0}^\textrm{(pot)} =
  v_\infty^{-\alpha}
  \left(\prod_{i=1}^{n}\int_{\bell_i} \frac{1}{(\mp \bell_i\cdot \bn - i0)^{\alpha_i}} \right) 
\frac{1}{\bD_1^{\beta_1} \bD_2^{\beta_2} \cdots \bD_m^{\beta_m}} + \cO(v_\infty^{-\alpha+1})\,,
\end{equation}
where $\alpha = \sum_{i=1}^n \alpha_i$. Moreover, it is always possible to choose the set of independent boundary integrals  such that one never needs to compute any explicit subleading contribution.\vskip 4pt
It is clear that, for potential integrals, the $i0$-prescription of the square propagators becomes irrelevant, such that the potential region contributes to the complete answer only through conservative terms.\footnote{Furthermore,
the scaling in $v_\infty$ is always integer-powered, which means that the potential region contributes to $\vec{h}_{n_1,n_2,0}(\epsilon)$ with $n_1=-\alpha$ and $n_2=0$.} Conveniently, these are Euclidean Feynman integrals (with linear and square propagators) and any modern integration tool can be straightforwardly applied to them.
In particular, they can be easily Feynman- or Schwinger-parametrized and, most importantly, IBP reduced \cite{4pmeft}.\footnote{Albeit unnecessary, it is also straightforward to compute higher-order contributions in the small-velocity expansion, which can be shown to belong to the same integral families as those in~\eqref{eq:potBCgeneric}. This allows us to PN expand the potential contributions to arbitrary order, with an answer that can always be IBP-reduced to a finite set of master integrals.}
The explicit computational methods and results for two- and three- loop boundary integrals is discussed in App.~\ref{sec:app-bc}.

\subsection{Radiation modes} \label{sec:radReg}
The radiation regions are characterized by propagators that can be on-shell.  In principle, the procedure and rules are similar as before, see \S\ref{sec:MoR}. For completeness, we discuss below some generic features of the procedure for the particular cases at two- and three-loop order. 
\vskip 4pt

\subsubsection{Two loops} It is easy to see that integrals having only one of the incoming velocities appearing in the Dirac-$\delta$ functions do not have radiative regions. That is the case because, upon going to the rest frame of the corresponding particle, we can eradicate all time components of the loop momenta. We thus only consider integrals of the type $\cM^{(12;\pm\pm)}_{\alpha_1\alpha_2;\beta_1\cdots\beta_5}$ with the set of square propagators given in ~\eqref{eq:intFam3PM}.\footnote{The ``mirrored'' version $\cM^{(21;\pm\pm)}_{\alpha_1\alpha_2;\beta_1\cdots\beta_5}$ is obtained by a simple relabeling symmetry.} For notational simplicity we drop all indices, propagator powers, and $i0$-prescriptions in the following paragraph, and reinstate them in the final result.\vskip 4pt
At 3PM order, the only radiation region is found by the shift of loop momenta $k=\ell_1+\ell_2-q$ and $\ell=\ell_1$ leading (schematically) to
\begin{equation}
  \begin{aligned}
    \cM
    &\sim \int_{\ell\,k} \frac{\delta(\ell\cdot u_1)\delta((k-\ell+q)\cdot u_2)}{[\pm\ell\cdot u_2][\pm(k-\ell+q)\cdot u_1][-\ell^2][-(k-\ell+q)^2][-k^2][-(\ell-q)^2][-(k-\ell)^2]}\,.
  \end{aligned}
\end{equation}
Upon going to the rest frame of particle 1, and applying the scaling of the radiation region,
\begin{align}
  k &\sim (v_\infty,v_\infty)|\bq|\,, \quad  \ell \sim (v_\infty,1)|\bq|\,,
\end{align}
we find
\begin{equation}
  \begin{aligned}
    \cM &\sim \int_{\ell\,k} \frac{v_\infty^{d+2}\delta(v_\infty\ell^0)\delta(\gamma v_\infty k^0-v_\infty^2 k^z+ v_\infty \ell^z))}{[\mp v_\infty \ell^z][\pm v_\infty k^0][\bell^2][-(v_\infty k^0)^2+(v_\infty\bk-\bell+\bq)^2][-(v_\infty k^0)^2+v_\infty^2\bk^2]}\\
    &\qquad\times\frac{1}{[(\bell-\bq)^2][-(v_\infty k^0)^2+(v_\infty\bk-\bell)^2]}\\
    &=\int_{\bell\,\bk} \frac{v_\infty^d}{[\mp v_\infty\bell\cdot\bn][\mp v_\infty\bell\cdot\bn][\bell^2][(\bell-\bq)^2][-v_\infty^2((-\bell\cdot\bn)^2-\bk^2)][(\bell-\bq)^2][\bell^2]}+\dots\,,
  \end{aligned}
\end{equation}
where the ellipsis stands for higher order terms in the small velocity expansion. Note that the $d$-dependent power of $v_\infty$ stems from the rescaling of the integral measure $\dd^{d+1}k\rightarrow v_\infty^{d+1}\dd^{d+1}k$. In the soft limit we then arrive at
\begin{equation}
  \begin{aligned}
  \left.\cM^{(12;\pm\pm)}_{\alpha_1\alpha_2;\beta_1\cdots\beta_5}\right|_{v_\infty\rightarrow 0}^\textrm{(1rad)} &=  \int_\bell \frac{v_\infty^{d-\alpha_1-\alpha_2-2\beta_3}}{[\mp \bell\cdot\bn-i0]^{\alpha_1}[\mp \bell\cdot\bn-i0]^{\alpha_2}[\bell^2]^{\beta_1+\beta_5}[(\bell-\bq)^2]^{\beta_2+\beta_4}}\\
  &\qquad\times\int_\bk \frac{1}{[-((-\bell\cdot\bn)^2-\bk^2)]^{\beta_3}} + \cO(v_\infty^{d-\alpha_1-\alpha_2-\beta_3+1})\,,
  \end{aligned}
\end{equation}
where the last (square) propagator must be
understood as follows:
\begin{equation}
  \frac{1}{[-((-\bell\cdot\bn)^2-\bk^2)]^{\beta_3}}
  = \begin{cases}
  \frac{1}{[\bk^2+(\bell\cdot\bn)^2-i0]^{\beta_3}} & \textrm{conservative (Feynman)}\,,\\
  \frac{1}{[\bk^2 - (-\bell\cdot\bn\pm i0)^2]^{\beta_3}} & \textrm{causal (ret/adv)}\,.
  \end{cases}
\end{equation}
These integrals contribute to boundary constants $\vec{h}_{n_1,n_2,0}$ with $n_1=3-\alpha_1-\alpha_2-2\beta_3$ and $n_2=-2$. Notice also the overall scaling in $v_\infty$ becomes $d$-dependent, which is a distinctive feature of on-shell propagators. Furthermore, since the result with the Feynman propagator is purely imaginary, there is no conservative contribution from radiation modes at 3PM order.  
\vskip 4pt

Let us add a few remarks. In comparison to common Euclidean Feynman integrals, these integrals have the following non-standard features:
{\it i)} The last propagator has a quartic term in the vectorial quantities, i.e.\ $(\bell\cdot \bn)^2$. This is a problem for all IBP implementations known to us, although in principle this should not be an obstacle. 
{\it ii)} The causal version can be easily computed recursively using the tadpole formula in~\eqref{eq:tadpoles} and the one-loop integral~\eqref{pm-1-loop}. However, the Feynman counterpart needs a somewhat careful handling of the different types of $i0$-prescriptions for linear propagators, which appear after performing the $\bk$ tadpole integral using~\eqref{eq:tadpoles}. In practice, the latter are more easily computed in parameterized form. \vskip 4pt

\subsubsection{Three loops}\label{sec:BC3Loops}
Since the $(111)$ and $(222)$ configurations are potential-only,  we can concentrate on the $(112)$, with all others, i.e.~$(122)$, $(212)$, $(121),\,\dots$, related by momentum relabeling and shift symmetries. These integrals feature to two on-shell square propagators, as we discuss~next.  
\vskip 4pt

\textbf{\textit{Propagator set I. 2rad.}} Integral families with the set of $\{D_i^I\}$ propagators, see~\eqref{eq:props4PM}, all have a region with two on-shell legs, associated to the shift $k_1=\ell_3-\ell_1$, $k_2=\ell_2-\ell_3$, and $\ell=\ell_3$, such that (ignoring indices and $i0$'s for simplicity):
\begin{equation}
  \begin{aligned}
    \cM^\textrm{I}
    &\sim\int_{k_1k_2\,\ell} 
      \frac{\delta((\ell-k_1)\cdot u_1)\delta((k_2+\ell)\cdot u_1)\delta(\ell\cdot u_2)}{[\pm(\ell-k_1)\cdot u_2][\pm(k_2+\ell)\cdot u_2][\pm\ell\cdot u_1][-(\ell-k_1)^2][-(k_2+\ell)^2][-\ell^2]}\\
    &\qquad\quad \times\frac{1}{[-(\ell-k_1-q)^2][-(k_2+\ell-q)^2][-(\ell-q)^2][-(-k_1-k_2)^2][-k_2^2][-k_1^2]}\,.
  \end{aligned}
\end{equation}
By going to the rest frame of particle 1 and applying the scaling for this region:
\begin{align}
  k_1 &\sim (v_\infty,v_\infty)|\bq|\,, & k_2 &\sim (v_\infty,v_\infty)|\bq|\,, & \ell &\sim (v_\infty,1)|\bq|\,,
\end{align}
we arrive, at the leading order, at  
\begin{equation}
  \begin{aligned}
    \cM^\textrm{I}&\sim\int_{\bell}
      \frac{v_\infty^{2d}}{[\mp v_\infty\bell\cdot\bn][\mp v_\infty\bell\cdot\bn][\pm v_\infty\bell\cdot\bn][\bell^2][\bell^2][\bell^2][(\bell-\bq)^2][(\bell-\bq)^2][(\bell-\bq)^2]}\\
      &\qquad\times\int_{\bk_1\bk_2}\frac{1}{[v_\infty^2(-\bk_1-\bk_2)^2][-v_\infty^2((-\bell\cdot\bn)^2-\bk_2^2)][-v_\infty^2((\bell\cdot\bn)^2-\bk_1^2)]}+\dots\,,
  \end{aligned}
\end{equation}
which yields, reinstating indices and prescriptions,
\begin{equation}
  \begin{aligned}
    \left.\cM^{\textrm{I},(112;\pm\pm\pm)}_{\alpha_1\alpha_2\alpha_3;\beta_1\cdots\beta_9}\right|_{v_\infty\rightarrow 0}^\textrm{(2rad)}&=\int_{\bell}
      \frac{v_\infty^{2d-\alpha_1-\alpha_2-\alpha_3-2\beta_7-2\beta_8-2\beta_9}}{[\mp\bell\cdot\bn-i0]^{\alpha_1}[\mp\bell\cdot\bn-i0]^{\alpha_2}[\pm\bell\cdot\bn-i0]^{\alpha_3}}\\
      &\quad\times\frac{1}{[\bell^2]^{\beta_1+\beta_2+\beta_3}[(\bell-\bq)^2]^{\beta_4+\beta_5+\beta_6}}\\
      &\quad\times\int_{\bk_1,\bk_2}\frac{1}{[(-\bk_1-\bk_2)^2]^{\beta_7}[-((-\bell\cdot\bn)^2-\bk_2^2)]^{\beta_8}[-((\bell\cdot\bn)^2-\bk_1^2)]^{\beta_9}}\,.
  \end{aligned}
\end{equation}
Similarly to what we discussed before the last two propagators must be understood as follows:
\begin{equation}\label{eq:3DpropFeynVsCausal}
  \frac{1}{[-((\sigma \bell\cdot\bn)^2-\bk_i^2)]}
  = \begin{cases}
  \frac{1}{[\bk_i^2+(\bell\cdot\bn)^2-i0]} & \textrm{conservative (Feynman)}\,,\\
  \frac{1}{[\bk_i^2 - (\sigma \bell\cdot\bn\pm i0)^2]} & \textrm{causal (ret/adv)}\,,
  \end{cases}
\end{equation}
with $\sigma\in\{-1,1\}$.
The scaling in $v_\infty$ tells us that this region contributes to boundary constants $\vec{h}_{n_1,n_2,0}$ with $n_1=6-\alpha_1-\alpha_2-\alpha_3-2\beta_7-2\beta_8-2\beta_9$ and $n_2=-4$. Unlike 3PM, radiation modes do contribute to both the conservative \cite{4pmeft2} and dissipative \cite{4pmeftot} sectors at 4PM order.\vskip 4pt 

\textbf{\textit{Propagator set I. 1rad.}} The same integral families have other regions with a single on-shell propagator. The first contribution emerges from the last one. After the shift $\ell_3=k+\ell_1$,
\begin{equation}
  \begin{aligned}
    \cM^{\textrm{I}} &\sim \int_{\ell_1\ell_2\,k} \begin{multlined}[t]
      \frac{\delta(\ell_1\cdot u_1)\delta(\ell_2\cdot u_1)\delta((k+\ell_1)\cdot u_2)}{[\pm\ell_1\cdot u_2][\pm\ell_2\cdot u_2][\pm(k+\ell_1)\cdot u_1][-\ell_1^2][-\ell_2^2][-(k+\ell_1)^2][-(\ell_1-q)^2]}\\
      \times\frac{1}{[-(\ell_2-q)^2][-(k+\ell_1-q)^2][-(\ell_1-\ell_2)^2][-(\ell_2-\ell_1-k)^2][-k^2]}\,,
    \end{multlined}
  \end{aligned}
\end{equation}
and the relevant scaling now becomes,
\begin{align}\label{eq:scalingI1rad}
k &\sim (v_\infty,v_\infty)|\bq|\,, & \ell_1 &\sim (v_\infty,1)|\bq|\,, & \ell_2 &\sim (v_\infty,1)|\bq|\,,
\end{align}
yielding, in the rest frame of particle 1, the leading order contribution 
\begin{equation}
\left.\cM^{\textrm{I},(112;\pm\pm\pm)}_{\alpha_1\alpha_2\alpha_3;\beta_1\cdots\beta_9}\right|_{v_\infty\rightarrow 0}^\textrm{(1rad,1)} =
\begin{multlined}[t]\int_{\bell_1\bell_2}
\frac{v_\infty^{d-\alpha_1-\alpha_2-\alpha_3-2\beta_9}}{[\mp \bell_1\cdot\bn-i0]^{\alpha_1}[\mp \bell_2\cdot\bn-i0]^{\alpha_2}[\pm \bell_1\cdot\bn-i0]^{\alpha_3}}\\
\times\frac{1}{[\bell_1^2]^{\beta_1+\beta_3}[\bell_2^2]^{\beta_2}[(\bell_1-\bq)^2]^{\beta_4+\beta_6}[(\bell_2-\bq)^2]^{\beta_5}[(\bell_1-\bell_2)^2]^{\beta_7+\beta_8}}\\
\times\int_\bk\frac{1}{[-((\bell_1\cdot\bn)^2-\bk^2)]^{\beta_9}}\,.
\end{multlined}
\end{equation}
Another radiative contribution arises when the second-to-last propagator goes on-shell. Applying the momentum shift $\ell_3=\ell_2-k$ and the same scaling \eqref{eq:scalingI1rad}, we arrive, at leading order in the velocity, at
\begin{equation}
\left.\cM^{\textrm{I},(112;\pm\pm\pm)}_{\alpha_1\alpha_2\alpha_3;\beta_1\cdots\beta_9}\right|_{v_\infty\rightarrow 0}^\textrm{(1rad,2)} =
\begin{multlined}[t]\int_{\bell_1\bell_2} 
\frac{v_\infty^{d-\alpha_1-\alpha_2-\alpha_3-2\beta_8}}{[\mp\bell_1\cdot\bn-i0]^{\alpha_1}[\mp\bell_2\cdot\bn-i0]^{\alpha_2}[\pm\bell_2\cdot\bn-i0]^{\alpha_3}}\\
\times\frac{1}{[\bell_1^2]^{\beta_1}[\bell_2^2]^{\beta_2+\beta_3}[(\bell_1-\bq)^2]^{\beta_4}[(\bell_2-\bq)^2]^{\beta_5+\beta_6}[(\bell_1-\bell_2)^2]^{\beta_7+\beta_9}}\\
\times\int_\bk\frac{1}{[-((-\bell_2\cdot\bn)^2-\bk^2)]^{\beta_8}}\,.
\end{multlined}
\end{equation}
The last propagator is to be understood, once more, according to~\eqref{eq:3DpropFeynVsCausal}.
The scaling in $v_\infty$ tells us to assign both of these contributions to boundary constants $\vec{h}_{n_1,n_2,0}$ with $n_1=3-\alpha_1-\alpha_2-\alpha_3-2\beta_{r}$ (where $r=8$ or $r=9$) and $n_2=-2$. \vskip 4pt

\textbf{\textit{Propagator set II. 1rad.}} In the case of integral families with the set of $\{D_i^{II}\}$ propagators, see~\eqref{eq:props4PM}, there is only one relevant region with a single on-shell propagator.\footnote{In principle, these integral families have more regions, however, none of the integrals that have them either appear in the computation of the impulse, or can be related to an integral with the set of $\{D_i^{I}\}$ propagators.} This region can be found following the shift $\ell_3=k-\ell_1-\ell_2+q$,
\begin{equation}
    \cM^{\textrm{II}} \sim  \begin{multlined}[t] \int_{\ell_1\ell_2\,k}
      \frac{\delta(\ell_1\cdot u_1)\delta(\ell_2\cdot u_1)\delta((k-\ell_1-\ell_2)\cdot u_2)}{[\pm\ell_1\cdot u_2][\pm\ell_2\cdot u_2][\pm(k-\ell_1-\ell_2)\cdot u_1][-\ell_1^2][-\ell_2^2][-(k-\ell_1-\ell_2+q)^2]}\\
      \times\frac{1}{[-(\ell_1-q)^2][-(\ell_2-q)^2][-(k-\ell_1-\ell_2)^2][-(\ell_1+\ell_2-q)^2][-(k-\ell_1)^2][-k^2]}\,,
    \end{multlined}
\end{equation}
so that, upon rescaling the loop momenta according to~\eqref{eq:scalingI1rad}, the leading term is given by
\begin{equation}
    \left.\cM^{\textrm{II},(112;\pm\pm\pm)}_{\alpha_1\alpha_2\alpha_3;\beta_1\cdots\beta_9}\right|_{v_\infty\rightarrow 0}^\textrm{(1rad)} =
     \begin{multlined}[t]\int_{\bell_1\bell_2}
      \frac{v_\infty^{d-\alpha_1-\alpha_2-\alpha_3-2\beta_9}}{[\mp\bell_1\cdot\bn]^{\alpha_1}[\mp\bell_2\cdot\bn]^{\alpha_2}[\mp(\bell_1+\bell_2)\cdot\bn]^{\alpha_3}[\bell_1^2]^{\beta_1+\beta_8}[\bell_2^2]^{\beta_2}}\\
      \times\frac{1}{[(\bell_1+\bell_2-\bq)^2]^{\beta_3+\beta_7}[(\bell_1-\bq)^2]^{\beta_4}[(\bell_2-\bq)^2]^{\beta_5}[(\bell_1+\bell_2)^2]^{\beta_6}}\\
      \times\int_\bk\frac{1}{[-((-(\bell_1+\bell_2)\cdot\bn)^2-\bk^2)]^{\beta_9}}\,.
    \end{multlined}
\end{equation}
Due to the $v_\infty$-scaling, these integrals contribute to the boundary constants $\vec{h}_{n_1,n_2,0}$ with $n_1=3-\alpha_1-\alpha_2-\alpha_3-2\beta_9$ and $n_2=-2$. Similarly to the 3PM case, contributions with a single radiative mode can be shown to be purely dissipative~\cite{4pmeftot}.\vskip 4pt The integration procedure for all of these boundary integrals and relevant results can be found in App.~\ref{sec:app-bc}.

\section{Scattering data }\label{sec:data}


In this section we collect the (spin-independent) results obtained in the EFT approach to 4PM order. See the ancillary file for a ready-to-use notebook including all the expressions. 

\subsection{Total impulse}

The impulse's coefficients for particle 1 (defined in \eqref{impulsetot}) are entirely conservative to 2PM order, and given by, 
\begin{equation}
  \begin{aligned}
    c_{1b}^{(1)\textrm{tot}} &= \frac{2 \left(2 \gamma ^2-1\right) m_1 m_2}{\sqrt{\gamma ^2-1}}\,, & \qquad\frac{c_{1b}^{(2)\textrm{tot}}}{\pi} &= -\frac{3 \left(5 \gamma ^2-1\right) m_1 m_2 M}{4 \sqrt{\gamma ^2-1}}\,,\\
    c_{1\check{u}_1}^{(1)\textrm{tot}} &= 0\,, & c_{1\check{u}_1}^{(2)\textrm{tot}} &= -\frac{2 \left(1-2 \gamma ^2\right)^2 m_1 m_2^2}{\gamma ^2-1}\,,\\
    c_{1\check{u}_2}^{(1)\textrm{tot}} &= 0\,, & c_{1\check{u}_2}^{(2)\textrm{tot}} &= \frac{2 \left(1-2 \gamma ^2\right)^2 m_1^2 m_2}{\gamma ^2-1}\,,
  \end{aligned}
\end{equation}
whereas at 3PM we have both conservative and dissipative terms, totaling \cite{3pmeft,eftrad}
\begin{equation}
  \begin{aligned}
    c_{1b}^{(3)\textrm{tot}} &= \frac{\left(-32 \gamma ^6+64 \gamma ^4-32 \gamma ^2+2\right) m_2 m_1(m_1^2+m_2^2)}{\left(\gamma ^2-1\right)^{5/2}}\\
    &\quad +m_2^2 m_1^2 \left[4 \left(\frac{8 \gamma ^4-24 \gamma ^2-6}{\gamma ^2-1}-\frac{\gamma  \left(1-2 \gamma ^2\right)^2 \left(2 \gamma^2-3\right)}{\left(\gamma ^2-1\right)^{5/2}}\right) \arccosh(\gamma)\right.\\
    &\qquad\qquad\qquad\quad\left.+\frac{4}{3} \left(\frac{\left(5 \gamma ^2-8\right) \left(1-2 \gamma ^2\right)^2}{\left(\gamma ^2-1\right)^2}+\frac{-20 \gamma ^7+90 \gamma ^5-120 \gamma ^3+53 \gamma }{\left(\gamma ^2-1\right)^{5/2}}\right)\right]\,,
     \end{aligned}
    \end{equation}
    \begin{equation}
  \begin{aligned}
    \frac{c_{1\check{u}_1}^{(3)\textrm{tot}}}{\pi} &= \frac{3 \left(2 \gamma ^2-1\right) \left(5 \gamma ^2-1\right) m_1 m_2^2 \left(m_1+m_2\right)}{2 \left(\gamma ^2-1\right)}\,,\\
    \frac{c_{1\check{u}_2}^{(3)\textrm{tot}}}{\pi} &=
    \begin{multlined}[t]
    -\frac{3 \left(2 \gamma ^2-1\right) \left(5 \gamma ^2-1\right) m_1^3 m_2}{2 \left(\gamma ^2-1\right)}
    - m_1^2 m_2^2 \left[
    \frac{3 \left(2 \gamma ^2-1\right) \left(5 \gamma ^2-1\right)}{2(\gamma ^2-1)}\right.\\
    +\frac{210 \gamma ^6-552 \gamma ^5+339 \gamma ^46912 \gamma ^3+3148 \gamma ^2-3336 \gamma +1151}{48\left(\gamma ^2-1\right)^{3/2}}\\
    +\frac{\gamma  \left(2 \gamma ^2-3\right) \left(35 \gamma ^4-30 \gamma ^2+11\right) }{16  \left(\gamma ^2-1\right)^2}\arccosh(\gamma)\\
    \left.-\frac{ \left(35 \gamma ^4+60 \gamma ^3-150 \gamma ^2+76 \gamma -5\right) }{8 \sqrt{\gamma  ^2-1}}\log \left(\frac{\gamma +1}{2}\right)
    \right]\,.
    \end{multlined}
  \end{aligned}
\end{equation}
The most intricate computation is at 4PM order, yielding the very state-of-the-art~\cite{4pmeft,4pmeft2,4pmeftot},
\begingroup
\allowdisplaybreaks
\begin{align}\label{imptot}
      \frac{c^{(4)\rm tot}_{1b}}{\pi} =&\,
        -\frac{3 h_1 m_1 m_2 (m_1^3+m_2^3)}{64 (\gamma ^2-1)^{5/2}}
        +m_1^2 m_2^2 (m_1+m_2)\Bigg[
          \frac{21 h_2 \mathrm{E}^2\left(\frac{\gamma -1}{\gamma +1}\right)}{32 (\gamma -1) \sqrt{\gamma ^2-1}}
          +\frac{3 h_3 \mathrm{K}^2\left(\frac{\gamma -1}{\gamma +1}\right)}{16 \left(\gamma ^2-1\right)^{3/2}}
          \nonumber\\
          &~
          -\frac{3 h_4 \mathrm{E}\left(\frac{\gamma -1}{\gamma +1}\right) \mathrm{K}\left(\frac{\gamma -1}{\gamma +1}\right)}{16 \left(\gamma ^2-1\right)^{3/2}}
          +\frac{\pi ^2 h_5}{8 \sqrt{\gamma ^2-1}}
          +\frac{h_6 \log \left(\frac{\gamma -1}{2}\right)}{16 \left(\gamma ^2-1\right)^{3/2}}
          +\frac{3 h_7 \text{Li}_2\left(\sqrt{\frac{\gamma -1}{\gamma +1}}\right)}{(\gamma -1) (\gamma +1)^2}
          \nonumber\\
          &~
          -\frac{3 h_7 \text{Li}_2\left(\frac{\gamma -1}{\gamma +1}\right)}{4 (\gamma -1) (\gamma +1)^2}
          \Bigg]
        +m_1^3 m_2^2\Bigg[
          \frac{h_8}{48 \left(\gamma ^2-1\right)^3}
          +\frac{\sqrt{\gamma ^2-1} h_9}{768 (\gamma -1)^3 \gamma ^9 (\gamma +1)^4}
          \nonumber\\
          &~
          +\frac{h_{10} \log \left(\frac{\gamma +1}{2}\right)}{8 \left(\gamma ^2-1\right)^2}
          -\frac{h_{11} \log \left(\frac{\gamma +1}{2}\right)}{32 \left(\gamma ^2-1\right)^{5/2}}
          +\frac{h_{12} \log (\gamma )}{16 \left(\gamma ^2-1\right)^{5/2}}
          -\frac{h_{13}\arccosh(\gamma )}{8 (\gamma -1) (\gamma +1)^4}
          \nonumber\\
          &~
          +\frac{h_{14}\arccosh(\gamma )}{16 \left(\gamma ^2-1\right)^{7/2}}
          -\frac{3 h_{15} \log \left(\frac{\gamma +1}{2}\right) \log \left(\frac{\gamma -1}{\gamma +1}\right)}{8 \sqrt{\gamma ^2-1}}
          +\frac{3 h_{16}\arccosh(\gamma ) \log \left(\frac{\gamma -1}{\gamma +1}\right)}{16 \left(\gamma ^2-1\right)^2}
          \nonumber\\
          &~
          -\frac{3 h_{17} \text{Li}_2\left(\frac{\gamma -1}{\gamma +1}\right)}{64 \sqrt{\gamma ^2-1}}
          -\frac{3}{32} \sqrt{\gamma ^2-1} h_{18} \text{Li}_2\left(\frac{1-\gamma }{\gamma +1}\right)
          \Bigg]
        +m_1^2 m_2^3 \Bigg[
          \frac{3 h_{15} \log \left(\frac{2}{\gamma -1}\right) \log \left(\frac{\gamma +1}{2}\right)}{8 \sqrt{\gamma ^2-1}}
          \nonumber\\
          &~
          +\frac{3 h_{16} \log \left(\frac{\gamma -1}{2}\right)\arccosh(\gamma )}{16 \left(\gamma ^2-1\right)^2}
          +\frac{h_{19}}{48 \left(\gamma ^2-1\right)^3}
          +\frac{h_{20}}{192 \gamma ^7 \left(\gamma ^2-1\right)^{5/2}}
          +\frac{h_{21} \log \left(\frac{\gamma +1}{2}\right)}{8 \left(\gamma ^2-1\right)^2}
          \nonumber\\
          &~
          +\frac{h_{22} \log \left(\frac{\gamma +1}{2}\right)}{16 \left(\gamma ^2-1\right)^{3/2}}
          +\frac{h_{23} \log (\gamma )}{2 \left(\gamma ^2-1\right)^{3/2}}
          -\frac{h_{24}\arccosh(\gamma )}{16 \left(\gamma ^2-1\right)^3}
          +\frac{h_{25}\arccosh(\gamma )}{16 \left(\gamma ^2-1\right)^{7/2}}
          -\frac{3 h_{26}\arccosh^2(\gamma )}{32 \left(\gamma ^2-1\right)^{7/2}}
          \nonumber\\
          &~
          +\frac{3 h_{27} \log ^2\left(\frac{\gamma +1}{2}\right)}{2 \sqrt{\gamma ^2-1}}
          +\frac{3 h_{28} \log \left(\frac{\gamma +1}{2}\right)\arccosh(\gamma )}{16 \left(\gamma ^2-1\right)^2}
          +\frac{h_{29} \text{Li}_2\left(\frac{1-\gamma }{\gamma +1}\right)}{4 \sqrt{\gamma ^2-1}}
          +\frac{3 h_{30} \text{Li}_2\left(\frac{\gamma -1}{\gamma +1}\right)}{8 \sqrt{\gamma ^2-1}}
          \Bigg]\,,
\nn\\[0.5 em]
      c^{(4)\rm tot}_{1\check{u}_1} =&\,
      \frac{9 \pi ^2 h_{31} m_1 m_2^2 \left(m_1+m_2\right){}^2}{32 \left(\gamma ^2-1\right)}
      +\frac{2 h_{32} m_1 m_2^2 \left(m_1^2+m_2^2\right)}{\left(\gamma ^2-1\right)^3}
      \nonumber\\
      &~~
      +m_1^2 m_2^3\Bigg[
        \frac{4 h_{33}}{3 \left(\gamma ^2-1\right)^3}
        -\frac{8 h_{34}}{3 \left(\gamma ^2-1\right)^{5/2}}
        +\frac{8 h_{35}\arccosh(\gamma )}{\left(\gamma ^2-1\right)^3}
        -\frac{16 h_{36}\arccosh(\gamma )}{\left(\gamma ^2-1\right)^{3/2}}
        \Bigg]\,,
\nonumber\\[0.5 em]
      c^{(4)\rm tot}_{1\check{u}_2}  =&\,
      -m_1^4 m_2 \left(\frac{9 \pi ^2 h_{31}}{32 \left(\gamma ^2-1\right)}+\frac{2 h_{32}}{\left(\gamma ^2-1\right)^3}\right)
      +m_1^3 m_2^2\Bigg[
        -\frac{4 h_{37}}{3 \left(\gamma ^2-1\right)^3}
        +\frac{h_{38}}{705600 \gamma ^8 \left(\gamma ^2-1\right)^{5/2}}
        \nonumber\\
        &
        +\frac{\pi ^2 h_{39}}{192 \left(\gamma ^2-1\right)^2}
        +\frac{h_{40}\arccosh(\gamma )}{6720 \gamma ^9 \left(\gamma ^2-1\right)^3}
        +\frac{32 h_{41}\arccosh(\gamma )}{3 \left(\gamma ^2-1\right)^{3/2}}
        -\frac{8 h_{42}\arccosh^2(\gamma )}{\left(\gamma ^2-1\right)^2}
        \nonumber\\
        &
        +\frac{32 h_{43}\arccosh^2(\gamma )}{\left(\gamma ^2-1\right)^{7/2}}
        +\frac{h_{44} \log (2)\arccosh(\gamma )}{8 \left(\gamma ^2-1\right)^2}
        +\frac{3 h_{45} \left(\text{Li}_2\left(\frac{\gamma -1}{\gamma +1}\right)-4 \text{Li}_2\left(\sqrt{\frac{\gamma -1}{\gamma +1}}\right)\right)}{16 \left(\gamma ^2-1\right)^2}
        \nonumber\\
        &
        +\frac{3 h_{46} \left(\log \left(\frac{\gamma +1}{2}\right)\arccosh(\gamma )-2 \text{Li}_2\left(\sqrt{\gamma ^2-1}-\gamma \right)\right)}{8 \left(\gamma ^2-1\right)^2}
        \nonumber\\
        &
        -\frac{h_{47} \left(\text{Li}_2\Big(-\big(\gamma -\sqrt{\gamma ^2-1}\big)^2\Big)-2 \log (\gamma )\arccosh(\gamma )\right)}{16 \left(\gamma ^2-1\right)^2}
        \Bigg]
\nn\\
&
      +m_1^2 m_2^3\Bigg[
        -\frac{2 h_{48}}{45 \left(\gamma ^2-1\right)^3}
        +\frac{h_{49}}{1440 \gamma ^7 \left(\gamma ^2-1\right)^{5/2}}
        +\frac{\pi ^2 h_{50}}{48 \left(\gamma ^2-1\right)^2}
        +\frac{h_{51}\arccosh(\gamma )}{480 \gamma ^8 \left(\gamma ^2-1\right)^3}
        \nonumber\\
        &
        -\frac{16 h_{52}\arccosh(\gamma )}{5 \left(\gamma ^2-1\right)^{3/2}}
        -\frac{16 h_{53}\arccosh^2(\gamma )}{\left(\gamma ^2-1\right)^2}
        -\frac{32 h_{54}\arccosh^2(\gamma )}{\left(\gamma ^2-1\right)^{7/2}}
        -\frac{h_{55} \log (2)\arccosh(\gamma )}{4 \left(\gamma ^2-1\right)^2}
        \nonumber\\
        &
        +\frac{h_{56} \left(\text{Li}_2\!\left(\frac{\gamma -1}{\gamma +1}\right)-4 \text{Li}_2\!\left(\sqrt{\frac{\gamma -1}{\gamma +1}}\right)\right)}{32 \left(\gamma ^2-1\right)^2}
        +\frac{h_{57} \left(\log\! \left(\frac{2}{\gamma +1}\right)\arccosh(\gamma )+2 \text{Li}_2\!\left(\sqrt{\gamma ^2-1}-\gamma \right)\right)}{4 \left(\gamma ^2-1\right)^2}
        \nonumber\\
        &
        +\frac{h_{58} \left(\text{Li}_2\Big(-\big(\gamma -\sqrt{\gamma ^2-1}\big)^2\Big)-2 \log (\gamma )\arccosh(\gamma )\right)}{8 \left(\gamma ^2-1\right)^2}
        \Bigg]\,,
\end{align}
\endgroup
which includes conservative, dissipative, as well as hereditary and nonlinear radiation-reaction effects. See App.~\ref{sec:app-pols} for the value of the $h_i$ polynomials. The impulse for the companion follows by  replacing $1\leftrightarrow 2$ in the masses, incoming velocities, and impact parameters. 
  
\subsubsection{Conservative}

A conservative part of the total impulse can be identified by replacing retarded propagators with Feynman's $i0$-prescription, while retaining the real part of the answer \cite{eftrad}. The first distinction appears at 3PM order, where we have~\cite{3pmeft}
\begin{align}
    c^{(3)\rm cons}_{1b} &=  -\frac{2 m_1 m_2(m_1^2+m_2^2)}{(\gamma^2-1)^{\frac{5}{2}}}\left(16\gamma^6-32\gamma^4+16\gamma^2-1\right)
    \nonumber\\
    &\quad -\frac{4m_1^2m_2^2\gamma}{3(\gamma^2-1)^{\frac{5}{2}}}\left(20\gamma^6-90\gamma^4+120\gamma^2-53\right)
    +\frac{8 m_1^2m_2^2}{\gamma^2-1}\left(4\gamma^4-12\gamma^2-3\right)\arccosh(\gamma)\,,
    \nonumber\\[0.5 em]
 c^{(3)\rm cons}_{1\check{u}_1} &=\frac{3\pi}{2} m_1m_2^2(m_1+m_2)\frac{(2\gamma^2-1)(5\gamma^2-1)}{\gamma^2-1}\,,
 \nonumber\\[0.5 em]
  c^{(3)\rm cons}_{1\check{u}_2} &=-\frac{3\pi}{2} m_1^2m_2(m_1+m_2)\frac{(2\gamma^2-1)(5\gamma^2-1)}{\gamma^2-1}\,,
\end{align}
from potential-only modes. At the next order, on the other hand, we find \cite{4pmeft,4pmeft2,4pmeftot}
\begingroup
\allowdisplaybreaks
\begin{align}
  \label{impcon}
    \frac{c^{(4)\rm cons}_{1b}}{\pi} =&\, 
      -\frac{3 h_1 m_1 m_2 \left(m_1^3+m_2^3\right)}{64 \left(\gamma ^2-1\right)^{5/2}}
      +m_1^2 m_2^2 \left(m_1+m_2\right)\Bigg[
        \frac{21 \sqrt{\gamma ^2-1} h_2 \mathrm{E}^2\left(\frac{\gamma -1}{\gamma +1}\right)}{32 (\gamma -1)^2 (\gamma +1)}
        \nn\\
        &~
        +\frac{3 h_3 \mathrm{K}^2\left(\frac{\gamma -1}{\gamma +1}\right)}{16 \left(\gamma ^2-1\right)^{3/2}}
        -\frac{3 h_4 \mathrm{E}\left(\frac{\gamma -1}{\gamma +1}\right) \mathrm{K}\left(\frac{\gamma -1}{\gamma +1}\right)}{16 \left(\gamma ^2-1\right)^{3/2}}
        +\frac{\pi ^2  h_5}{8 \sqrt{\gamma ^2-1}}
        +\frac{h_6 \log \left(\frac{\gamma -1}{2}\right)}{16 \left(\gamma ^2-1\right)^{3/2}}
        \nn\\
        &~
        +\frac{3 h_7 \text{Li}_2\left(\sqrt{\frac{\gamma -1}{\gamma +1}}\right)}{(\gamma -1) (\gamma +1)^2}
        -\frac{3 h_7 \text{Li}_2\left(\frac{\gamma -1}{\gamma +1}\right)}{4 (\gamma -1) (\gamma +1)^2}
        -\frac{3 h_{15} \log \left(\frac{\gamma -1}{2}\right) \log \left(\frac{\gamma +1}{2}\right)}{8 \sqrt{\gamma ^2-1}}
        \nn\\
        &~
        +\frac{3 h_{16} \log \left(\frac{\gamma -1}{2}\right) \arccosh(\gamma )}{16 \left(\gamma ^2-1\right)^2}
        +\frac{h_{20}}{192 \gamma ^7 \left(\gamma ^2-1\right)^{5/2}}
        +\frac{h_{22} \log \left(\frac{\gamma +1}{2}\right)}{16 \left(\gamma ^2-1\right)^{3/2}}
        \nn\\
        &~
        +\frac{h_{23} \log (\gamma )}{2 \left(\gamma ^2-1\right)^{3/2}}
        -\frac{h_{24} \arccosh(\gamma )}{16 \left(\gamma ^2-1\right)^3}
        -\frac{3 h_{26} \arccosh^2(\gamma )}{32 \left(\gamma ^2-1\right)^{7/2}}
        +\frac{3 h_{27} \log ^2\left(\frac{\gamma +1}{2}\right)}{2 \sqrt{\gamma ^2-1}}
        \nn\\
        &~
        +\frac{3 h_{28} \log \left(\frac{\gamma +1}{2}\right) \arccosh(\gamma )}{16 \left(\gamma ^2-1\right)^2}
        +\frac{h_{29} \text{Li}_2\left(\frac{1-\gamma }{\gamma +1}\right)}{4 \sqrt{\gamma ^2-1}}
        +\frac{3 \sqrt{\gamma ^2-1} h_{30} \text{Li}_2\left(\frac{\gamma -1}{\gamma +1}\right)}{8 (\gamma -1) (\gamma +1)}
        \Bigg]
    \nn\\[0.5 em]
    c^{(4)\rm cons}_{1\check{u}_1} =&\,
    \frac{9 \pi ^2 h_{31} m_1 m_2^2 \left(m_1+m_2\right){}^2}{32 \left(\gamma ^2-1\right)}
    +\frac{2 h_{32} m_1 m_2^2 \left(m_1^2+m_2^2\right)}{\left(\gamma ^2-1\right)^3}
    \nn\\
    &~
    +m_1^2 m_2^3\Bigg[
      \frac{4 h_{33}}{3 \left(\gamma ^2-1\right)^3}
      -\frac{16 h_{36} \arccosh(\gamma )}{\left(\gamma ^2-1\right)^{3/2}}
      \Bigg]
      \nn\\[0.5 em]
    c^{(4)\rm cons}_{1\check{u}_2} =&\,
    -\frac{9 \pi ^2 h_{31} m_1^2 m_2 \left(m_1+m_2\right){}^2}{32 \left(\gamma ^2-1\right)}
    -\frac{2 h_{32} m_1^2 m_2 \left(m_1^2+m_2^2\right)}{\left(\gamma ^2-1\right)^3}
    \nn\\
    &~
    +m_1^3 m_2^2\Bigg[
      -\frac{4 h_{33}}{3 \left(\gamma ^2-1\right)^3}
      +\frac{16 h_{36} \arccosh(\gamma )}{\left(\gamma ^2-1\right)^{3/2}}
    \Bigg]\,,
\end{align}
\endgroup
which includes not only potential contributions but as well (2rad) regions involving two radiation modes associated with hereditary-type effects. 

\subsubsection{Dissipative}

The non-conservative part of the impulse starts at 3PM order, and reads \cite{eftrad}
\begin{align}
  c^{(3)\rm diss}_{1b} &=\frac{4}{3} m_1^2m_2^2\frac{(2\gamma^2-1)^2}{(\gamma^2-1)^{5/2}}\left(\sqrt{\gamma^2-1}(5\gamma^2-8)+3\gamma(3-2\gamma^2)\arccosh(\gamma)\right)\,,
  \nn\\[0.5 em]
    c^{(3)\rm diss}_{1\check{u}_1} &= 0 \,,
    \nn\\
    c^{(3)\rm diss}_{1\check{u}_2} &=   \frac{\pi m_1^2m_2^2}{48(\gamma^2-1)^{3/2}}\bigg[\left(
    -210 \gamma^6 + 552 \gamma^5 -339 \gamma^4+912 \gamma^3-3148 \gamma^2+3336 \gamma-1151\right)
    \nn\\[0.5 em]
    &\quad +  6(35\gamma^6+60\gamma^5-185\gamma^4+16\gamma^3+145\gamma^2-76\gamma+5)\log\left(\frac{\gamma+1}{2}\right) 
    \nn\\
    &\quad -3\gamma(70\gamma^6-165\gamma^4+112\gamma^2-33)\frac{\arccosh(\gamma)}{\sqrt{\gamma^2-1}}\bigg]\,,
\end{align}
where only integrals with a single radiation mode contribute at two-loop order in the dissipative sector. At 4PM, on the other hand, we encounter two distinct contributions. For the (instantaneous) part, involving a single radiation mode, we find \cite{4pmeftot}
\begingroup
\allowdisplaybreaks
\begin{align}
  \label{impdis1}
  \frac{c^{(4) \rm diss}_{1b,\rm 1rad}}{\pi} =&\, 
    m_1^3 m_2^2\Bigg[
      \frac{h_8}{48 \left(\gamma ^2-1\right)^3}
      +\frac{h_{10} \log \left(\frac{\gamma +1}{2}\right)}{8 \left(\gamma ^2-1\right)^2}
      +\frac{h_{14} \arccosh(\gamma )}{16 \left(\gamma ^2-1\right)^{7/2}}
      \Bigg]
      \nonumber\\
      &~
    +m_1^2 m_2^3\Bigg[
      \frac{h_{19}}{48 \left(\gamma ^2-1\right)^3}
      +\frac{h_{21} \log \left(\frac{\gamma +1}{2}\right)}{8 \left(\gamma ^2-1\right)^2}
      +\frac{h_{25} \arccosh(\gamma )}{16 \left(\gamma ^2-1\right)^{7/2}}
      \Bigg]\,,
 \nn\\[0.7 em]
  c^{(4) \rm diss}_{1\check{u}_1,\rm 1rad} =&\,
  m_1^2 m_2^3 \Bigg[
    -\frac{8 h_{34}}{3 \left(\gamma ^2-1\right)^{5/2}}
    +\frac{8 h_{35} \arccosh(\gamma )}{\left(\gamma ^2-1\right)^3}
    \Bigg]\,,
\\[0.7 em]
  c^{(4)\rm diss}_{1\check{u}_2,\rm 1rad} =&\,
    m_1^3 m_2^2\Bigg[
      \frac{h_{38}}{705600 \gamma ^8 \left(\gamma ^2-1\right)^{5/2}}
      +\frac{\pi ^2 \left(108 h_{31}(\gamma^2-1)+h_{39}\right)}{192 \left(\gamma ^2-1\right)^2}
      +\frac{h_{40} \arccosh(\gamma )}{6720 \gamma ^9 \left(\gamma ^2-1\right)^3}
      \nonumber\\
      &~
      +\frac{32 h_{43} \arccosh^2(\gamma )}{\left(\gamma ^2-1\right)^{7/2}}
      +\frac{h_{44} \log (2) \arccosh(\gamma )}{8 \left(\gamma ^2-1\right)^2}
      +\frac{3 h_{45} \left(\text{Li}_2\left(\frac{\gamma -1}{\gamma +1}\right)-4 \text{Li}_2\left(\sqrt{\frac{\gamma -1}{\gamma +1}}\right)\right)}{16 \left(\gamma ^2-1\right)^2}
      \nonumber\\
      &~
      +\frac{3 h_{46} \left(\log \left(\frac{\gamma +1}{2}\right) \arccosh(\gamma )-2 \text{Li}_2\left(\sqrt{\gamma ^2-1}-\gamma \right)\right)}{8 \left(\gamma ^2-1\right)^2}
      \nonumber\\
      &~
      -\frac{h_{47} \left(\text{Li}_2\Big(-\big(\gamma -\sqrt{\gamma ^2-1}\big)^2\Big)-2 \log (\gamma ) \arccosh(\gamma )\right)}{16 \left(\gamma ^2-1\right)^2}
      +\frac{8 h_{60} \arccosh^2(\gamma )}{\left(\gamma ^2-1\right)^2}
      \Bigg]
      \nonumber\\
      &~
    +m_1^2 m_2^3\Bigg[
      \frac{h_{49}}{1440 \gamma ^7 \left(\gamma ^2-1\right)^{5/2}}
      +\frac{\pi ^2 \left(27 h_{31}(\gamma^2-1)+2 h_{50}\right)}{96 \left(\gamma ^2-1\right)^2}
      +\frac{h_{51} \arccosh(\gamma )}{480 \gamma ^8 \left(\gamma ^2-1\right)^3}
      \nonumber\\
      &~
      -\frac{32 h_{54} \arccosh^2(\gamma )}{\left(\gamma ^2-1\right)^{7/2}}
      -\frac{h_{55} \log (2) \arccosh(\gamma )}{4 \left(\gamma ^2-1\right)^2}
      +\frac{h_{56} \left(\text{Li}_2\left(\frac{\gamma -1}{\gamma +1}\right)-4 \text{Li}_2\left(\sqrt{\frac{\gamma -1}{\gamma +1}}\right)\right)}{32 \left(\gamma ^2-1\right)^2}
      \nonumber\\
      &~
      -\frac{h_{57} \left(\log \left(\frac{\gamma +1}{2}\right) \arccosh(\gamma )-2 \text{Li}_2\left(\sqrt{\gamma ^2-1}-\gamma \right)\right)}{4 \left(\gamma ^2-1\right)^2}
      \nonumber\\
      &~
      +\frac{h_{58} \left(\text{Li}_2\Big(-\big(\gamma -\sqrt{\gamma ^2-1}\big)^2\Big)-2 \log (\gamma ) \arccosh(\gamma )\right)}{8 \left(\gamma ^2-1\right)^2}
      +\frac{16 h_{59} \arccosh^2(\gamma )}{\left(\gamma ^2-1\right)^2}
      \Bigg]\,,
\nn
\end{align}
\endgroup
whereas regions having two radiation modes going on-shell, including hereditary as well as nonlinear radiation-reaction effects, yield
\begin{align}\label{impdis2}
    \frac{c^{(4) \rm diss}_{1b,\rm 2rad}}{\pi} =&\,
      m_1^3 m_2^2\bigg[
        \frac{\sqrt{\gamma ^2-1} \left(h_9-4 \gamma ^2 (\gamma +1) h_{20}\right)}{768 (\gamma -1)^3 \gamma ^9 (\gamma +1)^4}
        +\frac{\log (\gamma ) \left(h_{12}-8 \left(\gamma ^2-1\right) h_{23}\right)}{16 \left(\gamma ^2-1\right)^{5/2}}
        \nonumber\\
        &~
        +\frac{\arccosh(\gamma ) \left((\gamma +1) h_{24}-2 (\gamma -1)^2 h_{13}\right)}{16 (\gamma -1)^3 (\gamma +1)^4}
        \nonumber\\
        &~
        +\frac{3 h_{26} \arccosh^2(\gamma )}{32 \left(\gamma ^2-1\right)^{7/2}}
        +\frac{3 \left(h_{15}-4 h_{27}\right) \log ^2\left(\frac{\gamma +1}{2}\right)}{8 \sqrt{\gamma ^2-1}}
        \nonumber\\
        &~
        +\log \left(\frac{\gamma +1}{2}\right) \left(\frac{-2 \left(\gamma ^2-1\right) h_{22}-h_{11}}{32 \left(\gamma ^2-1\right)^{5/2}}-\frac{3 \left(h_{16}+h_{28}\right) \arccosh(\gamma )}{16 \left(\gamma ^2-1\right)^2}\right)
        \nonumber\\
        &~
        +\frac{\left(-3 \left(\gamma ^2-1\right) h_{18}-8 h_{29}\right) \text{Li}_2\left(\frac{1-\gamma }{\gamma +1}\right)}{32 \sqrt{\gamma ^2-1}}
        -\frac{3 \left(h_{17}+8 h_{30}\right) \text{Li}_2\left(\frac{\gamma -1}{\gamma +1}\right)}{64 \sqrt{\gamma ^2-1}}
        \bigg]\,,
\nonumber\\[0.7 em]
c^{(4) \rm diss}_{1\check{u}_1,\rm 2rad} =&\, 0\,,
\\[0.7 em]
c^{(4)\rm diss}_{1\check{u}_2,\rm 2rad} =&\,
  m_1^3 m_2^2\bigg[
    \frac{4 \left(h_{33}-h_{37}\right)}{3 \left(\gamma ^2-1\right)^3}
    -\frac{16 \left(3 h_{36}-2 h_{41}\right) \arccosh(\gamma )}{3 \left(\gamma ^2-1\right)^{3/2}}
    -\frac{8 \left(h_{42}+h_{60}\right) \arccosh^2(\gamma )}{\left(\gamma ^2-1\right)^2}
    \bigg]
    \nonumber\\
    &~
  +m_1^2 m_2^3\bigg[
    \frac{2 \left(45 h_{32}-h_{48}\right)}{45 \left(\gamma ^2-1\right)^3}
    -\frac{16 h_{52} \arccosh(\gamma )}{5 \left(\gamma ^2-1\right)^{3/2}}
    -\frac{16 \left(h_{53}+h_{59}\right) \arccosh^2(\gamma )}{\left(\gamma ^2-1\right)^2}
    \bigg]\,.
\nonumber
\end{align}
Notice that, since $c^{(4)}_{1\check{u}_1}$ is related by the mass-shell condition to lower order contributions at 3PM featuring at most 1rad terms,  the $c^{(4) \rm diss}_{1\check{u}_1,\rm 2rad}$ coefficient is consistently zero. We~also find that $c^{(4) \rm diss}_{1b,\rm 2rad} \propto m_1^3m_2^2$, with a vanishing ${\cal O}(m_1^2m_2^3)$ term. This means that the $m_1^2m_2^3$ contribution to the full 2rad-impulse in the $b$-direction comes entirely from the Feynman-only~part. This is not only consistent with PN results \cite{Bini:2022enm}, it is also expected from the fact that having such term in the dissipative sector would imply the existence of an additional---beyond the Feynman-only part---time-symmetric conservative-like contribution to the impulse, at first order in the self-force expansion~\cite{Barack:2018yvs}. This, however, would be in tension with known PN \cite{Bini:2021gat,Bini:2021qvf,Bini:2022yrk,Bini:2022xpp} and PM results~\cite{4pmeft2}. (See also \cite{Bini:2022enm} for additional PN-type constraints implying the vanishing of the ${\cal O}(m_1^2m_2^3)$ part of the dissipative 2rad piece of the impulse in the $b$-direction.)

\subsection{Radiated momentum}

The impulse allows us to derive the change in the mechanical momentum of the system, that gives us the total radiated momentum, \beq P^\mu_{\rm rad}=-(\Delta p^\mu_1+\Delta p_2^\mu)\,,\eeq  
from which we can derive a series of GW observables.

\subsubsection{Recoil}

The expression for the (space-like) recoil is somewhat lengthy, but it is instructive to look at the PN expansion. In particular, expanding in $v_\infty$ we find along the $b$-direction, 
\beq
\begin{aligned}
      \label{pb}
\frac{b^4 P^{\rm 4PM}_{b,\rm rad}}{\pi \Delta_m  G^4M^5 \nu^2}=\frac{37  }{30 }+\frac{1661   v_{\infty }^2}{560 }+\frac{1491   v_{\infty }^3}{400 }
+\frac{23563 v_{\infty }^4}{10080 } -\frac{26757   v_{\infty }^5}{5600 }+\frac{700793   v_{\infty }^6}{506880 } +\mathcal{O}(v_{\infty }^7)\,,
   \end{aligned}
\eeq 
in agreement with a recent PN derivation in~\cite{Bini:2022enm}. The result in \eqref{pb}, which affects the value of the relative deflection angle, has no parallel at 3PM order. 

\subsubsection{Total energy}

The radiated energy, in the incoming center-of-mass frame, in an hyperbolic-like scattering process can be obtained via
\beq
\Delta E_{\rm hyp} \equiv P_{\rm rad}\cdot \frac{m_1 u_1 + m_2 u_2 }{M \Gamma}\,.
\eeq
Inputting the value for the impulse(s), the ${\cal O}(G^4)$ contribution is given by
\begingroup
\allowdisplaybreaks
\begin{align}
    \label{de4}
    \Delta E_{\rm hyp}^{\rm 4PM} 
    =&\,
      -\frac{G^4 M^5 \nu ^2}{b^4 \Gamma }\Bigg\{
      \frac{15 \pi ^2 \left(\gamma ^2-1\right) \left(27 \left(\gamma ^2-1\right) h_{31}
        +2 h_{50}\right)
        +64 \left(45 h_{32}-h_{48}\right)}{1440 \left(\gamma ^2-1\right)^3}
        \nonumber\\
        &\,
      +\frac{h_{49}}{1440 \gamma ^7 \left(\gamma ^2-1\right)^{5/2}}
      -\arccosh^2(\gamma ) \left(\frac{16 h_{53}}{\left(\gamma ^2-1\right)^2}+\frac{32 h_{54}}{\left(\gamma ^2-1\right)^{7/2}}\right)
      \nonumber\\
      &\,
      -\frac{h_{55} \log (2)\arccosh(\gamma )}{4 \left(\gamma ^2-1\right)^2}
      +\frac{h_{57} \log \left(\frac{2}{\gamma +1}\right)\arccosh(\gamma )}{4 \left(\gamma ^2-1\right)^2}
      -\frac{h_{58} \log (\gamma )\arccosh(\gamma )}{4 \left(\gamma ^2-1\right)^2}
      \nonumber\\
      &\,
      +\arccosh(\gamma ) \left(\frac{h_{51}}{480 \gamma ^8 \left(\gamma ^2-1\right)^3}
      -\frac{16 h_{52}}{5 \left(\gamma ^2-1\right)^{3/2}}\right)
      -\frac{h_{56} \text{Li}_2\left(\sqrt{\frac{\gamma -1}{\gamma +1}}\right)}{8 \left(\gamma ^2-1\right)^2}
      \nonumber\\
      &\,
      +\frac{h_{56} \text{Li}_2\left(\frac{\gamma -1}{\gamma +1}\right)}{32 \left(\gamma ^2-1\right)^2}
      +\frac{h_{57} \text{Li}_2\left(\sqrt{\gamma ^2-1}-\gamma \right)}{2 \left(\gamma ^2-1\right)^2}
      +\frac{h_{58} \text{Li}_2\Big(-\big(\gamma -\sqrt{\gamma ^2-1}\big)^2\Big)}{8 \left(\gamma ^2-1\right)^2}
      \nonumber\\
      &\,
      +\nu\Bigg[
        \frac{4 \left(-45 h_{32}+30 h_{33}-30 h_{37}+h_{48}\right)}{45 \left(\gamma ^2-1\right)^3}+\frac{\pi ^2 \left(54 \left(\gamma ^2-1\right) h_{31}+h_{39}-4 h_{50}\right)}{96 \left(\gamma ^2-1\right)^2}
        \nonumber\\
        &\,
        -\arccosh^2(\gamma ) \left(\frac{16 \left(h_{42}-2 h_{53}\right)}{\left(\gamma ^2-1\right)^2}-\frac{64 (h_{43}+h_{54})}{\left(\gamma ^2-1\right)^{7/2}}\right)
        \\
        &\,
        +\frac{h_{38}-490 \gamma  \left(3840 \gamma ^7 h_{34}+h_{49}\right)}{352800 \gamma ^8 \left(\gamma ^2-1\right)^{5/2}}
        +\frac{(3 h_{46}+2h_{57}) \log \left(\frac{\gamma +1}{2}\right)\arccosh(\gamma )}{4 \left(\gamma ^2-1\right)^2}
        \nonumber\\
        &\,
        +\frac{\left(h_{44}+2 h_{55}\right) \log (2)\arccosh(\gamma )}{4 \left(\gamma ^2-1\right)^2}
        +\frac{\left(h_{47}+2 h_{58}\right) \log (\gamma )\arccosh(\gamma )}{4 \left(\gamma ^2-1\right)^2}
        \nonumber\\
        &\,
        +\arccosh(\gamma ) \left(\frac{53760 \gamma ^9 h_{35}-14 \gamma  h_{51}+h_{40}}{3360 \gamma ^9 \left(\gamma ^2-1\right)^3}-\frac{32 \left(15 h_{36}-10 h_{41}-3 h_{52}\right)}{15 \left(\gamma ^2-1\right)^{3/2}}\right)
        \nonumber\\
        &\,
        +\frac{\left(h_{56}-6 h_{45}\right) \text{Li}_2\left(\sqrt{\frac{\gamma -1}{\gamma +1}}\right)}{4 \left(\gamma ^2-1\right)^2}
        -\frac{\left(h_{56}-6 h_{45}\right) \text{Li}_2\left(\frac{\gamma -1}{\gamma +1}\right)}{16 \left(\gamma ^2-1\right)^2}
        \nonumber\\
        &\,
        -\frac{\left(3 h_{46}+2 h_{57}\right) \text{Li}_2\left(\sqrt{\gamma ^2-1}-\gamma \right)}{2 \left(\gamma ^2-1\right)^2}
        -\frac{\left(h_{47}+2 h_{58}\right) \text{Li}_2\Big(-\big(\gamma -\sqrt{\gamma ^2-1}\big)^2\Big)}{8 \left(\gamma ^2-1\right)^2}
        \Bigg]
      \Bigg\}\,,
      \nonumber
\end{align}
\endgroup
which, after PN-expanding, 
\begin{align}\label{DE}
 \frac{b^4 \Delta E_{\rm hyp}^{\rm 4PM}}{G^4M^5\nu^2} =&~ \frac{1568}{45 v_{\infty }}
 +\left(\frac{18608}{525}-\frac{1136 \nu }{45}\right) v_{\infty }+\frac{3136 v_{\infty }^2}{45} + \left(\frac{764 \nu ^2}{45}-\frac{356 \nu }{63}+\frac{220348}{11025}\right) v_{\infty }^3
 \nn\\
 &~+\left(\frac{1216}{105}-\frac{2272 \nu }{45}\right) v_{\infty }^4 + \left(-\frac{622 \nu  ^3}{45}+\frac{3028 \nu ^2}{1575}-\frac{199538 \nu }{33075}-\frac{151854}{13475}\right) v_{\infty }^5 
 \nn\\ 
 &~+ \left(\frac{1528 \nu ^2}{45}-\frac{8056 \nu }{1575}+\frac{117248}{1575}\right) v_{\infty }^6+\mathcal{O}(v_{\infty }^7)\,,
\end{align}
is in perfect agreement with the existent PN literature \cite{Damour:2020tta,Cho:1,Cho:2,Bini:2021gat,Bini:2021qvf,Bini:2022yrk,Bini:2022xpp,Bini:2022enm}, 
\subsubsection{GW flux}
The B2B map allows us to relate the total radiated energy for the hyperbolic-like motion to its counterpart over a period of an elliptic-like orbit  via the relation \cite{b2b3}
\beq \Delta E_{\rm ell}(j) = \Delta E_{\rm hyp}(j) - \Delta E_{\rm hyp}(-j)\,,\eeq   
which nicely agrees in the overlap with PN data at 3PM order \cite{Parra3}, but at the same time yields the expected vanishing contribution for even orders with bound states. It is then convenient to compute instead the GW flux, which can be used for generic orbits. The energy flux can be PM-expanded as follows (in an isotropic gauge) \cite{b2b3}
\beq
      \begin{aligned}
   \frac{dE}{dt} =  \frac{M}{r}\sum_n  {\cal F}^{(n)}_E(\gamma) \left(\frac{GM}{r}\right)^{(n+3)}\,,
    \end{aligned}
    \eeq
 and similarly for the total radiated energy  (recall $j = p_\infty b/(GM^2\nu)$) 
   \beq \Delta { E}_{\rm hyp} (j) \!= \!\sum_{n=0}^{\infty} \frac{\Delta E_{j\, \rm hyp}^{(n)}}{j^{n+3}}\,.\eeq
Hence, using the conservative-like part of the scattering trajectory, thus working within an adiabatic expansion, we find \cite{b2b3}
\beq
\begin{aligned}
\label{f12}
    M\pi\xi \,{\cal F}_E^{(0)} &= \frac{2 \Gamma  \nu \Delta E_{j\,\textrm{hyp}}^{(0)}}{\left(\gamma ^2-1\right) }\,,\\
    M\pi\xi \, {\cal F}_E^{(1)} &= \frac{3 \pi  \Gamma ^2 \nu \Delta E_{j\,\textrm{hyp}}^{(1)} }{4 \left(\gamma ^2-1\right)^{3/2}}-\frac{2 \Delta E_{j\,\textrm{hyp}}^{(0)} \nu ^3}{\left(\gamma ^2-1\right)^2 \Gamma ^6 \xi ^2} \bigg[(\gamma -1)^3 \left(10 \gamma ^3-10 \gamma ^2-9 \gamma +5\right) \nu ^2\\
      &\quad\quad+4 \left(5 \gamma ^5-8 \gamma ^4+\gamma ^3+4 \gamma ^2-3 \gamma +1\right) \nu +8 \gamma ^4-4 \gamma ^2-1\bigg]\,.
  \end{aligned}
\eeq
The result in \eqref{f12} can be readily applied to elliptic-like motion.\footnote{Nonlocal-in-time effects due to tail terms, e.g.~\cite{tail}, do not enter in the GW flux until higher PM orders.} See the ancillary file for explicit values.

\subsection{Scattering angle}

The relative impulse, $\Delta \bp$, can be obtained from the value of the total recoil, see e.g.~\cite{Damour:2020tta}, 
\beq
\label{relative}
\Delta \bp =  \Delta \bp_1 + \frac{E_1}{E} \bP_{\rm rad} + {\cal O}\big(\bP_{\rm rad}^2\big)\,,
\eeq
from which we can define a deflection angle, 
 \beq \frac{\chi_{\rm rel}}{2} \equiv \frac{1}{2} \mathrm{arccos}\left(\frac{\bp_+\cdot \bp_- }{ |{\bp}_-||\bp_+|}\right) = \sum_{n=1}^{\infty} \chi^{(n)}_{b,\rm rel} \left(\frac{GM}{b}\right)^n  = \sum_{n=1}^{\infty} \frac{\chi^{(n)}_{j,\rm rel} }{j^n}  \,, \eeq where $\bp_-$ and $\bp_+ \equiv \bp_- + \Delta \bp$ are the relative incoming/outgoing {$\bf 3$}-momenta. At 4PM order, we can decompose it as follows (in impact parameter space)
 \beq
\begin{aligned}
\chi^{(4)}_{b, \rm rel}(\gamma) &=  \chi^{(4)\rm cons}_{b,\rm rel}(\gamma) +   \chi^{(4)\rm rr}_{b,\rm rel}(\gamma)\,,
\\[0.35 em]
\chi^{(4)\rm rr}_{b,\rm rel}(\gamma) &= \chi^{(4)\rm rr, 1rad}_{b,\rm rel}(\gamma) +  \chi^{(4)\rm rr, 2rad}_{b,\rm rel}(\gamma)\,,
\end{aligned}
\eeq
where the (Feynman-only) conservative part, ignoring the recoil, is given by 
\begingroup
\allowdisplaybreaks
\begin{align}
\label{anglecons}
  \frac{ \chi^{(4)\rm cons}_{b,\rm rel}}{\pi\Gamma}
  =&\,
    \frac{3 h_{61}}{128 \left(\gamma ^2-1\right)^3}
    +\nu\Bigg[
      -\frac{3 h_3 \mathrm{K}^2\left(\frac{\gamma -1}{\gamma +1}\right)}{32 \left(\gamma ^2-1\right)^2}
      +\frac{3 h_4 \mathrm{E}\left(\frac{\gamma -1}{\gamma +1}\right) \mathrm{K}\left(\frac{\gamma -1}{\gamma +1}\right)}{32 \left(\gamma ^2-1\right)^2}
      +\frac{\pi ^2 h_5}{16(1- \gamma ^2)}
      \nonumber\\
      &~
      +\frac{3 h_{27} \log ^2\left(\frac{\gamma +1}{2}\right)}{4(1- \gamma ^2)}
      -\frac{h_6 \log \left(\frac{\gamma -1}{2}\right)}{32 \left(\gamma ^2-1\right)^2}
      +\frac{3 h_{15} \log \left(\frac{\gamma -1}{2}\right) \log \left(\frac{\gamma +1}{2}\right)}{16 \left(\gamma ^2-1\right)}
      -\frac{h_{22} \log \left(\frac{\gamma +1}{2}\right)}{32 \left(\gamma ^2-1\right)^2}
      \nonumber\\
      &~
      -\frac{h_{23} \log (\gamma )}{4 \left(\gamma ^2-1\right)^2}
      +\frac{3 h_{26} \arccosh^2(\gamma )}{64 \left(\gamma ^2-1\right)^4}
      +\frac{h_{24} \arccosh(\gamma )}{32 \left(\gamma ^2-1\right)^{7/2}}
      -\frac{3 h_{16} \log \left(\frac{\gamma -1}{2}\right) \arccosh(\gamma )}{32 \left(\gamma ^2-1\right)^{5/2}}
      \nonumber\\
      &~
      -\frac{3 h_{28} \log \left(\frac{\gamma +1}{2}\right) \arccosh(\gamma )}{32 \left(\gamma ^2-1\right)^{5/2}}
      -\frac{h_{62}}{384 \gamma ^7 \left(\gamma ^2-1\right)^3}
      -\frac{21 h_2 \mathrm{E}^2\left(\frac{\gamma -1}{\gamma +1}\right)}{64 (\gamma -1)^2 (\gamma +1)}
      \nonumber\\
      &~
      -\frac{3 \sqrt{\gamma ^2-1} h_7 \text{Li}_2\left(\sqrt{\frac{\gamma -1}{\gamma +1}}\right)}{2 (\gamma -1)^2 (\gamma +1)^3}
      +\frac{h_{29} \text{Li}_2\left(\frac{1-\gamma }{\gamma +1}\right)}{8(1- \gamma^2)}
      \nonumber\\
      &~
      +\left(\frac{3 \sqrt{\gamma ^2-1} h_7}{8 (\gamma -1)^2 (\gamma +1)^3}
      +\frac{3 h_{30}}{16-16 \gamma ^2}\right) \text{Li}_2\left(\frac{\gamma -1}{\gamma +1}\right)
      \Bigg]\,,
\end{align}
\endgroup
in agreement with the derivation in \cite{4pmeft,4pmeft2} (see also App.~\ref{sec:app-angle}); whereas for the remaining terms, we~find
\begingroup
\allowdisplaybreaks 
\begin{align}
  \frac{\Gamma\chi^{(4)\rm rr, 1rad}_{b,\rm rel}}{\pi\nu} =&
    \frac{h_{64}}{96 \left(\gamma ^2-1\right)^{7/2}}+\frac{h_{65} \log \left(\frac{\gamma +1}{2}\right)}{16 \left(\gamma ^2-1\right)^{5/2}}+\frac{h_{63} \arcsinh\left(\frac{\sqrt{\gamma -1}}{\sqrt{2}}\right)}{8 \left(\gamma ^2-1\right)^4}
    \nn\\
    &~
    -\frac{h_{25} \arccosh(\gamma )}{32 \left(\gamma ^2-1\right)^4}
    +\nu\Bigg[
      \frac{h_{67}}{96 \left(\gamma ^2-1\right)^{7/2}}+\frac{h_{68} \log \left(\frac{\gamma +1}{2}\right)}{16 \left(\gamma ^2-1\right)^{5/2}}
      \nn\\
      &~
      -\frac{\arccosh(\gamma ) \left((\gamma +1) h_{14}+(\gamma -3) h_{25}\right)}{32 \left(\gamma ^2-1\right)^4}
      +\frac{h_{66} \arcsinh \left(\frac{\sqrt{\gamma -1}}{\sqrt{2}}\right)}{8 (\gamma -1)^2 (\gamma +1)^4}
      \Bigg]\,,
\label{anglediss}
\\[0.6 em]
  \frac{\Gamma\chi^{(4)\rm rr, 2rad}_{b,\rm rel}}{\pi\nu^2} =&~
    \frac{\log \left(\frac{\gamma +1}{2}\right) \left(2 \left(\gamma ^2-1\right) h_{22}+h_{11}\right)}{64 (\gamma -1)^3 (\gamma +1)^2}
    -\frac{\log (\gamma ) \left(h_{12}-8 \left(\gamma ^2-1\right) h_{23}\right)}{32 (\gamma -1)^3 (\gamma +1)^2}
    \nn\\
    &~
    +\frac{\arccosh(\gamma ) \left(2 (\gamma -1)^2 h_{13}-(\gamma +1) h_{24}\right)}{32 \left(\gamma ^2-1\right)^{7/2}}
    -\frac{3 \left(h_{15}-4 h_{27}\right) \log ^2\left(\frac{\gamma +1}{2}\right)}{16 (\gamma -1)}
    \nn\\
    &~
    +\frac{3 \sqrt{\gamma ^2-1} \left(h_{16}+h_{28}\right) \log \left(\frac{\gamma +1}{2}\right) \arccosh(\gamma )}{32 (\gamma -1)^3 (\gamma +1)^2}
    -\frac{h_9-4 \gamma ^2 (\gamma +1) h_{20}}{1536 \gamma ^9 \left(\gamma ^2-1\right)^3}
    \nn\\
    &~
    -\frac{3 h_{26} \arccosh^2(\gamma )}{64 (\gamma -1)^4 (\gamma +1)^3}+\left(\frac{3}{64} (\gamma +1) h_{18}+\frac{h_{29}}{8 (\gamma -1)}\right) \text{Li}_2\left(\frac{1-\gamma }{\gamma +1}\right)
    \nn\\
    &~
    +\frac{3 \left(h_{17}+8 h_{30}\right) \text{Li}_2\left(\frac{\gamma -1}{\gamma +1}\right)}{128 (\gamma -1)}\,,
\label{anglediss2}
\end{align}
\endgroup
for the dissipative parts involving one (1rad) and two (2rad) propagators going on-shell. See App.~\ref{sec:app-pols} for the value of the $h_i$ polynomials. 

\subsection{Firsov resummation} 

Restricted to interactions which conserve energy and momentum, the scattering angle allows us to compute the PM components of the square of the momentum of each particle (or {\it impetus}) in the incoming center-of-mass frame,  
\beq
\label{impetus}
\bp^2 = p_\infty^2 + \sum_{n=1}^{\infty} P_n \left(\frac{G}{r}\right)^n = p_\infty^2 \left(1+ \sum_{n=1}^{\infty} f_n\left( \frac{GM}{r}\right)^n\right) \,,   
\eeq
via the Firsov parameterization (with $\bar p \equiv |\bp|/p_\infty$) \cite{paper1,paper2}
\beq
\label{firsov1}
\widebar\bp^2 = \exp\left[ \frac{2}{\pi} \int_{r|\widebar\bp|}^\infty \frac{\chi_b\,\dd b}{\sqrt{ b^2-r^2\widebar\bp^2}}\right]\,,
\eeq
yielding the general formula
\beq
\begin{aligned}
\label{eq:fi}
f_n = \sum_{\sigma\in\mathcal{P}(n)}g_\sigma^{(n)} \prod_{\ell} \left(\widehat{\chi}_b^{(\sigma_{\ell})}\right)^{\sigma^{\ell}}\,, \quad\quad P_n = p_\infty^2 M^n f_n\,,
\end{aligned}
\eeq
where
\beq
\begin{aligned}
g_\sigma^{(n)} &= \frac{2(2-n)^{\Sigma^{\ell} - 1}}{\prod_{\ell} (2\sigma^{\ell})!!}\,,\quad \quad
\widehat{\chi}_b^{(n)} = \frac{2}{\sqrt{\pi}}\frac{\Gamma(\frac{n}{2})}{\Gamma(\frac{n+1}{2})}\chi^{(n)}_b\,,\quad\quad \Sigma^\ell \equiv \sum_{\ell} \sigma^{\ell}\,.
\end{aligned}
\eeq
The $\mathcal{P}(n)$'s in \eqref{eq:fi} are the set of integer partitions of $n= \sigma_{\ell} \sigma^{\ell}$ (implicit summation), with mutually different $\sigma_{\ell}$'s. For instance, to 4PM we have 
\beq\label{fn}
\begin{aligned}
f_1 &= 2 \chi_b^{(1)}\,,\quad
f_2 = \frac{4}{\pi}\chi_b^{(2)}\,,\\
f_3 &= \frac{1}{3}\left(\chi_b^{(1)}\right)^3-\frac{4}{\pi}\chi_b^{(1)}\chi_b^{(2)}+\chi_b^{(3)}\,,\\ 
 f_4 &= -\frac{2}{3}\left(\chi_b^{(1)}\right)^4+\frac{8}{\pi}\left(\chi_b^{(1)}\right)^2\chi_b^{(2)}-\frac{8}{\pi^2}\left(\chi_b^{(2)}\right)^2-2\chi_b^{(1)}\chi_b^{(3)}+\frac{8}{3\pi}\chi_b^{(4)}\,.
    \end{aligned}
\eeq
We can also (recursively) reconstruct a Hamiltonian in an isotropic gauge,
\beq
H  (r,\bp^2)= \sum_{n=0}^{\infty} \frac{c^H_n(\bp^2) }{n!} \left(\frac{G}{r}\right)^n\,,
\eeq
where $c_0^H \equiv  E_1(\bp^2) + E_2(\bp^2) = \sqrt{\bp^2+m_1^2} + \sqrt{\bp^2+m_2^2}$. The relationship to the $f_n$ (and $P_n$) coefficients can be found in~\cite{paper1}, which we recommend to the reader for further details.\vskip 4pt

From the knowledge of the $f_k$'s, or equivalently the $P_k$'s, to an $n$PM value we can then obtain a `Firsov-resummed' deflection angle---which descends from what we called an `$f_n$-theory' in \cite{paper1}---by simply going back to the original relationship,  
\beq
\chi =  -\pi + 2b \int_{r_{\rm min}}^\infty \frac{\dd r}{r\sqrt{r^2\widebar\bp^2(r,E)-b^2}}\,,\label{chiF}
\eeq
and performing the integration, with $r_{\rm min}$ the distance of closest approach defined through the condition $p_r(r_{\rm min}) = 0$, or by explicitly finding a closed-form (resummed) expression. See~\cite{paper1} for a few specific examples.\vskip 4pt The $f_4$-theory approximation to the impetus formula in \eqref{impetus}, obtained using \eqref{fn} and the results in \cite{4pmeftot}, can also be compared directly against numerical simulations, for instance, by using the parameterization~\cite{paper1}
\beq
\begin{aligned}
r(\lambda) &= \lambda \, e^{-A(\lambda)}\,,\quad
\widebar\bp^2(r(\lambda))= e^{2A(\lambda)}\,,\\
A(\lambda) &\equiv \frac{1}{\pi} \int_\lambda^\infty \frac{\chi_b \dd b}{\sqrt{ b^2-\lambda^2}}\,,
\end{aligned}
\eeq
to match to data. See \cite{Damour:2022ybd} for this and other implementations and comparisons demonstrating incredible accuracy. 

\vskip 4pt \textbf{\textit{Conservative-like.}} Although, technically speaking, only the contributions from the impulse leading to \eqref{anglecons} conserve {\it both} the total energy and momentum in the incoming frame at 4PM order, the relative dynamics may still include other {\it conservative-like} interactions. In fact, since effects at second order in the radiation-reaction forces do not dissipate energy at this order\footnote{This is  supported by the comparison to the PN-expanded value of the total radiated energy \cite{Bini:2022enm}.} and, moreover, the total relative momentum is also conserved, we conclude that the (dissipative) 2rad term in \eqref{anglediss2} may be added to an (effectively) Hamiltonian-like description of the dynamics at ${\cal O}(G^4)$. Hence, introducing an effective angle,
\beq
\chi^{(4)\rm eff}_{b, \rm rel} = \chi^{(4)\rm cons}_{b, \rm rel} + \chi^{(4)\rm rr, 2rad}_{b, \rm rel}\,,
\eeq 
which would feature in the above formulae, we obtain an effective (relative) impetus formula,
\beq
\bp^2_{\rm eff} = p_{\infty}^2 \left(1 + \sum_{n=1}^{\infty} f_n^{\rm eff} \left(\frac{GM}{r}\right)^n\right) \,.  
\eeq
We can then input this expression to obtain an effective PM-resummed (conservative-like) deflection angle in \eqref{chiF}.\footnote{In principle, if one ignores recoil effects, for instance for the case of equal-mass scattering, one could also include even the 1rad term in \eqref{anglediss} to an effective Firsov representation. This was done in \cite{Damour:2022ybd} with spectacular success in matching to numerical simulations.}  The value of the coefficients can be found in the ancillary file.

\section{Conclusions \& Outlook}\label{sec:conc}

We elaborated here on the systematic framework to compute GW observables for compact binaries via the EFT approach \cite{pmeft,3pmeft,eftrad} in combination with modern integration techniques from particle physics~\cite{Smirnov:2012gma,Weinzierl:2022eaz,Kotikov:1991pm,Remiddi:1997ny,Henn:2013pwa,Prausa:2017ltv,Lee:2020zfb,Lee:2014ioa,Adams:2018yfj,Chetyrkin:1981qh,Tkachov:1981wb,Smirnov:2019qkx,Smirnov:2020quc,Lee:2012cn,Lee:2013mka,Beneke:1997zp,Jantzen:2012mw,Smirnov:2015mct,Meyer:2016zeb,Meyer:2016slj,Broedel:2019kmn,Primo:2017ipr,Hidding:2020ytt,Goncharov:2001iea,Chen:1977oja,Duhr:2014woa,Duhr:2019tlz,Dlapa:2020cwj,Smirnov:2021rhf,Lee:2019zop,Blumlein:2021pgo}. The EFT formalism was recently employed to derive the total spacetime impulse in the scattering of non-spinning compact bodies to 4PM order, reported in \cite{4pmeft2,4pmeftot}, which we have described in more detail in the present paper. Perfect agreement is found in the overlap with various partial results at ${\cal O}(G^4)$, within the PN \cite{Damour:2020tta,Cho:1,Cho:2,Bini:2021gat,Bini:2021qvf,Bini:2022yrk,Bini:2022xpp,Bini:2022enm} and PM \cite{4pmzvi2,Manohar:2022dea,DiVecchia:2022nna} expansions, respectively. A more recent analysis including a comparison between the ($f_4$-)Firsov-resummed deflection angle (see \S\ref{sec:data}) and numerical simulations was performed in~\cite{Damour:2022ybd}, finding an exquisite agreement between theory and numerical data. All these nontrivial checks give us confidence in the validity of our complete 4PM results~\cite{4pmeft2,4pmeftot}.\vskip 4pt   

Although the framework discussed here has been very successful to tackle both the conservative and dissipative dynamics of relativistic compact bodies in hyperbolic-like motion to 4PM order, there are various issues and subtleties which deserve further study:
\begin{itemize}
\item  \textbf{\textit{Local vs Nonlocal (in time).}} As it is well known, starting at ${\cal O}(G^4)$, tail-type hereditary interactions introduce nonlocal-in-time effects in the conservative dynamics \cite{tail,4pmeft,4pmeft2,b2b3}. Up to a given PM order, the latter can be described in terms of an (averaged) effectively local Hamiltonian or impetus formula. However, the coefficients depend on the trajectory, which implies $\bp^2_{\rm hyp} \neq \bp^2_{\rm ell}$. Hence, even though we have shown that it works for all the local-in-time as well as the trademark tail-type logarithmic corrections for generic orbits \cite{4pmeft,4pmeft2,b2b3}, the full B2B analytic continuation between radial actions for unbound and bound states only applies to the large-eccentricity limit. This means that, unlike local terms and logarithms, the remaining nonlocal pieces cannot be continued smoothly from hyperbolic motion to generic elliptic orbits, e.g. the circular case. Nevertheless, we expect the B2B map to approach the true solution for highly eccentric bound orbits. As it was argued in \cite{b2b3}, the analytic continuation may still be possible at the level of the integrand in the PN expansion, prior to performing the time integration over the trajectory. For the PM integrand, however, this would require a non-relativistic split into time and space, while identifying at the same time hereditary-type contributions. We are presently exploring this possibility. 

\item  \textbf{\textit{Memory effects.}} The conservative and dissipative part of the scattering angle at 5PN order appeared in \cite{Blumlein:2021txe} and \cite{Almeida:2022jrv}, respectively. Unfortunately, in the overlap at~${\cal O}(G^4)$, we find disagreement between the results in \cite{Blumlein:2021txe,Almeida:2022jrv} and ours \cite{4pmeft,4pmeft2,4pmeftot}. Among the relevant PN corrections are {\it memory} contributions which, unlike tails described by $QQM$ terms in the effective action (with $Q$ some generic multipole moment and $M$ the total mass/energy), are described by $QQQ$ corrections in PN theory.\footnote{In addition, nonlinear gravitational contributions to the radiation-reaction force, as well as tail-like interactions beyond the monopole coupling, also yield terms of this type. As a consequence, these are often all piled up together in the PN effective action.}
On the one hand, as expected (see e.g.~\cite{Bini:2021gat}), agreement between PN and PM computations in the conservative sector requires nonzero  $QQQ$-type terms. On the other hand, according to the derivations in \cite{Blumlein:2021txe,Almeida:2022jrv}, the PN side includes regions with three (on-shell) radiation modes, that are absent in our PM calculation. One reason for the mismatch of regions may be simply due to the use of IBP relations.\footnote{After using IBP relations, the relevant PN integrals can be reduced to masters also featuring two radiation modes \cite{Blumlein:2021txe}. However, unlike the PM derivation, there is no constrain on their individual frequencies.} Or it may also be related to the fact that we are computing the total impulse, inserting the trajectories yielding Dirac-$\delta$ constraints; whereas the PN calculation involves the worldline Hamiltonian/Lagrangian. Had the PM and PN results agreed, we would then simply conclude that memory terms in the latter are ultimately described by 2rad-type regions in the impulse to ${\cal O}(G^4)$ in the former. However, the disagreement in the total value, and in particular the mass scaling of the result \cite{Bini:2021gat}, suggests that additional issues are still present in the comparison.\footnote{One possible issue is the following. The PM derivation of the impulse involves the incoming center-of-mass, while the PN action is obtained in the relative frame. Because of effects at second order in the radiation-reaction forces, the mismatch may be due to recoil-type contributions which have not been accounted for on the PN side. Let us  also point out that the relative deflection angle, see~\S\ref{sec:data}, has not resolved the discrepancy.} This is currently under investigation.
\item \textbf{\textit{High-energy limit.}} Another puzzling feature of the solution at 4PM order is the behavior of the spacetime impulse in the limit $\gamma \to \infty$, where we find
\begin{equation}
\begin{aligned}
c_{1b}^{(4)\textrm{tot}} &\xrightarrow{\gamma\rightarrow\infty} \frac{35\pi\gamma^3(7-12 \log(2)^2+\log(256))m_1^2 m_2^3}{8}\,,\\
c_{1\check{u}_1}^{(4)\textrm{tot}} &\xrightarrow{\gamma\rightarrow\infty} 64\gamma^3 m_1^2 m_2^3\,,
\qquad\qquad\qquad ~ c_{1u_1}^{(4)\textrm{tot}} \xrightarrow{\gamma\rightarrow\infty} -\frac{6848\gamma^2\log(2\gamma)m_1^3m_2^2}{105}\,,\\
c_{1\check{u}_2}^{(4)\textrm{tot}} &\xrightarrow{\gamma\rightarrow\infty} -\frac{6848\gamma^3\log(2\gamma)m_1^3 m_2^2}{105}\,,\quad
c_{1u_2}^{(4)\textrm{tot}} \xrightarrow{\gamma\rightarrow\infty} 64\gamma^2 m_1^2m_2^3\,,\\
\end{aligned}
\end{equation}
with the $c_{1u_a}$ coefficients defined w.r.t.~the (unchecked) incoming velocities; as well as for the total radiated energy, yielding
\beq
\frac{b^4 \Gamma \Delta E_{\rm hyp}^{\rm 4PM}}{G^4M^5\nu^2} \xrightarrow{\gamma\rightarrow \infty} \frac{13696}{105} \gamma ^3 \nu \log (2 \gamma)\,.
\eeq
If we then take, concurrently, the massless limit, $m_a \to 0$, while keeping $s \equiv \gamma m_1 m_2$ fixed, we discover that 
$c_{1b}^{(4)\textrm{tot}}  \sim   s^3/m_1$, whereas the (unchecked) velocity components go to zero.\footnote{We should, however, make the following observation. While the $c^{(n)}_{1u_a}$'s may remain finite in the high-energy/massless limit, the velocities themselves scale as  $u_a = p_a/m_a$, which diverge.} Similarly, for the radiated energy we find a logarithmic divergence,\footnote{In fact, retaining subleading pieces, we find that the $m_1 \to 0$ limit also generates terms proportional to $m_2/m_1$, which diverge if we keep $m_2$ finite.} 
\beq
\frac{\Delta E_{\rm hyp}^{\rm 4PM}}{\sqrt{s}} \sim \left(\chi^{(1)}\right)^4 \log\left(\frac{s}{m_1m_2}\right)\,,
\eeq
where we used $\chi^{(1)} \simeq  \frac{G\sqrt{s}}{b}$ for the leading order value of the deflection angle. In~principle, because of analyticity argument, it is believed that all even coefficients should vanish in the massless case, see e.g. \cite{DiVecchia:2019kta}. Hence, in the perturbative PM regime, we conclude that the massive theory (with $u_a^2=1$) does not extrapolate smoothly to the massless case ($u_a^2=0$).\footnote{The divergence in the $c^{(4)\rm tot}_{1b}$ coefficient is also intimately connected to the lack of ${\cal O}(m_2^3m_1^2)$ contribution to $c_{1b,\rm 2rad}^{(4)\rm diss}$ (see  \S\ref{sec:data}). In fact, it  turns out impossible to cancel the unwanted high-energy behavior while remaining consistent with all known PN as well as self-force data \cite{Bini:2022enm}.} We expect these issues to be resolved by a non-perturbative understanding of the high-energy limit, see e.g.~\cite{Gruzinov:2014moa,DiVecchia:2022nna} for some recent developments.

\item  \textbf{\textit{Scalability.}}
\begin{outline}
\1 \textbf{\textit{Feynman diagrams.}}
The integrand in the EFT approach is constructed via Feynman rules featuring (bulk) graviton vertices that grow factorially in the number of legs. Even after optimization~\cite{pmeft}, performing the tensor contractions and collecting contributions with equal integral structure becomes time consuming as well as memory critical. For instance, the construction of the 5PM integrand, already known, has taken the order of a few days (so far ignoring spin and tidal degrees of freedom). In this case, a dedicated code within \texttt{FORM} \cite{Ruijl:2017dtg}, together with heavy use of parallelization techniques, turned out to be sufficient. However, at 6PM and beyond, new ideas may be necessary to tame the factorial growth, as well as to produce more manageable integrand representations.

\1  \textbf{\textit{IBP reduction.}}
The IBP algorithms play a key role in PM computations. However, there are a few issues which will require further development in order to move forward to higher loop orders. Firstly, to our knowledge, none of the known programs in the literature can automatically handle symmetries of retarded propagators.
As a result, we have only implemented some symmetry relations at the level of the master integrals.
While this was sufficient to tackle all the three-loop computations at 4PM, higher order effects will present new challenges.
Secondly, the number of equations that need to be generated, and reduced according to Laporta's algorithm \cite{Laporta:2000dsw}, grows rapidly with every loop order and numerator power. This poses a serious challenge for state-of-the-art IBP programs already at four-loop order. It is clear that reductions of the most complicated integral families beyond 5PM will not be possible without significant (hardware and software) improvements.\footnote{We should also mention here some possible alternatives, such as the use of algebraic geometry~\cite{Tarasov:2004ks,DBLP:journals/corr/abs-cs-0509070,Smirnov:2005ky,Smirnov:2006tz,Smirnov:2006wh,Lee:2008tj,Barakat:2022ttc,Barakat:2022qlc} and intersection theory~\cite{Mastrolia:2018uzb,Mizera:2019gea,Mizera:2019vvs,Frellesvig:2019uqt,Frellesvig:2019kgj,Frellesvig:2020qot,Chestnov:2022alh,Cacciatori:2022mbi,Chestnov:2022xsy}. However, these are still at an early stage of development and therefore need to mature further in order to display their full potential.} Needless to say, research in precision gravity greatly benefits from advances in these areas and we thus encourage the community to extend the existent programs to systematize more general scenarios, including both Dirac-$\delta$ functions and retarded Green's functions, as well as explore new venues to tackle large(r) systems.

\1 \textbf{\textit{Integration.}} Challenges in the integration problem can be divided into two categories. Firstly, there is the methodology of differential equations. Assuming~the IBP-reduction step is successfully done, the problem is reduced to finding a form where a solution can be obtained, ideally in the $\epsilon$-factorized form. However, both the size of the differential equations and function space are known to become drastically more complicated the higher the loop order. For sectors involving polylogarithms, we can compare to other similar cases, e.g.~\cite{Bruser:2020bsh}, and also make use of the \texttt{INITIAL} algorithm \cite{Dlapa:2020cwj}.\footnote{It was already observed in \cite{Bruser:2020bsh} that going from three to four loops introduced additional poles at higher roots of unity into the differential equations. This, in turn, led to the development of the \texttt{INITIAL} algorithm \cite{Dlapa:2020cwj}, which we expect to handle the polylogarithmic sectors at 5PM.} At the same time, the non-polylogarithmic sectors are also expected to become more complicated at four- and higher loop orders, where experience shows that each extra loop increases by one the order of the corresponding Picard-Fuchs equation. Hence, at 5PM we might encounter an object satisfying a fourth-order Picard-Fuchs equation, for which we may not have a representation in terms of elliptic integrals. In this regard, recent advances in the particle physics community, see e.g.~\cite{Pogel:2022ken,Pogel:2022vat,Duhr:2022dxb,Dlapa:2022wdu}, may lead to a solution of this problem. Alternatively, we may proceed without having a fully $\epsilon$-factorized form. This was possible at 4PM order, notably because the final result is free of iterated integrals of modular forms, only containing complete elliptic integrals~\cite{4pmeftot}. Whether this pattern continues at 5PM, and beyond, is still unclear.\footnote{The principle of `maximal transcendentality' suggests that the three-loop result has maximal weight two (modulo overall factors of $\pi$), such that the four-loop answer should have maximal transcendental weight three. If we assign weight one to $\mathrm{K}(z)$, this would explain the absence of integrals of the type $\int\mathrm{K}^2(z)\mathrm{d} z$ at 4PM, but in principle they could appear at the next order.} Secondly, we have the evaluation of the boundary constants. Presently, all the necessary potential-only boundary integrals can be obtained with known techniques to four-loop order, see e.g.\,\cite{Lee:2015eva}. That is also the case for integrals involving the 2rad- and anticipated 3rad-type regions. Yet, integrals involving a single radiation mode at 5PM appear to be slightly more difficult than their three-loop counterparts. Although the 5PM order might not be the stumbling block---since we expect that extensions of the strategies we discussed here will succeed---going beyond four loops may require new strategies for the computations of the boundary conditions. In~this regard, the use of numerical tools may ultimately prove useful \cite{Jinno:2022sbr}. 

\end{outline}

\end{itemize}

In addition to addressing all of the above challenges, there are several other directions in which progress can still be achieved without significant obstacles. For instance, including spin and tidal effects, see e.g. \cite{tidaleft,pmefts}, to 4PM order. Moreover, so far we have concentrated on the impulse, however, there has been also a widespread interest in the change of angular momentum,~e.g.~\cite{Veneziano:2022zwh,Bini:2022wrq,Riva:2023xxm}. At the mechanical level, this can be readily computed within the EFT framework. New results in these directions will appear shortly.
\vskip 4pt
Last but not least, our computations are unveiling the structure of the PM expansion in general relativity, with the methodology of differential equations unraveling a very rich space of possible functions, e.g.~MPLs, $\mathrm{K}(z)$, $\mathrm{E}(z)$, perhaps featuring iterated integrals on elliptic curves at higher PM orders. One interesting venue, once more PM data is known, is the possibility to perform a resummation in Newton's constant, plausibly through an {\it ansatz} fulfilling various constraints.  At this stage, and already envisioning the complexity of the 5PM order and beyond, is clear that having any additional (non)perturbative knowledge---for instance at the level of the space of functions or particular limits such as, e.g., the high-energy or specific $\gamma$ values other than the soft region---would be extremely useful to help us {\it bootstrap} even further, e.g.~\cite{Almelid:2017qju},  the relativistic two-body problem. We are presently exploring these possibilities.

\begin{acknowledgments}
Throughout this PM journey---from \cite{pmeft} to \cite{4pmeftot}---we have benefited (one way or the other) from discussions with various people, in particular Babis Anastasiou, Zvi Bern, Donato Bini, Luc Blanchet, Johannes Bl\"umlein, Ruth Britto, Alessandra Buonanno, Clifford Cheung, Gihyuk Cho, Thibault Damour, Paolo Di Vecchia, Lance Dixon, Claude Duhr, Stefano Foffa, Walter Goldberger, Carlo Heissenberg, Harald Ita, Gustav Jakobsen, Henrik Johansson, David Kosower, Francois Larrouturou, Gustav Mogull, Rourou Ma, Jakob Neef, Donal O'Connell, Ben Page, Georgios Papathanasiou, Julio Parra-Martinez, Jan Plefka, Massimiliano Riva, Radu Roiban, Ira Rothstein, Michael Ruf, Rodolfo Russo, Volker Schomerus, Chia-Hsien Shen, Mikhail Solon, Jan Steinhoff, Riccardo Sturani, Gabriele Veneziano, Justin Vines, Zixin Yang, Mao Zeng, Yang Zhang and Xiaoran Zhao. We would like to thank also the participants of the KITP program ``High-Precision Gravitational Waves" (supported in part by the National Science Foundation under Grant No. NSF PHY-1748958) for discussions.
The work of C.D., G.K. and R.A.P. is funded by the ERC Consolidator Grant ``Precision Gravity: From the LHC to LISA"  provided by the European Research Council (ERC) under the European Union's H2020 research and innovation programme (grant agreement No. 817791). Z.L.~is supported partially by DFF grant 1026-00077B, the Carlsberg Foundation, and the European Union's Horizon 2020 research and innovation program under the Marie Sk\l{}odowska-Curie grant agreement No.\,847523 `INTERACTIONS'.

\end{acknowledgments}


\appendix

\section{Integral parametrization}\label{sec:param}
We discuss here a generic strategy to parametrize PM integrals with Feynman propagators (which also includes the potential-only case), and in particular the systematic treatment of the Dirac-$\delta$ functions leading to novel parametrized forms.

\subsection{Elementary integrals}

Using the formula for a complex Gaussian intregral, having introduced a small real part,
\begin{equation}\label{eq:gaussianC1}
  \int_{-\infty}^\infty \dd \eta\, e^{i(a + i0)\eta^2}  = e^{i\pi/4} \sqrt{\pi \over a+i0}\,, \quad a\in\mathbb{R},
\end{equation}
we can derive the generic Gaussian formula in Minkowski space:
\begin{equation}\label{eq:genericGaussian}
  \begin{aligned}
  &\int \dd^{d+1}\ell_1\cdots \dd^{d+1}\ell_L e^{i\left(a_{ij}\ell_i \cdot \ell_j + 2 b_i\cdot \ell_i\right)}\\
    &\qquad =e^{\frac{i\pi}{4}L(1-d)}\pi^{\frac{L(d+1)}{2}}\prod_i\left[\left(a_i + i0\right)^{-\frac{1}{2}}\left(a_i-i0\right)^{-\frac{d}{2}}\right] e^{-i\,a_{ij}^{-1} b_i\cdot b_j}\,,
  \end{aligned}
\end{equation}
where $\{a_i\}$ is the collection of eigenvalues of the matrix $a_{ij}$. Let us stress that it is crucial to keep the $i0$'s in order to properly account for phase factors.
\vskip 4pt

Another useful formulae is given by
\begin{equation}\label{eq:feynmanInt}
  \int_0^\infty \dd\eta\, \eta^{\sigma-1} e^{i\eta(a + y\,i0)}=\Gamma(\sigma)e^{\frac{i\pi}{2}\sigma}(a+y\, i0)^{-\sigma}\,,
\end{equation}
which holds for real $a$ and real $y>0$. Similarly, we also use the representation\footnote{Note that the integration goes over all real numbers, and not only over the positive ones. This formula can also be generalized to the case of derivatives acting on the delta function.}
\begin{equation}
  \delta(a) = {1 \over 2\pi}\int_{-\infty}^{\infty} \dd \beta\, e^{i\,\beta a}\,.
\end{equation}

\subsection{Feynman form}

For propagators with Feynman's $i0$-prescription we use the common Schwinger parametrization, such that
\begin{equation}\label{eq:paramBasic}
  \frac{1}{(a\pm i0)^\nu} = \frac{e^{\mp \frac{i\pi \nu}{2}}}{\Gamma(\nu)}\int_0^\infty \dd \alpha\, \alpha^{\nu-1}e^{\pm i\,\alpha (a\pm i0)}\,,
\end{equation}
which holds for $\nu>0$ and real $a$. In general we encounter the following type of parametrized integrals
\begin{equation}
  I \propto \int_{-\infty}^\infty \prod_{i=0}^L \dd \beta_i \int_0^\infty \prod_{i=1}^{\tilde{m}} \dd \alpha_i\, \alpha_i^{\nu_i-1} \int_\ell e^{i\left(a_{ij}\ell_i\cdot \ell_j+2 b_i\cdot \ell_i+c\right)}\,,
\end{equation}
where the $\beta_i$-parameters are associated to the parametrization of the $L$ Dirac-$\delta$ functions and the $x_i$-parameters to the $\tilde{m}=L+m$ (linear and square) propagators.
The factors of $a_{ij}$, $b_i$, and $c$, are also functions of the $\alpha_i$- and $\beta_i$-parameters.
All these objects are homogeneous of degree one in these parameters. The $\nu_i$'s are the powers of the propagators, and note we do not distinguish between square and linear types. For simplicity, below we assume that the matrix $a_{ij}$ is positive definite. For the actual computations one must be careful when using the Gaussian integral formulae in~\eqref{eq:genericGaussian} due to the appearance of various phases, due to definiteness of the minors of the $a$ matrix.\vskip 4pt
Performing the Gaussian integrals over the loop momenta~$\ell_i$, the result takes the schemat\-ic form
\begin{equation}
  I \propto \int_{-\infty}^\infty \prod_{i=0}^L \dd \beta_i\, \int_0^\infty \prod_{i=1}^{\tilde{m}} \dd \alpha_i\, \alpha_i^{\nu_i-1} \det(a)^{-(d+1)/2}e^{-i\frac{\adj(a)_{ij}b_i\cdot b_j-\det(a)c}{\det(a)}}\,,
\end{equation}
where, for later convenience, we pulled out an inverse factor of $\det(a)$ in the exponent by expressing the inverse matrix as the adjugate\footnote{\url{https://en.wikipedia.org/wiki/Adjugate_matrix}} divided by the determinant.
Since the arguments of Dirac-$\delta$ functions are linear in the loop momenta, and therefore only contribute to the vector $b$, the matrix $a_{ij}$ is independent of $\beta_i$-parameters. Hence, the $\beta_i$-integrals are in Gaussian form,
\begin{equation}
  \int_{-\infty}^\infty\left(\prod_{i=1}^L \dd \beta_i \right) e^{i\frac{\vec{\beta} \cdot A \cdot\vec{\beta} +2\vec{\beta} \cdot\vec B + C}{\det(a)}}\,,
\end{equation}
where $\vec{\beta} = (\beta_1,\dots,\beta_L)$. In order to determine the $A$ matrix, the vector $\vec{B}$, and the scalar $C$ as functions of the $\alpha_i$-parameters, we use the identity: 
\begin{equation}
  -\adj(a)_{ij}b_i\cdot b_j+\det(a) c = \vec{\beta}\cdot A \cdot \vec{\beta}+2\vec{\beta}\cdot\vec{B}+C\,.
\end{equation}
We can then use standard formulae to perform the resulting integrals, leading to the following Schwinger-type form 
\begin{equation}
  I \propto \int_0^\infty\prod_{i=1}^{\tilde{m}} \dd \alpha_i\, \alpha_i^{\nu_i-1} \det(a)^{-(d+1)/2+L/2}\det(A)^{-1/2}e^{-i\frac{\tilde{\cF}}{\det(A)\det(a)}}\,,
\end{equation}
where we defined the (Symanzik) polynomial $\tilde\cF=\vec{B}\cdot \adj(A) \cdot \vec{B}-\det(A)C$.
The various factors are all homogeneous polynomials in the $\alpha_i$-parameters: $\det(a)$ is of degree $L$; an entry of the $L\times L$ dimensional matrix $A$ is of degree $L-1$, such that its determinant is of degree $L(L-1)$ and its adjugate of degree $(L-1)^2$; an entry of the vector $\vec{B}$ is of degree $L$; and $C$ is of degree $L+1$.
\vskip 4pt

We continue by using the trick of inserting the identity,
\begin{equation}
  1 = \int_0^\infty\dd\eta\, \delta\left(\eta - \sum_{i=1}^{\tilde{m}} \alpha_i\right)\,,
\end{equation}
followed by a rescaling of all parameters $\alpha_i \rightarrow \eta \alpha_i$, yielding
\begin{equation}
  \begin{aligned}
    I &\propto \int_0^\infty\prod_{i=1}^{\tilde{m}} \dd \alpha_i\, \alpha_i^{\nu_i-1}\delta(1-\alpha) \int_0^\infty\dd\eta\,\eta^{\nu-1-Ld/2}  \det(a)^{-(d+1)/2+L/2}\det(A)^{-1/2}e^{\frac{-i\eta\tilde{\cF}}{\det(A)\det(a)}}
    \\
    &\propto\int_0^\infty\prod_{i=1}^{\tilde{m}} \dd \alpha_i\, \alpha_i^{\nu_i-1}\delta(1-\alpha) \det(a)^{-(d+1)/2-Ld/2+L/2+\nu}\det(A)^{-1/2-Ld/2+\nu}\tilde{\cF}^{Ld/2-\nu}
    \\
    &=\int_0^\infty\prod_{i=1}^{\tilde{m}} \dd \alpha_i\, \alpha_i^{\nu_i-1}\delta(1-\alpha)\frac{\cU^{\nu+L/2-(d+1)/2-Ld/2}\cU_\delta^{\nu-Ld/2-1/2}}{\tilde{\cF}^{\nu-Ld/2}}\,,
  \end{aligned}
\end{equation}
with $\alpha=\sum_{i=1}^{\tilde{m}}\alpha_i$, $\nu=\sum_{i=1}^{\tilde{m}} \nu_i$, $\cU=\det(a)$, and $\cU_\delta=\det(A)$, and we have used the integral formula~\eqref{eq:feynmanInt} with $\sigma=\nu-L(d+1)/2+L/2$, while we have ignored all the $i0$'s that we emphasize must be properly kept in order to correctly incorporate crucial phase factors.\vskip 4pt

In the absence of (uncut) linear propagators, we have $\vec{B}=0$ and we can rescale the first Symanzik polynomial by an inverse of $\cU_\delta=\det(A)$, i.e. $\cF=\tilde{\cF}/\cU_\delta=C$, leading to the simpler, and also more common, form
\begin{equation}
  I \propto \int_0^\infty\prod_{i=1}^m \dd\alpha_i \alpha_i^{\nu_i-1}\delta(1-\alpha)\frac{\cU^{\nu+L/2-(d+1)/2-Ld/2}\cU_\delta^{-1/2}}{\cF^{\nu-Ld/2}}\,.
\end{equation}
\vskip 4pt
Notice that the degree of homogeneity of $\cU_\delta=\det(A)$ vanishes for $L=1$, in which case the final parametrization consists of the more standard two polynomials.


\newpage
\section{Elliptic off-diagonal blocks}\label{sec:app-de}

\subsection{General method}

After bringing the elliptic diagonal block to $\epsilon$-form~\eqref{eq:ellEformDiag}, one still has to deal with sectors having the elliptic sector, $S_{\text{ell}}$, as a subsector. Let $S_{i}$ be such sector.
The differential equations, restricted to the two sectors, are given by
\begin{align}
\frac{\partial}{\partial x}\begin{pmatrix}
\vec{g}_{S_{\text{ell}}} \\
\vec{g}^{\text{diag}}_{S_i}
\end{pmatrix}
=\begin{pmatrix}
\epsilon\tilde{D}_{\text{ell}} & 0 \\
\hat{C}_{i,\text{ell}} & \epsilon\tilde{D}_{i}
\end{pmatrix}
\begin{pmatrix}
\vec{g}_{S_{\text{ell}}} \\
\vec{g}^{\text{diag}}_{S_i}
\end{pmatrix}+\ldots\,,
\end{align}
where the diagonal blocks have already been transformed into $\epsilon$-form and the ellipsis indicate that other subsectors have been omitted.
Similar to the elliptic diagonal block $\tilde{D}_{\text{ell}}$, the off-diagonal block $\hat{C}_{i,\text{ell}}$ will also depend on elliptic integrals.
While one can again use an ansatz for this block, it turns out that, after a simplification, it is often possible to manually transform $\hat{C}_{i,\text{ell}}$ to $\epsilon$-form.
Before we discuss this procedure in more detail, let us analyze why the off-diagonal block is generally complicated after the transformation of the diagonal blocks into $\epsilon$-form.\vskip 4pt

As discussed in \S\ref{sec:eps-form}, a transformation of the diagonal block of $S_i$ into $\epsilon$-form
\begin{equation}
\label{eq:diag-can-transf-2}
 \vec{g}^{\text{diag}}_{S_i}=T_i\vec{f}_{S_i}\,,
\end{equation}
can be found e.g.\ with \texttt{Libra} or \texttt{epsilon}. The resulting basis elements will be a linear combination of master integrals $\mathcal{M}^{(a_1 \cdots a_n; \pm\cdots\pm)}_{\,\,\alpha_1 \cdots \alpha_n; \beta_1 \cdots \beta_m}$ defined in Eq.~\eqref{PMint}:
\begin{equation}
\label{eq:I-lin-comb}
 g^{\text{diag}}_{S_i,j}=\sum_{\vec{\alpha}}h_{j,\vec{\alpha}}(x,\epsilon)\mathcal{M}_{\vec{\alpha}},
\end{equation}
where we collected the indices $\alpha_1,\ldots,\alpha_n;\beta_1,\ldots,\beta_m$ into a multi-index $\vec{\alpha}$ and omitted the family superscripts.
The $h_{j,\vec{\alpha}}(x,\epsilon)$ are coefficient functions.
On the other hand, a basis $\vec{g}_{S_i}$ that also brings the off-diagonal block $\hat{C}_{i,\text{ell}}$ into $\epsilon$-form, i.e.
\begin{align}
\frac{\partial}{\partial x}\begin{pmatrix}
\vec{g}_{S_\text{ell}} \\
\vec{g}_{S_i}
\end{pmatrix}
=\begin{pmatrix}
\epsilon\tilde{D}_{\text{ell}} & 0 \\
\epsilon\tilde{C}_{i,\text{ell}} & \epsilon\tilde{D}_{i}
\end{pmatrix}
\begin{pmatrix}
\vec{g}_{S_\text{ell}} \\
\vec{g}_{S_i}
\end{pmatrix}+\ldots\,,
\end{align}
will need corrections from the elliptic subsector:
\begin{align}
 \vec{g}_{S_i}=\vec{g}^{\text{diag}}_{S_i}+T_{i,\text{ell}}\,\vec{g}_{S_\text{ell}}\,.
\end{align}
Our main observation is that if the integrals in $\vec{g}^{\text{diag}}_{S_i}$ are complicated expressions in terms of $\mathcal{M}_{\vec{\alpha}}$, i.e. many $h_{j,\vec{\nu}}(x,\epsilon)$ are non-zero high-degree rational (or algebraic) functions in $x$ and $\epsilon$, then the off-diagonal block $\hat{C}_{i,\text{ell}}$ will likewise be very complicated.
This will, in turn, require an equally complicated $T_{i,\text{ell}}$ that is difficult to compute.\vskip 4pt

The goal is therefore to search for an alternative basis for the $\epsilon$-form of the diagonal block, $\vec{g}^{\text{diag,simple}}_{S_i}$, such that its elements are algebraically simple when written in terms of master integrals $\mathcal{M}_{\vec{\alpha}}$.
Since we explicitly know a basis $\vec{g}^{\text{diag}}_{S_i}$, a better suited version can be found with the following strategy:
From the differential equations it can be easily seen that any constant (in $x$ and $\epsilon$) transformation on $\vec{g}^{\text{diag}}_{S_i}$ will still result in $\tilde{D}_{i}$ being in $\epsilon$-form. We can therefore use IBP relations to search for Feynman integrals $\mathcal{M}_{\vec{\alpha}}$ which can be written as
\begin{equation}
\label{eq:cS-form}
 \mathcal{M}_{\vec{\alpha}}=\mathcal{N}_{\vec{\alpha}}(x,\epsilon)\ \vec{c}_{\vec{\alpha}}\cdot\vec{g}^{\text{diag}}_{S_i}+\ldots\,,
\end{equation}
where $\vec{c}_{\vec{\alpha}}$ is a vector of constants and the ellipsis indicate subsector integrals. Further, out of the found integrals, one can choose those with as simple prefactors $\mathcal{N}_{\vec{\alpha}}(x,\epsilon)$ as possible.
Then it is clear that
\begin{equation}
\label{eq:new-off-simple}
 \frac{1}{\mathcal{N}_{\vec{\alpha}}(x,\epsilon)}\mathcal{M}_{\vec{\alpha}}=\vec{c}_{\vec{\alpha}}\cdot\vec{g}^{\text{diag}}_{S_i}+\ldots\,,
\end{equation}
has to be an integral which we can choose as a member of $\vec{g}^{\text{diag,simple}}_{S_i}$, because, up to subsector integrals, it is simply a constant linear combination of the integrals $\vec{g}^{\text{diag}}_{S_i}$. Note however that this new integral is not simply a rotation of the basis $\vec{f}_{S_i}$ as in \eqref{eq:diag-can-transf-2} but already includes transformations on the off-diagonal part, which is indicated through the ellipsis in \eqref{eq:new-off-simple}. In practice, to find a complete basis, we found it sufficient to perform the test \eqref{eq:cS-form} on all integrals inside the sector $S_i$ with at most three doubled propagators or numerator power equal to three.\vskip 4pt

To summarize, one can often find an alternative basis $\vec{g}^{\text{diag,simple}}_{S_i}$, which consists of 'simple' expressions and which still transforms the diagonal block into canonical form. 
We then observe that this leads to a significant simplification of the off-diagonal blocks $\hat{C}_{i,j}$. Often, one can then manually transform $\hat{C}_{i,j}$ to $\epsilon$-form. An example of this will be given in the next section.

\subsection{Three-loop example}
\label{sec:app-ell-off-diag}

After doing the diagonal transformations described in \S\ref{sec:DE-examples-4PM}, there will be off-diagonal blocks which depend on the elliptic sector and therefore also on complete elliptic integrals.
An example of such a sector $S$, within in the family $\mathcal{M}^{(112;+++)}$, is characterized by the set of positive propagator powers $\{\beta_2,\beta_4,\beta_6,\beta_7,\beta_8,\beta_9\}$. It has the elliptic sector as one of its subsectors.
The off-diagonal block is too complicated to be displayed here, and, as discussed in the previous section, we observed that the main reason for this is that $\vec{g}^\text{diag}_{S}$, the basis which transforms the diagonal block into canonical form, is a complicated expression in terms of integrals $\mathcal{M}^{(112; +++)}_{\alpha_1 \alpha_2 \alpha_3; \beta_1 \cdots \beta_9}$.
We therefore use IBP reduction to search for integrals that are linear combinations of the basis $\vec{g}^\text{diag}_{S}$ with constant coefficients up to subsector integrals, see Eq.\ \eqref{eq:cS-form}. We find e.g.\
\begin{equation}
\begin{aligned}
\mathcal{M}^{(112;+++)}_{\shortminusS10\shortminus1;010101111}&=\epsilon^{-1}\ \vec{c}_1\cdot\vec{g}^\text{diag}_{S}+\ldots, \\
\mathcal{M}^{(112;+++)}_{0\shortminus1\shortminus1;010101111}&=\epsilon^{-1}\ \vec{c}_2\cdot\vec{g}^\text{diag}_{S}+\ldots, \\
\mathcal{M}^{(112;+++)}_{00\shortminus2;010101111}&=(1-\epsilon)^{-1}\ \vec{c}_3\cdot\vec{g}^\text{diag}_{S}+\ldots, \\
\mathcal{M}^{(112;+++)}_{000;\shortminusS1201\shortminus11111}&=\frac{x}{1-x^2}\ \vec{c}_4\cdot\vec{g}^\text{diag}_{S}+\ldots,
\end{aligned}
\end{equation}
where the ellipsis indicate subsector integrals.
Therefore, an alternative basis is given by
\begin{equation}
\vec{g}^{\text{diag,simple}}_{S}=\begin{pmatrix}
\epsilon\, \mathcal{M}^{(112;+++)}_{\shortminusS10\shortminus1;010101111} \\
\epsilon\, \mathcal{M}^{(112;+++)}_{0\shortminus1\shortminus1;010101111} \\
(1-\epsilon)\,\mathcal{M}^{(112;+++)}_{00\shortminus2;010101111} \\
\frac{1-x^2}{x}\,\mathcal{M}^{(112;+++)}_{000;\shortminusS1201\shortminus11111}
\end{pmatrix}
\end{equation}
and, as expected, the off-diagonal block $C_{\text{ell}}(x,\epsilon)$ is considerably simpler when using this basis.
Specifically, it is of the form
\begin{align}
C_{\text{ell}}(x,\epsilon)=C_{\text{ell}}^{(0)}(x)+\epsilon C_{\text{ell}}^{(1)}(x)\,,
\end{align}
where
\begin{equation}
\label{eq:exC}
C_{\text{ell}}^{(0)}(x)=\begin{pmatrix}
\frac{\mathrm{K}(1-x^2)[2x^2\mathrm{K}(1-x^2)-(1+x^2)\mathrm{E}(1-x^2)]}{2\pi^2x(1-x^2)} & 0 & 0 \\
0 & 0 & 0 \\
\frac{\mathrm{K}(1-x^2)[(1+x^2)\mathrm{K}(1-x^2)-2\mathrm{E}(1-x^2)]}{\pi^2(1-x^2)} & 0 & 0 \\
-\frac{4\mathrm{K}(1-x^2)\mathrm{E}(1-x^2)}{\pi^2x} & 0 & 0
\end{pmatrix}\, .
\end{equation}
To get rid of the remaining $C_{\text{ell}}^{(0)}(x)$ one can use the transformation
\begin{equation}
\begin{aligned}
T_{\text{ell}}&=\int\dd x\, C_{\text{ell}}^{(0)}(x) =\begin{pmatrix}
\frac{(1+x^2)\mathrm{K}^2(1-x^2)}{4\pi^2} & 0 & 0 \\
0 & 0 & 0 \\
\frac{x\mathrm{K}^2(1-x^2)}{\pi^2} & 0 & 0 \\
\frac{2(1-x^2)\mathrm{K}^2(1-x^2)}{\pi^2} & 0 & 0 \\
\end{pmatrix}\, .
\end{aligned}
\end{equation}
This results in the $\epsilon$-form
\begin{equation}
\begin{aligned}
 \epsilon\tilde{C}_{\text{ell}}&=C_{\text{ell}}-\frac{\partial}{\partial x}T_{\text{ell}}+\epsilon\left(\tilde{D}T_{\text{ell}}-T_{\text{ell}}\tilde{D}_{\text{ell}}\right) \\
 &=\epsilon\left(C_{\text{ell}}^{(1)}(x)+\tilde{D}T_{\text{ell}}-T_{\text{ell}}\tilde{D}_{\text{ell}}\right) \\
 &=\epsilon \begin{pmatrix}
    -\frac{2x\mathrm{K}^2(1-x^2)}{(1-x^2)} & 0 & 0 \\
    0 & 0 & 0 \\
    -\frac{2(1+x^2)\mathrm{K}^2(1-x^2)}{(1-x^2)} & 0 & 0 \\
    0 & 0 & 0 
   \end{pmatrix}\,.
\end{aligned}
\end{equation}
The other off-diagonal blocks can be treated in an analogue way.

\subsection{The $\epsilon$-form at three loops}\label{sec:app-4PM-final-DE}

The final differential equations in $\epsilon$-form are given by
\begin{equation}
 \partial_x\vec{g}(x,\epsilon)=\epsilon\tilde{A}(x)\vec{g}(x,\epsilon)\,,
\end{equation}
where the matrix $\tilde{A}(x)$ is
\begin{equation}
\label{eq:GW-aeps}
 \begin{aligned}
  \tilde{A}(x)&=M_1\frac{\pi^2}{x(1-x^2)\mathrm{K}^2(1-x^2)}+M_2\frac{1}{1-x}+M_3\frac{1}{x}+M_4\frac{1}{1+x}\\
  &\quad+M_5\frac{x}{1+x^2}+M_6\frac{\mathrm{K}^2(1-x^2)}{\pi^2x(1-x^2)}+M_7\frac{\mathrm{K}^2(1-x^2)}{\pi^2(1-x^2)}\\
  &\quad+M_8\frac{\mathrm{K}^2(1-x^2)}{\pi^2x}+M_9\frac{\mathrm{K}^2(1-x^2)}{\pi^2}+M_{10}\frac{(1-x^2)\mathrm{K}^2(1-x^2)}{\pi^2x}\\
  &\quad+M_{11}\frac{\mathrm{K}^4(1-x^2)}{\pi^4x(1-x^2)}+M_{12}\frac{\mathrm{K}^4(1-x^2)}{\pi^4x}\\
  &\quad+M_{13}\frac{(1-x^2)\mathrm{K}^4(1-x^2)}{\pi^4x}+M_{14}\frac{(1-x^2)^2\mathrm{K}^4(1-x^2)}{\pi^4x}\,,
 \end{aligned}
\end{equation}
and the $M_i$ are matrices with rational constant entries.
Clearly, the result for the integrals cannot be written in terms of MPLs anymore. Instead, we use the more general iterated integrals
\begin{equation}
 I(h_1,h_2,\ldots,h_n;z,z_0)=\int_{z_0}^z\mathrm{d}t\,h_1(t)I(h_2,\ldots,h_n;t,z_0),\qquad I(;x,x_0)=1.
\end{equation}
We note that the integration kernels $h_i$ in Eq.\ \eqref{eq:GW-aeps} are modular forms \cite{Adams:2017ejb} and can be written as polynomials in the functions $E_{2,8,1,1,2}(\tau),E_{2,8,1,1,4}(\tau)$ and $E_{2,8,1,1,8}(\tau)$ defined in~\cite{Walden:2020odh}, where $\tau=i \mathrm{K}(x^2)/ \mathrm{K}(1 - x^2)$.
This is true for all kernels except $x/(1+x^2)$, which however appears only together with other polylogarithmic kernels, meaning that all affected iterated integrals can be written as multiple polylogarithms.
The representation of the kernels as modular forms and polylogarithmic kernels makes the numeric evaluation of the ensuing iterated integrals very efficient \cite{Walden:2020odh}.
However, it turns out that in our case the final result is free of elliptic integration kernels and only involves multiple polylogarithms.


\newpage
\section{Boundary integrals}\label{sec:app-bc}

\subsection{Two-loop boundary relations}
\label{sec:bcrels-3PM}

We now apply the methods discussed in \S\ref{sec:bc-exp} to the example of \S\ref{sec:3PM-de-example}. To this end, we first use the differential equations in $\epsilon$-form to expand the master integrals in $v_\infty$,
\begin{align}\label{eq:f-exp-app}
 \vec{f}(v_\infty,\epsilon)\simeq T^{-1}P(v_\infty,\epsilon)v_\infty^{\epsilon\tilde{M}_{1}}\vec{g}_0(\epsilon)=\sum_{n_1,n_2,k}v_\infty^{n_1+n_2\epsilon}\log^kv_\infty\,H_{n_1,n_2,k}(\epsilon)\vec{g}_0(\epsilon)\,,
\end{align}
where $\vec{g}=T\vec{f}$ and $P(v_\infty,\epsilon)=\mathbb{1}+\sum_{i=1}^\infty v_\infty^iP^{(i)}(\epsilon)$ can be computed recursively. The matrix exponential evaluates to
\begin{equation}
 v_\infty^{\epsilon \tilde{M}_1}=\begin{pmatrix}
                           v_\infty^{-2\epsilon} & 0 & 0 & 0 & 0 & 0 & 0 \\
                           0 & v_\infty^{-2\epsilon} & 0 & 0 & 0 & 0 & 0 \\
                           0 & 0 & 1 & 0 & 0 & 0 & 0 \\
                           0 & 0 & 0 & v_\infty^{6\epsilon} & 0 & 0 & 0 \\
                           0 & 0 & 0 & 0 & 1 & 0 & 0 \\
                           -\epsilon v_\infty^{-2\epsilon}\log v_\infty & \epsilon v_\infty^{-2\epsilon}\log v_\infty & \tfrac{1}{2}(v_\infty^{-2\epsilon}-1) & \tfrac{1}{8}(v_\infty^{6\epsilon}-v_\infty^{-2\epsilon}) & 0 & v_\infty^{-2\epsilon} & 0 \\
                           \tfrac{1}{2}(v_\infty^{2\epsilon}-v_\infty^{-2\epsilon}) & 0 & 0 & \tfrac{1}{4}(v_\infty^{2\epsilon}-v_\infty^{6\epsilon}) & \tfrac{1}{2}(v_\infty^{2\epsilon}-1) & 0 & v_\infty^{2\epsilon} 
                          \end{pmatrix}\, .
\end{equation}
We then compare this result to the explicit expansion of the master integrals found through the \texttt{asy2.m} code:
\begin{equation}
 \begin{aligned}
  \vec{f}(v_\infty)&\simeq \sum_{n_1,n_2,k}v_\infty^{n_1+n_2\epsilon}\log^kv_\infty\,\vec{h}_{n_1,n_2,k}(\epsilon)\, .
 \end{aligned}
\end{equation}
An example for eight of the (infinitely many) coefficients is given by
\begin{equation}
 \begin{pmatrix}
  h_{0,0,0,2}(\epsilon) \\
  h_{-1,6,0,2}(\epsilon) \\
  h_{-1,-2,0,4}(\epsilon) \\
  h_{0,0,0,5}(\epsilon) \\
  h_{0,2,0,6}(\epsilon) \\
  h_{-3,-2,0,7}(\epsilon) \\
  h_{-2,-2,0,7}(\epsilon) \\
  h_{-3,-2,1,7}(\epsilon)
 \end{pmatrix}=
 \begin{pmatrix}
  H_{0,0,0,2}(\epsilon) \\
  H_{-1,6,0,2}(\epsilon) \\
  H_{-1,-2,0,4}(\epsilon) \\
  H_{0,0,0,5}(\epsilon) \\
  H_{0,2,0,6}(\epsilon) \\
  H_{-3,-2,0,7}(\epsilon) \\
  H_{-2,-2,0,7}(\epsilon) \\
  H_{-3,-2,1,7}(\epsilon)
 \end{pmatrix}\vec{g}_0(\epsilon)\equiv H_{\text{depend}}(\epsilon)\,\vec{g}_0(\epsilon)\, ,
\end{equation}
where the last index $j$ of $h_{n_1,n_2,k,j}$ and $H_{n_1,n_2,k,j}$ refers to the $f_j$ master integral.
Since there are seven master integrals in the considered example, likewise $\vec{g}_0(\epsilon)$ has seven entries. Therefore, the matrix $H_{\text{depend}}(\epsilon)$ is an $8\times 7$ matrix given by:
\begin{equation}
 \hspace*{-0.5cm}\begin{pmatrix}
 0 & 0 & \frac{1+\epsilon}{16 (1+2 \epsilon) (6 \epsilon-1)} & 0 & 0 & 0 & 0 \\
 0 & 0 & 0 & \frac{1+\epsilon}{64 \epsilon (1+2 \epsilon)} & 0 & 0 & 0 \\
 0 & \frac{-1-\epsilon}{16 (1+2 \epsilon)} & 0 & 0 & 0 & 0 & 0 \\
 0 & 0 & 0 & 0 & \frac{-1-\epsilon}{16 \epsilon (1+2 \epsilon)} & 0 & 0 \\
 \frac{-1-\epsilon}{4 (1+2 \epsilon) (1+4 \epsilon)} & 0 & 0 & \frac{-1-\epsilon}{8 (1+2 \epsilon) (1+4 \epsilon)} & \frac{-1-\epsilon}{4 (1+2 \epsilon) (1+4 \epsilon)} & 0 & \frac{-1-\epsilon}{2 (1+2 \epsilon) (1+4 \epsilon)} \\
 \frac{2 \epsilon-1}{8 (1+\epsilon)} & -\frac{(1+2 \epsilon) (2+5 \epsilon)}{8 \epsilon (1+\epsilon)} & \frac{-1-4 \epsilon}{8 \epsilon} & \frac{1+4 \epsilon}{32 \epsilon} & 0 & \frac{-1-4 \epsilon}{4 \epsilon} & 0 \\
 \frac{5+6 \epsilon+4 \epsilon^2}{16 (1+\epsilon)} & \frac{(2+3 \epsilon) (3+10 \epsilon+4 \epsilon^2)}{16 \epsilon (1+\epsilon)} & \frac{(3+2 \epsilon) (1+4 \epsilon)}{16 \epsilon} & -\frac{(3+2 \epsilon) (1+4 \epsilon)}{64 \epsilon} & 0 & \frac{(3+2 \epsilon) (1+4 \epsilon)}{8 \epsilon} & 0 \\
 \frac{1}{4} (1+4 \epsilon) & \frac{1}{4} (-1-4 \epsilon) & 0 & 0 & 0 & 0 & 0
               \end{pmatrix}
\end{equation}
From the above matrix we find that, indeed, seven of the $h_{n_1,n_2,k,j}$ are linearly independent. If this was not the case we would have to consider different sets until a complete and independent one is found. In the case of $H_{\text{depend}}$, it turns out that the last row is not linearly independent from the first seven,
\begin{equation}
 h_{-3,-2,1,7}(\epsilon)=(3+2\epsilon)h_{-3,-2,0,7}(\epsilon) + 2h_{-2,-2,0,7}(\epsilon),
\end{equation}
hence we choose
\begin{equation}
 \vec{h}_{\text{indep}}(\epsilon)\equiv\begin{pmatrix}
  h_{0,0,0,2}(\epsilon) \\
  h_{-1,6,0,2}(\epsilon) \\
  h_{-1,-2,0,4}(\epsilon) \\
  h_{0,0,0,5}(\epsilon) \\
  h_{0,2,0,6}(\epsilon) \\
  h_{-3,-2,0,7}(\epsilon) \\
  h_{-2,-2,0,7}(\epsilon)
 \end{pmatrix}=H_{\text{indep}}(\epsilon)\,\vec{g}_0(\epsilon)\,,
\end{equation}
as a linearly independent set, such that $H_{\text{indep}}$ is built from the corresponding rows. The result then corresponds to an invertible square matrix. Finally, we compute $\vec{g}_0=H^{-1}_{\text{indep}}\vec{h}_{\text{indep}}$ from the knowledge of $\vec{h}_{\text{indep}}$. The latter are determined through the asymptotic expansion of the master integrals in the soft limit. Note that, by the method of region, we only find boundary integrals with $n_2=0,-2$ at two-loop order. Hence, those with $n_2=6$ and $n_2=2$ can be disregarded.
\vskip 4pt

The above procedure translates directly in the same way to the three-loop case. That is because the expansion of the new elliptic elements of the differential equations still produce terms of the form $v_\infty^{n_1+n_2\epsilon}\log^kv_\infty$, as in~\eqref{eq:f-exp-app}. In what follows we provide some details of the derivations of the master integrals in the near-static limit.

\subsection{Two-loop potential region}
We present here the results for potential-only boundary integrals at two-loop order, identified in eq.~\eqref{eq:potBCgeneric},
\begin{equation}\label{eq:potBC3PM}
\boldsymbol\cM^{(\pm\pm),\textrm{pot}}_{\alpha_1\alpha_2;\beta_1\cdots\beta_5} = \int_{\bell_1\bell_2}\frac{1}{(\pm\bell_1\cdot\bn-i0)^{\alpha_1}(\pm\bell_2\cdot\bn-i0)^{\alpha_2}\bD_1^{\beta_1}\bD_2^{\beta_2}\bD_3^{\beta_3}\bD_4^{\beta_4}\bD_5^{\beta_5}}\,,
\end{equation}
where we have used a bolded notation to emphasize that these are 3d-type integrals. The propagators we encounter take the form
\begin{align}\label{}
\bD_1 &= \vecbf{\ell}_1^2\,, &
\bD_2 &= \vecbf{\ell}_2^2\,, &
\bD_3 &= (\vecbf{\ell}_1 + \vecbf{\ell}_2 - \vecbf{q})^{2}\,, &
\bD_4 &= (\vecbf{\ell}_1 - \vecbf{q})^2\,, &
\bD_5 &= (\vecbf{\ell}_2 - \vecbf{q})^2\,.
\end{align}
Using IBP relations, we identify nine master integrals with all possible sign configurations of linear propagators, yielding 
\begingroup
\allowdisplaybreaks
\begin{align}
\boldsymbol\cM^{(\pm\pm),\textrm{pot}}_{00;11100} &= e^{2\epsilon \gamma_\textrm{E}}\frac{\Gamma^3(\tfrac{1}{2} -\epsilon)\, \Gamma(2\epsilon)}{\Gamma(\tfrac{3}{2} -3\epsilon)} 
\label{eq:K2_1}\,,\\
\boldsymbol\cM^{(\pm\pm),\textrm{pot}}_{00;11011} &= e^{2\epsilon \gamma_\textrm{E}}\frac{\Gamma^4(\tfrac{1}{2} -\epsilon)\,\Gamma^2(\tfrac{1}{2} +\epsilon)}{\Gamma^2(1-2\epsilon)}
\label{eq:K2_2}\,,\\
\boldsymbol\cM^{(\pm\pm),\textrm{pot}}_{01;11100} 
&= ie^{2\epsilon\gamma_\textrm{E}}\,\frac{\sqrt{\pi}\, \Gamma(\tfrac{1}{2} -2\epsilon)\,\Gamma^2(\tfrac{1}{2} -\epsilon)\,\Gamma(-\epsilon)\,\Gamma(\tfrac{1}{2} +2\epsilon)}{\Gamma(\tfrac{1}{2} -3\epsilon)\,\Gamma(1-2\epsilon)}
\label{eq:K2_3}\,,\\
\boldsymbol\cM^{(\pm\pm),\textrm{pot}}_{01;11011} &= 
i e^{2\epsilon\gamma_\textrm{E}}\, \frac{\sqrt{\pi}\, \Gamma^2(\tfrac{1}{2}-\epsilon)\,\Gamma^2(-\epsilon)\,\Gamma(\tfrac{1}{2}+\epsilon)\,\Gamma(1+\epsilon)}{\Gamma(1-2\epsilon)\,\Gamma(-2\epsilon)}
\label{eq:K2_4}\,,\\
\boldsymbol\cM^{(\pm\pm),\textrm{pot}}_{01;10110} &= 
ie^{2\epsilon\gamma_\textrm{E}}\frac{2^{6\epsilon}\,\pi\, \Gamma(\epsilon)\,\Gamma(\tfrac{1}{2}-2\epsilon)\,\Gamma(\tfrac{1}{2}+2\epsilon)}{\Gamma(1-\epsilon)}
\label{eq:K2_5}\,,\\
\boldsymbol\cM^{(++),\textrm{pot}}_{11;11100} &= 2 \boldsymbol\cM^{(+-),\textrm{pot}}_{11;11100}
=  -e^{2\epsilon\gamma_\textrm{E}}\frac{4\pi}{3}\frac{\Gamma^3(-\epsilon)\,\Gamma(2\epsilon+1)}{\Gamma(-3\epsilon)}\,,\\
\boldsymbol\cM^{(++),\textrm{pot}}_{11;11011} &= - e^{2\epsilon\gamma_\textrm{E}} \frac{\pi\,\Gamma^4(-\epsilon)\,\Gamma^2(\epsilon+1)}{\Gamma^2(-2\epsilon)}\,,\\
\boldsymbol\cM^{(\pm\pm),\textrm{pot}}_{02;10110} &= 
- e^{2\epsilon\gamma_\textrm{E}}\, \frac{4\epsilon\, \Gamma(2\epsilon)\,\Gamma^2(-2\epsilon)\, \Gamma(\tfrac{1}{2}-\epsilon)\,\Gamma(\tfrac{1}{2}+\epsilon)}{\Gamma(-4\epsilon)}
\label{eq:K2_9}\,.
\end{align}
\endgroup
As expected, each of them has either a double-bubble or sunrise topology for the square propagators. We present below three different strategies we used to derive these results.\vskip 8pt

\textbf{\textit{Recursive integration.}} While we are mainly interested in the two- and three-loop computations, it turns out to be useful to analyze first the case of one-loop integrals. That is because integration at higher loop orders can be evaluated, at least partially, recursively via lower order integrals. Starting at $\mathcal{O}(G^2)$, all scalar integrals appearing in the computation of the conservative dynamics of non-spinning \cite{pmeft} and spinning \cite{pmefts} binary systems can be immersed into the following form
\begin{align}\label{static-1-loop-def}
\boldsymbol\cM^{(\pm),\textrm{pot}}_{\alpha_1;\beta_1\beta_2} =
\int_\bell \frac{1}{[\pm\bell\cdot\bn-i0]^{\alpha_1} [\vecbf{\ell}^2]^{\beta_1}\, [(\vecbf{\ell}- \vecbf{q})^2]^{\beta_2}},
\end{align}
where $\bn = (0,0,1)$ and we use a coordinate systems such that $q^z=0$ in the rest frame. The result can be obtained via direct evaluation \cite{Smirnov:2012gma}
\begin{align}\label{pm-1-loop}
\boldsymbol\cM^{(\pm),\textrm{pot}}_{\alpha_1;\beta_1\beta_3} &= e^{\epsilon\gamma_E}
\frac{2^{\alpha_1 -1} i^{\alpha_1}\, \Gamma(\frac{\alpha_1}{2})\, \Gamma(\frac{d - \alpha_1}{2} - \beta_1)\,
 \Gamma(\frac{d- \alpha_1}{2} - \beta_2)\, \Gamma(\frac{\alpha_1 - d}{2} + \beta_1 + \beta_2)}
{\Gamma(\alpha_1) \Gamma(\beta_1)\Gamma(\beta_2)\,\Gamma(d - \alpha_1 - \beta_1 - \beta_2)}(\bq^2)^{(d-\alpha_1)/2-\beta_1-\beta_2}\,,
\end{align}
for $\beta_1>0$, $\beta_2>0$, and $\bn\cdot\bq=0$, whereas in dim.~reg.~the integral vanishes for $\beta_1\leq 0$ or $\beta_2\leq 0$.\vskip 4pt

The reader will now simply observe that a number of the above boundary integrals can be evaluated by making use of the one-loop (bubble) formula in \eqref{pm-1-loop}, {\it loop by loop}. In particular, $\boldsymbol\cM^{(\pm\pm),\textrm{pot}}_{00;11100}$, $\boldsymbol\cM^{(\pm\pm),\textrm{pot}}_{00;11011}$, $\boldsymbol\cM^{(\pm\pm),\textrm{pot}}_{01;11100}$, $\boldsymbol\cM^{(\pm\pm),\textrm{pot}}_{01;11011}$, and $\boldsymbol\cM^{(\pm\pm),\textrm{pot}}_{11;11011}$, can be straightforwardly computed in this fashion.
\vskip 8pt
\textbf{\textit{Symmetrization trick.}} Next we consider integrals with a sunrise topology (for the square propagators) together with two linear propagators, e.g. $\boldsymbol\cM^{(+\pm),\textrm{pot}}_{11;11100}$. We first introduce an auxiliary loop integration, such that the three square propagators can be written in fully-symmetric form in all loop momenta (including the new auxiliary loop variable), i.e.,
\begin{align}
\boldsymbol\cM^{(+\pm),\textrm{pot}}_{11;11100} &=
 \int_{\bell_1\bell_2\bell_3}  \frac{\pi^{-d/2}e^{-\epsilon \gamma_E}}{[\bell_1\cdot\bn-i0][\pm\bell_2\cdot\bn-i0]}\,
 \frac{\delta^{(d)}(\vecbf{\ell}_{123} {-} \vecbf{q})}{\vecbf{\ell}_1^2\, \vecbf{\ell}_2^2\, \vecbf{\ell}_{3} ^2}.
\end{align}
(Notice that the $\gamma_E$ and $\pi$'s appear due to our conventions and the fact that the integral over $\bell_3$ is artificial.)
The above integral is invariant under the permutation of the three loop momenta. Hence, it can be written as \begin{align}
\boldsymbol\cM^{(+\pm),\textrm{pot}}_{11;11100} &=
\frac{1}{3!}
 \int_{\bell_1\bell_2\bell_3}
\bigg( \frac{\pi^{-d/2}e^{-\epsilon\gamma_E}}{[\bell_1\cdot\bn-i0][\pm\bell_2\cdot\bn-i0]} + \operatorname{perm}(1,2,3) \bigg)
 \frac{\delta^{(d)}(\bell_{123} {-} \vecbf{q})}{\bell_1^2\, \bell_2^2\, \bell_{3} ^2}\,,
\end{align}
which becomes,
\begin{align}
\boldsymbol\cM^{(+\pm),\textrm{pot}}_{11;11100}
&=  \rho_\pm \frac{(2\pi i)^2}{ 6}  \int {\dd^{d-1} \bell^\perp_1 \dd^{d-1} \bell^\perp_2 \over \pi^d}
\frac{e^{2\epsilon\gamma_E}} {(\bell_1^\perp)^2\, (\bell_2^\perp)^2\, (\bell_{12}^\perp - \bq)^2}\\
&=
- \frac{2\rho_\pm\,\pi}{3} e^{2\epsilon\gamma_E}\,
\frac{\Gamma^3(-\epsilon)\, \Gamma(1 + 2\epsilon)}{\Gamma(-3\epsilon)}\,,
\end{align}
with $\rho_+=2$ and $\rho_-=1$. To compute the above integral, we have used \eqref{pm-1-loop} in the last line, recursively, together with the following identities \cite{Saotome:2012vy}
\begin{align}
\label{delta-id-2-1}
\delta(z_1 + z_2 + z_3) \left( \frac{1}{z_1 - i0}\frac{1}{z_{12} - i0} + \text{perms} \right) 
&= {(2\pi i)^2}\, \delta(z_1)\delta(z_2)\delta(z_3),
\\[0.35 em]
\label{delta-id-2-2}
\delta(z_1 + z_2 {+} z_3) \left(\frac{1}{z_1 - i0}\frac{1}{z_{2} - i0} + \text{perms} \right) 
&= 2{(2\pi i)^2}\, \delta(z_1)\delta(z_2)\delta(z_3),
\end{align}
where $z_{12} = z_1 + z_2$, $z_i\in\mathbb{R}$, and ``perms'' stands for a sum over all permutations in all three variables $z_i$. Notice that these two expressions differ simply by a factor of 2. This is due to the fact that their sum corresponds to the cut of the two linear propagators, i.e.,
\begin{align}
\boldsymbol\cM^{(++),\textrm{pot}}_{11;11100} + \boldsymbol\cM^{(+-),\textrm{pot}}_{11;11100}
&= {(2\pi i)^2 \over 2}
 \int_{\bell_1\bell_2}\delta(\bell_1\cdot\bn)\, \delta(\bell_2\cdot\bn)\,
{(\vecbf{q}^2)^{4 - d} \over \vecbf{\ell}_1^2\, \vecbf{\ell}_2^2\, (\vecbf{\ell}_{12} {-} q)^2}
=  3\boldsymbol\cM^{(+-),\textrm{pot}}_{11;11100},
\end{align}
which implies
\begin{align}\label{2loop-hard-ratio}
\boldsymbol\cM^{(++),\textrm{pot}}_{11;11100} = 2\, \boldsymbol\cM^{(+-),\textrm{pot}}_{11;11100}.
\end{align}

\vskip 8pt
\textbf{\textit{Direct integration.}} There are two other master integrals that cannot be computed with the above trick, i.e., $\boldsymbol\cM^{(\pm\pm),\textrm{pot}}_{01;10110}$ and $\boldsymbol\cM^{(\pm\pm),\textrm{pot}}_{02;10110}$. Let us start with the former, and concentrate on the integration over the $\vecbf{\ell}_2$ momentum.
In the parametric representation we have
\begin{align}
\int_{\bell_2} & \frac{1}{(\pm\bell_2\cdot\bn-i0)\, (\bell_1 + \bell_2)^2}
\\
&=2^{4-d} e^{i \pi  d+\epsilon\gamma_E}\, \Gamma\big(2-\tfrac{d}{2}\big)
\int_0^\infty \dd \alpha_1 \dd \alpha_2\,\delta(1 {-} \alpha_1 {-} \alpha_2)\,
\alpha_1^{2-d} \alpha_2^{d/2-2}  \big(\mp 4 \alpha_1 \bell_1\cdot\bn - \alpha_2\big)^{\frac{d-4}{2}}.
\nonumber
\end{align}
Using the Cheng-Wu theorem~\cite{Cheng:1987ga} to remove one of the integration variables, followed by a direct integration of the leftover integral, we obtain
\begin{align}\label{}
\int_{\bell_2}  \frac{1}{(\pm\bell_2\cdot\bn-i0)\, (\bell_1 + \bell_2)^2}
= e^{\epsilon\gamma_E}\frac{(2i)^{1-2\epsilon}\, \Gamma(2\epsilon)\,\Gamma\big(\tfrac{1}{2} - \epsilon\big)}{(\mp\bell_1\cdot\bn - i0)^{2\epsilon}}\,.
\end{align}
The remaining integral over $\vecbf{\ell}_1$ becomes~$\boldsymbol\cM^{(\pm),\textrm{pot}}_{(2\epsilon) 1 1}$, which can be evaluated using \eqref{pm-1-loop}. The final result is given by
\begin{align}\label{}
\boldsymbol\cM^{(\pm\pm),\textrm{pot}}_{01;10110} &= 
ie^{2\epsilon\gamma_E} \frac{2^{6\epsilon}\,\pi\, \Gamma(\epsilon)\,\Gamma(\tfrac{1}{2}-2\epsilon)\,\Gamma(\tfrac{1}{2}+2\epsilon)}{\Gamma(1-\epsilon)}\,.
\end{align}

The same procedure can be directly applied to the evaluation of $\boldsymbol\cM^{(\pm\pm),\textrm{pot}}_{02;10110}$. Alternatively, the integration over the various momenta (in both of these integrals) can be related by the following IBP relation
\begin{align}\label{}
\int_{\bell_2} \frac{1}{(\pm\bell_2\cdot\bn-i0)^2\, (\bell_1 + \bell_2)^2}
=
\frac{3-d}{\mp \bell_1\cdot\bn}
\int_{\bell_2}  \frac{1}{(\pm\bell_2\cdot\bn-i0)\, (\bell_1 + \bell_2)^2}\,.
\end{align}
which would allow us to also use the one-loop recursive formula, obtaining
\begin{align}
\boldsymbol\cM^{(\pm\pm),\textrm{pot}}_{02;10110} &= 
- e^{2\epsilon\gamma_E}\frac{4\epsilon\, \Gamma(2\epsilon)\,\Gamma^2(-2\epsilon)\, \Gamma(\tfrac{1}{2}-\epsilon)\,\Gamma(\tfrac{1}{2}+\epsilon)}{\Gamma(-4\epsilon)}\,.
\end{align}

\subsection{Three-loop potential region}
We present here the results for potential boundary integrals at 4PM order:\begin{equation}
\boldsymbol\cM^{(\pm\pm\pm),\textrm{pot}}_{\alpha_1\alpha_2\alpha_3;\beta_1\cdots\beta_9} = \int_{\bell_1\bell_2\bell_3}\frac{1}{(\pm\bell_1\cdot\bn-i0)^{\alpha_1}(\pm\bell_2\cdot\bn-i0)^{\alpha_2}(\pm\bell_3\cdot\bn-i0)^{\alpha_3}\bD_1^{\beta_1}\cdots\bD_9^{\beta_9}}\,,
\end{equation}
with denominators 
\begin{align}
\bD_1 &= \bell_1^2\,, &
\bD_2 &= \bell_2^2\,, &
\bD_3 &= \bell_3^2\,, &
\bD_4 &= (\bell_1-\bq)^2\,, &
\bD_5 &= (\bell_2-\bq)^2\,, &
\bD_6 &= (\bell_3-\bq)^2\,, 
\end{align}
which are common to both sets introduced in~\eqref{eq:props4PM}, and the distinct ones given by
\begin{align}
\bD_7^{\textrm{I}} &= (\bell_1-\bell_2)^2\,, &
\bD_8^{\textrm{I}} &= (\bell_2-\bell_3)^2\,, &
\bD_9^{\textrm{I}} &= (\bell_3-\bell_1)^2\,, \\
\bD_7^{\textrm{II}} &= (\bell_{12}-\bq)^2\,, &
\bD_8^{\textrm{II}} &= (\bell_{23}-\bq)^2\,, &
\bD_9^{\textrm{II}} &= (\bell_{123}-\bq)^2\,.
\end{align}

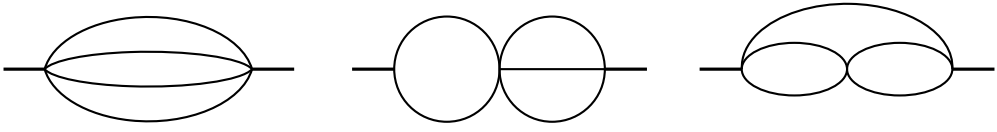
\begin{figure}
\centering
\begin{tikzpicture}[scale=0.7] 
  \draw[line width=0.8pt] (0,0) arc (10:170:2 and 1.2);
  \draw[line width=0.8pt] (0,0) arc (10:170:2 and 0.4);
  \draw[line width=0.8pt] (0,0) arc (-10:-170:2 and 0.4);
  \draw[line width=0.8pt] (0,0) arc (-10:-170:2 and 1.2);
  \draw[xshift=0cm, -, line width=1.2pt] (0,0) -- (0.8,0);
  \draw[xshift=-39.2mm, -, line width=1.2pt] (0,0) -- (-0.8,0); 
  \draw[xshift=47mm, xshift=10mm, line width=0.8pt] (0,0) circle (1.0);
  \draw[xshift=47mm, xshift=-10mm, line width=0.8pt](0,0) circle (1.0);
  \draw[xshift=47mm, -, line width=0.8pt] (0,0) -- (2.0,0);  
  \draw[xshift=47mm, xshift=20mm, -, line width=1.2pt] (0,0) -- (0.8,0);
  \draw[xshift=47mm, xshift=-20mm, -, line width=1.2pt] (0,0) -- (-0.8,0); 
  \draw[xshift=113mm, xshift=-10mm, line width=0.8pt] (0,0) ellipse (1.0 and 0.5);
  \draw[xshift=113mm, xshift=10mm, line width=0.8pt] (0,0) ellipse (1.0 and 0.5);
  \draw[xshift=113mm, xshift=20mm, line width=0.8pt] (0,0) arc (-2:182:2 and 1.2);
  \draw[xshift=113mm, xshift=20mm, -, line width=1.2pt] (0,0) -- (0.8,0);
  \draw[xshift=113mm, xshift=-20mm, -, line width=1.2pt] (0,0) -- (-0.8,0); 
\end{tikzpicture}
\caption{Topologies (of square propagators) of potential three-loop boundary integrals.}
\label{fig:pot3Loop}
\end{figure}

All three-loop static integrals (for the non-spinning case) can be reduced to the three topologies (of square propagators) shown in Fig.~\ref{fig:pot3Loop} \cite{4pmeft,4pmeft2,4pmeftot}.
The values for the relevant master integrals are given by
\begingroup
\allowdisplaybreaks
\begin{align}
\label{eq:D0}
\boldsymbol\cM^{\textrm{I},(\pm\pm\pm),\textrm{pot}}_{000;001110011} &=
e^{3\epsilon\gamma_E}\frac{\Gamma({1/2} - 3\epsilon)\Gamma({1/2} + 3\epsilon)\Gamma^5({1/2} - \epsilon)\Gamma^2({1/2} + \epsilon)}{\Gamma(1-4\epsilon)\Gamma^2(1-2\epsilon)\Gamma(1 + 2\epsilon)}\,,\\
\boldsymbol\cM^{\textrm{I},(\pm\pm\pm),\textrm{pot}}_{000;011101100} &=
\boldsymbol\cM^{\textrm{II},(\pm\pm\pm),\textrm{pot}}_{000;001101011} =
e^{3\epsilon\gamma_E}\frac{\Gamma (2\epsilon) \Gamma(\epsilon + {1/2}) \Gamma^5({1/2}-\epsilon)}{\Gamma({3/2} - 3\epsilon)\Gamma(1-2\epsilon)}\,,\\
\boldsymbol\cM^{\textrm{II},(\pm\pm\pm),\textrm{pot}}_{000;001100110} &=
e^{3\epsilon\gamma_E}\frac{\Gamma^4({1/2}-\epsilon) \Gamma(3 \epsilon - {1/2})}{\Gamma (2-4 \epsilon)}\,,\\
\boldsymbol\cM^{\textrm{I},(\pm\pm\pm),\textrm{pot}}_{001;001110011} &= e^{3\epsilon\gamma_E}\frac{i \sqrt{\pi}\Gamma^4(1/2-\epsilon)\Gamma(-3\epsilon)\Gamma(-\epsilon)\Gamma(1/2+\epsilon)}{\Gamma^2(1-2\epsilon)\Gamma(-4\epsilon)\Gamma(1+2\epsilon)}\,,\\
\boldsymbol\cM^{\textrm{I},(\pm\pm\pm),\textrm{pot}}_{001;001100110} &=
e^{3\epsilon\gamma_E}\frac{i\sqrt{\pi}\Gamma^2(1/2-2\epsilon)\Gamma^4(1/2-\epsilon)\Gamma(3\epsilon)}{\Gamma(1-4\epsilon)\Gamma^2(1-2\epsilon)}\,,\\
\label{eq:B1}
\boldsymbol\cM^{\textrm{II},(\pm\pm\pm),\textrm{pot}}_{001;001100110} &=
e^{3\epsilon\gamma_E}\frac{i\sqrt{\pi}\Gamma(1-3\epsilon)\Gamma^3(1/2-\epsilon)\Gamma(-\epsilon)\Gamma(3\epsilon)}{\Gamma(1-4\epsilon)\Gamma(3/2-3\epsilon)}\,,\\
\label{eq:B4m}
\boldsymbol\cM^{\textrm{I},(\pm+-),\textrm{pot}}_{011;001100110} &=
 \begin{multlined}[t]
 - {3 \over 2} \pi^{5/2} \bigg[
   \frac{1}{\epsilon^2} - \frac{6}{\epsilon}\log(2) - \frac{3}{4}(\pi^2-24\log^2(2))\\
   -\frac{1}{6}\left(-27\pi^2\log(2)+216\log^3(2)+1370\zeta(3)\right)\epsilon\bigg] + \mathcal{O}(\epsilon^2)\,,
 \end{multlined}\\
\boldsymbol\cM^{\textrm{I},(\pm++),\textrm{pot}}_{011;001100110} &=
 \begin{multlined}[t]
 \frac{1}{2}\boldsymbol\cM^{\textrm{II},(\pm++),\textrm{pot}}_{011;001110001} 
 =-\frac{\pi^{5/2}}{2}\bigg[
   \frac{1}{\epsilon^2} - \frac{6\log(2)}{\epsilon} + \frac{1}{12}(7 \pi^2 + 216 \log^2(2))\\
	-\frac{1}{2}\left(7\pi^2\log(2)+72\log^3(2)+158\zeta(3)\right)\epsilon\bigg]+\mathcal{O}(\epsilon^2)\,,
 \end{multlined}\\
\label{eq:B3m}
\boldsymbol\cM^{\textrm{II},(\pm+-),\textrm{pot}}_{011001110001} &=
-\pi^{5/2}\bigg[
 \begin{multlined}[t]
   \frac{1}{\epsilon^2} - \frac{6\log(2)}{\epsilon} - \frac{1}{12} (17 \pi^2 -216 \log^2(2))\\
   +\frac{1}{2}\left(17\pi^2\log(2)-72\log^3(2)-606\zeta(3)\right)\epsilon\bigg] + \mathcal{O}(\epsilon^2)\,,
 \end{multlined}\\
\label{eq:B5B6}
\boldsymbol\cM^{\textrm{I},(+-+),\textrm{pot}}_{111;001010101} &=
3\boldsymbol\cM^{\textrm{I},(+++),\textrm{pot}}_{111;001010101}
=\boldsymbol\cM^{\textrm{I},(++-),\textrm{pot}}_{111;001010101} 
=\frac{3}{2}\boldsymbol\cM^{\textrm{II},(++-),\textrm{pot}}_{111;001110001}\\
&=\frac{1}{2}\boldsymbol\cM^{\textrm{II},(+++),\textrm{pot}}_{111;001110001} =
-e^{3\epsilon\gamma_E}\frac{i\pi^{3/2}\Gamma^4(-\epsilon)\Gamma(1+3\epsilon)}{\Gamma(-4\epsilon)}.
\nn
\end{align}
\endgroup
For most of these integrals the strategies that we discussed in the previous section apply. In particular, all integrals with one or no linear propagator, \eqref{eq:D0} to \eqref{eq:B1}, can be computed through a recursive application of the one-loop formula in \eqref{pm-1-loop}. For all the integrals with three linear propagators, \eqref{eq:B5B6}, a version of the symmetrization trick  can be implemented. However, for the integrals with exactly two linear propagators, all the strategies discussed in the previous section fail. We give an explicit derivation below. The full dependence on $\epsilon$ includes hypergeometric functions.
We have therefore quoted here only their values to fourth order in $\epsilon$, which is sufficient to compute the total impulse. \vskip 4pt
Using~\eqref{pm-1-loop}, the four integrals in question can be reduced to the following form:
\begin{align}\label{}
\boldsymbol\cM^{\textrm{II},(\,\,\,+\pm),\textrm{pot}}_{011001110001} &=
\frac{\Gamma^2(\tfrac{1}{2} - \epsilon)\, \Gamma(\tfrac{1}{2}  + \epsilon)}{\Gamma(1 - 2\epsilon)} \int_{\bell_1\bell_2}\frac{e^{\epsilon\gamma_E}}{(\ell_1^z) (\pm\ell_2^z)}
\frac{1}{\vecbf{\ell}_1^2 \vecbf{\ell}_2^2 [(\vecbf{\ell}_{12} - \vecbf{q})^2]^{(4-d)/2}},
\\
\boldsymbol\cM^{\textrm{I},(\,\,\,+\pm),\textrm{pot}}_{011;001100110} &=
\frac{\Gamma^2(\tfrac{1}{2}  - \epsilon)\, \Gamma(\tfrac{1}{2}  + \epsilon)}{\Gamma(1-2\epsilon)} \int_{\bell_1\bell_2}  \frac{e^{\epsilon\gamma_E}}{(\ell_1^z)(\pm\ell_{12}^z)}
\frac{1}{\vecbf{\ell}_1^2 \vecbf{\ell}_2^2 [(\vecbf{\ell}_{12} - \vecbf{q})^2]^{(4-d)/2}},
\end{align}
where the first entry in the $(\,\,\, +\pm)$ upper index is left blank since it is independent of that propagator. We then only need to concentrate on the following type of (two-loop) generalized integrals:
\begin{align}
\mathcal{I}_1^\pm
&\equiv
\int_{\bell_1\bell_2} \frac{e^{-2\epsilon\gamma_E}}{(\ell_1^z)(\pm\ell_2^z)} \frac{1}{ \vecbf{\ell}_1^2 \vecbf{\ell}_2^2 [(\vecbf{\ell}_{12} - \vecbf{q})^2]^\alpha}\,,
\\
\mathcal{I}_2^\pm
&\equiv
\int_{\bell_1\bell_2} \frac{e^{-2\epsilon\gamma_E}}{(\ell_1^z) (\pm\ell_{12}^z)} \frac{1}{\vecbf{\ell}_1^2 \vecbf{\ell}_2^2 [(\vecbf{\ell}_{12} - \vecbf{q})^2]^\alpha}\,,
\end{align}
with $\alpha\equiv (4 {-} d)/2$, which obey the relation
\begin{align}
\mathcal{I}_2^+
&=
\frac{1}{2} \int_{\bell_1,\bell_2} \left[ \frac{1}{(\ell_1^z - i 0)} + \frac{1}{(\ell_2^z - i 0)} \right] \frac{e^{-2\epsilon\gamma_E}}{(\ell_{12}^z - i 0)} \frac{1}{\vecbf{\ell}_1^2 \vecbf{\ell}_2^2 [(\vecbf{\ell}_{12} - \vecbf{q})^2]^\alpha}
\nonumber \\
&=
\frac{1}{2} \int_{\bell_1,\bell_2} \frac{e^{-2\epsilon\gamma_E}}{(\ell_1^z - i 0) (\ell_2^z - i 0)}\frac{1}{\vecbf{\ell}_1^2\vecbf{\ell}_2^2 [(\vecbf{\ell}_{12} - \vecbf{q})^2]^\alpha}=
\frac{1}{2}\, \mathcal{I}_1^+\,.\label{i1i2rel}
\end{align}
When $\alpha=1$ this reduces to the two-loop identity in~\eqref{2loop-hard-ratio}.\vskip 4pt
We compute $\mathcal{I}_1$ and $\mathcal{I}_2$ by employing a  parametric representation. Firstly, we have \begin{align}
\mathcal{I}_1^\pm = 
\frac{i^{\nu } e^{-i\pi d/2}}{\Gamma(\alpha)}
\left(\prod_{i=1}^{5 }\int_0^\infty \dd x_i\right) x_3^{\alpha - 1}\, \mathcal{U}^{-d/2}  e^{-i\mathcal{F}^\pm/\mathcal{U}}\,,
\end{align}
where $\mathcal{U}$ and $\mathcal{F}$ are given by
\begin{align}
\mathcal{U} &= x_1 x_2+x_1 x_3+x_2 x_3\,,
\\
\mathcal{F}^\pm &= x_1 x_2 x_3 + \frac{1}{4} \big(\pm 2 x_3 x_4 x_5 - (x_2{+}x_3) x_4^2 - (x_1{+}x_3) x_5^2\big)\,.
\end{align}
Notice that $\mathcal{U}$ is independent of $x_4$ and $x_5$ and $\mathcal{F}$ is a quadratic polynomial in these two variables. Therefore, we can simply integrate them out, yielding \begin{align}
\mathcal{I}_1^\pm = 
\frac{i^{1+\alpha} e^{-i\pi d/2}}{\Gamma(\alpha)}
\left(\prod_{i=1}^{3}\int_0^\infty \dd x_i\right)
x_3^{\alpha-1}\,
\mathcal{U}^{\frac{1-d}{2}}
e^{- i x_1 x_2 x_3/\mathcal{U}}
\left(\pi \,\pm\, 2 \Arctan{x_3 \over \sqrt{\cal U}}\right)\,.
\end{align}
At this point, we note the appearance of the factor $x_3^{\alpha-1}$ with a non-integer exponent and the square root in the $\Arctan$ term. In order to remove it, we introduce the Feynman form,
\begin{align}
\mathcal{I}_1^\pm = 
- \frac{\Gamma(3-d+\alpha)}{\Gamma(\alpha)}
\Bigg(\prod_{i=1}^{3}\int_0^\infty \dd x_i\Bigg) \delta(1 {-} X)
{\mathcal{U}^{\alpha + (7-3 d)/2} \over x_3^{4-d} (x_1 x_2)^{\alpha+3-d} } \bigg(\pi \pm 2\Arctan{x_3 \over \sqrt{\cal U}}\bigg),
\end{align}
with $X= \sum_{i\in I} x_i$ and $I=\{1,2,3\}$. As in the usual Feynman parametric representation, one is free to choose any non-empty subset $I\subset\{1,2,3\}$~\cite{Cheng:1987ga}. The choice $X=x_3$ turns out to be convenient. The next step is to rationalize the argument of the $\Arctan$. We find this can be realized via the substitution
\beq
x_1 \to \frac{z^2-y z^2}{1 + y z^2}\,, \quad
x_2 \to y z^2\,,
\eeq
leading to
\begin{align}\label{app-I1-m3}
\mathcal{I}_1^\pm =
- {2\Gamma(3 - d + \alpha) \over \Gamma(\alpha)}
\int_0^\infty & \dd z\,  z^{-2-3 d+4 (d-\alpha)+2 \alpha}\, \Big[\pi  \,\pm\, 2 \operatorname{arccot}(z) \Big]
\\
&
\times\int_0^1 \dd y\, y^{-3+d-\alpha }\, (1-y)^{-3+d-\alpha }\, \big(1+y z^2\big)^{2-d+\alpha}.
\nonumber
\end{align}
An observation is that the first term in the square bracket is independent of the signs of the integral.
For this part, performing the integrals for $z$ and $y$ successively gives
\begin{align}\label{}
- \frac{\pi\, \Gamma^2(-\epsilon) \Gamma(1 - \alpha - \epsilon) \Gamma(\alpha + 2\epsilon)}{\Gamma(\alpha)\Gamma(1 - \alpha - 3\epsilon)}\,,
\end{align}
which is in perfect agreement with the result obtained by using the cutting rules in the two linear propagators,
\begin{align}
\mathcal{I}_1^{+} + \mathcal{I}_1^{-}
&= \frac{(2\pi i)^2}{2}
\int{\dd^{d-1}\bell_1^\perp \dd^{d-1}\bell_2^\perp \over \pi^{d}}
\frac{1}{[\vecbf{\ell}_1^\perp]^2 [\vecbf{\ell}_2^\perp]^2 [(\vecbf{\ell}_{12}^\perp - \vecbf{q})^2]^\alpha}
\nonumber\\
&=
- \frac{2\pi \Gamma^2(-\epsilon) \Gamma(1 - \alpha - \epsilon)\Gamma(\alpha + 2\epsilon)}{\Gamma(\alpha) \Gamma(1 - \alpha - 3\epsilon)}\,.
\end{align}
To evaluate the second term in \eqref{app-I1-m3}, we rewrite the $\operatorname{arccot}$ as
\begin{align}\label{}
\operatorname{arccot}(z) = \int_0^1 \dd w\,\frac{z}{w^2+z^2},
\end{align}
so that the integrand becomes a rational function in $y$, $z$ and $w$.
We find it convenient to use the integral representation of the hypergeometric function,
\begin{align}
{}_2F_1(a,b;c;z) 
:=  {\Gamma(c) \over \Gamma(b)\, \Gamma(c-b)}
\int_0^1 \dd t\, t^{b-1} (1-t z)^{-a} (1-t)^{-b+c-1},
\end{align}
and the iterated-integration relation between higher- and lower-order terms \cite{slater1966generalized}
\begin{align}\label{}
{}_{p+1}F_{q+1}\!\big(\vec{a}_{p},a_0; \vec{b}_{q},b_0; z\big) 
= \frac{\Gamma(b_0)}{\Gamma(a_0)\, \Gamma(b_{0} {-} a_{0})}
\int_{0}^{1} dt\,  t^{a_0 - 1}(1-t)^{b_0-a_0-1}\,
{}_{p}F_{q}\big(\vec{a}_{p};\vec{b}_{q}; tz\big),
\end{align}
where $\vec{a}_{p} \equiv (a_1,\ldots, a_p)$, and similarly for $\vec{b}_{q}$. Performing the integrals for $y$, $z$ and $w$ successively, we finally find that the second term in \eqref{app-I1-m3} can be evaluated to the following hypergeometric functions: ${}_3F_2$ and ${}_4F_3$.
The final analytic expression for $\mathcal{I}_1^\pm$ then becomes
\begin{align}\label{I1-pFq}
\mathcal{I}_1^\pm =
& - {\pi  \Gamma(-\epsilon)^2\,\Gamma(1 {-} \alpha {-} \epsilon)\, \Gamma(\alpha {+} 2\epsilon) \over \Gamma(\alpha)\, \Gamma(1 {-} \alpha {-} 3\epsilon)}
\mp  {\Gamma\big(\tfrac{3}{2} {-} \alpha {-} \epsilon\big)\, \Gamma\big(\alpha {+} \epsilon {-} \tfrac{1}{2}\big)\, \Gamma(\alpha {+} 2\epsilon) 
\over \Gamma(\alpha)}
\nonumber\\
&~~\times
\bigg(
\Gamma^2(1 {-} \alpha {-} 2\epsilon)\, \Gamma(1 {-} \alpha {-} \epsilon)\, 
{}_3\tilde{F}_2\big(1 {-} \alpha {-} \epsilon, 1 {-} \alpha {-} 2\epsilon, 1 {-}\alpha {-} 2 \epsilon; 2 {-} 2\alpha {-} 4\epsilon, 2 {-} \alpha {-} \epsilon; 1\big)
\nonumber\\
& \qquad
-\sqrt{\pi}\,\Gamma^2\big(\tfrac{1}{2}-\epsilon\big)\,
{}_4\tilde{F}_3\Big(\tfrac{1}{2},1,\tfrac{1}{2}-\epsilon ,\tfrac{1}{2}-\epsilon; \tfrac{3}{2}, \tfrac{3}{2} -\alpha -3 \epsilon, \alpha +\epsilon+\tfrac{1}{2};1\Big)
\bigg),
\end{align}
where ${}_p\tilde{F}_q$,
\begin{align}\label{}
{}_{p}\tilde{F}_{q} \big(\vec{a}_{p}; \vec{b}_{q}; z\big) 
:= {{}_{p}F_{q}\big(\vec{a}_{p}; \vec{b}_{q}; z\big) \over \Gamma(b_1)\cdots \Gamma(b_q)}\,,
\end{align}
is the regularized hypergeometric function associated with ${}_pF_q$.
Following the same procedure, one can derive the full analytical form for $\mathcal{I}_2$,
\begin{align}\label{I2-pFq}
\mathcal{I}_2^\pm =
&
- {\pi\,\Gamma^2(-\epsilon)\, \Gamma(1 {-} \alpha {-} \epsilon)\, \Gamma(\alpha {+} 2\epsilon) \over \Gamma(\alpha)\, \Gamma(1 {-} \alpha {-} 3\epsilon)}
\mp {\Gamma\big(\tfrac{3}{2} {-} \epsilon\big)\, \Gamma\big(\epsilon {-} \tfrac{1}{2}\big)  \over \Gamma(\alpha)}
\\
&~\times
\bigg[\Gamma(-2\epsilon)\, \Gamma(-\epsilon)\, \Gamma(3 {-} \alpha {-} 2\epsilon)\, \Gamma(\alpha {+} 2\epsilon {-} 2) \, 
{}_3\tilde{F}_2\big(-\epsilon, -2\epsilon, 1 {-} \alpha {-} 2\epsilon; 1 {-} \epsilon, 1 {-} \alpha {-} 4\epsilon; 1\big)
\nonumber\\
&\qquad 
-\sqrt{\pi}\,\Gamma\big(\tfrac{1}{2} {-} \epsilon\big)\, \Gamma\big(\tfrac{3}{2} {-} \alpha {-} \epsilon\big)\, \Gamma(\alpha {+} 2\epsilon)\,
{}_4\tilde{F}_3\Big(\tfrac{1}{2}, 1, \tfrac{1}{2} {-} \epsilon, \tfrac{3}{2} {-} \alpha {-} \epsilon; \tfrac{3}{2}, \epsilon {+} \tfrac{3}{2}, \tfrac{3}{2} {-} \alpha {-} 3\epsilon; 1\Big)
\bigg]\,,
\nonumber
\end{align}
which obeys the relation in \eqref{i1i2rel}. A numerical check for \eqref{I1-pFq} and \eqref{I2-pFq} has been performed in \cite{Jinno:2022sbr}.
Furthermore, we have checked that our analytic results satisfy non-trivial dimensional recurrence relations \cite{Tarasov:1996br,Laporta:2000dsw,Lee:2009dh}.\vskip 4pt In order to match these results to the boundary conditions appearing in the method of differential expansions, an expansion in $\epsilon$ is required. It is straightforward to expand $\Gamma$ functions, however, it is difficult to do so for hypergeometric functions, in particular in the presence of half-integer parameters \cite{Huber:2005yg,Huber:2007dx}, such as the ${}_3F_2$ and ${}_4F_3$ in \eqref{I1-pFq} and \eqref{I2-pFq}.
While we can obtain the first one or two orders via a series expansion in \texttt{Mathematica}, we had to rely on a different approach by {\it bootstrapping} analytic constants (see also \cite{Blumlein:2021pgo} for alternative options). We performed a formal Laurent expansion and numerically evaluated the coefficients to high precision. It is then possible to reconstruct analytic expressions based on an ansatz using \textit{Mathematica's} built-in function \texttt{FindIntegerNullVector},\footnote{See e.g. \cite{bailey1991polynomial,Bailey:1999nv} for the theoretical idea behind such integer relation algorithms.} leading to the results cited in~\eqref{eq:B4m} - \eqref{eq:B3m}.\vskip 4pt

We have also checked that the results satisfy the relation
\begin{equation}\label{eq:B3B4id}
\begin{aligned}
\boldsymbol\cM^{\textrm{II},(\pm++),\textrm{pot}}_{011;001110001}+\boldsymbol\cM^{\textrm{II},(\pm+-),\textrm{pot}}_{011001110001}
&=\boldsymbol\cM^{\textrm{I},(\pm++),\textrm{pot}}_{011;001100110}+\boldsymbol\cM^{\textrm{I},(\pm+-),\textrm{pot}}_{011;001100110}\\
&=
-\frac{2\pi e^{3\gamma_E \epsilon} \Gamma({1}/{2}-2\epsilon) \Gamma^2({1/2} - \epsilon) \Gamma^2(-\epsilon) \Gamma({1/2} + 3\epsilon)}{\Gamma({1/2} - 4\epsilon) \Gamma(1-2\epsilon)}\,,
\end{aligned}
\end{equation}
which can be derived by rewriting the sum of the linear propagators as Dirac-$\delta$ functions, reducing the integral to a (lower-dimensional) version with one fewer linear propagator. The integral can then be performed with the recursive one-loop strategy.

\subsection{Three-loop radiative I: Conservative}

The conservative part is determined by using Feynman's $i0$-prescription for the graviton propagators, while retaining the real part of the total impulse \cite{eftrad}. At 4PM order, only regions with two radiation modes contribute \cite{4pmeft,4pmeft2,4pmeftot}. Following \S\ref{sec:bc-exp}, we find a total of 39 independent boundary conditions with the relevant scaling $v_\infty^{-4\epsilon}$. As we discussed earlier, see~\eqref{eq:momConservation}, the $b$-direction, together with the complete 3PM result \cite{3pmeft}, is sufficient to reconstruct the full answer. Hence, ignoring 21 contributions needed for the $u$-direction, reduces the problem to 18 boundary integrals to be computed.\vskip 4pt

Following the method of regions in parameter space discussed in~\S\ref{sec:MoR}, the relevant integration regions correspond to either two or three $\alpha$-parameters being rescaled by $\alpha_i\rightarrow v_\infty^{-2}\alpha_i$. For boundary integrals with three simultaneously rescaled $\alpha$-parameters, a total of five, they start at an order in $v_\infty$ beyond what is required from the boundary relations, and therefore can be set to zero.\vskip 4pt

The remaining 13 boundary integrals, having exactly two $\alpha$-parameters rescaled, have one of the two following parametric representation, which only differ on the $x_4$ integration; either
\begin{align}
\boldsymbol\cI_{1}^{\textrm{2rad,cons}} &= 
e^{\frac{3}{4}i\pi(2 \epsilon-1)+3\gamma_\textrm{E}\epsilon}
v_\infty^{2-4 \epsilon}
\!\int_0^{\infty}\! \dd x_1 \dd x_2 \dd x_3 \dd x_4\, e^{- i \frac{x_1 x_2}{x_1+x_2}}
\frac{(x_1 + x_2)^{\epsilon-1} (x_3x_4)^{\epsilon-\frac{3}{2}}}{\sqrt{x_3+x_4-x_1-x_2}}\,,
\end{align}
or
\begin{align}
\boldsymbol\cI_{2}^{\textrm{2rad,cons}} &= 
e^{\frac{1}{4}i\pi(3 \epsilon-1)+3\gamma_\textrm{E}\epsilon}
v_\infty^{-4 \epsilon}
\!\int_0^{\infty}\! \dd x_1 \dd x_2 \dd x_3 \dd x_4\, e^{- i \frac{x_1 x_2}{x_1+x_2}}
\frac{x_4(x_1 + x_2)^{\epsilon-1} (x_3x_4)^{\epsilon-\frac{3}{2}}}{\sqrt{x_3+x_4-x_1-x_2}}\,.
\end{align}
We present the evaluation of the former. The latter integral can easily be performed following similar steps.\vskip 4pt
By successively performing the $\alpha_3$ and $\alpha_4$ integrals, realizing that they evaluate to $\beta$-functions, we find the following combination of $\Gamma$-functions:
\begin{align}\label{}
\boldsymbol\cI_{1}^{\textrm{2rad,cons}} = 
\frac{i e^{\frac{1}{4}i\pi(14\epsilon-3)}v_\infty^{2-4\epsilon}}{\sqrt{\pi}}
\Gamma(\tfrac{3}{2} -2\epsilon)\,\Gamma^2(\epsilon-\tfrac{1}{2})
\int_0^{\infty} \dd x_1 \dd x_2\, e^{-\frac{i x_1 x_2}{x_1+x_2}}\, (x_1 + x_2)^{3 \epsilon - \frac{5}{2}}\,.
\end{align}
The remaining integral corresponds to an ordinary (mass-independent) bubble. We then find
\begin{align}\label{}
\boldsymbol\cI_{1}^{\textrm{2rad,cons}}
&=-v_\infty^{2-4 \epsilon}\,e^{2 i \pi \epsilon +3\gamma_\textrm{E}\epsilon}
\frac{\sqrt{\pi}\, 2^{4\epsilon}\, \sin(\pi\epsilon)\, \Gamma(\tfrac{3}{2}-2\epsilon)\, \Gamma(1-\epsilon)\, \Gamma(\epsilon - \tfrac{1}{2})\, \Gamma(2\epsilon - 1) }{ \cos(3\pi\epsilon)\, \Gamma (2-3\epsilon)}\,.
\end{align}
Following the same procedure the second integral evaluates to
\begin{equation}
\boldsymbol\cI_{2}^{\textrm{2rad,cons}}
=-v_\infty^{-4\epsilon}\,e^{2 i \pi \epsilon +3\gamma_\textrm{E}\epsilon}
\frac{\pi 2^{6\epsilon}\Gamma(\tfrac{1}{2}-2\epsilon)\Gamma(\epsilon-\tfrac{1}{2})\Gamma(\tfrac{1}{2}+\epsilon)}{\cos(3\pi\epsilon)\Gamma(1-3\epsilon)}\,.
\end{equation}

\subsection{Three-loop radiative II: Dissipative}

We now consider the case of causal graviton propagators, dictated by retarded boundary conditions in the $i0$-prescription. In \S\ref{sec:BC3Loops} we identified two types of radiative integrals which, unlike the conservative sector, now can have either one or two graviton propagators going on-shell,
\begin{equation}\label{eq:radIIdiss}
\begin{aligned}
\boldsymbol\cI^{\textrm{2rad}} &= \int_\bell \frac{1}{[\pm\bell\cdot\bn-i0]^{\alpha}[\bell^2]^{\beta_1}[(\bell-\bq)^2]^{\beta_2}}\\
&\quad\times\int_{\bk_1\bk_2}\frac{1}{[(\bk_1+\bk_2)^2]^{\beta_3}[(\bk_1^2-(\bell\cdot\bn\pm i0)^2)]^{\beta_4}[(\bk_2^2-(\bell\cdot\bn\pm i0)^2)]^{\beta_5}}\,,\\
\boldsymbol\cI^{\textrm{1rad}} &= \int_{\bell_1\bell_2}\frac{1}{[\pm\bell_1\cdot\bn-i0]^{\alpha_1}[\pm\bell_2\cdot\bn-i0]^{\alpha_2}[\bell_1^2]^{\beta_1}[\bell_2^2]^{\beta_2}[(\bell_1+\bell_2-\bq)^2]^{\beta_3}}\\
&\quad\times\frac{1}{[(\bell_1-\bq)^2]^{\beta_4}[(\bell_2-\bq)^2]^{\beta_5}}\int_{\bk}\frac{1}{[(\bk^2-(\bell_1\cdot\bn\pm i0)^2)]^{\beta_6}}
\end{aligned}
\end{equation}
The 2rad integral can be computed straightforwardly using formula~\eqref{eq:inner2rad} for the inner two-loop integral, followed by a simple application of the result~\eqref{pm-1-loop} for the remaining one-loop integral~\eqref{static-1-loop-def}. The 1rad region, on the other hand, is a tad more involved.\vskip 4pt

After having performed the inner integral of $\boldsymbol\cI^{\textrm{1rad}}$, one is left with a two-loop potential-only boundary integral of the type in~\eqref{eq:potBC3PM}.
Unfortunately, due to the presence of a non-integer exponent, we cannot perform an IBP reduction for the leftover two-loop integral, which would reduce it to the set of masters in \eqref{eq:K2_1} to \eqref{eq:K2_9}.
In the following we discuss a few paradigmatic cases and describe the different strategies we used to compute them.\vskip 4pt

Let us start by elaborating on a special property for some of the integrals in \eqref{eq:radIIdiss}. It is often the case that a linear propagator appears twice, yet with a different sign, e.g.
\begin{equation}
  \int_\bell \frac{1}{[\bell\cdot\bn-i0]^{\alpha_1}[-\bell\cdot\bn-0]^{\alpha_2}} [\cdots]\,.
\end{equation}
By performing an explicit parametrization we can show that these propagators can be then combined into a single factor. More generally, we have
\begin{equation}
  \begin{aligned}
  \int_\bell \frac{1}{[\bell\cdot\bn-i0]^{\alpha_1}[-\bell\cdot\bn-i0]^{\alpha_2}} [\cdots] &=
  e^{i\pi \phi_1} \int_\bell \frac{1}{[\bell\cdot\bn-i0]^{\alpha_1+\alpha_2}} [\cdots]\\
  &\quad+e^{i\pi \phi_2} \int_\bell \frac{1}{[-\bell\cdot\bn-i0]^{\alpha_1+\alpha_2}} [\cdots]\,,
  \end{aligned}
\end{equation}
where $\phi_1$ and $\phi_2$ are $d$-dependent phases, which become integers for the case of $d$-independent exponents. Using the above decomposition, at the end of the day we arrive at integrals with either one or two independent linear propagators.\vskip 4pt

\textbf{\textit{One linear propagator.}} 
Most integrals with a single linear propagator either have a scaleless sub-integral, or can be performed by iteratively using the one-loop formula~\eqref{pm-1-loop}. However, one type integral cannot be computed using an iterative strategy, namely
\begin{equation}
  \int_{\bell_1\bell_2} \frac{1}{[-\bell_1\cdot\bn]^{-1+2\epsilon}[\bell_1^2][\bell_2^2]^{-1}[(\bell_2-\bq)^2][(\bell_1-\bell_2)^2]}\,,
\end{equation}
carrying an implicit $i0$-prescription in all propagators.
For this specific case we use an IBP reduction on the $\bell_2$ integration, yielding
\begin{equation}\label{eq:subloopIBP}
  \int_{\bell_2} \frac{1}{[\bell_2^2]^{-1}[(\bell_2-\bq)^2][(\bell_1-\bell_2)^2]} = \bell_1\cdot\bn \int_{\bell_2} \frac{1}{[(\bell_2-\bq)^2][(\bell_1-\bell_2)^2]}\,.
\end{equation}
Even though, in general, IBPs are not particularly helpful on partial integration, since they often introduce rational coefficients that depend on the other loop momenta, in this case the integral can be easily performed by iteratively using the standard one-loop formula in~\eqref{pm-1-loop}.\vskip 4pt

\textbf{\textit{Two linear propagators.}} An IBP of the type in\eqref{eq:subloopIBP} can also be applied to many of the integrals with two linear propagators. For all the remaining ones which do not fit this pattern, we performed a direct integration in parameter space.
We demonstrate the strategy below with an explicit example.\vskip 4pt

Consider the integral
\begin{equation}
  I = \int_{\bell_1\bell_2} \frac{1}{[-\bell_1\cdot\bn]^{2\epsilon}[\bell_2\cdot\bn][\bell_1^2][(\bell_2-\bq)^2][(\bell_1-\bell_2)^2]}\,,
\end{equation}
 which in a Schwinger parametrization takes the form
\begin{equation}
  I \propto \int_0^\infty \dd x_1\cdots \dd x_5 \,x_1^{2 \epsilon -1} \left(x_4 x_5+x_3 \left(x_4+x_5\right)\right)^{\epsilon -\frac{3}{2}} e^{-\frac{\left(x_4+x_5\right) x_1^2-2 x_2 x_5 x_1+4 x_3 x_4 x_5+x_2^2 \left(x_3+x_5\right)}{4 \left(x_4 x_5+x_3 \left(x_4+x_5\right)\right)}}\,,
\end{equation}
where we have dropped some $\epsilon$-dependent overall factors.
The integration over the parameters coming from the linear propagators,  i.e. $x_1$ and $x_2$, evaluate to a hypergeometric function.
Upon going to Feynman form, and dropping some additional $\epsilon$-dependent overall factors, we find
\begin{equation}
  \begin{aligned}
  I &\propto \int_0^\infty \!\dd x_3\dd x_4 \dd x_4 \,\delta(1-x_{345})\left(x_3 x_4 x_5\right){}^{-3 \epsilon -\frac{1}{2}} \left(x_3+x_5\right){}^{\epsilon -\frac{1}{2}} \left(x_4 x_5+x_3 \left(x_4+x_5\right)\right){}^{4 \epsilon -1}\\
  &\quad\times\left[2 x_5 \Gamma \big(\epsilon +\tfrac{1}{2}\big) \, _2F_1\big(\tfrac{1}{2},\epsilon +\tfrac{1}{2};\tfrac{3}{2};-\tfrac{x_5^2}{x_4 x_5+x_3 \left(x_4+x_5\right)}\big)+ \sqrt{\pi}\Gamma (\epsilon )\sqrt{x_3 x_4+ \left(x_3+x_4\right) x_5}\right]\,,
  \end{aligned}
\end{equation}
where $x_{345}=x_3+x_4+x_5$. We use, once again, the Cheng-Wu theorem~\cite{Cheng:1987ga} to perform the $x_5$ integral, by localizing $x_5\rightarrow 1$. The final two-fold integration can then be performed after a change of variables
\beq  y = \frac{x_4+x_3 (x_4+1)}{x_4+1} \,, \quad z=\frac{1}{x_4+x_3 (x_4+1)}\,,
\eeq
such that the final answer becomes
\begin{align}
I = 
 -&\frac{i \pi ^2 4^{\epsilon } e^{2 \gamma  \epsilon +i \pi  \epsilon } \sec (3 \pi  \epsilon )}{\Gamma (2-4 \epsilon )  \Gamma \left(\frac{1}{2}-3 \epsilon \right) \Gamma \left(\epsilon +\frac{1}{2}\right) \Gamma (2 \epsilon +1)}\Big[\pi  2^{4 \epsilon +1} \Gamma (2-4 \epsilon ) \Gamma (2 \epsilon )
 \nonumber\\
& +\pi ^{5/2} (1-4 \epsilon ) \epsilon  \csc (2 \pi  \epsilon ) \sec ^2(\pi  \epsilon ) \, _3\tilde{F}_2\left(\tfrac{1}{2},\tfrac{1}{2}-\epsilon ,\epsilon +\tfrac{1}{2};\tfrac{3}{2},2 \epsilon +1;1\right)
\\
&-\sec (\pi  \epsilon ) \Gamma\big(\tfrac{1}{2}-3 \epsilon \big)\, \Gamma \big(\tfrac{1}{2}-\epsilon \big)\, \Gamma (2 \epsilon +1) \, _3F_2\left(\tfrac{1}{2}-3 \epsilon ,\tfrac{1}{2}-2 \epsilon  ,\tfrac{1}{2}-\epsilon ;1-2 \epsilon ,\tfrac{3}{2}-2 \epsilon ;1\right)\Big]
\nonumber
\end{align}
after reinstating the overall factors. A similar strategy applies to all integrals we encounter at 4PM order.


\section{Conservative angle}\label{sec:app-angle}

\subsection{From the impulse}
The (Feynman-only) conservative scattering angle in the center-of-mass frame, ignoring the recoil, is computed from the impulse via the relation, e.g. \cite{paper1,pmeft},
\begin{equation}
  2 \sin\left(\frac{\chi^\textrm{cons}}{2}\right) = \frac{\sqrt{-(\Delta p_1^\textrm{cons})^2}}{p_\infty}\,.
\end{equation}
Using the decomposition in~\eqref{eq:DpConsStructure}, the above expression becomes
\begin{equation}
  \chi^\textrm{cons} = \arcsin \left[
    \sum_{n=1}^\infty \frac{1}{(j\Gamma)^n}\sum_{i=1}^{\lceil n/2\rceil} \tilde{c}_{b,\rm cons}^{(n,i)}
    \right]\,,
\end{equation}
with
\begin{equation}
  \tilde{c}_{b,\rm cons}^{(n,i)} = -\Gamma\nu^{i-1}N^{n,i}(\nu)(\gamma^2-1)^{(n-1)/2}c_{b,\textrm{cons}}^{(n,i)}\,.
\end{equation}
After PM-expanding,
\begin{equation}
  \frac{\chi}{2} = \sum_{n=1}^\infty \chi_b^{(n)} \left(\frac{G M}{b}\right)^n = \sum_{n=1}^\infty \frac{\chi_j^{(n)}}{j^n}= \sum_{n=1}^\infty \frac{\tilde\chi_j^{(n)}}{(j\Gamma)^n}\,,
\end{equation}
and using the series expansion of the $\arcsin$, we can then easily obtain a closed-form expression for the $\tilde{\chi}_j^{(n),\textrm{cons}}$ coefficients,
\begin{equation}
  \tilde{\chi}_j^{(n),\textrm{cons}} = \sum_{\sigma\in\cP(n)} \frac{2^{\Sigma(\sigma) - 3}(1-(-1)^{\Sigma(\sigma)})\Gamma(\Sigma(\sigma)/2)^2}{\pi}\prod_i \frac{\left(\sum_{j=1}^{\lceil \sigma_i/2\rceil} \tilde{c}_{b,\rm cons}^{(\sigma_i,j)}\right)^{\sigma^i}}{(\sigma^i)!}\,,
\end{equation}
with $\Sigma(\sigma) = \sum_i \sigma^i$, for a partition~$\sigma\in\cP(n)$ of the integer number~$n$, i.e. $\sum_i \sigma_i\sigma^i=n$, where $\sigma_i$ are unique integers.\footnote{The factor $(1-(-1)^{\Sigma(\sigma)})$ is filtering for all integer partitions where $\Sigma(\sigma)$ is an odd number. One could in principle restrict the sum to only such partitions.}
For instance, to 4PM we find
\begin{equation}
  \begin{aligned}
    \tilde\chi_j^{(1),\textrm{cons}} &= \frac{1}{2}\tilde{c}_{b,\rm cons}^{(1,1)}\,,\\
    \tilde\chi_j^{(2),\textrm{cons}} &= \frac{1}{2}\tilde{c}_{b,\rm cons}^{(2,1)}\,,\\
    \tilde\chi_j^{(3),\textrm{cons}} &= \frac{1}{12}\left(\left(\tilde{c}_{b,\rm cons}^{(1,1)}\right)^3+6\tilde{c}_{b,\rm cons}^{(3,1)}+6\tilde{c}_{b,\rm cons}^{(3,2)}\right)\,,\\
    \tilde\chi_j^{(4),\textrm{cons}} &= \frac{1}{4}\left(\left(\tilde{c}_{b,\rm cons}^{(1,1)}\right)^2\tilde{c}_{b,\rm cons}^{(2,1)}+2\tilde{c}_{b,\rm cons}^{(4,1)}+2\tilde{c}_{b,\rm cons}^{(4,2)}\right)\,.
  \end{aligned}
\end{equation}
\vskip 4pt

\subsection{Probe limit} The test-particle limit plays a significant role in determining the conservative dynamics of compact binary systems with generic mass ratios. For instance,
at the first two orders in the PM expansion, it captures the complete information \cite{Vines:2018gqi,paper1}. At higher orders, it remains a crucial part of the full result by providing consistency checks, both for the total answer as well as the master integrals, which often also contribute at higher orders in the mass-ratio, e.g. \cite{3pmeft}. In this appendix we give an  analytic derivation of the test-particle scattering angle of non-spinning bodies to arbitrary orders in $G$.\vskip 4pt

In the probe limit we have
\begin{align}
m_2\gg m_1 \quad\Longrightarrow\quad
\mu \simeq m_1,~~
M \simeq m_2, ~~
\nu = {\mu \over M} \ll 1,
\end{align}
describing a particle of mass $\mu \simeq m_1$ is moving in a Schwarzschild background with a mass parameter $M \simeq m_2$\,,
\begin{align}
ds^{2} \equiv g_{\mu\nu} dx^\mu dx^\nu = \left(1 - {2GM \over r} \right) dt^2 - \left(1 - {2GM \over r}\right)^{-1} \,dr^2 
- r^2 \left(d\theta^2 + \sin^2\theta \, d\varphi^2\right).
\end{align}
For simplicity, but without loss of generality, we consider motion along the equatorial plane, i.e. $\theta=\tfrac{\pi}{2}$ and $p_\theta=0$. Starting from the on-shell condition,
\begin{align}\label{app-schw-shell}
g_{\mu\nu} p^\mu p^\nu -  \mu^2  \,=\, 0\,,
\end{align}
with $p_{\mu} = (\mu\mathcal{E}_0, p_r, p_\theta, p_\phi)$ the four-momentum, we obtain
\begin{align}
p_r^2 \,=\,  {G^2 M^2 \mu^2 \big({\cal E}^2 + 2u-1\big)+J^2 u^2 (2u - 1)  \over  G^2 M^2 (1-2 u)^2},
\end{align}
where $J=p_\phi$ and $u=GM/r$, such that the scattering angle becomes \begin{align}
\chi + \pi \,=\, \int_{-\infty}^{\infty} dr\,{\partial p_r(J,u,\mathcal{E}_0) \over \partial J}\,.
\end{align}
Introducing the following (dimensionless) variables, similarly to the comparable-mass case, 
\begin{align}
y &\equiv {J \over \mu r}\,,\quad j
\equiv \frac{J}{Gm_1m_2}\,,\quad c_0 \,\equiv\, {\mathcal{E}_0}^2 - 1\,,
\end{align}
we have \cite{damour1,damour2}
\begin{align}\label{chi-integral-rep}
{\chi \over 2} \,=\, - {\pi \over 2}  + \int_{0}^{\sqrt{c_0}} {dy \over \sqrt{c_0 - y^2 + 2y(1+y^2) j^{-1}}}\,.
\end{align}
Expanding the integrand in \eqref{chi-integral-rep} in powers of $1/j$, we arrive at
\begin{align}
{\chi \over 2} \,=\, \sum_{n=1}^\infty {(-2)^n \over j^n}\, {\Gamma(n{+}1/2) \over \sqrt{\pi}\, \Gamma(n{+}1)}\,
\int_{0}^{\sqrt{c_0}} dy\, {(y^3 + y)^{n}  \over  (c_0 - y^2)^{(2n+1)/2}}\,,
\end{align}
where we used
\begin{align}
 \int_{0}^{\sqrt{c_0}}  {dy  \over  \sqrt{c_0 - y^2}} \,=\, {\pi \over 2}\,.
\end{align}
Hence, rescaling $y^2  \to c_0 z$, we have
\begin{align}
\chi \,&=\, \sum_{n=1}^\infty {(-2)^n \over j^n (\gamma^2 {-} 1)^{n/2}}\,{\Gamma(n{+}1/2) \over \sqrt{\pi}\, \Gamma(n{+}1)}\,
\int_{0}^{1} dz\, {z^{(n+1)/2} 
\big((\gamma^2 {-} 1)z + 1\big)^{n}  \over  (1 - z)^{n+1/2}},
\end{align}
after identifying ${\cal E}_0 \to \gamma$ for the two-body problem. Somewhat surprisingly, we find that the PM coefficients can be written in a compact form,
\begin{align}
\chi = \sum_{n=1}^\infty {1 \over j^n}\,
{(-2)^n  \over (\gamma^2 - 1)^{n/2}}\,
{\Gamma\left(\frac{1}{2} {-} n\right) \Gamma\left(n {+} \frac{1}{2}\right) \Gamma\left(\frac{n+1}{2}\right) \, 
\over \sqrt{\pi} \Gamma(n+1)}\,
_2\tilde{F}_1\!\left(-n,\tfrac{n+1}{2};1 {-} \tfrac{n}{2}; 1-\gamma^2\right),
\end{align}
with ${}_2\tilde{F}_1$ the regularized hypergeometric function associated to ${}_2{F}_1$.


\newpage
\section{Polynomials in the impulse/angle}\label{sec:app-pols}

\begingroup
\allowdisplaybreaks
\begin{align*}
            h_1 &= 515 \gamma ^6-1017 \gamma ^4+377 \gamma ^2-3\\
            h_2 &= 380 \gamma ^2+169\\
            h_3 &= 1200 \gamma ^2+2095 \gamma +834\\
            h_4 &= 1200 \gamma ^3+2660 \gamma ^2+2929 \gamma +1183\\
            h_5 &= -25 \gamma ^6+30 \gamma ^4+60 \gamma ^3-129 \gamma ^2+76 \gamma -12\\
            h_6 &= 210 \gamma ^6-552 \gamma ^5+339 \gamma ^4-912 \gamma ^3+3148 \gamma ^2-3336 \gamma +1151\\
            h_7 &= -\gamma  \left(2 \gamma ^2-3\right) \left(15 \gamma ^2-15 \gamma +4\right)\\
            h_8 &= 420 \gamma ^9+3456 \gamma ^8-1338 \gamma ^7-15822 \gamma ^6+13176 \gamma ^5+9563 \gamma ^4-16658 \gamma ^3\\
            &\quad+8700 \gamma ^2-496 \gamma -1049\\
            h_9 &= -22680 \gamma ^{21}+11340 \gamma ^{20}+116100 \gamma ^{19}-34080 \gamma ^{18}-216185 \gamma ^{17}+74431 \gamma ^{16}\\
            &\quad+232751 \gamma ^{15}-304761 \gamma ^{14}+333545 \gamma ^{13}-32675 \gamma ^{12}-500785 \gamma ^{11}+535259 \gamma ^{10}\\
            &\quad-181493 \gamma ^9+3259 \gamma ^8+9593 \gamma ^7+9593 \gamma ^6-3457 \gamma ^5-3457 \gamma ^4\\
            &\quad+885 \gamma ^3+885 \gamma ^2-210 \gamma -210\\
            h_{10} &= -280 \gamma ^7+50 \gamma ^6+970 \gamma ^5+27 \gamma ^4-1432 \gamma ^3+444 \gamma ^2+366 \gamma -129\\
            h_{11} &= 2835 \gamma ^{11}-10065 \gamma ^9-700 \gamma ^8+13198 \gamma ^7+1818 \gamma ^6-9826 \gamma ^5+5242 \gamma ^4\\
            &\quad+11391 \gamma ^3+18958 \gamma ^2+10643 \gamma +2074\\
            h_{12} &= \gamma  \left(945 \gamma ^{10}-2955 \gamma ^8+4874 \gamma ^6-5014 \gamma ^4+8077 \gamma ^2+5369\right)\\
            h_{13} &= \gamma  \left(280 \gamma ^7+580 \gamma ^6+90 \gamma ^5-856 \gamma ^4-2211 \gamma ^3+1289 \gamma ^2+2169 \gamma -1965\right)\\
            h_{14} &= \gamma  \left(2 \gamma ^2-3\right) \left(280 \gamma ^7-890 \gamma ^6-610 \gamma ^5+1537 \gamma ^4+380 \gamma ^3-716 \gamma ^2-82 \gamma +85\right)\\
            h_{15} &= 35 \gamma ^4+60 \gamma ^3-150 \gamma ^2+76 \gamma -5\\
            h_{16} &= \gamma  \left(2 \gamma ^2-3\right) \left(35 \gamma ^4-30 \gamma ^2+11\right)\\
            h_{17} &= 315 \gamma ^8-860 \gamma ^6+690 \gamma ^4-960 \gamma ^3+1732 \gamma ^2-1216 \gamma +299\\
            h_{18} &= 315 \gamma ^6-145 \gamma ^4+65 \gamma ^2+21\\
            h_{19} &= 840 \gamma ^9+1932 \gamma ^8+234 \gamma ^7-17562 \gamma ^6+20405 \gamma ^5-2154 \gamma ^4-11744 \gamma ^3\\
            &\quad+12882 \gamma ^2-4983 \gamma +102\\
            h_{20} &= 3600 \gamma ^{16}+4320 \gamma ^{15}-23840 \gamma ^{14}+7824 \gamma ^{13}+14128 \gamma ^{12}+16138 \gamma ^{11}-9872 \gamma ^{10}\\
            &\quad-47540 \gamma ^9+63848 \gamma ^8-37478 \gamma ^7+13349 \gamma ^6-1471 \gamma ^4+207 \gamma ^2-45\\
            h_{21} &= -350 \gamma ^7+1425 \gamma ^5-400 \gamma ^4-1480 \gamma ^3+660 \gamma ^2+285 \gamma -124\\
            h_{22} &= -300 \gamma ^7+210 \gamma ^6+1112 \gamma ^5+2787 \gamma ^4+2044 \gamma ^3+3692 \gamma ^2+6744 \gamma +1759\\
            h_{23} &= \gamma  \left(75 \gamma ^6-140 \gamma ^4-283 \gamma ^2-852\right)\\
            h_{24} &= \gamma  \left(2 \gamma ^2-3\right) \left(210 \gamma ^6-720 \gamma ^5+339 \gamma ^4-576 \gamma ^3+3148 \gamma ^2-3504 \gamma +1151\right)\\
            h_{25} &= \gamma  \left(2 \gamma ^2-3\right) \left(350 \gamma ^7-960 \gamma ^6-705 \gamma ^5+1632 \gamma ^4+432 \gamma ^3-768 \gamma ^2-93 \gamma +96\right)\\
            h_{26} &= \gamma ^2 \left(3-2 \gamma ^2\right)^2 \left(35 \gamma ^4-30 \gamma ^2+11\right)\\
            h_{27} &= 15 \gamma ^3+60 \gamma ^2+19 \gamma +8\\
            h_{28} &= \gamma  \left(70 \gamma ^6-645 \gamma ^4+768 \gamma ^2+63\right)\\
            h_{29} &= -75 \gamma ^6+90 \gamma ^4+333 \gamma ^2+60
\\
            h_{30} &= 25 \gamma ^6-30 \gamma ^4+60 \gamma ^3+129 \gamma ^2+76 \gamma +12\\
            h_{31} &= \left(1-5 \gamma ^2\right)^2\\
            h_{32} &= 80 \gamma ^8-192 \gamma ^6+152 \gamma ^4-44 \gamma ^2+3\\
            h_{33} &= \gamma  \left(2 \gamma ^2-1\right) \left(64 \gamma ^6-216 \gamma ^4+258 \gamma ^2-109\right)\\
            h_{34} &= \left(2 \gamma ^2-1\right)^3 \left(5 \gamma ^2-8\right)\\
            h_{35} &= \gamma  \left(2 \gamma ^2-3\right) \left(2 \gamma ^2-1\right)^3\\
            h_{36} &= 8 \gamma ^6-28 \gamma ^4+6 \gamma ^2+3\\
            h_{37} &= \gamma  \left(384 \gamma ^8-1528 \gamma ^6+384 \gamma ^4+2292 \gamma ^2-1535\right)\\
            h_{38} &= 393897472 \gamma ^{16}-791542442 \gamma ^{14}-3429240286 \gamma ^{12}+3966858415 \gamma ^{10}\\
            &\quad+767410066 \gamma ^8-21241500 \gamma ^6+7188300 \gamma ^4-1837500 \gamma ^2+385875\\
            h_{39} &= 1575 \gamma ^7-2700 \gamma ^6-3195 \gamma ^5+3780 \gamma ^4+4993 \gamma ^3-1188 \gamma ^2-1485 \gamma +108\\
            h_{40} &= -3592192 \gamma ^{18}+2662204 \gamma ^{16}+46406238 \gamma ^{14}-37185456 \gamma ^{12}-25426269 \gamma ^{10}\\
            &\quad+222810 \gamma ^8-246540 \gamma ^6+79800 \gamma ^4-19950 \gamma ^2+3675\\
            h_{41} &= 44 \gamma ^6-32 \gamma ^4-425 \gamma ^2-82\\
            h_{42} &= \gamma  \left(16 \gamma ^6+24 \gamma ^4-226 \gamma ^2-151\right)\\
            h_{43} &= \gamma ^2 \left(4 \gamma ^8-59 \gamma ^4+35 \gamma ^2+60\right)\\
            h_{44} &= -525 \gamma ^7+1065 \gamma ^5-3883 \gamma ^3+1263 \gamma\\
            h_{45} &= 175 \gamma ^7-150 \gamma ^6-355 \gamma ^5+210 \gamma ^4+185 \gamma ^3-66 \gamma ^2-37 \gamma +6\\
            h_{46} &= -175 \gamma ^7+355 \gamma ^5-185 \gamma ^3+37 \gamma\\
            h_{47} &= \gamma  \left(525 \gamma ^6-1065 \gamma ^4-2773 \gamma ^2+1041\right)\\
            h_{48} &= 96 \gamma ^{10}-8464 \gamma ^8+54616 \gamma ^6-70104 \gamma ^4+9916 \gamma ^2+13895\\
            h_{49} &= 6144 \gamma ^{16}-587336 \gamma ^{14}+4034092 \gamma ^{12}-417302 \gamma ^{10}-5560073 \gamma ^8-142640 \gamma ^6\\
            &\quad+35710 \gamma ^4-8250 \gamma ^2+1575\\
            h_{50} &= -3747 \gamma ^6+3249 \gamma ^4+8535 \gamma ^2+1051\\
            h_{51} &= 24576 \gamma ^{18}+213480 \gamma ^{16}-1029342 \gamma ^{14}-1978290 \gamma ^{12}+3752006 \gamma ^{10}\\
            &\quad+816595 \gamma ^8-55260 \gamma ^6+13690 \gamma ^4-3100 \gamma ^2+525\\
            h_{52} &= \gamma  \left(16 \gamma ^6+204 \gamma ^4-496 \gamma ^2-869\right)\\
            h_{53} &= \gamma ^2 \left(8 \gamma ^4-6 \gamma ^2-9\right)\\
            h_{54} &= \gamma  \left(2 \gamma ^2-3\right) \left(8 \gamma ^6-6 \gamma ^4-51 \gamma ^2-8\right)\\
            h_{55} &= -4321 \gamma ^6+3387 \gamma ^4+15261 \gamma ^2+2057\\
            h_{56} &= 2100 \gamma ^7-4996 \gamma ^6+1755 \gamma ^5+4332 \gamma ^4-6422 \gamma ^3+4212 \gamma ^2-1209 \gamma +36\\
            h_{57} &= -1249 \gamma ^6+1083 \gamma ^4+1053 \gamma ^2+9\\
            h_{58} &= -1823 \gamma ^6+1221 \gamma ^4+13155 \gamma ^2+2039\\
            h_{59} &= -24 \gamma ^6+18 \gamma ^4+111 \gamma ^2+16\\
            h_{60} &= \gamma  \left(26 \gamma ^2-9\right)
\\
    h_{61} &= 35 (\gamma -1) (\gamma +1) \left(33 \gamma ^4-18 \gamma ^2+1\right)\\
    h_{62} &= 3600 \gamma ^{16}+4320 \gamma ^{15}-35360 \gamma ^{14}+33249 \gamma ^{13}+27952 \gamma ^{12}-25145 \gamma ^{11}-15056 \gamma ^{10}\\
    &\quad-32177 \gamma ^9+64424 \gamma ^8-38135 \gamma ^7+13349 \gamma ^6-1471 \gamma ^4  +207 \gamma ^2-45\\
    h_{63} &= \gamma ^2 \left(2 \gamma ^2-3\right) \left(2 \gamma ^2-1\right) \left(35 \gamma ^4-30 \gamma ^2+11\right)\\
    h_{64} &= -4140 \gamma ^8+702 \gamma ^7+15018 \gamma ^6-8491 \gamma ^5-9366 \gamma ^4+10052 \gamma ^3-6210 \gamma ^2+2681 \gamma -102\\
    h_{65} &= 210 \gamma ^7-240 \gamma ^6-755 \gamma ^5+216 \gamma ^4+1200 \gamma ^3-508 \gamma ^2-295 \gamma +124\\
    h_{66} &= \gamma  \left(2 \gamma ^2-3\right) \left(2 \gamma ^2-1\right) \left(35 \gamma ^4-30 \gamma ^2+11\right)\\
    h_{67} &= -(\gamma -1) \big(420 \gamma ^9+7596 \gamma ^8-2040 \gamma ^7-30840 \gamma ^6+21667 \gamma ^5+18929 \gamma ^4-26710 \gamma ^3\\
    &\quad+14910 \gamma ^2-3177 \gamma -947\big)\\
    h_{68} &= (\gamma -1) \left(490 \gamma ^7-290 \gamma ^6-1725 \gamma ^5+189 \gamma ^4+2632 \gamma ^3-952 \gamma ^2-661 \gamma +253\right)\,.
\end{align*}
\endgroup

\newpage
\bibliographystyle{JHEP}
\bibliography{refs}

\providecommand{\href}[2]{#2}\begingroup\raggedright\begin{thebibliography}{100}

\bibitem{LIGOScientific:2021djp}
{\scshape LIGO Scientific, VIRGO, KAGRA} collaboration, \emph{{GWTC-3: Compact
  Binary Coalescences Observed by LIGO and Virgo During the Second Part of the
  Third Observing Run}},  \href{https://arxiv.org/abs/2111.03606}{{\ttfamily
  2111.03606}}.

\bibitem{buosathya}
A.~Buonanno and B.~Sathyaprakash, \emph{{Sources of Gravitational Waves: Theory
  and Observations}},  \href{https://arxiv.org/abs/1410.7832}{{\ttfamily
  1410.7832}}.

\bibitem{tune}
R.~A. Porto, \emph{{The Tune of Love and the Nature(ness) of Spacetime}},
  \href{https://doi.org/10.1002/prop.201600064}{\emph{Fortsch. Phys.}
  {\bfseries 64} (2016) 723}
  [\href{https://arxiv.org/abs/1606.08895}{{\ttfamily 1606.08895}}].

\bibitem{music}
R.~A. Porto, \emph{{The Music of the Spheres: The Dawn of Gravitational Wave
  Science}},  \href{https://arxiv.org/abs/1703.06440}{{\ttfamily 1703.06440}}.

\bibitem{Maggiore:2019uih}
M.~Maggiore et~al., \emph{{Science Case for the Einstein Telescope}},
  \href{https://doi.org/10.1088/1475-7516/2020/03/050}{\emph{JCAP} {\bfseries
  03} (2020) 050} [\href{https://arxiv.org/abs/1912.02622}{{\ttfamily
  1912.02622}}].

\bibitem{Barausse:2020rsu}
E.~Barausse et~al., \emph{{Prospects for Fundamental Physics with LISA}},
  \href{https://doi.org/10.1007/s10714-020-02691-1}{\emph{Gen. Rel. Grav.}
  {\bfseries 52} (2020) 81} [\href{https://arxiv.org/abs/2001.09793}{{\ttfamily
  2001.09793}}].

\bibitem{Bernitt:2022aoa}
S.~Bernitt et~al., \emph{{Fundamental Physics in the Gravitational-Wave Era}},
  \href{https://doi.org/10.1080/10619127.2021.1988473}{\emph{Nucl. Phys. News}
  {\bfseries 32} (2022) 16}.

\bibitem{Ajith:2012az}
P.~Ajith et~al., \emph{{The NINJA-2 catalog of hybrid
  post-Newtonian/numerical-relativity waveforms for non-precessing black-hole
  binaries}},
  \href{https://doi.org/10.1088/0264-9381/29/12/124001}{\emph{Class. Quant.
  Grav.} {\bfseries 29} (2012) 124001}
  [\href{https://arxiv.org/abs/1201.5319}{{\ttfamily 1201.5319}}].

\bibitem{Szilagyi:2015rwa}
B.~Szil\'agyi et~al., \emph{{Approaching the Post-Newtonian Regime with
  Numerical Relativity: A Compact-Object Binary Simulation Spanning 350
  Gravitational-Wave Cycles}},
  \href{https://doi.org/10.1103/PhysRevLett.115.031102}{\emph{Phys. Rev. Lett.}
  {\bfseries 115} (2015) 031102}
  [\href{https://arxiv.org/abs/1502.04953}{{\ttfamily 1502.04953}}].

\bibitem{Dietrich:2018phi}
T.~Dietrich et~al., \emph{{CoRe database of binary neutron star merger
  waveforms}}, \href{https://doi.org/10.1088/1361-6382/aaebc0}{\emph{Class.
  Quant. Grav.} {\bfseries 35} (2018) 24LT01}
  [\href{https://arxiv.org/abs/1806.01625}{{\ttfamily 1806.01625}}].

\bibitem{Damour:2008yg}
T.~Damour, \emph{{Introductory lectures on the Effective One Body formalism}},
  \href{https://doi.org/10.1142/S0217751X08039992}{\emph{Int. J. Mod. Phys. A}
  {\bfseries 23} (2008) 1130}
  [\href{https://arxiv.org/abs/0802.4047}{{\ttfamily 0802.4047}}].

\bibitem{blanchet}
L.~Blanchet, \emph{{Gravitational Radiation from Post-Newtonian Sources and
  Inspiralling Compact Binaries}},
  \href{https://doi.org/10.12942/lrr-2014-2}{\emph{Living Rev.Rel.} {\bfseries
  17} (2014) 2} [\href{https://arxiv.org/abs/1310.1528}{{\ttfamily
  1310.1528}}].

\bibitem{Schafer:2018kuf}
G.~Sch{\"a}fer and P.~Jaranowski, \emph{{Hamiltonian formulation of general
  relativity and post-Newtonian dynamics of compact binaries}},
  \href{https://doi.org/10.1007/s41114-018-0016-5}{\emph{Living Rev. Rel.}
  {\bfseries 21} (2018) 7} [\href{https://arxiv.org/abs/1805.07240}{{\ttfamily
  1805.07240}}].

\bibitem{Barack:2018yvs}
L.~Barack and A.~Pound, \emph{{Self-force and radiation reaction in general
  relativity}}, \href{https://doi.org/10.1088/1361-6633/aae552}{\emph{Rept.
  Prog. Phys.} {\bfseries 82} (2019) 016904}
  [\href{https://arxiv.org/abs/1805.10385}{{\ttfamily 1805.10385}}].

\bibitem{walterLH}
W.~D. Goldberger, \emph{{Les Houches lectures on effective field theories and
  gravitational radiation}},  in \emph{Les Houches Summer School - Session 86},
  1, 2007, \href{https://arxiv.org/abs/hep-ph/0701129}{{\ttfamily
  hep-ph/0701129}}.

\bibitem{iragrg}
I.~Rothstein, \emph{{Progress in Effective Field Theory Approach to the Binary
  Inspiral Problem}},
  \href{https://doi.org/10.1007/s10714-014-1726-y}{\emph{Gen. Rel. Grav.}
  {\bfseries 46} (2014) 1726}.

\bibitem{foffa}
S.~Foffa and R.~Sturani, \emph{{Effective field theory methods to model compact
  binaries}}, \href{https://doi.org/10.1088/0264-9381/31/4/043001}{\emph{Class.
  Quant. Grav.} {\bfseries 31} (2014) 043001}
  [\href{https://arxiv.org/abs/1309.3474}{{\ttfamily 1309.3474}}].

\bibitem{review}
R.~A. Porto, \emph{{The effective field theorist's approach to gravitational
  dynamics}}, \href{https://doi.org/10.1016/j.physrep.2016.04.003}{\emph{Phys.
  Rept.} {\bfseries 633} (2016) 1}
  [\href{https://arxiv.org/abs/1601.04914}{{\ttfamily 1601.04914}}].

\bibitem{Goldberger:2022ebt}
W.~D. Goldberger, \emph{{Effective field theories of gravity and compact binary
  dynamics: A Snowmass 2021 whitepaper}},
  \href{https://arxiv.org/abs/2206.14249}{{\ttfamily 2206.14249}}.

\bibitem{Buonanno:2022pgc}
A.~Buonanno, M.~Khalil, D.~O'Connell, R.~Roiban, M.~P. Solon and M.~Zeng,
  \emph{{Snowmass White Paper: Gravitational Waves and Scattering Amplitudes}},
   in \emph{{2022 Snowmass Summer Study}}, 4, 2022,
  \href{https://arxiv.org/abs/2204.05194}{{\ttfamily 2204.05194}}.

\bibitem{LISA}
{\scshape LISA} collaboration, \emph{{Laser Interferometer Space Antenna}},
  \href{https://arxiv.org/abs/1702.00786}{{\ttfamily 1702.00786}}.

\bibitem{CE}
D.~Reitze et~al., \emph{{Cosmic Explorer: The U.S. Contribution to
  Gravitational-Wave Astronomy beyond LIGO}}, {\emph{Bull. Am. Astron. Soc.}
  {\bfseries 51} (2019) 035}
  [\href{https://arxiv.org/abs/1907.04833}{{\ttfamily 1907.04833}}].

\bibitem{ET}
M.~Punturo et~al., \emph{The einstein telescope: A third-generation
  gravitational wave observatory},
  \href{https://doi.org/10.1088/0264-9381/27/19/194002}{\emph{Class. Quant.
  Grav.} {\bfseries 27} (2010) 194002}.

\bibitem{4pndjs}
T.~Damour, P.~Jaranowski and G.~Schafer, \emph{{Nonlocal-in-time action for the
  fourth post-Newtonian conservative dynamics of two-body systems}},
  \href{https://doi.org/10.1103/PhysRevD.89.064058}{\emph{Phys. Rev.}
  {\bfseries D89} (2014) 064058}
  [\href{https://arxiv.org/abs/1401.4548}{{\ttfamily 1401.4548}}].

\bibitem{4pnbla}
L.~Bernard, L.~Blanchet, A.~Boh\'e, G.~Faye and S.~Marsat, \emph{{Fokker action
  of nonspinning compact binaries at the fourth post-Newtonian approximation}},
  \href{https://doi.org/10.1103/PhysRevD.93.084037}{\emph{Phys. Rev.}
  {\bfseries D93} (2016) 084037}
  [\href{https://arxiv.org/abs/1512.02876}{{\ttfamily 1512.02876}}].

\bibitem{4pnbla2}
L.~Bernard, L.~Blanchet, A.~Boh\'e, G.~Faye and S.~Marsat, \emph{{Energy and
  periastron advance of compact binaries on circular orbits at the fourth
  post-Newtonian order}}, {\emph{Phys. Rev.} {\bfseries D95} (2017) 044026}
  [\href{https://arxiv.org/abs/1610.07934}{{\ttfamily 1610.07934}}].

\bibitem{Marchand:2017pir}
T.~Marchand, L.~Bernard, L.~Blanchet and G.~Faye, \emph{{Ambiguity-Free
  Completion of the Equations of Motion of Compact Binary Systems at the Fourth
  Post-Newtonian Order}},
  \href{https://doi.org/10.1103/PhysRevD.97.044023}{\emph{Phys. Rev. D}
  {\bfseries 97} (2018) 044023}
  [\href{https://arxiv.org/abs/1707.09289}{{\ttfamily 1707.09289}}].

\bibitem{damour3n}
D.~Bini, T.~Damour and A.~Geralico, \emph{{Novel approach to binary dynamics:
  application to the fifth post-Newtonian level}},
  \href{https://doi.org/10.1103/PhysRevLett.123.231104}{\emph{Phys.\,Rev.\,Lett.}
  {\bfseries 123} (2019) 231104}
  [\href{https://arxiv.org/abs/1909.02375}{{\ttfamily 1909.02375}}].

\bibitem{binidam1}
D.~Bini, T.~Damour and A.~Geralico, \emph{{Sixth post-Newtonian local-in-time
  dynamics of binary systems}},
  \href{https://doi.org/10.1103/PhysRevD.102.024061}{\emph{Phys. Rev. D}
  {\bfseries 102} (2020) 024061}
  [\href{https://arxiv.org/abs/2004.05407}{{\ttfamily 2004.05407}}].

\bibitem{binidam2}
D.~Bini, T.~Damour and A.~Geralico, \emph{{Sixth post-Newtonian
  nonlocal-in-time dynamics of binary systems}},
  \href{https://doi.org/10.1103/PhysRevD.102.084047}{\emph{Phys. Rev. D}
  {\bfseries 102} (2020) 084047}
  [\href{https://arxiv.org/abs/2007.11239}{{\ttfamily 2007.11239}}].

\bibitem{binit}
D.~Bini, T.~Damour and A.~Geralico, \emph{{Scattering of tidally interacting
  bodies in post-Minkowskian gravity}},
  \href{https://doi.org/10.1103/PhysRevD.101.044039}{\emph{Phys. Rev. D}
  {\bfseries 101} (2020) 044039}
  [\href{https://arxiv.org/abs/2001.00352}{{\ttfamily 2001.00352}}].

\bibitem{Bini:2021gat}
D.~Bini, T.~Damour and A.~Geralico, \emph{{Radiative contributions to
  gravitational scattering}},
  \href{https://doi.org/10.1103/PhysRevD.104.084031}{\emph{Phys. Rev. D}
  {\bfseries 104} (2021) 084031}
  [\href{https://arxiv.org/abs/2107.08896}{{\ttfamily 2107.08896}}].

\bibitem{Bini:2021qvf}
D.~Bini and A.~Geralico, \emph{{Higher-order tail contributions to the energy
  and angular momentum fluxes in a two-body scattering process}},
  \href{https://doi.org/10.1103/PhysRevD.104.104020}{\emph{Phys. Rev. D}
  {\bfseries 104} (2021) 104020}
  [\href{https://arxiv.org/abs/2108.05445}{{\ttfamily 2108.05445}}].

\bibitem{Bini:2022yrk}
D.~Bini and A.~Geralico, \emph{{Momentum recoil in the relativistic two-body
  problem: Higher-order tails}},
  \href{https://doi.org/10.1103/PhysRevD.105.084028}{\emph{Phys. Rev. D}
  {\bfseries 105} (2022) 084028}
  [\href{https://arxiv.org/abs/2202.03037}{{\ttfamily 2202.03037}}].

\bibitem{Bini:2022xpp}
D.~Bini and A.~Geralico, \emph{{Multipolar invariants and the eccentricity
  enhancement function parametrization of gravitational radiation}},
  \href{https://doi.org/10.1103/PhysRevD.105.124001}{\emph{Phys. Rev. D}
  {\bfseries 105} (2022) 124001}
  [\href{https://arxiv.org/abs/2204.08077}{{\ttfamily 2204.08077}}].

\bibitem{Bini:2022enm}
D.~Bini, T.~Damour and A.~Geralico, \emph{{Radiated momentum in gravitational
  two-body scattering including time-asymmetric effects}},
  \href{https://arxiv.org/abs/2210.07165}{{\ttfamily 2210.07165}}.

\bibitem{Damour:2020tta}
T.~Damour, \emph{{Radiative contribution to classical gravitational scattering
  at the third order in $G$}},
  \href{https://doi.org/10.1103/PhysRevD.102.124008}{\emph{Phys. Rev. D}
  {\bfseries 102} (2020) 124008}
  [\href{https://arxiv.org/abs/2010.01641}{{\ttfamily 2010.01641}}].

\bibitem{Marchand:2020fpt}
T.~Marchand, Q.~Henry, F.~Larrouturou, S.~Marsat, G.~Faye and L.~Blanchet,
  \emph{{The mass quadrupole moment of compact binary systems at the fourth
  post-Newtonian order}},
  \href{https://doi.org/10.1088/1361-6382/ab9ce1}{\emph{Class. Quant. Grav.}
  {\bfseries 37} (2020) 215006}
  [\href{https://arxiv.org/abs/2003.13672}{{\ttfamily 2003.13672}}].

\bibitem{Larrouturou:2021dma}
F.~Larrouturou, Q.~Henry, L.~Blanchet and G.~Faye, \emph{{The Quadrupole Moment
  of Compact Binaries to the Fourth post-Newtonian Order: I. Non-Locality in
  Time and Infra-Red Divergencies}},
  \href{https://arxiv.org/abs/2110.02240}{{\ttfamily 2110.02240}}.

\bibitem{Larrouturou:2021gqo}
F.~Larrouturou, L.~Blanchet, Q.~Henry and G.~Faye, \emph{{The Quadrupole Moment
  of Compact Binaries to the Fourth post-Newtonian Order: II. Dimensional
  Regularization and Renormalization}},
  \href{https://arxiv.org/abs/2110.02243}{{\ttfamily 2110.02243}}.

\bibitem{nrgr}
W.~D. Goldberger and I.~Z. Rothstein, \emph{{An Effective field theory of
  gravity for extended objects}},
  \href{https://doi.org/10.1103/PhysRevD.73.104029}{\emph{Phys. Rev.}
  {\bfseries D73} (2006) 104029}
  [\href{https://arxiv.org/abs/hep-th/0409156}{{\ttfamily hep-th/0409156}}].

\bibitem{nrgrs}
R.~A. Porto, \emph{{Post-Newtonian Corrections to the Motion of Spinning Bodies
  in NRGR}}, \href{https://doi.org/10.1103/PhysRevD.73.104031}{\emph{Phys. Rev.
  D} {\bfseries 73} (2006) 104031}
  [\href{https://arxiv.org/abs/gr-qc/0511061}{{\ttfamily gr-qc/0511061}}].

\bibitem{dis1}
W.~Goldberger and I.~Rothstein, \emph{{Dissipative Effects in the Worldline
  Approach to Black Hole Dynamics}},
  \href{https://doi.org/10.1103/PhysRevD.73.104030}{\emph{Phys. Rev. D}
  {\bfseries 73} (2006) 104030}
  [\href{https://arxiv.org/abs/hep-th/0511133}{{\ttfamily hep-th/0511133}}].

\bibitem{dis2}
R.~A. Porto, \emph{{Absorption Effects due to Spin in the Worldline Approach to
  Black Hole Dynamics}},
  \href{https://doi.org/10.1103/PhysRevD.77.064026}{\emph{Phys. Rev. D}
  {\bfseries 77} (2008) 064026}
  [\href{https://arxiv.org/abs/0710.5150}{{\ttfamily 0710.5150}}].

\bibitem{prl}
R.~A. Porto and I.~Z. Rothstein, \emph{The hyperfine
  {E}instein-{I}nfeld-{H}offmann potential},
  \href{https://doi.org/10.1103/PhysRevLett.97.021101}{\emph{Phys. Rev. Lett.}
  {\bfseries 97} (2006) 021101}
  [\href{https://arxiv.org/abs/gr-qc/0604099}{{\ttfamily gr-qc/0604099}}].

\bibitem{nrgrss}
R.~A. Porto and I.~Z. Rothstein, \emph{{Spin(1)Spin(2) Effects in the Motion of
  Inspiralling Compact Binaries at Third Order in the Post-Newtonian
  Expansion}},
  \href{https://doi.org/10.1103/PhysRevD.78.044012}{\emph{Phys.Rev.} {\bfseries
  D78} (2008) 044012} [\href{https://arxiv.org/abs/0802.0720}{{\ttfamily
  0802.0720}}].

\bibitem{nrgrs2}
R.~A. Porto and I.~Z. Rothstein, \emph{{Next to Leading Order Spin(1)Spin(1)
  Effects in the Motion of Inspiralling Compact Binaries}},
  \href{https://doi.org/10.1103/PhysRevD.78.044013}{\emph{Phys.Rev.} {\bfseries
  D78} (2008) 044013} [\href{https://arxiv.org/abs/0804.0260}{{\ttfamily
  0804.0260}}].

\bibitem{nrgrso}
R.~A. Porto, \emph{{Next-to-Leading Order Spin-Orbit Effects in the Motion of
  Inspiralling Compact Binaries}},
  \href{https://doi.org/10.1088/0264-9381/27/20/205001}{\emph{Class. Quant.
  Grav.} {\bfseries 27} (2010) 205001}
  [\href{https://arxiv.org/abs/1005.5730}{{\ttfamily 1005.5730}}].

\bibitem{andirad}
W.~D. Goldberger and A.~Ross, \emph{{Gravitational radiative corrections from
  effective field theory}},
  \href{https://doi.org/10.1103/PhysRevD.81.124015}{\emph{Phys. Rev.}
  {\bfseries D81} (2010) 124015}
  [\href{https://arxiv.org/abs/0912.4254}{{\ttfamily 0912.4254}}].

\bibitem{andirad2}
A.~Ross, \emph{{Multipole expansion at the level of the action}},
  \href{https://doi.org/10.1103/PhysRevD.85.125033}{\emph{Phys. Rev.}
  {\bfseries D85} (2012) 125033}
  [\href{https://arxiv.org/abs/1202.4750}{{\ttfamily 1202.4750}}].

\bibitem{amps}
R.~A. Porto, A.~Ross and I.~Z. Rothstein, \emph{{Spin induced multipole moments
  for the gravitational wave amplitude from binary inspirals to 2.5
  Post-Newtonian order}},
  \href{https://doi.org/10.1088/1475-7516/2012/09/028}{\emph{JCAP} {\bfseries
  1209} (2012) 028} [\href{https://arxiv.org/abs/1203.2962}{{\ttfamily
  1203.2962}}].

\bibitem{srad}
R.~A. Porto, A.~Ross and I.~Z. Rothstein, \emph{{Spin induced multipole moments
  for the gravitational wave flux from binary inspirals to third Post-Newtonian
  order}}, \href{https://doi.org/10.1088/1475-7516/2011/03/009}{\emph{JCAP}
  {\bfseries 1103} (2011) 009}
  [\href{https://arxiv.org/abs/1007.1312}{{\ttfamily 1007.1312}}].

\bibitem{chadRR}
C.~R. Galley and M.~Tiglio, \emph{Radiation reaction and gravitational waves in
  the effective field theory approach},
  \href{https://doi.org/10.1103/PhysRevD.79.124027}{\emph{Phys. Rev. D}
  {\bfseries 79} (2009) 124027}.

\bibitem{chadbr2}
C.~R. Galley and A.~K. Leibovich, \emph{Radiation reaction at 3.5
  post-{N}ewtonian order in effective field theory},
  \href{https://doi.org/10.1103/PhysRevD.86.044029}{\emph{Phys. Rev. D}
  {\bfseries 86} (2012) 044029}
  [\href{https://arxiv.org/abs/1205.3842}{{\ttfamily 1205.3842}}].

\bibitem{tail}
C.~Galley, A.~Leibovich, R.~A. Porto and A.~Ross, \emph{{Tail Effect in
  Gravitational Radiation Reaction: Time Nonlocality and Renormalization Group
  Evolution}}, \href{https://doi.org/10.1103/PhysRevD.93.124010}{\emph{Phys.
  Rev. D} {\bfseries 93} (2016) 124010}
  [\href{https://arxiv.org/abs/1511.07379}{{\ttfamily 1511.07379}}].

\bibitem{natalia1}
N.~T. Maia, C.~R. Galley, A.~K. Leibovich and R.~A. Porto, \emph{{Radiation
  reaction for spinning bodies in effective field theory I: Spin-orbit
  effects}}, \href{https://doi.org/10.1103/PhysRevD.96.084064}{\emph{Phys.
  Rev.} {\bfseries D96} (2017) 084064}
  [\href{https://arxiv.org/abs/1705.07934}{{\ttfamily 1705.07934}}].

\bibitem{natalia2}
N.~T. Maia, C.~R. Galley, A.~K. Leibovich and R.~A. Porto, \emph{{Radiation
  reaction for spinning bodies in effective field theory II: Spin-spin
  effects}}, \href{https://doi.org/10.1103/PhysRevD.96.084065}{\emph{Phys.
  Rev.} {\bfseries D96} (2017) 084065}
  [\href{https://arxiv.org/abs/1705.07938}{{\ttfamily 1705.07938}}].

\bibitem{dis3}
W.~D. Goldberger, J.~Li and I.~Z. Rothstein, \emph{{Non-conservative effects on
  spinning black holes from world-line effective field theory}},
  \href{https://doi.org/10.1007/JHEP06(2021)053}{\emph{JHEP} {\bfseries 06}
  (2021) 053} [\href{https://arxiv.org/abs/2012.14869}{{\ttfamily
  2012.14869}}].

\bibitem{withchad}
C.~Galley and R.~A. Porto, \emph{{Gravitational Self-Force in the
  Ultra-Relativistic Limit: the ``Large-$N$" Expansion}},
  \href{https://doi.org/10.1007/JHEP11(2013)096}{\emph{JHEP} {\bfseries 11}
  (2013) 096} [\href{https://arxiv.org/abs/1302.4486}{{\ttfamily 1302.4486}}].

\bibitem{apparent}
R.~A. Porto and I.~Rothstein, \emph{{Apparent Ambiguities in the Post-Newtonian
  Expansion for Binary Systems}},
  \href{https://doi.org/10.1103/PhysRevD.96.024062}{\emph{Phys. Rev. D}
  {\bfseries 96} (2017) 024062}
  [\href{https://arxiv.org/abs/1703.06433}{{\ttfamily 1703.06433}}].

\bibitem{nrgr4pn1}
S.~Foffa and R.~Sturani, \emph{{Dynamics of the gravitational two-body problem
  at fourth post-Newtonian order and at quadratic order in the Newton
  constant}}, \href{https://doi.org/10.1103/PhysRevD.87.064011}{\emph{Phys.
  Rev. D} {\bfseries 87} (2013) 064011}
  [\href{https://arxiv.org/abs/1206.7087}{{\ttfamily 1206.7087}}].

\bibitem{nrgr4pn2}
S.~Foffa, R.~A. Porto, I.~Rothstein and R.~Sturani, \emph{{Conservative
  dynamics of binary systems to fourth Post-Newtonian order in the EFT approach
  II: Renormalized Lagrangian}},
  \href{https://doi.org/10.1103/PhysRevD.100.024048}{\emph{Phys. Rev.}
  {\bfseries D100} (2019) 024048}
  [\href{https://arxiv.org/abs/1903.05118}{{\ttfamily 1903.05118}}].

\bibitem{5pn1}
S.~Foffa, P.~Mastrolia, R.~Sturani, C.~Sturm and W.~J. Torres~Bobadilla,
  \emph{{Static two-body potential at fifth post-Newtonian order}},
  \href{https://doi.org/10.1103/PhysRevLett.122.241605}{\emph{Phys. Rev. Lett.}
  {\bfseries 122} (2019) 241605}
  [\href{https://arxiv.org/abs/1902.10571}{{\ttfamily 1902.10571}}].

\bibitem{5pn2}
J.~Bl{\"u}mlein, A.~Maier and P.~Marquard, \emph{{Five-Loop Static Contribution
  to the Gravitational Interaction Potential of Two Point Masses}},
  {\emph{Phys. Lett. B} {\bfseries 800} (2020) 135100}
  [\href{https://arxiv.org/abs/1902.11180}{{\ttfamily 1902.11180}}].

\bibitem{hered1}
S.~Foffa and R.~Sturani, \emph{{Hereditary terms at next-to-leading order in
  two-body gravitational dynamics}},
  \href{https://doi.org/10.1103/PhysRevD.101.064033}{\emph{Phys. Rev. D}
  {\bfseries 101} (2020) 064033}
  [\href{https://arxiv.org/abs/1907.02869}{{\ttfamily 1907.02869}}].

\bibitem{hered2}
G.~L. Almeida, S.~Foffa and R.~Sturani, \emph{{Tail contributions to
  gravitational conservative dynamics}},
  \href{https://doi.org/10.1103/PhysRevD.104.124075}{\emph{Phys. Rev. D}
  {\bfseries 104} (2021) 124075}
  [\href{https://arxiv.org/abs/2110.14146}{{\ttfamily 2110.14146}}].

\bibitem{tail3}
L.~Blanchet, S.~Foffa, F.~Larrouturou and R.~Sturani, \emph{{Logarithmic tail
  contributions to the energy function of circular compact binaries}},
  \href{https://doi.org/10.1103/PhysRevD.101.084045}{\emph{Phys. Rev. D}
  {\bfseries 101} (2020) 084045}
  [\href{https://arxiv.org/abs/1912.12359}{{\ttfamily 1912.12359}}].

\bibitem{blum}
J.~Bl{\"u}mlein, A.~Maier, P.~Marquard and G.~Sch{\"a}fer, \emph{{Testing
  binary dynamics in gravity at the sixth post-Newtonian level}},
  \href{https://doi.org/10.1016/j.physletb.2020.135496}{\emph{Phys. Lett. B}
  {\bfseries 807} (2020) 135496}
  [\href{https://arxiv.org/abs/2003.07145}{{\ttfamily 2003.07145}}].

\bibitem{blum2}
J.~Bl\"umlein, A.~Maier, P.~Marquard and G.~Sch\"afer, \emph{{The 6th
  post-Newtonian potential terms at $O(G_N^4)$}},
  \href{https://doi.org/10.1016/j.physletb.2021.136260}{\emph{Phys. Lett. B}
  {\bfseries 816} (2021) 136260}
  [\href{https://arxiv.org/abs/2101.08630}{{\ttfamily 2101.08630}}].

\bibitem{Blumlein:2020pyo}
J.~Bl\"umlein, A.~Maier, P.~Marquard and G.~Sch\"afer, \emph{{The fifth-order
  post-Newtonian Hamiltonian dynamics of two-body systems from an effective
  field theory approach: potential contributions}},
  \href{https://doi.org/10.1016/j.nuclphysb.2021.115352}{\emph{Nucl. Phys. B}
  {\bfseries 965} (2021) 115352}
  [\href{https://arxiv.org/abs/2010.13672}{{\ttfamily 2010.13672}}].

\bibitem{Blumlein:2021txe}
J.~Bl\"umlein, A.~Maier, P.~Marquard and G.~Sch\"afer, \emph{{The fifth-order
  post-Newtonian Hamiltonian dynamics of two-body systems from an effective
  field theory approach}},
  \href{https://doi.org/10.1016/j.nuclphysb.2022.115900}{\emph{Nucl. Phys. B}
  {\bfseries 983} (2022) 115900}
  [\href{https://arxiv.org/abs/2110.13822}{{\ttfamily 2110.13822}}].

\bibitem{radnrgr}
A.~K. Leibovich, N.~T. Maia, I.~Z. Rothstein and Z.~Yang, \emph{{Second
  post-Newtonian order radiative dynamics of inspiralling compact binaries in
  the Effective Field Theory approach}},
  \href{https://doi.org/10.1103/PhysRevD.101.084058}{\emph{Phys. Rev. D}
  {\bfseries 101} (2020) 084058}
  [\href{https://arxiv.org/abs/1912.12546}{{\ttfamily 1912.12546}}].

\bibitem{pardo}
B.~A. Pardo and N.~T. Maia, \emph{{Next-to-leading order spin-orbit effects in
  the equations of motion, energy loss and phase evolution of binaries of
  compact bodies in the effective field theory approach}},
  \href{https://doi.org/10.1103/PhysRevD.102.124020}{\emph{Phys. Rev. D}
  {\bfseries 102} (2020) 124020}
  [\href{https://arxiv.org/abs/2009.05628}{{\ttfamily 2009.05628}}].

\bibitem{Almeida:2022jrv}
G.~L. Almeida, S.~Foffa and R.~Sturani, \emph{{Gravitational radiation
  contributions to the two-body scattering angle}},
  \href{https://arxiv.org/abs/2209.11594}{{\ttfamily 2209.11594}}.

\bibitem{Cho:2021mqw}
G.~Cho, B.~Pardo and R.~A. Porto, \emph{{Gravitational radiation from
  inspiralling compact objects: Spin-spin effects completed at the
  next-to-leading post-Newtonian order}},
  \href{https://doi.org/10.1103/PhysRevD.104.024037}{\emph{Phys. Rev. D}
  {\bfseries 104} (2021) 024037}
  [\href{https://arxiv.org/abs/2103.14612}{{\ttfamily 2103.14612}}].

\bibitem{Cho2022}
G.~Cho, R.~A. Porto and Z.~Yang, \emph{{Gravitational radiation from
  inspiralling compact objects: Spin effects to fourth Post-Newtonian order}},
  \href{https://arxiv.org/abs/2201.05138}{{\ttfamily 2201.05138}}.

\bibitem{Cho:1}
G.~Cho, S.~Dandapat and A.~Gopakumar, \emph{{Third order post-Newtonian
  gravitational radiation from two-body scattering: Instantaneous energy and
  angular momentum radiation}},
  \href{https://doi.org/10.1103/PhysRevD.105.084018}{\emph{Phys. Rev. D}
  {\bfseries 105} (2022) 084018}
  [\href{https://arxiv.org/abs/2111.00818}{{\ttfamily 2111.00818}}].

\bibitem{Cho:2}
G.~Cho, \emph{{Third post-Newtonian gravitational radiation from two-body
  scattering. II. Hereditary energy radiation}},
  \href{https://doi.org/10.1103/PhysRevD.105.104035}{\emph{Phys. Rev. D}
  {\bfseries 105} (2022) 104035}
  [\href{https://arxiv.org/abs/2203.10872}{{\ttfamily 2203.10872}}].

\bibitem{Kim:2022bwv}
J.-W. Kim, M.~Levi and Z.~Yin, \emph{{N$^3$LO Quadratic-in-Spin Interactions
  for Generic Compact Binaries}},
  \href{https://arxiv.org/abs/2209.09235}{{\ttfamily 2209.09235}}.

\bibitem{Kim:2022pou}
J.-W. Kim, M.~Levi and Z.~Yin, \emph{{N$^3$LO Spin-Orbit Interaction via the
  EFT of Spinning Gravitating Objects}},
  \href{https://arxiv.org/abs/2208.14949}{{\ttfamily 2208.14949}}.

\bibitem{Mandal:2022nty}
M.~K. Mandal, P.~Mastrolia, R.~Patil and J.~Steinhoff, \emph{{Gravitational
  Spin-Orbit Hamiltonian at NNNLO in the post-Newtonian framework}},
  \href{https://arxiv.org/abs/2209.00611}{{\ttfamily 2209.00611}}.

\bibitem{Mandal:2022ufb}
M.~K. Mandal, P.~Mastrolia, R.~Patil and J.~Steinhoff, \emph{{Gravitational
  Quadratic-in-Spin Hamiltonian at NNNLO in the post-Newtonian framework}},
  \href{https://arxiv.org/abs/2210.09176}{{\ttfamily 2210.09176}}.

\bibitem{Damour:2022ybd}
T.~Damour and P.~Rettegno, \emph{{Strong-field scattering of two black holes:
  Numerical Relativity meets Post-Minkowskian gravity}},
  \href{https://arxiv.org/abs/2211.01399}{{\ttfamily 2211.01399}}.

\bibitem{damour1}
T.~Damour, \emph{{Gravitational scattering, post-Minkowskian approximation and
  Effective One-Body theory}},
  \href{https://doi.org/10.1103/PhysRevD.94.104015}{\emph{Phys. Rev.}
  {\bfseries D94} (2016) 104015}
  [\href{https://arxiv.org/abs/1609.00354}{{\ttfamily 1609.00354}}].

\bibitem{damour2}
T.~Damour, \emph{{High-energy gravitational scattering and the general
  relativistic two-body problem}},
  \href{https://doi.org/10.1103/PhysRevD.97.044038}{\emph{Phys. Rev.}
  {\bfseries D97} (2018) 044038}
  [\href{https://arxiv.org/abs/1710.10599}{{\ttfamily 1710.10599}}].

\bibitem{Damour:2019lcq}
T.~Damour, \emph{{Classical and quantum scattering in post-Minkowskian
  gravity}}, \href{https://doi.org/10.1103/PhysRevD.102.024060}{\emph{Phys.
  Rev. D} {\bfseries 102} (2020) 024060}
  [\href{https://arxiv.org/abs/1912.02139}{{\ttfamily 1912.02139}}].

\bibitem{paper1}
G.~K{\"a}lin and R.~A. Porto, \emph{{From Boundary Data to Bound States}},
  \href{https://doi.org/10.1007/JHEP01(2020)072}{\emph{JHEP} {\bfseries 01}
  (2020) 072} [\href{https://arxiv.org/abs/1910.03008}{{\ttfamily
  1910.03008}}].

\bibitem{paper2}
G.~K{\"a}lin and R.~A. Porto, \emph{{From boundary data to bound states. Part
  II: Scattering angle to dynamical invariants (with twist)}},
  \href{https://doi.org/10.1007/JHEP02(2020)120}{\emph{JHEP} {\bfseries 02}
  (2020) 120} [\href{https://arxiv.org/abs/1911.09130}{{\ttfamily
  1911.09130}}].

\bibitem{b2b3}
G.~Cho, G.~K\"alin and R.~A. Porto, \emph{{From Boundary Data to Bound States
  III: Radiative Effects}},  \href{https://arxiv.org/abs/2112.03976}{{\ttfamily
  2112.03976}}.

\bibitem{pmeft}
G.~K\"alin and R.~A. Porto, \emph{{Post-Minkowskian Effective Field Theory for
  Conservative Binary Dynamics}},
  \href{https://doi.org/10.1007/JHEP11(2020)106}{\emph{JHEP} {\bfseries 11}
  (2020) 106} [\href{https://arxiv.org/abs/2006.01184}{{\ttfamily
  2006.01184}}].

\bibitem{3pmeft}
G.~K\"alin, Z.~Liu and R.~A. Porto, \emph{{Conservative Dynamics of Binary
  Systems to Third Post-Minkowskian Order from the Effective Field Theory
  Approach}}, \href{https://doi.org/10.1103/PhysRevLett.125.261103}{\emph{Phys.
  Rev. Lett.} {\bfseries 125} (2020) 261103}
  [\href{https://arxiv.org/abs/2007.04977}{{\ttfamily 2007.04977}}].

\bibitem{tidaleft}
G.~K\"alin, Z.~Liu and R.~A. Porto, \emph{{Conservative Tidal Effects in
  Compact Binary Systems to Next-to-Leading Post-Minkowskian Order}},
  \href{https://doi.org/10.1103/PhysRevD.102.124025}{\emph{Phys. Rev. D}
  {\bfseries 102} (2020) 124025}
  [\href{https://arxiv.org/abs/2008.06047}{{\ttfamily 2008.06047}}].

\bibitem{pmefts}
Z.~Liu, R.~A. Porto and Z.~Yang, \emph{{Spin Effects in the Effective Field
  Theory Approach to Post-Minkowskian Conservative Dynamics}},
  \href{https://doi.org/10.1007/JHEP06(2021)012}{\emph{JHEP} {\bfseries 06}
  (2021) 012} [\href{https://arxiv.org/abs/2102.10059}{{\ttfamily
  2102.10059}}].

\bibitem{eftrad}
G.~K\"alin, J.~Neef and R.~A. Porto, \emph{{Radiation-Reaction in the Effective
  Field Theory Approach to Post-Minkowskian Dynamics}},
  \href{https://arxiv.org/abs/2207.00580}{{\ttfamily 2207.00580}}.

\bibitem{4pmeft}
C.~Dlapa, G.~K\"alin, Z.~Liu and R.~A. Porto, \emph{{Dynamics of Binary Systems
  to Fourth Post-Minkowskian Order from the Effective Field Theory Approach}},
  \href{https://doi.org/10.1016/j.physletb.2022.137203}{\emph{Phys. Lett. B}
  {\bfseries 831} (2022) 137203}
  [\href{https://arxiv.org/abs/2106.08276}{{\ttfamily 2106.08276}}].

\bibitem{4pmeft2}
C.~Dlapa, G.~K\"alin, Z.~Liu and R.~A. Porto, \emph{{Conservative Dynamics of
  Binary Systems at Fourth Post-Minkowskian Order in the Large-Eccentricity
  Expansion}},
  \href{https://doi.org/10.1103/PhysRevLett.128.161104}{\emph{Phys. Rev. Lett.}
  {\bfseries 128} (2022) 161104}
  [\href{https://arxiv.org/abs/2112.11296}{{\ttfamily 2112.11296}}].

\bibitem{4pmeftot}
C.~Dlapa, G.~K\"alin, Z.~Liu, J.~Neef and R.~A. Porto, \emph{{Radiation
  Reaction and Gravitational Waves at Fourth Post-Minkowskian Order}},
  \href{https://doi.org/10.1103/PhysRevLett.130.101401}{\emph{Phys. Rev. Lett.}
  {\bfseries 130} (2023) 101401}
  [\href{https://arxiv.org/abs/2210.05541}{{\ttfamily 2210.05541}}].

\bibitem{janmogul}
G.~Mogull, J.~Plefka and J.~Steinhoff, \emph{{Classical black hole scattering
  from a worldline quantum field theory}},
  \href{https://doi.org/10.1007/JHEP02(2021)048}{\emph{JHEP} {\bfseries 02}
  (2021) 048} [\href{https://arxiv.org/abs/2010.02865}{{\ttfamily
  2010.02865}}].

\bibitem{janmogul2}
G.~U. Jakobsen, G.~Mogull, J.~Plefka and J.~Steinhoff, \emph{{Classical
  Gravitational Bremsstrahlung from a Worldline Quantum Field Theory}},
  \href{https://doi.org/10.1103/PhysRevLett.126.201103}{\emph{Phys. Rev. Lett.}
  {\bfseries 126} (2021) 201103}
  [\href{https://arxiv.org/abs/2101.12688}{{\ttfamily 2101.12688}}].

\bibitem{Jakobsen:2021zvh}
G.~U. Jakobsen, G.~Mogull, J.~Plefka and J.~Steinhoff, \emph{{SUSY in the sky
  with gravitons}}, \href{https://doi.org/10.1007/JHEP01(2022)027}{\emph{JHEP}
  {\bfseries 01} (2022) 027}
  [\href{https://arxiv.org/abs/2109.04465}{{\ttfamily 2109.04465}}].

\bibitem{Jakobsen:2022fcj}
G.~U. Jakobsen and G.~Mogull, \emph{{Conservative and Radiative Dynamics of
  Spinning Bodies at Third Post-Minkowskian Order Using Worldline Quantum Field
  Theory}}, \href{https://doi.org/10.1103/PhysRevLett.128.141102}{\emph{Phys.
  Rev. Lett.} {\bfseries 128} (2022) 141102}
  [\href{https://arxiv.org/abs/2201.07778}{{\ttfamily 2201.07778}}].

\bibitem{Mougiakakos:2021ckm}
S.~Mougiakakos, M.~M. Riva and F.~Vernizzi, \emph{{Gravitational Bremsstrahlung
  in the post-Minkowskian effective field theory}},
  \href{https://doi.org/10.1103/PhysRevD.104.024041}{\emph{Phys. Rev. D}
  {\bfseries 104} (2021) 024041}
  [\href{https://arxiv.org/abs/2102.08339}{{\ttfamily 2102.08339}}].

\bibitem{Riva:2021vnj}
M.~M. Riva and F.~Vernizzi, \emph{{Radiated momentum in the post-Minkowskian
  worldline approach via reverse unitarity}},
  \href{https://doi.org/10.1007/JHEP11(2021)228}{\emph{JHEP} {\bfseries 11}
  (2021) 228} [\href{https://arxiv.org/abs/2110.10140}{{\ttfamily
  2110.10140}}].

\bibitem{Mougiakakos:2022sic}
S.~Mougiakakos, M.~M. Riva and F.~Vernizzi, \emph{{Gravitational Bremsstrahlung
  with tidal effects in the post-Minkowskian expansion}},
  \href{https://arxiv.org/abs/2204.06556}{{\ttfamily 2204.06556}}.

\bibitem{Riva:2022fru}
M.~M. Riva, F.~Vernizzi and L.~K. Wong, \emph{{Gravitational Bremsstrahlung
  from Spinning Binaries in the Post-Minkowskian Expansion}},
  \href{https://arxiv.org/abs/2205.15295}{{\ttfamily 2205.15295}}.

\bibitem{Jakobsen:2022psy}
G.~U. Jakobsen, G.~Mogull, J.~Plefka and B.~Sauer, \emph{{All Things Retarded:
  Radiation-Reaction in Worldline Quantum Field Theory}},
  \href{https://arxiv.org/abs/2207.00569}{{\ttfamily 2207.00569}}.

\bibitem{Jinno:2022sbr}
R.~Jinno, G.~K\"alin, Z.~Liu and H.~Rubira, \emph{{Machine Learning
  Post-Minkowskian Integrals}},
  \href{https://arxiv.org/abs/2209.01091}{{\ttfamily 2209.01091}}.

\bibitem{Damgaard:2019lfh}
P.~H. Damgaard, K.~Haddad and A.~Helset, \emph{{Heavy Black Hole Effective
  Theory}}, \href{https://doi.org/10.1007/JHEP11(2019)070}{\emph{JHEP}
  {\bfseries 11} (2019) 070}
  [\href{https://arxiv.org/abs/1908.10308}{{\ttfamily 1908.10308}}].

\bibitem{Brandhuber:2023hhy}
A.~Brandhuber, G.~R. Brown, G.~Chen, S.~De~Angelis, J.~Gowdy and G.~Travaglini,
  \emph{{One-loop Gravitational Bremsstrahlung and Waveforms from a Heavy-Mass
  Effective Field Theory}},  \href{https://arxiv.org/abs/2303.06111}{{\ttfamily
  2303.06111}}.

\bibitem{ira1}
D.~Neill and I.~Z. Rothstein, \emph{{Classical Space-Times from the S Matrix}},
  \href{https://doi.org/10.1016/j.nuclphysb.2013.09.007}{\emph{Nucl. Phys.}
  {\bfseries B877} (2013) 177}
  [\href{https://arxiv.org/abs/1304.7263}{{\ttfamily 1304.7263}}].

\bibitem{Vaidya:2014kza}
V.~Vaidya, \emph{{Gravitational spin Hamiltonians from the S matrix}},
  \href{https://doi.org/10.1103/PhysRevD.91.024017}{\emph{Phys. Rev.}
  {\bfseries D91} (2015) 024017}
  [\href{https://arxiv.org/abs/1410.5348}{{\ttfamily 1410.5348}}].

\bibitem{Walter}
W.~D. Goldberger and A.~K. Ridgway, \emph{{Bound states and the classical
  double copy}}, \href{https://doi.org/10.1103/PhysRevD.97.085019}{\emph{Phys.
  Rev.} {\bfseries D97} (2018) 085019}
  [\href{https://arxiv.org/abs/1711.09493}{{\ttfamily 1711.09493}}].

\bibitem{Goldberger:2016iau}
W.~D. Goldberger and A.~K. Ridgway, \emph{{Radiation and the classical double
  copy for color charges}},
  \href{https://doi.org/10.1103/PhysRevD.95.125010}{\emph{Phys. Rev. D}
  {\bfseries 95} (2017) 125010}
  [\href{https://arxiv.org/abs/1611.03493}{{\ttfamily 1611.03493}}].

\bibitem{cheung}
C.~Cheung, I.~Z. Rothstein and M.~P. Solon, \emph{{From Scattering Amplitudes
  to Classical Potentials in the Post-Minkowskian Expansion}},
  \href{https://doi.org/10.1103/PhysRevLett.121.251101}{\emph{Phys. Rev. Lett.}
  {\bfseries 121} (2018) 251101}
  [\href{https://arxiv.org/abs/1808.02489}{{\ttfamily 1808.02489}}].

\bibitem{bohr}
N.~E.~J. Bjerrum-Bohr, P.~H. Damgaard, G.~Festuccia, L.~Plante and P.~Vanhove,
  \emph{{General Relativity from Scattering Amplitudes}},
  \href{https://doi.org/10.1103/PhysRevLett.121.171601}{\emph{Phys. Rev. Lett.}
  {\bfseries 121} (2018) 171601}
  [\href{https://arxiv.org/abs/1806.04920}{{\ttfamily 1806.04920}}].

\bibitem{Guevara:2018wpp}
A.~Guevara, A.~Ochirov and J.~Vines, \emph{{Scattering of Spinning Black Holes
  from Exponentiated Soft Factors}},
  \href{https://doi.org/10.1007/JHEP09(2019)056}{\emph{JHEP} {\bfseries 09}
  (2019) 056} [\href{https://arxiv.org/abs/1812.06895}{{\ttfamily
  1812.06895}}].

\bibitem{cristof1}
A.~Cristofoli, N.~E.~J. Bjerrum-Bohr, P.~H. Damgaard and P.~Vanhove,
  \emph{{Post-Minkowskian Hamiltonians in general relativity}},
  \href{https://doi.org/10.1103/PhysRevD.100.084040}{\emph{Phys. Rev. D}
  {\bfseries 100} (2019) 084040}
  [\href{https://arxiv.org/abs/1906.01579}{{\ttfamily 1906.01579}}].

\bibitem{donal}
D.~A. Kosower, B.~Maybee and D.~O'Connell, \emph{{Amplitudes, Observables, and
  Classical Scattering}},
  \href{https://doi.org/10.1007/JHEP02(2019)137}{\emph{JHEP} {\bfseries 02}
  (2019) 137} [\href{https://arxiv.org/abs/1811.10950}{{\ttfamily
  1811.10950}}].

\bibitem{donalvines}
B.~Maybee, D.~O'Connell and J.~Vines, \emph{{Observables and amplitudes for
  spinning particles and black holes}},
  \href{https://doi.org/10.1007/JHEP12(2019)156}{\emph{JHEP} {\bfseries 12}
  (2019) 156} [\href{https://arxiv.org/abs/1906.09260}{{\ttfamily
  1906.09260}}].

\bibitem{zvi1}
Z.~Bern, C.~Cheung, R.~Roiban, C.-H. Shen, M.~P. Solon and M.~Zeng,
  \emph{{Scattering Amplitudes and the Conservative Hamiltonian for Binary
  Systems at Third Post-Minkowskian Order}},
  \href{https://doi.org/10.1103/PhysRevLett.122.201603}{\emph{Phys. Rev. Lett.}
  {\bfseries 122} (2019) 201603}
  [\href{https://arxiv.org/abs/1901.04424}{{\ttfamily 1901.04424}}].

\bibitem{Haddad:2020que}
K.~Haddad and A.~Helset, \emph{{Tidal effects in quantum field theory}},
  \href{https://doi.org/10.1007/JHEP12(2020)024}{\emph{JHEP} {\bfseries 12}
  (2020) 024} [\href{https://arxiv.org/abs/2008.04920}{{\ttfamily
  2008.04920}}].

\bibitem{Aoude:2022thd}
R.~Aoude, K.~Haddad and A.~Helset, \emph{{Classical gravitational
  spinning-spinless scattering at $\mathcal{O}(G^{2} S^{\infty})$}},
  \href{https://arxiv.org/abs/2205.02809}{{\ttfamily 2205.02809}}.

\bibitem{Bjerrum-Bohr:2021din}
N.~E.~J. Bjerrum-Bohr, P.~H. Damgaard, L.~Plant\'e and P.~Vanhove, \emph{{The
  amplitude for classical gravitational scattering at third Post-Minkowskian
  order}}, \href{https://doi.org/10.1007/JHEP08(2021)172}{\emph{JHEP}
  {\bfseries 08} (2021) 172}
  [\href{https://arxiv.org/abs/2105.05218}{{\ttfamily 2105.05218}}].

\bibitem{andres2}
D.~Kosmopoulos and A.~Luna, \emph{{Quadratic-in-spin Hamiltonian at $
  \mathcal{O} $(G$^{2}$) from scattering amplitudes}},
  \href{https://doi.org/10.1007/JHEP07(2021)037}{\emph{JHEP} {\bfseries 07}
  (2021) 037} [\href{https://arxiv.org/abs/2102.10137}{{\ttfamily
  2102.10137}}].

\bibitem{4pmzvi}
Z.~Bern, J.~Parra-Martinez, R.~Roiban, M.~S. Ruf, C.-H. Shen, M.~P. Solon
  et~al., \emph{{Scattering Amplitudes and Conservative Binary Dynamics at
  ${\cal O}(G^4)$}},
  \href{https://doi.org/10.1103/PhysRevLett.126.171601}{\emph{Phys. Rev. Lett.}
  {\bfseries 126} (2021) 171601}
  [\href{https://arxiv.org/abs/2101.07254}{{\ttfamily 2101.07254}}].

\bibitem{4pmzvi2}
Z.~Bern, J.~Parra-Martinez, R.~Roiban, M.~S. Ruf, C.-H. Shen, M.~P. Solon
  et~al., \emph{{Scattering Amplitudes, the Tail Effect, and Conservative
  Binary Dynamics at O(G4)}},
  \href{https://doi.org/10.1103/PhysRevLett.128.161103}{\emph{Phys. Rev. Lett.}
  {\bfseries 128} (2022) 161103}
  [\href{https://arxiv.org/abs/2112.10750}{{\ttfamily 2112.10750}}].

\bibitem{Gabriele}
P.~Di~Vecchia, C.~Heissenberg, R.~Russo and G.~Veneziano, \emph{{Radiation
  Reaction from Soft Theorems}},
  \href{https://doi.org/10.1016/j.physletb.2021.136379}{\emph{Phys. Lett. B}
  {\bfseries 818} (2021) 136379}
  [\href{https://arxiv.org/abs/2101.05772}{{\ttfamily 2101.05772}}].

\bibitem{Gabriele2}
P.~Di~Vecchia, C.~Heissenberg, R.~Russo and G.~Veneziano, \emph{{The eikonal
  approach to gravitational scattering and radiation at $ \mathcal{O}
  $(G$^{3}$)}}, \href{https://doi.org/10.1007/JHEP07(2021)169}{\emph{JHEP}
  {\bfseries 07} (2021) 169}
  [\href{https://arxiv.org/abs/2104.03256}{{\ttfamily 2104.03256}}].

\bibitem{Parra}
J.~Parra-Martinez, M.~S. Ruf and M.~Zeng, \emph{{Extremal black hole scattering
  at $\mathcal{O}(G^3)$: graviton dominance, eikonal exponentiation, and
  differential equations}},
  \href{https://doi.org/10.1007/JHEP11(2020)023}{\emph{JHEP} {\bfseries 11}
  (2020) 023} [\href{https://arxiv.org/abs/2005.04236}{{\ttfamily
  2005.04236}}].

\bibitem{Parra3}
E.~Herrmann, J.~Parra-Martinez, M.~S. Ruf and M.~Zeng, \emph{{Radiative
  Classical Gravitational Observables at $\mathcal O(G^3)$ from Scattering
  Amplitudes}}, \href{https://doi.org/10.1007/JHEP10(2021)148}{\emph{JHEP}
  {\bfseries 10} (2021) 148}
  [\href{https://arxiv.org/abs/2104.03957}{{\ttfamily 2104.03957}}].

\bibitem{Brandhuber:2021eyq}
A.~Brandhuber, G.~Chen, G.~Travaglini and C.~Wen, \emph{{Classical
  gravitational scattering from a gauge-invariant double copy}},
  \href{https://doi.org/10.1007/JHEP10(2021)118}{\emph{JHEP} {\bfseries 10}
  (2021) 118} [\href{https://arxiv.org/abs/2108.04216}{{\ttfamily
  2108.04216}}].

\bibitem{Cristofoli:2021vyo}
A.~Cristofoli, R.~Gonzo, D.~A. Kosower and D.~O'Connell, \emph{{Waveforms from
  amplitudes}}, \href{https://doi.org/10.1103/PhysRevD.106.056007}{\emph{Phys.
  Rev. D} {\bfseries 106} (2022) 056007}
  [\href{https://arxiv.org/abs/2107.10193}{{\ttfamily 2107.10193}}].

\bibitem{FebresCordero:2022jts}
F.~Febres~Cordero, M.~Kraus, G.~Lin, M.~S. Ruf and M.~Zeng, \emph{{Conservative
  Binary Dynamics with a Spinning Black Hole at $\mathcal{O}(G^3)$ from
  Scattering Amplitudes}},  \href{https://arxiv.org/abs/2205.07357}{{\ttfamily
  2205.07357}}.

\bibitem{Manohar:2022dea}
A.~V. Manohar, A.~K. Ridgway and C.-H. Shen, \emph{{Radiated Angular Momentum
  and Dissipative Effects in Classical Scattering}},
  \href{https://arxiv.org/abs/2203.04283}{{\ttfamily 2203.04283}}.

\bibitem{DiVecchia:2022nna}
P.~Di~Vecchia, C.~Heissenberg, R.~Russo and G.~Veneziano, \emph{{The eikonal
  operator at arbitrary velocities I: the soft-radiation limit}},
  \href{https://doi.org/10.1007/JHEP07(2022)039}{\emph{JHEP} {\bfseries 07}
  (2022) 039} [\href{https://arxiv.org/abs/2204.02378}{{\ttfamily
  2204.02378}}].

\bibitem{chadprl}
C.~R. Galley, \emph{{Classical Mechanics of Nonconservative Systems}},
  \href{https://doi.org/10.1103/PhysRevLett.110.174301}{\emph{Phys. Rev. Lett.}
  {\bfseries 110} (2013) 174301}
  [\href{https://arxiv.org/abs/1210.2745}{{\ttfamily 1210.2745}}].

\bibitem{iraTasi}
I.~Z. Rothstein, \emph{{TASI lectures on effective field theories}},
  \href{https://arxiv.org/abs/hep-ph/0308266}{{\ttfamily hep-ph/0308266}}.

\bibitem{beneke}
M.~Beneke and V.~A. Smirnov, \emph{{Asymptotic expansion of Feynman integrals
  near threshold}}, {\emph{Nucl. Phys.} {\bfseries B522} (1998) 321}
  [\href{https://arxiv.org/abs/hep-ph/9711391}{{\ttfamily hep-ph/9711391}}].

\bibitem{nrgrG5}
S.~Foffa, P.~Mastrolia, R.~Sturani and C.~Sturm, \emph{{Effective field theory
  approach to the gravitational two-body dynamics, at fourth post-Newtonian
  order and quintic in the Newton constant}},
  \href{https://doi.org/10.1103/PhysRevD.95.104009}{\emph{Phys. Rev. D}
  {\bfseries 95} (2017) 104009}
  [\href{https://arxiv.org/abs/1612.00482}{{\ttfamily 1612.00482}}].

\bibitem{Smirnov:2012gma}
V.~A. Smirnov, \emph{{Analytic tools for Feynman integrals}}. {Springer}, 2012,
  \href{https://doi.org/10.1007/978-3-642-34886-0}{10.1007/978-3-642-34886-0}.

\bibitem{Weinzierl:2022eaz}
S.~Weinzierl, \emph{{Feynman Integrals}}. 1, 2022,
  \href{https://doi.org/10.1007/978-3-030-99558-4}{10.1007/978-3-030-99558-4},
  [\href{https://arxiv.org/abs/2201.03593}{{\ttfamily 2201.03593}}].

\bibitem{Kotikov:1991pm}
A.~Kotikov, \emph{{Differential equation method: The Calculation of N point
  Feynman diagrams}},
  \href{https://doi.org/10.1016/0370-2693(91)90536-Y}{\emph{Phys. Lett. B}
  {\bfseries 267} (1991) 123}.

\bibitem{Remiddi:1997ny}
E.~Remiddi, \emph{{Differential equations for Feynman graph amplitudes}},
  {\emph{Nuovo Cim. A} {\bfseries 110} (1997) 1435}
  [\href{https://arxiv.org/abs/hep-th/9711188}{{\ttfamily hep-th/9711188}}].

\bibitem{Henn:2013pwa}
J.~M. Henn, \emph{{Multiloop integrals in dimensional regularization made
  simple}}, \href{https://doi.org/10.1103/PhysRevLett.110.251601}{\emph{Phys.
  Rev. Lett.} {\bfseries 110} (2013) 251601}
  [\href{https://arxiv.org/abs/1304.1806}{{\ttfamily 1304.1806}}].

\bibitem{Prausa:2017ltv}
M.~Prausa, \emph{{epsilon: A tool to find a canonical basis of master
  integrals}}, \href{https://doi.org/10.1016/j.cpc.2017.05.026}{\emph{Comput.
  Phys. Commun.} {\bfseries 219} (2017) 361}
  [\href{https://arxiv.org/abs/1701.00725}{{\ttfamily 1701.00725}}].

\bibitem{Lee:2020zfb}
R.~N. Lee, \emph{{Libra: A package for transformation of differential systems
  for multiloop integrals}},
  \href{https://doi.org/10.1016/j.cpc.2021.108058}{\emph{Comput. Phys. Commun.}
  {\bfseries 267} (2021) 108058}
  [\href{https://arxiv.org/abs/2012.00279}{{\ttfamily 2012.00279}}].

\bibitem{Lee:2014ioa}
R.~N. Lee, \emph{{Reducing differential equations for multiloop master
  integrals}}, \href{https://doi.org/10.1007/JHEP04(2015)108}{\emph{JHEP}
  {\bfseries 04} (2015) 108} [\href{https://arxiv.org/abs/1411.0911}{{\ttfamily
  1411.0911}}].

\bibitem{Adams:2018yfj}
L.~Adams and S.~Weinzierl, \emph{{The $\varepsilon$-form of the differential
  equations for Feynman integrals in the elliptic case}},
  \href{https://doi.org/10.1016/j.physletb.2018.04.002}{\emph{Phys. Lett. B}
  {\bfseries 781} (2018) 270}
  [\href{https://arxiv.org/abs/1802.05020}{{\ttfamily 1802.05020}}].

\bibitem{Chetyrkin:1981qh}
K.~Chetyrkin and F.~Tkachov, \emph{{Integration by Parts: The Algorithm to
  Calculate beta Functions in 4 Loops}},
  \href{https://doi.org/10.1016/0550-3213(81)90199-1}{\emph{Nucl. Phys. B}
  {\bfseries 192} (1981) 159}.

\bibitem{Tkachov:1981wb}
F.~Tkachov, \emph{{A Theorem on Analytical Calculability of Four Loop
  Renormalization Group Functions}},
  \href{https://doi.org/10.1016/0370-2693(81)90288-4}{\emph{Phys. Lett. B}
  {\bfseries 100} (1981) 65}.

\bibitem{Smirnov:2019qkx}
A.~V. Smirnov and F.~S. Chuharev, \emph{{FIRE6: Feynman Integral REduction with
  Modular Arithmetic}},
  \href{https://doi.org/10.1016/j.cpc.2019.106877}{\emph{Comput. Phys. Commun.}
  {\bfseries 247} (2020) 106877}
  [\href{https://arxiv.org/abs/1901.07808}{{\ttfamily 1901.07808}}].

\bibitem{Smirnov:2020quc}
A.~Smirnov and V.~Smirnov, \emph{{How to choose master integrals}},
  \href{https://doi.org/10.1016/j.nuclphysb.2020.115213}{\emph{Nucl. Phys. B}
  {\bfseries 960} (2020) 115213}
  [\href{https://arxiv.org/abs/2002.08042}{{\ttfamily 2002.08042}}].

\bibitem{Lee:2012cn}
R.~Lee, \emph{{Presenting LiteRed: a tool for the Loop InTEgrals REDuction}},
  \href{https://arxiv.org/abs/1212.2685}{{\ttfamily 1212.2685}}.

\bibitem{Lee:2013mka}
R.~N. Lee, \emph{{LiteRed 1.4: a powerful tool for reduction of multiloop
  integrals}}, \href{https://doi.org/10.1088/1742-6596/523/1/012059}{\emph{J.
  Phys. Conf. Ser.} {\bfseries 523} (2014) 012059}
  [\href{https://arxiv.org/abs/1310.1145}{{\ttfamily 1310.1145}}].

\bibitem{Beneke:1997zp}
M.~Beneke and V.~A. Smirnov, \emph{{Asymptotic expansion of Feynman integrals
  near threshold}},
  \href{https://doi.org/10.1016/S0550-3213(98)00138-2}{\emph{Nucl. Phys.}
  {\bfseries B522} (1998) 321}
  [\href{https://arxiv.org/abs/hep-ph/9711391}{{\ttfamily hep-ph/9711391}}].

\bibitem{Jantzen:2012mw}
B.~Jantzen, A.~V. Smirnov and V.~A. Smirnov, \emph{{Expansion by regions:
  revealing potential and Glauber regions automatically}},
  \href{https://doi.org/10.1140/epjc/s10052-012-2139-2}{\emph{Eur. Phys. J. C}
  {\bfseries 72} (2012) 2139}
  [\href{https://arxiv.org/abs/1206.0546}{{\ttfamily 1206.0546}}].

\bibitem{Smirnov:2015mct}
A.~V. Smirnov, \emph{{FIESTA4: Optimized Feynman integral calculations with GPU
  support}}, \href{https://doi.org/10.1016/j.cpc.2016.03.013}{\emph{Comput.
  Phys. Commun.} {\bfseries 204} (2016) 189}
  [\href{https://arxiv.org/abs/1511.03614}{{\ttfamily 1511.03614}}].

\bibitem{Meyer:2016zeb}
C.~Meyer, \emph{{Evaluating multi-loop Feynman integrals using differential
  equations: automatizing the transformation to a canonical basis}},
  \href{https://doi.org/10.22323/1.260.0028}{\emph{PoS} {\bfseries LL2016}
  (2016) 028}.

\bibitem{Meyer:2016slj}
C.~Meyer, \emph{{Transforming differential equations of multi-loop Feynman
  integrals into canonical form}},
  \href{https://doi.org/10.1007/JHEP04(2017)006}{\emph{JHEP} {\bfseries 04}
  (2017) 006} [\href{https://arxiv.org/abs/1611.01087}{{\ttfamily
  1611.01087}}].

\bibitem{Broedel:2019kmn}
J.~Broedel, C.~Duhr, F.~Dulat, R.~Marzucca, B.~Penante and L.~Tancredi,
  \emph{{An analytic solution for the equal-mass banana graph}},
  \href{https://doi.org/10.1007/JHEP09(2019)112}{\emph{JHEP} {\bfseries 09}
  (2019) 112} [\href{https://arxiv.org/abs/1907.03787}{{\ttfamily
  1907.03787}}].

\bibitem{Primo:2017ipr}
A.~Primo and L.~Tancredi, \emph{{Maximal cuts and differential equations for
  Feynman integrals. An application to the three-loop massive banana graph}},
  \href{https://doi.org/10.1016/j.nuclphysb.2017.05.018}{\emph{Nucl. Phys. B}
  {\bfseries 921} (2017) 316}
  [\href{https://arxiv.org/abs/1704.05465}{{\ttfamily 1704.05465}}].

\bibitem{Hidding:2020ytt}
M.~Hidding, \emph{{DiffExp, a Mathematica package for computing Feynman
  integrals in terms of one-dimensional series expansions}},
  \href{https://doi.org/10.1016/j.cpc.2021.108125}{\emph{Comput. Phys. Commun.}
  {\bfseries 269} (2021) 108125}
  [\href{https://arxiv.org/abs/2006.05510}{{\ttfamily 2006.05510}}].

\bibitem{Goncharov:2001iea}
A.~Goncharov, \emph{{Multiple polylogarithms and mixed Tate motives}},
  \href{https://arxiv.org/abs/math/0103059}{{\ttfamily math/0103059}}.

\bibitem{Chen:1977oja}
K.-T. Chen, \emph{{Iterated path integrals}},
  \href{https://doi.org/10.1090/S0002-9904-1977-14320-6}{\emph{Bull. Am. Math.
  Soc.} {\bfseries 83} (1977) 831}.

\bibitem{Duhr:2014woa}
C.~Duhr, \emph{{Mathematical aspects of scattering amplitudes}},
  \href{https://arxiv.org/abs/1411.7538}{{\ttfamily 1411.7538}}.

\bibitem{Duhr:2019tlz}
C.~Duhr and F.~Dulat, \emph{{PolyLogTools \textemdash{} polylogs for the
  masses}}, \href{https://doi.org/10.1007/JHEP08(2019)135}{\emph{JHEP}
  {\bfseries 08} (2019) 135}
  [\href{https://arxiv.org/abs/1904.07279}{{\ttfamily 1904.07279}}].

\bibitem{Dlapa:2020cwj}
C.~Dlapa, J.~Henn and K.~Yan, \emph{{Deriving canonical differential equations
  for Feynman integrals from a single uniform weight integral}},
  \href{https://doi.org/10.1007/JHEP05(2020)025}{\emph{JHEP} {\bfseries 05}
  (2020) 025} [\href{https://arxiv.org/abs/2002.02340}{{\ttfamily
  2002.02340}}].

\bibitem{Smirnov:2021rhf}
A.~V. Smirnov, N.~D. Shapurov and L.~I. Vysotsky, \emph{{FIESTA5: numerical
  high-performance Feynman integral evaluation}},
  \href{https://arxiv.org/abs/2110.11660}{{\ttfamily 2110.11660}}.

\bibitem{Lee:2019zop}
R.~N. Lee, A.~V. Smirnov, V.~A. Smirnov and M.~Steinhauser, \emph{{Four-loop
  quark form factor with quartic fundamental colour factor}},
  \href{https://doi.org/10.1007/JHEP02(2019)172}{\emph{JHEP} {\bfseries 02}
  (2019) 172} [\href{https://arxiv.org/abs/1901.02898}{{\ttfamily
  1901.02898}}].

\bibitem{Blumlein:2021pgo}
J.~Bl\"umlein, \emph{{Analytic Integration Methods in Quantum Field Theory: An
  Introduction}},  in \emph{{Antidifferentiation and the Calculation of Feynman
  Amplitudes}}, 3, 2021, \href{https://arxiv.org/abs/2103.10652}{{\ttfamily
  2103.10652}}, \href{https://doi.org/10.1007/978-3-030-80219-6_1}{DOI}.

\bibitem{Pak:2010pt}
A.~Pak and A.~Smirnov, \emph{{Geometric approach to asymptotic expansion of
  Feynman integrals}},
  \href{https://doi.org/10.1140/epjc/s10052-011-1626-1}{\emph{Eur. Phys. J. C}
  {\bfseries 71} (2011) 1626}
  [\href{https://arxiv.org/abs/1011.4863}{{\ttfamily 1011.4863}}].

\bibitem{safi}
S.~Rafie-Zinedine, \emph{{Simplifying Quantum Gravity Calculations}},
  \href{https://arxiv.org/abs/1808.06086}{{\ttfamily 1808.06086}}.

\bibitem{Ruijl:2017dtg}
B.~Ruijl, T.~Ueda and J.~Vermaseren, \emph{{FORM version 4.2}},
  \href{https://arxiv.org/abs/1707.06453}{{\ttfamily 1707.06453}}.

\bibitem{xactpackage}
J.~M. Martín-García, \emph{{xAct: Efficient tensor computer algebra for the
  Wolfram Language}},  2002-2021.

\bibitem{Passarino:1978jh}
G.~Passarino and M.~J.~G. Veltman, \emph{{One Loop Corrections for $e^+ e^-$
  Annihilation Into $\mu^+ \mu^-$ in the Weinberg Model}},
  \href{https://doi.org/10.1016/0550-3213(79)90234-7}{\emph{Nucl. Phys. B}
  {\bfseries 160} (1979) 151}.

\bibitem{Cutkosky:1960sp}
R.~E. Cutkosky, \emph{{Singularities and discontinuities of Feynman
  amplitudes}}, \href{https://doi.org/10.1063/1.1703676}{\emph{J. Math. Phys.}
  {\bfseries 1} (1960) 429}.

\bibitem{Anastasiou:2002yz}
C.~Anastasiou and K.~Melnikov, \emph{{Higgs boson production at hadron
  colliders in NNLO QCD}},
  \href{https://doi.org/10.1016/S0550-3213(02)00837-4}{\emph{Nucl. Phys. B}
  {\bfseries 646} (2002) 220}
  [\href{https://arxiv.org/abs/hep-ph/0207004}{{\ttfamily hep-ph/0207004}}].

\bibitem{Anastasiou:2003ds}
C.~Anastasiou, L.~J. Dixon, K.~Melnikov and F.~Petriello, \emph{{High precision
  QCD at hadron colliders: Electroweak gauge boson rapidity distributions at
  NNLO}}, \href{https://doi.org/10.1103/PhysRevD.69.094008}{\emph{Phys. Rev. D}
  {\bfseries 69} (2004) 094008}
  [\href{https://arxiv.org/abs/hep-ph/0312266}{{\ttfamily hep-ph/0312266}}].

\bibitem{Anastasiou:2004vj}
C.~Anastasiou and A.~Lazopoulos, \emph{{Automatic integral reduction for higher
  order perturbative calculations}},
  \href{https://doi.org/10.1088/1126-6708/2004/07/046}{\emph{JHEP} {\bfseries
  07} (2004) 046} [\href{https://arxiv.org/abs/hep-ph/0404258}{{\ttfamily
  hep-ph/0404258}}].

\bibitem{Laporta:2000dsw}
S.~Laporta, \emph{{High precision calculation of multiloop Feynman integrals by
  difference equations}},
  \href{https://doi.org/10.1142/S0217751X00002159}{\emph{Int. J. Mod. Phys. A}
  {\bfseries 15} (2000) 5087}
  [\href{https://arxiv.org/abs/hep-ph/0102033}{{\ttfamily hep-ph/0102033}}].

\bibitem{Smirnov:2006wh}
A.~V. Smirnov and V.~A. Smirnov, \emph{{S-bases as a tool to solve reduction
  problems for Feynman integrals}},
  \href{https://doi.org/10.1016/j.nuclphysbps.2006.09.032}{\emph{Nucl. Phys. B
  Proc. Suppl.} {\bfseries 160} (2006) 80}
  [\href{https://arxiv.org/abs/hep-ph/0606247}{{\ttfamily hep-ph/0606247}}].

\bibitem{Smirnov:2006tz}
A.~V. Smirnov, \emph{{An Algorithm to construct Grobner bases for solving
  integration by parts relations}},
  \href{https://doi.org/10.1088/1126-6708/2006/04/026}{\emph{JHEP} {\bfseries
  04} (2006) 026} [\href{https://arxiv.org/abs/hep-ph/0602078}{{\ttfamily
  hep-ph/0602078}}].

\bibitem{Lee:2008tj}
R.~N. Lee, \emph{{Group structure of the integration-by-part identities and its
  application to the reduction of multiloop integrals}},
  \href{https://doi.org/10.1088/1126-6708/2008/07/031}{\emph{JHEP} {\bfseries
  07} (2008) 031} [\href{https://arxiv.org/abs/0804.3008}{{\ttfamily
  0804.3008}}].

\bibitem{vonManteuffel:2014ixa}
A.~von Manteuffel and R.~M. Schabinger, \emph{{A novel approach to integration
  by parts reduction}},
  \href{https://doi.org/10.1016/j.physletb.2015.03.029}{\emph{Phys. Lett. B}
  {\bfseries 744} (2015) 101}
  [\href{https://arxiv.org/abs/1406.4513}{{\ttfamily 1406.4513}}].

\bibitem{Peraro:2016wsq}
T.~Peraro, \emph{{Scattering amplitudes over finite fields and multivariate
  functional reconstruction}},
  \href{https://doi.org/10.1007/JHEP12(2016)030}{\emph{JHEP} {\bfseries 12}
  (2016) 030} [\href{https://arxiv.org/abs/1608.01902}{{\ttfamily
  1608.01902}}].

\bibitem{Klappert:2019emp}
J.~Klappert and F.~Lange, \emph{{Reconstructing rational functions with
  FireFly}}, \href{https://doi.org/10.1016/j.cpc.2019.106951}{\emph{Comput.
  Phys. Commun.} {\bfseries 247} (2020) 106951}
  [\href{https://arxiv.org/abs/1904.00009}{{\ttfamily 1904.00009}}].

\bibitem{Klappert:2020aqs}
J.~Klappert, S.~Y. Klein and F.~Lange, \emph{{Interpolation of dense and sparse
  rational functions and other improvements in FireFly}},
  \href{https://doi.org/10.1016/j.cpc.2021.107968}{\emph{Comput. Phys. Commun.}
  {\bfseries 264} (2021) 107968}
  [\href{https://arxiv.org/abs/2004.01463}{{\ttfamily 2004.01463}}].

\bibitem{Peraro:2019svx}
T.~Peraro, \emph{{FiniteFlow: multivariate functional reconstruction using
  finite fields and dataflow graphs}},
  \href{https://doi.org/10.1007/JHEP07(2019)031}{\emph{JHEP} {\bfseries 07}
  (2019) 031} [\href{https://arxiv.org/abs/1905.08019}{{\ttfamily
  1905.08019}}].

\bibitem{Boehm:2018fpv}
J.~B\"ohm, A.~Georgoudis, K.~J. Larsen, H.~Sch\"onemann and Y.~Zhang,
  \emph{{Complete integration-by-parts reductions of the non-planar hexagon-box
  via module intersections}},
  \href{https://doi.org/10.1007/JHEP09(2018)024}{\emph{JHEP} {\bfseries 09}
  (2018) 024} [\href{https://arxiv.org/abs/1805.01873}{{\ttfamily
  1805.01873}}].

\bibitem{Boehm:2020ijp}
J.~Boehm, M.~Wittmann, Z.~Wu, Y.~Xu and Y.~Zhang, \emph{{IBP reduction
  coefficients made simple}},
  \href{https://doi.org/10.1007/JHEP12(2020)054}{\emph{JHEP} {\bfseries 12}
  (2020) 054} [\href{https://arxiv.org/abs/2008.13194}{{\ttfamily
  2008.13194}}].

\bibitem{Usovitsch:2020jrk}
J.~Usovitsch, \emph{{Factorization of denominators in integration-by-parts
  reductions}},  \href{https://arxiv.org/abs/2002.08173}{{\ttfamily
  2002.08173}}.

\bibitem{Bendle:2019csk}
D.~Bendle, J.~B\"ohm, W.~Decker, A.~Georgoudis, F.-J. Pfreundt, M.~Rahn et~al.,
  \emph{{Integration-by-parts reductions of Feynman integrals using Singular
  and GPI-Space}}, \href{https://doi.org/10.1007/JHEP02(2020)079}{\emph{JHEP}
  {\bfseries 02} (2020) 079}
  [\href{https://arxiv.org/abs/1908.04301}{{\ttfamily 1908.04301}}].

\bibitem{Maierhofer:2017gsa}
P.~Maierh\"ofer, J.~Usovitsch and P.~Uwer, \emph{{Kira\textemdash{}A Feynman
  integral reduction program}},
  \href{https://doi.org/10.1016/j.cpc.2018.04.012}{\emph{Comput. Phys. Commun.}
  {\bfseries 230} (2018) 99}
  [\href{https://arxiv.org/abs/1705.05610}{{\ttfamily 1705.05610}}].

\bibitem{Klappert:2020nbg}
J.~Klappert, F.~Lange, P.~Maierh\"ofer and J.~Usovitsch, \emph{{Integral
  reduction with Kira 2.0 and finite field methods}},
  \href{https://doi.org/10.1016/j.cpc.2021.108024}{\emph{Comput. Phys. Commun.}
  {\bfseries 266} (2021) 108024}
  [\href{https://arxiv.org/abs/2008.06494}{{\ttfamily 2008.06494}}].

\bibitem{Meyer:2017joq}
C.~Meyer, \emph{{Algorithmic transformation of multi-loop master integrals to a
  canonical basis with CANONICA}},
  \href{https://doi.org/10.1016/j.cpc.2017.09.014}{\emph{Comput. Phys. Commun.}
  {\bfseries 222} (2018) 295}
  [\href{https://arxiv.org/abs/1705.06252}{{\ttfamily 1705.06252}}].

\bibitem{Dlapa:2022wdu}
C.~Dlapa, J.~M. Henn and F.~J. Wagner, \emph{{An algorithmic approach to
  finding canonical differential equations for elliptic Feynman integrals}},
  \href{https://arxiv.org/abs/2211.16357}{{\ttfamily 2211.16357}}.

\bibitem{wasow1965asymptotic}
W.~Wasow, \emph{Asymptotic simplification of self-adjoint differential
  equations with a parameter},
  \href{https://doi.org/https://doi.org/10.1016/0022-0396(66)90048-9}{\emph{Journal
  of Differential Equations} {\bfseries 2} (1966) 378}.

\bibitem{Bruser:2018jnc}
R.~Br\"user, S.~Caron-Huot and J.~M. Henn, \emph{{Subleading Regge limit from a
  soft anomalous dimension}},
  \href{https://doi.org/10.1007/JHEP04(2018)047}{\emph{JHEP} {\bfseries 04}
  (2018) 047} [\href{https://arxiv.org/abs/1802.02524}{{\ttfamily
  1802.02524}}].

\bibitem{Becher:2014oda}
T.~Becher, A.~Broggio and A.~Ferroglia, \emph{{Introduction to Soft-Collinear
  Effective Theory}}, vol.~896. Springer, 2015,
  \href{https://doi.org/10.1007/978-3-319-14848-9}{10.1007/978-3-319-14848-9},
  [\href{https://arxiv.org/abs/1410.1892}{{\ttfamily 1410.1892}}].

\bibitem{DiVecchia:2019kta}
P.~Di~Vecchia, S.~G. Naculich, R.~Russo, G.~Veneziano and C.~D. White, \emph{{A
  tale of two exponentiations in $ \mathcal{N} $ = 8 supergravity at subleading
  level}}, \href{https://doi.org/10.1007/JHEP03(2020)173}{\emph{JHEP}
  {\bfseries 03} (2020) 173}
  [\href{https://arxiv.org/abs/1911.11716}{{\ttfamily 1911.11716}}].

\bibitem{Gruzinov:2014moa}
A.~Gruzinov and G.~Veneziano, \emph{{Gravitational Radiation from Massless
  Particle Collisions}},
  \href{https://doi.org/10.1088/0264-9381/33/12/125012}{\emph{Class. Quant.
  Grav.} {\bfseries 33} (2016) 125012}
  [\href{https://arxiv.org/abs/1409.4555}{{\ttfamily 1409.4555}}].

\bibitem{Tarasov:2004ks}
O.~V. Tarasov, \emph{{Computation of Grobner bases for two loop propagator type
  integrals}}, \href{https://doi.org/10.1016/j.nima.2004.07.104}{\emph{Nucl.
  Instrum. Meth. A} {\bfseries 534} (2004) 293}
  [\href{https://arxiv.org/abs/hep-ph/0403253}{{\ttfamily hep-ph/0403253}}].

\bibitem{DBLP:journals/corr/abs-cs-0509070}
V.~P. Gerdt and D.~Robertz, \emph{A maple package for computing groebner bases
  for linear recurrence relations}, {\emph{CoRR} {\bfseries abs/cs/0509070}
  (2005) } [\href{https://arxiv.org/abs/cs/0509070}{{\ttfamily cs/0509070}}].

\bibitem{Smirnov:2005ky}
A.~V. Smirnov and V.~A. Smirnov, \emph{{Applying Grobner bases to solve
  reduction problems for Feynman integrals}},
  \href{https://doi.org/10.1088/1126-6708/2006/01/001}{\emph{JHEP} {\bfseries
  01} (2006) 001} [\href{https://arxiv.org/abs/hep-lat/0509187}{{\ttfamily
  hep-lat/0509187}}].

\bibitem{Barakat:2022ttc}
M.~Barakat, R.~Br\"user, T.~Huber and J.~Piclum, \emph{{IBP reduction via
  Gr\"obner bases in a rational double-shift algebra}},
  \href{https://doi.org/10.22323/1.416.0043}{\emph{PoS} {\bfseries LL2022}
  (2022) 043} [\href{https://arxiv.org/abs/2207.09275}{{\ttfamily
  2207.09275}}].

\bibitem{Barakat:2022qlc}
M.~Barakat, R.~Br\"user, C.~Fieker, T.~Huber and J.~Piclum, \emph{{Feynman
  integral reduction using Gr\"obner bases}},
  \href{https://arxiv.org/abs/2210.05347}{{\ttfamily 2210.05347}}.

\bibitem{Mastrolia:2018uzb}
P.~Mastrolia and S.~Mizera, \emph{{Feynman Integrals and Intersection Theory}},
  \href{https://doi.org/10.1007/JHEP02(2019)139}{\emph{JHEP} {\bfseries 02}
  (2019) 139} [\href{https://arxiv.org/abs/1810.03818}{{\ttfamily
  1810.03818}}].

\bibitem{Mizera:2019gea}
S.~Mizera, \emph{{Aspects of Scattering Amplitudes and Moduli Space
  Localization}}, Ph.D. thesis, Princeton, Inst. Advanced Study, 2020.
\newblock \href{https://arxiv.org/abs/1906.02099}{{\ttfamily 1906.02099}}.
\newblock 10.1007/978-3-030-53010-5.

\bibitem{Mizera:2019vvs}
S.~Mizera and A.~Pokraka, \emph{{From Infinity to Four Dimensions: Higher
  Residue Pairings and Feynman Integrals}},
  \href{https://doi.org/10.1007/JHEP02(2020)159}{\emph{JHEP} {\bfseries 02}
  (2020) 159} [\href{https://arxiv.org/abs/1910.11852}{{\ttfamily
  1910.11852}}].

\bibitem{Frellesvig:2019uqt}
H.~Frellesvig, F.~Gasparotto, M.~K. Mandal, P.~Mastrolia, L.~Mattiazzi and
  S.~Mizera, \emph{{Vector Space of Feynman Integrals and Multivariate
  Intersection Numbers}},
  \href{https://doi.org/10.1103/PhysRevLett.123.201602}{\emph{Phys. Rev. Lett.}
  {\bfseries 123} (2019) 201602}
  [\href{https://arxiv.org/abs/1907.02000}{{\ttfamily 1907.02000}}].

\bibitem{Frellesvig:2019kgj}
H.~Frellesvig, F.~Gasparotto, S.~Laporta, M.~K. Mandal, P.~Mastrolia,
  L.~Mattiazzi et~al., \emph{{Decomposition of Feynman Integrals on the Maximal
  Cut by Intersection Numbers}},
  \href{https://doi.org/10.1007/JHEP05(2019)153}{\emph{JHEP} {\bfseries 05}
  (2019) 153} [\href{https://arxiv.org/abs/1901.11510}{{\ttfamily
  1901.11510}}].

\bibitem{Frellesvig:2020qot}
H.~Frellesvig, F.~Gasparotto, S.~Laporta, M.~K. Mandal, P.~Mastrolia,
  L.~Mattiazzi et~al., \emph{{Decomposition of Feynman Integrals by
  Multivariate Intersection Numbers}},
  \href{https://doi.org/10.1007/JHEP03(2021)027}{\emph{JHEP} {\bfseries 03}
  (2021) 027} [\href{https://arxiv.org/abs/2008.04823}{{\ttfamily
  2008.04823}}].

\bibitem{Chestnov:2022alh}
V.~Chestnov, F.~Gasparotto, M.~K. Mandal, P.~Mastrolia, S.~J. Matsubara-Heo,
  H.~J. Munch et~al., \emph{{Macaulay matrix for Feynman integrals: linear
  relations and intersection numbers}},
  \href{https://doi.org/10.1007/JHEP09(2022)187}{\emph{JHEP} {\bfseries 09}
  (2022) 187} [\href{https://arxiv.org/abs/2204.12983}{{\ttfamily
  2204.12983}}].

\bibitem{Cacciatori:2022mbi}
S.~L. Cacciatori and P.~Mastrolia, \emph{{Intersection Numbers in Quantum
  Mechanics and Field Theory}},
  \href{https://arxiv.org/abs/2211.03729}{{\ttfamily 2211.03729}}.

\bibitem{Chestnov:2022xsy}
V.~Chestnov, H.~Frellesvig, F.~Gasparotto, M.~K. Mandal and P.~Mastrolia,
  \emph{{Intersection Numbers from Higher-order Partial Differential
  Equations}},  \href{https://arxiv.org/abs/2209.01997}{{\ttfamily
  2209.01997}}.

\bibitem{Bruser:2020bsh}
R.~Br\"user, C.~Dlapa, J.~M. Henn and K.~Yan, \emph{{Full Angle Dependence of
  the Four-Loop Cusp Anomalous Dimension in QED}},
  \href{https://doi.org/10.1103/PhysRevLett.126.021601}{\emph{Phys. Rev. Lett.}
  {\bfseries 126} (2021) 021601}
  [\href{https://arxiv.org/abs/2007.04851}{{\ttfamily 2007.04851}}].

\bibitem{Pogel:2022ken}
S.~P\"ogel, X.~Wang and S.~Weinzierl, \emph{{Taming Calabi-Yau Feynman
  Integrals: The Four-Loop Equal-Mass Banana Integral}},
  \href{https://doi.org/10.1103/PhysRevLett.130.101601}{\emph{Phys. Rev. Lett.}
  {\bfseries 130} (2023) 101601}
  [\href{https://arxiv.org/abs/2211.04292}{{\ttfamily 2211.04292}}].

\bibitem{Pogel:2022vat}
S.~P\"ogel, X.~Wang and S.~Weinzierl, \emph{{Bananas of equal mass: any loop,
  any order in the dimensional regularisation parameter}},
  \href{https://arxiv.org/abs/2212.08908}{{\ttfamily 2212.08908}}.

\bibitem{Duhr:2022dxb}
C.~Duhr, A.~Klemm, C.~Nega and L.~Tancredi, \emph{{The ice cone family and
  iterated integrals for Calabi-Yau varieties}},
  \href{https://doi.org/10.1007/JHEP02(2023)228}{\emph{JHEP} {\bfseries 02}
  (2023) 228} [\href{https://arxiv.org/abs/2212.09550}{{\ttfamily
  2212.09550}}].

\bibitem{Lee:2015eva}
R.~N. Lee and K.~T. Mingulov, \emph{{Introducing SummerTime: a package for
  high-precision computation of sums appearing in DRA method}},
  \href{https://doi.org/10.1016/j.cpc.2016.02.018}{\emph{Comput. Phys. Commun.}
  {\bfseries 203} (2016) 255}
  [\href{https://arxiv.org/abs/1507.04256}{{\ttfamily 1507.04256}}].

\bibitem{Veneziano:2022zwh}
G.~Veneziano and G.~A. Vilkovisky, \emph{{Angular momentum loss in
  gravitational scattering, radiation reaction, and the Bondi gauge
  ambiguity}},
  \href{https://doi.org/10.1016/j.physletb.2022.137419}{\emph{Phys. Lett. B}
  {\bfseries 834} (2022) 137419}
  [\href{https://arxiv.org/abs/2201.11607}{{\ttfamily 2201.11607}}].

\bibitem{Bini:2022wrq}
D.~Bini and T.~Damour, \emph{{Radiation-reaction and angular momentum loss at
  the second post-Minkowskian order}},
  \href{https://doi.org/10.1103/PhysRevD.106.124049}{\emph{Phys. Rev. D}
  {\bfseries 106} (2022) 124049}
  [\href{https://arxiv.org/abs/2211.06340}{{\ttfamily 2211.06340}}].

\bibitem{Riva:2023xxm}
M.~M. Riva, F.~Vernizzi and L.~K. Wong, \emph{{Angular momentum balance in
  gravitational two-body scattering: Flux, memory, and supertranslation
  invariance}},  \href{https://arxiv.org/abs/2302.09065}{{\ttfamily
  2302.09065}}.

\bibitem{Almelid:2017qju}
O.~Almelid, C.~Duhr, E.~Gardi, A.~McLeod and C.~D. White, \emph{{Bootstrapping
  the QCD soft anomalous dimension}},
  \href{https://doi.org/10.1007/JHEP09(2017)073}{\emph{JHEP} {\bfseries 09}
  (2017) 073} [\href{https://arxiv.org/abs/1706.10162}{{\ttfamily
  1706.10162}}].

\bibitem{Adams:2017ejb}
L.~Adams and S.~Weinzierl, \emph{{Feynman integrals and iterated integrals of
  modular forms}},
  \href{https://doi.org/10.4310/CNTP.2018.v12.n2.a1}{\emph{Commun. Num. Theor.
  Phys.} {\bfseries 12} (2018) 193}
  [\href{https://arxiv.org/abs/1704.08895}{{\ttfamily 1704.08895}}].

\bibitem{Walden:2020odh}
M.~Walden and S.~Weinzierl, \emph{{Numerical evaluation of iterated integrals
  related to elliptic Feynman integrals}},
  \href{https://doi.org/10.1016/j.cpc.2021.108020}{\emph{Comput. Phys. Commun.}
  {\bfseries 265} (2021) 108020}
  [\href{https://arxiv.org/abs/2010.05271}{{\ttfamily 2010.05271}}].

\bibitem{Saotome:2012vy}
R.~Saotome and R.~Akhoury, \emph{{Relationship Between Gravity and Gauge
  Scattering in the High Energy Limit}},
  \href{https://doi.org/10.1007/JHEP01(2013)123}{\emph{JHEP} {\bfseries 01}
  (2013) 123} [\href{https://arxiv.org/abs/1210.8111}{{\ttfamily 1210.8111}}].

\bibitem{Cheng:1987ga}
H.~Cheng and T.~T. Wu, \emph{{Expanding Protons:~Scattering at High Energies}}.
  The MIT Press, 1987.

\bibitem{slater1966generalized}
L.~J. Slater, \emph{Generalized hypergeometric functions}. Cambridge University
  Press, 1966.

\bibitem{Tarasov:1996br}
O.~V. Tarasov, \emph{{Connection between Feynman integrals having different
  values of the space-time dimension}},
  \href{https://doi.org/10.1103/PhysRevD.54.6479}{\emph{Phys. Rev. D}
  {\bfseries 54} (1996) 6479}
  [\href{https://arxiv.org/abs/hep-th/9606018}{{\ttfamily hep-th/9606018}}].

\bibitem{Lee:2009dh}
R.~N. Lee, \emph{{Space-time dimensionality D as complex variable: Calculating
  loop integrals using dimensional recurrence relation and analytical
  properties with respect to D}},
  \href{https://doi.org/10.1016/j.nuclphysb.2009.12.025}{\emph{Nucl. Phys. B}
  {\bfseries 830} (2010) 474}
  [\href{https://arxiv.org/abs/0911.0252}{{\ttfamily 0911.0252}}].

\bibitem{Huber:2005yg}
T.~Huber and D.~Maitre, \emph{{HypExp: A Mathematica package for expanding
  hypergeometric functions around integer-valued parameters}},
  \href{https://doi.org/10.1016/j.cpc.2006.01.007}{\emph{Comput. Phys. Commun.}
  {\bfseries 175} (2006) 122}
  [\href{https://arxiv.org/abs/hep-ph/0507094}{{\ttfamily hep-ph/0507094}}].

\bibitem{Huber:2007dx}
T.~Huber and D.~Maitre, \emph{{HypExp 2, Expanding Hypergeometric Functions
  about Half-Integer Parameters}},
  \href{https://doi.org/10.1016/j.cpc.2007.12.008}{\emph{Comput. Phys. Commun.}
  {\bfseries 178} (2008) 755}
  [\href{https://arxiv.org/abs/0708.2443}{{\ttfamily 0708.2443}}].

\bibitem{bailey1991polynomial}
D.~Bailey and H.~Ferguson, \emph{A polynomial time, numerically stable integer
  relation algorithm}, {\emph{NASA Technical Report RNR-91-032} (1991) }.

\bibitem{Bailey:1999nv}
D.~H. Bailey and D.~J. Broadhurst, \emph{{Parallel integer relation detection:
  Techniques and applications}},
  \href{https://doi.org/10.1090/S0025-5718-00-01278-3}{\emph{Math. Comput.}
  {\bfseries 70} (2001) 1719}
  [\href{https://arxiv.org/abs/math/9905048}{{\ttfamily math/9905048}}].

\bibitem{Vines:2018gqi}
J.~Vines, J.~Steinhoff and A.~Buonanno, \emph{{Spinning-black-hole scattering
  and the test-black-hole limit at second post-Minkowskian order}},
  \href{https://doi.org/10.1103/PhysRevD.99.064054}{\emph{Phys. Rev. D}
  {\bfseries 99} (2019) 064054}
  [\href{https://arxiv.org/abs/1812.00956}{{\ttfamily 1812.00956}}].

\end{thebibliography}\endgroup

\end{document}